%% file: main.tex
\documentclass[
    paper=A4,               
    twoside=false,          
    fontsize=11pt,          
    titlepage=true,         
    listof=totoc,           
    bibliography=totoc,     
    listof=left,            
    open=right,             
    headsepline=true,       
    footsepline=false,      
    captions=tableheading,  
    numbers=noendperiod,    
    parskip=half-,          
    headings=normal         
]{scrbook}

\usepackage{scrhack}    
\usepackage{float}
\usepackage{graphicx} 
\usepackage{caption}  
\usepackage{longtable}
\usepackage{amsmath}
\usepackage{scrlayer-scrpage}
\automark{chapter}
\clearpairofpagestyles
\setkomafont{pagehead}{\sffamily\small\upshape}
\ohead{\headmark}
\usepackage[a4paper]{geometry}
\usepackage[utf8]{inputenc}
\usepackage[english]{babel}
\usepackage[T1]{fontenc}
\usepackage{csquotes} 
\usepackage{xcolor} 
\usepackage{setspace}
\usepackage{array}
\usepackage{enumitem}

\usepackage{graphicx} 
\usepackage[backend=biber, style=authoryear-icomp, citestyle=authoryear]{biblatex} 

\usepackage{tabularx} 
\usepackage{ifthen}

\usepackage{url}  

\usepackage{caption}    
\usepackage{subcaption}

\usepackage{hyperref}
\hypersetup{colorlinks=true,linkcolor=blue, citecolor=magenta}
\usepackage{booktabs}
\usepackage{acronym} 
\usepackage{siunitx}[=v2]   
\begin{document}

\frontmatter
\input{cover-information/HWI-cover}
\cleardoublepage
\pagestyle{headings}
\tableofcontents
\listoffigures
\listoftables

\input{content-folder/abbreviations}

\mainmatter
\input{content-folder/body}

\printbibliography 

\let\oldpagestyle=\thepage 
\AddToHook{cmd/chapter/before}{
    \clearpage
    \setcounter{page}{1}
    \renewcommand*{\thepage}{\thechapter-\arabic{page}}
}

\AtEndDocument{
    \renewcommand*{\thepage}{\oldpagestyle}
}

\appendix
 \input{content-folder/appendix}

\end{document}

%% file: cover-information/HWI-cover.tex
\begin{titlepage}

\rmfamily
\normalfont

\newcommand{\orcid}[1]{\href{https://orcid.org/#1}{\includegraphics[width=8pt]{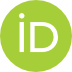}}}

\includegraphics[width=0.4\linewidth]{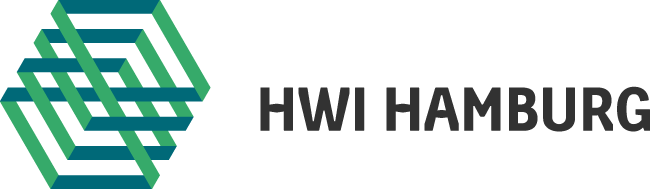}

\vspace{1cm}

{\huge\sffamily\bfseries
\setstretch{10}
Applying Computational Engineering Modelling\\
to Analyse the Social Impact of Conflict and\\
Violent Events
}

\vspace{0.5cm}

Felix Schwebel\textsuperscript{1, 2}\orcid{0009-0002-0864-9966}, 
Sebastian Meynen\textsuperscript{1}, 
and Manuel García-Herranz\textsuperscript{2}\orcid{0000-0002-4252-4975}\\
\textsuperscript{1}HWI, Universtiy of Hamburg, Hamburg, Germany\\
\textsuperscript{2}UNICEF, New York, NY, USA

\vspace{0.5cm}
\textbf{Abstract}\\[0.5em]
This thesis presents a novel framework for analysing the societal impacts of armed conflict by applying principles from engineering and material science. Building on the idea of a "social fabric", it recasts communities as plates with properties—such as resilience and vulnerability—analogous to material parameters like thickness or elasticity. Conflict events are treated as external forces that deform this fabric, revealing how repeated shocks and local weaknesses can compound over time. Using a custom Python-based Finite Element Analysis implementation, the thesis demonstrates how data on socioeconomic indicators (e.g., infrastructure, health, and demographics) and conflict incidents can be translated into a single computational model. Preliminary tests validate that results align with expected physical behaviours, and a proof-of-concept highlights how this approach can capture indirect or spillover effects and illuminate the areas most at risk of long-term harm. By bridging social science insights with computational modelling, this work offers an adaptable frame to inform both academic research and on-the-ground policy decisions for communities affected by violence.

\end{titlepage}

%% file: content-folder/abbreviations.tex
\chapter*{Abbreviations}
\addcontentsline{toc}{chapter}{Abbreviations}
\begin{acronym}

   \acro{ACLED}{Armed Conflict Location and Event Data}
   \acro{AI}{Artificial Intelligence}
   \acro{BRIC}{Baseline Resilience Indicators for Communities}
   \acro{CCVI}{Climate Conflict Vulnerability Index}
   \acro{CISI}{Critical Infrastructure Spatial Index}
    \acro{DOFs}{degrees of freedom}
    \acro{DRLRL}{Disaster Resilience of "Loss-Response" of Location}
    \acro{DROP}{Disaster Resilience of Place}
    \acro{FEA}{Finite Element Analysis}
    \acro{FEM}{Finite Element Method}
    \acro{GBD}{Global Burden of Disease}
    \acro{GDP}{Gross Domestic Product}
    \acro{GIS}{Geographic Information System}
    \acro{ND-GAIN}{Notre Dame Global Adaptation Initiative}
    \acro{PPP}{purchasing power parity}
    \acro{PTSD}{Post-traumatic stress disorder}
    \acro{SAR}{Spatial Autoregressive}
    \acro{SPI}{Standard Precipitation Index}
    \acro{SSA}{Sub-Saharan Africa}
    \acro{UCDP}{Uppsala Conflict Data Program}
    \acro{UN}{United Nations}
    \acro{V-Dem}{Varieties of Democracy}
    \acro{WASH}{Water, Sanitation and Hygiene}

\end{acronym}

%% file: content-folder/body.tex
\thispagestyle{plain}
\pagenumbering{arabic}

\chapter{Introduction}

\section{Background}

Armed conflicts—defined by \textcite{Kadir2018TheChildren} as any organised dispute involving weapons, violence, or force, whether within or across national borders—have escalated in frequency and intensity in recent decades as the number of major conflicts has almost tripled, and battle-related deaths have increased six-folded \parencite{Le2022TheCountries}. This trend poses significant challenges to global peace and long-term development \parencite{Azanaw2023EffectsMeta-analysis}. Especially as projections suggest that by 2030, over 60 \% of the world's poor will be concentrated in fragile and conflict-affected countries \parencite{Verwimp2019TheConflict, Kaila2023TheNigeria}. 

Contemporary warfare rarely remains confined to clear battlefields, resulting in wide-ranging damage to state institutions, economic structures, and community networks \parencite{InternationalBankforReconstructionandDevelopment/TheWorldBank2020BuildingAfrica, Slone2021ChildrensSupport}. \ac{SSA}, in particular, has witnessed a disproportionately high number of conflicts over the past three decades, with adverse effects persisting even long after the conflicts ended \parencite{Amberg2023ExaminingAnalysis, Kadir2018TheChildren, Wagner2018ArmedAnalysis}. These include large-scale displacement, destruction of critical infrastructure (e.g. hospitals, schools, \ac{WASH} facilities), and significant disruptions in supply chains and governance systems \parencite{Kadir2018TheChildren, Le2022TheCountries}. 

A 2017 review highlights that civilian casualties significantly outnumber those directly involved in conflicts, with a ratio exceeding five to one \parencite{Le2022TheCountries, Wagner2018ArmedAnalysis}. Children, in particular, experience an outsized share of these hardships \parencite{Bendavid2021TheChildren}. In 2017, roughly 420 million children worldwide were estimated to be living in conflict-affected areas, with a fifth to a sixth of all children in Africa, Asia, and the Americas exposed to conflict \parencite{Bendavid2021TheChildren}. The harm encompasses not only direct impacts, like deaths and injuries, but also extends to survivors and those indirectly affected, who may experience stunting, chronic illnesses, and significant mental health issues. \parencite{Carpiniello2023TheZones, Kadir2018TheChildren, Wagner2018ArmedAnalysis}. Between 1995 and 2015 alone, an estimated 4.9 to 5.5 million deaths of children under the age of five were linked to conflict, corresponding to 6.6 to 7.4 \% of all deaths in that age group \parencite{Wagner2018ArmedAnalysis}. 

Additionally, it is important to note that these threats are not limited to "frontline" areas or, as \textcite{Tapsoba2023TheConflict} indicated, "it is not only in areas where bullets are flying around that people need help coping with the adverse effects of conflict". Communities located tens or hundreds of kilometres away from active violence can still experience psychological distress, nutritional deficiencies, and breakdowns in community support \parencite{Tapsoba2023TheConflict}. Therefore, it is essential for policymakers and humanitarian organisations to develop effective strategies that not only protect and support communities directly affected but also recognise and enhance the resilience and coping abilities of those indirectly impacted. \parencite{Tapsoba2023TheConflict}. 

Moreover, the simultaneous challenges posed by climate hazards and conflict risks can further exacerbate development setbacks, force migration, and aggravate hunger and food insecurity \parencite{Hanze2024WhenDisasters, Mittermaier2024TheV1.1}. This is particularly problematic as those countries that are already the most vulnerable to climate hazards and have the least opportunity to improve their resilience appear to be disproportionately impacted by conflicts \parencite{Rosvold2023DisasterResilience}. In 2021, 23 of the bottom 30 countries on the \ac{ND-GAIN} index—those most vulnerable to climate change—had experienced at least one armed conflict, suggesting a vicious cycle of fragility, climate vulnerability, and violence \parencite{Rosvold2023DisasterResilience}. 

Against this backdrop, fragility and violence are recognised as central impediments to sustainable development in the 21st century \parencite{Kaila2023TheNigeria}. Modern warfare today is characterised by prolonged conflicts, extensive destruction due to powerful weaponry, and intentional assaults on civilian areas. This reality has led to demands for new, data-informed approaches to analyse not just direct violence but also the ways in which conflicts change across different social settings and how recurring or interconnected incidents can have a cumulative impact \parencite{Amberg2023ExaminingAnalysis, Cappelli2024LocalAfrica, Kadir2018TheChildren, Slone2021ChildrensSupport}. Such understanding is crucial for developing more effective strategies to prevent and mitigate the complex impacts of contemporary conflicts.


\section{Problem Statement}

Despite the increasing availability of datasets (e.g., \ac{UCDP}, \ac{ACLED}) and studies on armed conflict, there remains a fundamental gap in understanding how violence evolves within particular communities and why similar conflict events lead to vastly different outcomes \parencite{Bendavid2021TheChildren, Hanze2024WhenDisasters}.  Large-scale conflict datasets relying on media sources and field reports only capture direct violence data (e.g. fatalities and locations) but do not give any information about indirect harms, such as breakdown of critical services or serious health and mental health impacts \parencite{Bendavid2021TheChildren}. 

Furthermore, conventional analytical techniques frequently depend on aggregating data at the country or regional level or utilising uniform exposure buffers around conflict incidents \parencite{Amberg2023ExaminingAnalysis, Tapsoba2023TheConflict}. These approaches may mask the localised and evolving characteristics of conflict exposure. By relying on averages or coarse spatial units, these approaches risk significantly underestimating the real impact of conflicts, especially when indirect and cumulative effects fail to register as "major" events \parencite{Bendavid2021TheChildren, Wagner2018ArmedAnalysis}. A study by \parencite{Wagner2018ArmedAnalysis} even suggested that the number of deaths of infants and children younger than 5 years is approximately ten times higher than the estimations of the \ac{GBD} study when accounting not only for direct combat-related killings but also deaths caused indirectly by the violent event (e.g. by impacting personal health and health facility access).     

On the opposite end, micro-level surveys can offer in-depth insights into how conflict impacts household welfare, child health, mental well-being, and the capacity of local health services \parencite{Bruck2013MeasuringSurveys}. Yet conducting such surveys in conflict-affected settings can be dangerous, resource-intensive, and ethically complex, leading to potential gaps or biases in the data \parencite{Bendavid2021TheChildren, Norris2007CommunityReadiness, Verwimp2019TheConflict}. Insecure regions may be excluded from sampling, and respondents—fearing reprisals—might not disclose sensitive information \parencite{Bruck2013MeasuringSurveys, Norris2007CommunityReadiness}. Because of these complexities and challenges, it might be difficult for these surveys to capture long-term processes like post-conflict shifts or spatial shifts over an intermediate time span \parencite{Bendavid2021TheChildren}. 

Scholars have thus called for more holistic methods that integrate local vulnerability factors, event-based conflict data, and possible climate-related risk into multi-indicator analyses \parencite{Bendavid2021TheChildren, Cappelli2024LocalAfrica}. However, existing multi-indicator approaches often produce conservative or incomplete estimates because they rely on high-level aggregations (e.g. country means) or simple "exposure radius" definitions, which overlook fine-grained spillover and contagion effects as well as relevant heterogeneity in exposure \parencite{Amberg2023ExaminingAnalysis, Tapsoba2023TheConflict}. For instance, conflicts can spread geospatially to neighbouring areas, creating "hotspots" that reflect not just event intensity but also pre-existing social vulnerabilities and resource scarcities that might be responsible for location-specific conflict outcomes \parencite{Bendavid2021TheChildren, Cappelli2024LocalAfrica, Hanze2024WhenDisasters}. 

This also echoes insights from the field of disaster risk science, as indicated by \textcite{Hanze2024WhenDisasters}, that disasters are not discrete events but processes deeply embedded in socio-economic contexts. The occurrence of disasters is therefore not only a matter of the hazardous event itself but also of the exposure and vulnerability, which themselves can deteriorate over time, fuelling vicious cycles of conflict risk, climate hazards, and socio-institutional fragility \parencite{Cappelli2024LocalAfrica, Hanze2024WhenDisasters}. 

Despite these acknowledged shortcomings, no comprehensive computational framework seems to systematically integrate these varied layers (conflict events and local socio-economic conditions) into a unified model that is able to assess the impacts of conflict events and the characteristics of the local social systems in a dynamic, location-specific manner \parencite{Amberg2023ExaminingAnalysis}.

Consequently, there remains a fundamental need for an approach that not only compiles these indicators but also simulates how they interact over time and across space. In other words, the impact of a conflict event depends not just on \textbf{what} happens (e.g. a bombing versus a protest) but also on \textbf{where} it happens—and crucially, \textbf{in what kind of (social) environment}. Understanding these interactions and shifts could help guide humanitarian responses more effectively to regions most in need and advance the theoretical grasp of conflict dynamics by bridging the gap between micro-level granular insights and macro-level event data as well as providing better contexts for disaster, migration and conflict policies \parencite{Idler2024ConflictViolence, InternationalBankforReconstructionandDevelopment/TheWorldBank2020BuildingAfrica, Rosvold2023DisasterResilience}.


\section{Research Objectives}

In the past few decades, bridging quantitative physical models with the qualitative depth of the social sciences has emerged as a promising route for understanding complex societal phenomena \parencite{Castellano2009StatisticalDynamics, Weidlich1991PhysicsSynergetics}. Early work in synergetics underscored that while social systems demand nuanced, qualitative concepts, carefully formulated mathematical models can reveal hidden structures and emergent patterns. Moreover, as \parencite{Castellano2009StatisticalDynamics} suggests, global behaviours in large-scale systems often depend on higher level "universal" features rather than the microscopic details, making it feasible to adapt physical theories for social contexts. 

Building on these principles and the identified gaps in conflict analysis, this thesis proposes a physics-informed framework that moves beyond specific social indicators to provide a scalable approach to model and combine conflict and community characteristics. By reframing social vulnerabilities as "material properties" and conflict events as external "forces", this framework seeks to provide the foundation for further research and analyses that shed light on how violence propagates, how local vulnerabilities compound, and where critical thresholds lie. Ultimately, the goal is to catalyse more integrative research that combines social scientists' qualitative insights with physics-based modelling tools—thus expanding the current analytical repertoire for understanding and mitigating the impacts of armed conflicts. 

Specifically, this thesis pursues three interconnected objectives: 

\begin{enumerate}
    \item \textbf{Formulate a conceptual "social fabric" model}
    \begin{description}
        \item[Goal:] Establish a comprehensive modelling approach that enables the translation of both social characteristics and conflict events into physical analogues within a unified framework. 
        \item[Key Questions:] How can diverse social characteristics be conceptualised as "material properties" (e.g., thickness, elasticity) in a plate structure? In what ways can conflict events be incorporated as "forces" acting upon this social fabric?
        \item[Rationale:] As noted by \textcite{Weidlich1991PhysicsSynergetics}, a careful qualitative analysis must come before and inform quantitative modelling. By offering a flexible blueprint, domain experts can choose the variables that best reflect local contexts, ensuring the framework can adapt to various socio-political realities. 
    \end{description}

    \item \textbf{Demonstrate the framework's capabilities in a proof-of-concept}
    \begin{description}
        \item[Goal:] Present an illustrative application that shows how the framework simulates compound or protracted conflict scenarios. 
        \item[Key Question:] What insights can such simulations provide about the interaction between conflict events and local social conditions?
        \item[Rationale:] While a full empirical validation is beyond the scale of this thesis, offering simple conceptual examples helps underscore the potential of this interdisciplinary approach. 
    \end{description}
    
    \item \textbf{Establish foundations for future validation \& collaboration}
    \begin{description}
        \item[Goal:] Identifying key conceptual and practical gaps between this physics-based approach and potential real-world implementation needs.
        \item[Key Question:] What conceptual or practical gaps remain between this physics-based modelling approach and real-world data and validation needs, and how might these gaps be addressed through interdisciplinary collaboration?
        \item[Rationale:] As \textcite{Utsumi2022Armed2007} points out, no single initiative has solved the fundamental problems of merging survey data and \ac{GIS}-based conflict metrics. This framework, by contrast, aims to be a springboard for interdisciplinary cooperation among physicists, social scientists, and policymakers seeking more robust conflict analyses \parencite{InternationalBankforReconstructionandDevelopment/TheWorldBank2020BuildingAfrica}.
    \end{description}
    
\end{enumerate}


\section{Methodology}

To achieve the three objectives outlined in the previous section, a mixed approach is used that blends conceptual development with practical implementation and testing. The methodology was, therefore, designed to systematically survey the existing literature, develop and customise a computational workflow, and verify the feasibility of the proposed framework via pilot experiments and a proof-of-concept. This section describes these steps and their relevance to the subsequent chapters.

\subsection{Literature Review}

A targeted review of existing research was conducted to identify direct and indirect conflict impacts and evaluate current methods of measuring and assessing these impacts. The main objective of the review was to extract key features and mechanisms that any physics-based framework should incorporate to include the straightforward (e.g., fatalities, destruction) and more subtle (e.g., spatial distribution, spillovers) consequences of violence. 

Academic databases—including \textit{Google Scholar}, \textit{ScienceDirect}, \textit{Springer Nature}, and \textit{Sage Journals}—were searched using keywords such as "conflict impact", "conflict vulnerability", "conflict resilience", "impact of violence", "conflict exposure", and "conflict analyses". Additional keywords included "disaster impact" and "social science physics". \textit{\href{https://www.connectedpapers.com}{connectedpapers.com}} was used to explore derivative works from highly relevant papers and the most important prior works on the respective topics, enabling a structured exploration via visual graphs. By focusing on recent publications and influential earlier studies, the search achieved a balance of contemporary advancements and foundational concepts. This informed the framework's requirements regarding vulnerability indices, conflict-event handling, and indirect effects.

\subsection{Python-Based Implementation}

Following the requirements and dimensions of impacts identified in the literature review, a custom Python workflow was developed to support large-scale geospatial analyses and the flexible inclusion of socio-economic indicators as "material" parameters. This framework was constructed within a Jupyter Notebook environment to streamline the import of geospatial data (e.g., \texttt{.tif} files) and to allow direct import and translation of conflict events into external forces with user-defined distributions and temporal changes. While existing commercial structural analysis tools could theoretically be employed, an open-source, custom-built solution was chosen for its adaptability to conflict-specific modifications and to facilitate easier future modifications.

Additionally, the open-source Python setup facilitates interdisciplinary collaboration, allowing domain experts to modify the mapping of indicators to parameters or the handling of event definitions without needing to reconfigure a proprietary pipeline. This design also positions the projects to leverage emerging possibilities in advanced \ac{FEA} and machine-learning-assisted solvers, which have been shown to handle highly non-linear material behaviours and enable faster computing times \parencite{Maurizi2022PredictingNetworks}. Even though the current work exclusively employs a linear-elastic approach, the open-source framework simplifies potential future integrations with \ac{AI}-powered methods that could accelerate simulations and advance concepts even further.

\subsection{Preliminary Tests \& Proof-of-Concept}

Several experiments were carried out to verify both computational feasibility and conceptual validity. These experiments tested different material property configurations, boundary conditions, and force configurations, thereby revealing how the results change for different parameter magnitudes, overlapping loads and spatially heterogeneous parameters. The purpose was not to conduct any validation at this point but to confirm that the custom solver behaved as expected from the physical theory—for instance, verifying that plate thickness strongly influences peak displacement—and that displacement can be used as a meaningful proxy for compounding social impact in the presence of multiple events. 

After these baseline checks, a brief proof-of-concept application was developed to illustrate how the framework might be applied in an actual analysis setting. Although empirical validation exceeded the scope of this thesis, demonstrating the system's capacity to combine socio-economic and conflict data with this physics-inspired approach and visualise compounding and spatially spreading effects laid the groundwork for more extensive future work.


\chapter{Review of Existing Literature \& Approaches}

Conflict disrupts societies in ways that go far beyond battlefield casualties. Diverse research highlights not only direct losses such as deaths, injuries, and displacement but also indirect effects that compromise public health, erode social cohesion, destroy livelihoods, and undermine developmental progress. This chapter explores these multifaceted impacts by examining the direct and indirect consequences of conflict on communities, the spatial and temporal patterns through which conflicts spread and persist, and the factors that determine community vulnerability and resilience. It then surveys current methodological approaches for measuring and assessing conflict impacts, highlighting both progress and persistent gaps in the field. In doing so, the chapter establishes key foundations for the physics-based modelling framework presented in the following chapters.

\section{The Multifaceted Nature of Conflict Impacts}

\subsection{Direct \& Indirect Harm to Health \& Livelihoods}

Direct impacts include fatalities and injuries among combatants and civilians, forced displacement, and the destruction of critical infrastructure such as hospitals, schools, \ac{WASH} facilities, and roads \parencite{Bendavid2021TheChildren, Bernal2024ImpactChallenges, Bruck2013MeasuringSurveys, Kadir2018TheChildren, Le2022TheCountries, Tapsoba2023TheConflict, Wagner2018ArmedAnalysis}. These initial shocks often trigger further deterioration of health services, the fracturing of communal ties, and impediments to economic livelihoods, effectively eroding a society's capacity to recover \parencite{Amberg2023ExaminingAnalysis, Bendavid2021TheChildren, Bernal2024ImpactChallenges, Bruck2013MeasuringSurveys, Hanze2024WhenDisasters}. Some of the most vulnerable groups are children and women of reproductive age, who face heightened risks of direct injury, sexual violence, and forced recruitment \parencite{Bendavid2021TheChildren, Bernal2024ImpactChallenges, Carpiniello2023TheZones, Kadir2018TheChildren}. 

Indirect consequences are profoundly complex since they often unfold over extended periods but even often overshadow immediate combat mortality \parencite{Bernal2024ImpactChallenges, Wagner2018ArmedAnalysis}. Attacks on agriculture and infrastructure, for example, can cause significant disruption to supply chains, driving up hunger and malnutrition \parencite{Azanaw2023EffectsMeta-analysis, Le2022TheCountries}. Malnutrition and stunting, in turn, amplify mortality risks, while increased maternal stress and reduced healthcare access translate into neonatal complications, further reinforcing negative feedback loops in already fragile communities \parencite{Akresh2016DetailedImpacts, Amberg2023ExaminingAnalysis, Azanaw2023EffectsMeta-analysis, Bendavid2021TheChildren, Bernal2024ImpactChallenges, Le2022TheCountries, Wagner2018ArmedAnalysis}. As \textcite{Wagner2018ArmedAnalysis} observed, conflict in Africa significantly contributes to infant and under-five mortality, sometimes two to four times higher than the immediate combat-related death toll would indicate. Such finding underscores that conflict’s reach extends well beyond battle-related mortality, shaping long-term socioeconomic trajectories for entire generations \parencite{Cappelli2024LocalAfrica}.

The breakdown of health facilities and public health systems can additionally amplify the spread of diseases such as tuberculosis, measles, and malaria \parencite{Bendavid2021TheChildren, Kadir2018TheChildren}. In parallel, doctors, nurses, or other health personnel often leave conflict zones, limiting maternal health services, immunisations, and basic clinical care \parencite{Bendavid2021TheChildren, Bernal2024ImpactChallenges}. The mental health toll can similarly escalate, with \ac{PTSD}, anxiety, and depression rising two- or threefold in conflict-affected populations \parencite{Bendavid2021TheChildren, Bernal2024ImpactChallenges, Carpiniello2023TheZones, Kadir2018TheChildren, Slone2021ChildrensSupport}.

\subsection{Socioeconomic Disruption}

Beyond severe health consequences, conflicts can also distort local communities, leading households, firms and entire regions to adapt to a state of what \textcite{Verwimp2019TheConflict} describe as "war economy" \parencite{Bernal2024ImpactChallenges, Kaila2023TheNigeria, Le2022TheCountries}. Early economic literature treated conflicts as short-term shocks, but later research recognised that such violent events trigger structural changes in daily life and markets, sometimes stretching over decades \parencite{Cappelli2024LocalAfrica, Verwimp2019TheConflict}. According to \textcite{Tapsoba2023TheConflict}, individuals may adopt risk-coping strategies—even before or without any manifestation of violence—based solely on perceived risks. 

Inequality often widens under protracted violence, as wealthier or politically connected groups can secure better protection or migrate elsewhere, leaving poorer households with few coping strategies \parencite{Cappelli2024LocalAfrica, Le2022TheCountries, Norris2007CommunityReadiness}. The destruction of important infrastructure, such as roads, power lines, and water systems, cripples local industries, deters investment and restricts population mobility \parencite{Cappelli2024LocalAfrica, Kadir2018TheChildren, Sinha2023WaterApproach}. Agricultural yields suffer when farmers cannot safely access land or irrigation channels \parencite{Cappelli2024LocalAfrica, Le2022TheCountries}. Meanwhile, state funds are often diverted from development projects to military spending, stalling infrastructure repair and social services \parencite{Bernal2024ImpactChallenges}. Such upheavals can endure long after formal ceasefires, manifesting as fragile governance systems, heavily disrupted local markets, and chronic reliance on external humanitarian assistance \parencite{Bernal2024ImpactChallenges, Bruck2013MeasuringSurveys, Hanze2024WhenDisasters, Kaila2023TheNigeria}.

\subsection{Education \& Child Development}

As previously mentioned, children are among those who experience the severest impact of conflicts. In addition to the severe health and mental health impacts, violent events also cause major disruption of schooling and child development \parencite{Azanaw2023EffectsMeta-analysis, Le2022TheCountries, Slone2021ChildrensSupport, Utsumi2022Armed2007}. As conflict intensifies, schools may be destroyed, turned into shelters, or abandoned out of fear \parencite{Utsumi2022Armed2007}. Importantly, even the perception of a nearby conflict can lead parents to withdraw children from school \parencite{Tapsoba2023TheConflict, Utsumi2022Armed2007}. In Afghanistan, for instance, parental decisions to withdraw children—especially girls—were not solely tied to actual combat presence but to the perceived spillover effect, illustrating how rumour and fear also shape outcomes \parencite{Utsumi2022Armed2007}. 

Prolonged interruption of education undermines long-term human capital, reinforcing cycles of poverty and insecurity \parencite{Le2022TheCountries}. Like health and economic disruptions, these impacts may persist even after hostilities cease, as returning to school might be impossible due to teachers being killed or having fled or if families have depleted their resources to send their children to school. \parencite{Amberg2023ExaminingAnalysis}. The combination of exposure to armed actors, environmental destruction, disrupted schooling, and chronic fear produces long-term developmental setbacks for children \parencite{Le2022TheCountries, Slone2021ChildrensSupport, Wagner2018ArmedAnalysis}.


\section{Spatial \& Temporal Dynamics of Conflict}

\subsection{Shifting Hotspots \& Spillover Effects}
 
Assessing the effects of conflict on communities involves not only the direct and indirect impacts but also the spread and spillover effects that may not stem from direct destruction or being classified as an active conflict zone \parencite{Idler2024ConflictViolence, InternationalBankforReconstructionandDevelopment/TheWorldBank2020BuildingAfrica, Utsumi2022Armed2007}. \textcite{Schutte2011DiffusionWars} distinguishes between two main diffusion processes of conflict: relocation, where conflict "jumps" to new locations, and escalation, where violence radiates outwards from an existing hotspot. However, empirical studies find that escalation is much more frequent, creating "contagion" effects that pull neighbouring communities into violence \parencite{Cappelli2024LocalAfrica, Schutte2011DiffusionWars, Tapsoba2023TheConflict}. 

Multiple studies (e.g., \textcite{Amberg2023ExaminingAnalysis, Kaila2023TheNigeria, Stojetz2024ShockingNigeria, Tapsoba2023TheConflict, Utsumi2022Armed2007, Wagner2018ArmedAnalysis}) use distance-based or kernel-density measures to account for the geographical spread. \textcite{Utsumi2022Armed2007} demonstrates that education outcomes in Afghanistan were adversely affected, even in districts without actual clashes, because of parents' fear and rumours. Similarly, \textcite{Wagner2018ArmedAnalysis} found elevated child mortality up to 100 km away from conflict sites, suggesting that mere adjacency can threaten safety and diminish service accessibility. Such broad zones of impact challenge simplified "exposure radii" assumptions, such as 50 km or 100 km, that may underestimate the intangible spread of fear, rumour, or population displacement that typically crosses administrative boundaries \parencite{Utsumi2022Armed2007}. Recognising this spillover is essential to understanding the risks faced by vulnerable groups like children. It is vital to deliver important services and implement protective measures before the situation deteriorates into active conflict zones \parencite{Idler2024ConflictViolence, Utsumi2022Armed2007, Wagner2018ArmedAnalysis}.

\subsection{Protracted Conflicts \& Cumulative Effects}

As already described in the direct and indirect effects, conflicts also often recur or persist over multi-year periods, imposing repeated stress on fragile communities \parencite{Bendavid2021TheChildren, Mittermaier2024TheV1.1, Verwimp2019TheConflict, Wagner2018ArmedAnalysis}. Each new wave of violence further erodes the health and education systems, with child and maternal mortality rates climbing even between conflicts \parencite{Verwimp2019TheConflict, Wagner2018ArmedAnalysis}. \textcite{Wagner2018ArmedAnalysis} illustrates that the devastating effects on child mortality from consecutive violence lasting at least five years were four times greater than an unstained conflict lasting less than a year. 

Such repeated shocks also discourage development or infrastructure projects, as external investors or humanitarian programmes face uncertainty or heightened security risks \parencite{Bernal2024ImpactChallenges}. Over time, these recurrent shocks complicated any straightforward path to recovery. Those families most exposed to earlier violent events might have a diminished capacity to cope with subsequent crises, further exacerbating inequality \parencite{Amberg2023ExaminingAnalysis, Bendavid2021TheChildren, Wagner2018ArmedAnalysis}.


\section{Social Resilience \& Vulnerability}

Understanding how communities respond to and recover from armed conflict depends heavily on the concepts of vulnerability and resilience \parencite{Cutter2014TheResilience}. However, there are several concepts and definitions which require some clarification for this purpose of conflict and violent effect-specific properties \parencite{Boon2012BronfenbrennersDisasters, Cutter2008ADisasters, Cutter2014TheResilience}. 

\subsection{Vulnerability in Conflict-Affected Contexts}

In disaster risk science and social-ecological studies, vulnerability typically denotes the pre-existing characteristics—economic, demographic, institutional—that elevate a group's susceptibility to harm \parencite{Cutter2008ADisasters, Hanze2024WhenDisasters, Mittermaier2024TheV1.1}. In conflict zones, these factors may be tied to income equality, weak governance, fragile health systems, and ethnic divisions \parencite{Cappelli2024LocalAfrica, Cutter2008ADisasters, Hanze2024WhenDisasters}. \textcite{Fatima2022EvaluatingPakistan} described vulnerability as an invisible, multidimensional phenomenon that a single indicator cannot capture. Rather, it must be measured as a composite of exposure, sensitivity, and adaptive capacity using multiple indicators as proxies \parencite{Cutter2008ADisasters, Fatima2022EvaluatingPakistan, Zhou2009ResiliencePerspective}:

\begin{itemize}
    \item \textbf{Exposure}: Who or what is at risk? For instance, communities in close proximity to active violence or those already battling climate  (e.g., drought, floods) often face elevated stress \parencite{Cutter2008ADisasters}. 
    \item \textbf{Sensitivity}: How severely can people be harmed if exposed? Low maternal education, inadequate nutrition, and weak infrastructure might magnify even low-intensity attacks \parencite{Cutter2008ADisasters, Zhou2009ResiliencePerspective}. 
    \item \textbf{Adaptive Capacity}: To what extent can a system (e.g., a family, a village) adjust to or cope with stress? In conflict contexts, adaptation often involves displacement, reliance on aid, or alternative livelihood, but each strategy, or the lack of it, can also bring new vulnerabilities \parencite{Cutter2008ADisasters}. 
\end{itemize}

As \textcite{Tapsoba2023TheConflict} indicates, exposure does not necessarily involve the manifestation of a violent event, but just the perceived risk alone can undermine and impact vulnerable communities. Moreover, armed conflict frequently coexists with other stressors—such as climate shocks—leading to compounding vulnerabilities \parencite{Cappelli2024LocalAfrica, Caso2023The2018, Fatima2022EvaluatingPakistan, Stojetz2024ShockingNigeria}.

\subsection{Resilience in Conflict-Affects Contexts}

Resilience generally refers to a system's ability to absorb, adapt, or transform in response to stress without collapsing \parencite{Boon2012BronfenbrennersDisasters, Cutter2008ADisasters, Sinha2023WaterApproach}. In conflict settings, resilience captures the concept that while violence may inflict damage, communities can reorganise, maintain core functions (like health care or education), and eventually recover—or even "bounce forward" to a new, more robust state or equilibrium \parencite{Boon2012BronfenbrennersDisasters, Cutter2014TheResilience, Sinha2023WaterApproach}. 

Multiple models within this field offer various ways of understanding resilience as a concept. The \ac{DROP} model by \textcite{Cutter2008ADisasters} suggests that both inherent resilience and inherent vulnerabilities arise from "antecedent conditions" resulting from the interaction of human and environmental systems and the built environment. Similarly, the later \ac{BRIC} framework \parencite{Cutter2014TheResilience} identifies measurable factors across social, economic, infrastructural, institutional, community, and environmental domains that contribute to a community's overall resilience. \textcite{Zhou2009ResiliencePerspective}'s \ac{DRLRL} model focuses on the interplay of risk, loss potential, and local factors, highlighting how biophysical resilience and social resilience combine to produce the "resilience of action", which in turn feeds back into future risk levels. 

Other researchers emphasise the dynamic nature of resilience. \textcite{Norris2007CommunityReadiness} discusses four adaptive capabilities—economic development, social capital, information and communication, and community competence—that may offer resistance up to a certain point but can wear down if shocks become recurrent or too intense. In addition, community resources must be sufficiently robust, redundant, rapidly available, and equally distributed to buffer or counteract immediate effects and avoid dysfunction \parencite{Norris2007CommunityReadiness}. \textcite{Sinha2023WaterApproach} broadens this view with the lenses of absorbing shocks, self-organisation, and learning, highlighting how interdependencies (e.g., water infrastructure relying on electrical power for pumps) can either boost or undermine resilience. Meanwhile, \textcite{Boon2012BronfenbrennersDisasters} examines individual-level resilience through bioecological systems theory, arguing that personal well-being depends on social context and the quality of relationships among families, neighbours, and institutions. 

Overall, the interplay between vulnerability and resilience is essential to understanding communities' responses in conflict settings. Strong social structures and resources can absorb or limit the impact of violence, whereas communities lacking these capacities may experience the effects of conflict greatly magnified  \parencite{Cutter2008ADisasters, Cutter2014TheResilience, Stojetz2024ShockingNigeria}.


\section{Existing Approaches to Measure \& Assess Conflict Impacts}

This section of the literature review provides an overview of the approaches and frameworks already developed to quantify or interpret conflict impacts, providing additional insights into their strengths and shortcomings.

\subsection{Macro-Level Spatial Overlays}

One frequently used method uses spatial overlays to analyse conflict incidence at larger geographic scales, typically using georeferenced conflict events from databases like \ac{ACLED} or \ac{UCDP}. Rather than simply aggregating events to an entire region or country, many studies define a buffer around each violent event to approximate the local population "exposed" to that event \parencite{Utsumi2022Armed2007}. 

For instance, \textcite{Amberg2023ExaminingAnalysis} apply a 50 km buffer around conflict sites to determine which maternal and child health indicators might be affected by proximity to violence. Others may use a 20 km or 100 km, depending on the context, the type of conflict and the presumed spatial extent of the conflict \parencite{Kaila2023TheNigeria, Stojetz2024ShockingNigeria, Utsumi2022Armed2007, Wagner2018ArmedAnalysis}. 

While it provides a simple and scalable approach, binary exposure classification misses any spillover effects that spread beyond fixed boundaries or that do not necessarily follow symmetrical circular distance patterns, as well as compounding effects of multiple events at the same or partially overlapping location \parencite{Rosvold2023DisasterResilience, Utsumi2022Armed2007}. The size of the buffer radii might also be somewhat arbitrary and risk underestimating particularly indirect effects  \parencite{Stojetz2024ShockingNigeria, Utsumi2022Armed2007}. Additionally, these approaches usually do not account for location-specific vulnerabilities and compounding effects \parencite{Stojetz2024ShockingNigeria}.

\subsection{Micro-Level Surveys}

In contrast to the macro-level overlay approaches, micro-level surveys focus on direct data from affected communities, households, or individuals \parencite{Verwimp2019TheConflict}. The emphasis is on ground-level realities, including health outcomes, mental well-being, coping strategies, and socioeconomic disruptions \parencite{Verwimp2019TheConflict}. Individuals and households who have personally experienced, witnessed, or been forced to adapt to violence in some way are surveyed directly, revealing nuanced insights into indirect impacts, including psychological harm, child nutrition deficits, and forced displacement patterns \parencite{Verwimp2019TheConflict}.

Nevertheless, gathering detailed surveys in active conflict zones can pose severe ethical and practical challenges \parencite{Norris2007CommunityReadiness, Verwimp2019TheConflict}. Field teams might be barred from the most dangerous areas, while fear of reprisals can discourage honest reporting \parencite{Norris2007CommunityReadiness}. Some respondents may not recall or choose not to disclose sensitive events, leading to underestimation of genuine impacts \parencite{Bruck2013MeasuringSurveys, Norris2007CommunityReadiness}. Survey coverage might also be less uniform across large geographic areas, complicating attempts to draw consistent national or multi-national conclusions \parencite{Bendavid2021TheChildren}.

\subsection{Multi-Indicator or Composite Frameworks}

Another category includes multi-indicator or composite indices that integrate conflict data with socio-economic, demographic, or environmental metrics. The \ac{CCVI} by \textcite{Mittermaier2024TheV1.1}, for example, maps global climate and conflict hazards and vulnerability factors on a grid, resulting in a cell-by-cell risk assessment by combining the indicators using (weighted) generalised means. \textcite{Cutter2008ADisasters} and \textcite{Cutter2014TheResilience} have developed comprehensive models like \ac{DROP} and \ac{BRIC}, which combine structural attributes—health infrastructure, socio-economic status, governance, etc.—into aggregated scores. However, these models do not include conflict hazards as such and focus on resilience and vulnerability assessment in these specific cases \parencite{Cutter2008ADisasters}. 

Such frameworks offer specific standardised measures that can be tracked over time or compared across locations, acknowledging that risk arises from both hazards and the underlying vulnerability of populations \parencite{Mittermaier2024TheV1.1}. However, when using direct aggregation or weighted averaging, these methods mathematically fuse conflict and vulnerability measures, treating conflict impacts and community characteristics as parts of the same expression. This lack of mathematical separation between the "impact" (conflict events) and the "impacted" (community characteristics) constrains their ability to model sophisticated interaction patterns or emergent effects that might arise when multiple events compound over space and time \parencite{Amberg2023ExaminingAnalysis}. Furthermore, these indices often rely on subjective indicator selection and weighting decisions \parencite{Fatima2022EvaluatingPakistan, Hanze2024WhenDisasters}.

\subsection{Advanced Statistical or Spatial Models}

A growing body of work seeks to capture the fluid, often diffuse nature of conflict. \textcite{Utsumi2022Armed2007} uses a kernel density estimation to represent how conflict "spills over" beyond strict boundaries. Others, like \textcite{Tapsoba2023TheConflict}, employ space-time stochastic processes, estimating underlying risk distributions based on observed events and factoring in "perceived conflict risk" to explain preemptive coping strategies. 

Spatial econometric techniques—such as \ac{SAR} models—allow for "contagion" or "neighbour effects", where conflict in one cell raises the likelihood of conflict in the neighbouring cells \parencite{Cappelli2024LocalAfrica}. These advanced models aim to reflect how violence escalates or "diffuses" geographically,  capturing the dynamic interplay among repeated events over time \parencite{Cappelli2024LocalAfrica}. \textcite{Wang2024SpatiotemporalPerceptions} also uses Shannon entropy measures to quantify the spread of conflict as it expands or shifts. 

Despite their sophistication, advanced spatial and stochastic methods pose certain hurdles: they demand high-resolution spatiotemporal data (e.g., precise coordinates, consistent event reporting), risk parameter sensitivities (e.g., choosing bandwidths in kernel density), and may be complex to interpret \parencite{Utsumi2022Armed2007}. Additionally, location inaccuracies can degrade model accuracy, potentially underestimating or misrepresenting spillovers \parencite{Wagner2018ArmedAnalysis}.

\subsection{Mixed-Methods Approaches}

Finally, some studies integrate qualitative (interviews, focus groups, archival research) and quantitative (\ac{GIS}, surveys, or advanced statistical models) approaches to study historically, economically, and socially embedded conflict processes \parencite{Bernal2024ImpactChallenges, Idler2024ConflictViolence}. \textcite{Idler2024ConflictViolence} adopts this strategy to investigate "spatial shift" phenomena in Colombia and Iraq, mapping out how new armed actors enter territories, displace incumbents, and reshape patterns of violence. Combining interviews and focus groups with network or spatial analysis fosters a deeper understanding of context, rumour dynamics, power relations, and community resilience factors \parencite{Bernal2024ImpactChallenges, Idler2024ConflictViolence}. 

Mixed-methods research can reveal causal processes that might remain hidden in purely numeric approaches \parencite{Idler2024ConflictViolence}. However, such work is often resource-intensive, hard to scale, and contingent on the availability of secure access. Findings may also be case-specific or reliant on interpretive frameworks that limit cross-country comparisons.


\section{Summary of Literature Review Findings}

The previous sections have demonstrated that conflict impacts are multifaceted, directly and indirectly affecting health, education, and socio-economic structures. They have also shown that violence evolves through complex spatial and temporal patterns, with spillovers, protracted recurrence, and cumulative effects often hitting the most vulnerable communities the hardest. Beyond these immediate harms, community resilience and vulnerability emerge as critical factors in determining how severely conflict undermines local systems. Additionally, a wide range of existing methods for measuring and interpreting conflict impacts were examined. These methods differ in scale, data requirements, and their ability to capture indirect or dynamic processes.

Tables \ref{tab:summary-of-dimensions-of-conflict-impact} and \ref{tab:summary-of-current-approaches} provide concise overviews of these dimensions of conflict impact and commonly used approaches used to study them. Table \ref{tab:summary-of-dimensions-of-conflict-impact} outlines the categories of direct and indirect impacts, the spatial and temporal complexities of conflict, and the vulnerability/resilience factors that can magnify or mitigate harm. Table \ref{tab:summary-of-current-approaches} summarises the range of existing methods, from macro-level spatial overlays to micro-level surveys and advanced spatial models, highlighting the strengths and limitations of each approach. 

Both tables underscore the variety of challenges in conflict analysis. On the one hand, researchers must cope with direct and indirect harms, spatial spillovers, and protracted conflict timelines (Table \ref{tab:summary-of-dimensions-of-conflict-impact}). On the other, the approaches adopted to measure or interpret conflict exhibit distinctive strengths and drawbacks (Table \ref{tab:summary-of-current-approaches}). Spatial overlays, for example, handle large-scale data but often neglect intangible spread or local vulnerabilities, while micro-level surveys yield rich detail but can be difficult to implement in insecure environments. Composite frameworks and advanced models capture multiple facets yet risk complexity or require extensive data and parameter calibration. Mixed-methods research can identify deeper causal mechanisms but tends to be resource-intensive and context-specific. 
 
\begin{table}[H]
    \centering
    \begin{tabular}{>{\raggedright\arraybackslash}p{110pt}|>{\raggedright\arraybackslash}p{330pt}} \hline
         \textbf{Dimensions} & \textbf{Key Findings} \\ \hline \hline
         \textbf{Direct Impacts} & 
         \begin{itemize}[nosep]
             \item Fatalities and injuries
             \item Forced displacement
             \item Infrastructure destruction
             \item Immediate service disruption
         \end{itemize}
         \\ \hline
         \textbf{Indirect Impacts} &  
         \begin{itemize}[nosep]
             \item Health system breakdown
             \item Education system disruption
             \item Agriculture/supply chain disruption
             \item Economic system transformation
             \item Social cohesion breakdown
             \item Psychological impacts
         \end{itemize}
         \\ \hline
         \textbf{Spatial Dynamics} &
         \begin{itemize}[nosep]
             \item Direct zone of violence (extension, relocation)
             \item Spillover effects
             \item Non-uniform propagation
             \item Cross-boundary impacts
         \end{itemize}
         \\ \hline
         \textbf{Temporal Dynamics} & 
         \begin{itemize}[nosep]
             \item Immediate vs. long-term effects
             \item Compounding over time
             \item Anticipatory effects
             \item Post-conflict persistence
         \end{itemize}
         \\ \hline
         \textbf{Vulnerability Factors} & 
         \begin{itemize}[nosep]
             \item Pre-existing conditions
             \item Enhanced vulnerability of children/women
             \item Environmental context
         \end{itemize}
         \\ \hline
          \textbf{Resilience Factors} & 
         \begin{itemize}[nosep]
             \item Adaptive capacities
             \item Community resources and social capital
             \item System robustness and redundancy
             \item Recovery and transformation potential
         \end{itemize}
         \\ \hline
    \end{tabular}
    \caption[Summary of the key dimensions of conflict impact]{Summary the of key dimensions of conflict impact, including direct/indirect harm, spatial/temporal dynamics, and vulnerability/resilience factors.}
    \label{tab:summary-of-dimensions-of-conflict-impact}
\end{table}

\begin{table}[H]
    \centering
    \begin{tabular}{>{\raggedright\arraybackslash}p{100pt}|>{\raggedright\arraybackslash}p{170pt}|>{\raggedright\arraybackslash}p{170pt}} \hline
         \textbf{Method}& \textbf{Key Capabilties}&\textbf{Limitations}\\ \hline \hline
         \textbf{Spatial Overlays}& 
         \begin{itemize}[nosep]
             \item Large-scale analysis possible
             \item Clear visualisation
             \item Standardised measurement
             \item Good for direct impacts
         \end{itemize}
          &
          \begin{itemize}[nosep]
             \item Miss spillover effects
             \item Assume uniform propagation
             \item Ignore local context
             \item Static view
         \end{itemize}
          \\ \hline
         \textbf{Household Surveys}&  
         \begin{itemize}[nosep]
             \item Capture indirect effects
             \item Document local experiences
             \item Track vulnerability changes 
             \item Reveal coping strategies 
         \end{itemize}
          &
          \begin{itemize}[nosep]
             \item Limited coverage/access
             \item Resource intensive 
             \item Reporting biases
             \item Difficult to scale
         \end{itemize}
          \\ \hline
         \textbf{Multi-Indicator Frameworks}&
         \begin{itemize}[nosep]
             \item Integrate multiple factors 
             \item Compareable across regions
             \item Track multiple dimensions
         \end{itemize}&
          \begin{itemize}[nosep]
             \item Limited mathematical separation between "impact" and "impacted"
             \item Subjective weighting
             \item Static snapshots
             \item Risk oversimplification of variations 
         \end{itemize}
          \\ \hline
         \textbf{Advanced Spatial Models}& 
         \begin{itemize}[nosep]
             \item Model diffusion patterns
             \item Handle non-linear effects
             \item Account for spatial relations
             \item Track temporal changes
         \end{itemize}
          &
          \begin{itemize}[nosep]
             \item High data requirements
             \item Complex implementations
             \item Parameter sensitivity 
         \end{itemize}
          \\ \hline
         \textbf{Mixed-Methods Approaches}& 
         \begin{itemize}[nosep]
             \item Combine qualitative and quantitative insights
             \item Reveal causal mechanics 
             \item Document complex processes
         \end{itemize}
          &
          \begin{itemize}[nosep]
             \item Resource intensive
             \item Hard to scale 
             \item Context dependent
             \item Limited comparability
         \end{itemize}
          \\ \hline
    \end{tabular}
    \caption[Overview of five major methods used to quantify conflict impacts]{Overview of five major methods used to quantify conflict impacts, highlighting key strengths and limitations.}
    \label{tab:summary-of-current-approaches}
\end{table}

These findings form a foundation for the modelling approach proposed in this thesis. They highlight the need for a method that can integrate repeated or overlapping events, account for socio-economic vulnerabilities in a location-specific dynamic, and preserve a clear distinction between the "impact" and the "impacted". The next chapter introduces a physics-inspired analogy that offers a novel mechanism for combining these aspects into a single, computational framework, building on—and addressing key gaps within—the methods reviewed here.


\chapter{The "Social Fabric" Concept}

The literature review in the previous chapter showed that conflict impacts are multidimensional and can produce various direct and indirect consequences whose severity depends on local vulnerabilities and resilience capacities (Table \ref{tab:summary-of-dimensions-of-conflict-impact}). Existing analytical approaches (Table \ref{tab:summary-of-current-approaches}) each address parts of this complexity but also face distinct limitations. Spatial overlays capture direct violence but may miss spillovers and local socio-economic differences, while micro-level surveys offer rich insights but might lack scalability and uniform coverage in conflict contexts. Multi-indicator frameworks and advanced spatial models can tackle certain dynamic or indirect dimensions but often require extensive data and risk parameter sensitivities or do not allow for differentiation between conflict "impact" and socio-economic "context". 

In light of these constraints, three key dynamics emerge that existing methods typically struggle to represent simultaneously: 

\begin{enumerate}
    \item \textbf{Anticipatory and spillover effects}: Spillover effects from nearby conflict might also cause indirect impacts and drive communities to adopt coping strategies.
    \item \textbf{Compounding effects of repeated or protracted violence}: Each new shock acts on a social system already weakened by previous impacts, amplifying long-term harm. 
    \item \textbf{Spatial variation in vulnerability and resilience}: Community susceptibility to harm and capacity for recovery emerge from multiple interlocking factors (e.g., governance, infrastructure, and social networks) that vary across locations and influence how conflict impacts spread or amplify. 
\end{enumerate}

Against this backdrop, this chapter introduces a framework that reconceptualises conflict-affected regions as a "social fabric" with distinct properties analogous to those found in material science. By drawing on established principles of continuum mechanics and \ac{FEA}, the approach aims to capture how conflict impact propagates through communities and how local "material properties" (representing resilience or weakness) shape that propagation. While not intending to replicate every nuance of social systems, the framework provides a structured foundation for analysing how repeated violence accumulates, spreads, or intensifies in different settings—and thus offers a pathway beyond the limitations identified in earlier approaches.

\section{"Social Fabric" Analogy and Its Rationale}

The concept of a "social fabric" appears already in social science literature (e.g., mentioned by \textcite{Slone2021ChildrensSupport, Zhou2009ResiliencePerspective}) as a metaphor for how communities are woven together through relationships, institutions, and shared resources. This framework takes this metaphor one step further by drawing on physics and material science to provide mathematical tools for analysing how such social fabrics respond to impacts. The basic premise is drawn from an analogy to how plates in physics deform when forces are applied and how certain properties—thickness, elasticity, etc.—govern the spread of deformations and stress.

A simple illustration is the difference between steel and rubber in mechanical terms. Steel retains shape under substantial force because of its high stiffness and only yields if the load exceeds a certain threshold. Rubber, on the other side, deforms more easily under the same force but may also be more capable of rebounding once the stress is removed. These variations in response are determined by well-defined material parameters that govern how each material distributes and absorbs external forces. 

This physical framework offers valuable insights for analysing conflict-affected communities. Just as materials exhibit characteristic responses to loads based on their internal properties, communities demonstrate varying capacities to absorb and recover from conflict impacts. Some regions might function similarly to "rigid" steel, maintaining critical function even under significant pressure through strong institutions and robust infrastructure. Others might behave more like rubber, showing notable disruption to the same "force". Critically, when force is applied to either material, the effect is not confined to the point of impact, as stress waves or deformation patterns emerge, but is shaped by local characteristics. 

Likewise, communities experiencing conflict rarely confine the impact to a single location. Fear, displacement, or economic ripples might spread over large areas, just as physical stress might spread through a plate. Material science has developed mathematical tools for analysing exactly these kinds of dynamics of how materials with varying properties respond to stress, how effects propagate through non-uniform media, and how repeated loading leads to progressive changes in material behaviour \parencite{Bhaskar2021Plates, Rao2017SimulationMechanics}. Translated into social context, these tools can illuminate how recurrent violence interacts with local vulnerabilities, how disruptions spread to neighbouring areas, and how seemingly similar conflict events may yield drastically different outcomes based on underlying "social-fabric" differences. The following sections, therefore, describe and explain the basic physical parameters and principles, the used \ac{FEA} method and details about its specific implementation.

\section{Physical Parameters and Basic Principles}

To operationalise this analogy between social systems and physical materials, the framework adapts a linear-elastic approach as a first approximation. In engineering contexts, "linear elasticity" implies that deformation is directly proportional to the applied force and that the plate returns to its original state once the load is removed. Real communities, of course, might also exhibit "plastic" or irreversible damage—such as institutional breakdown or permanent displacement—but beginning with the linear-elastic theory offers a manageable starting point for modelling the superposition and spatial diffusion of conflict stress.

Under these simplified conditions, just three parameters are sufficient to characterise how a plate-like structure responds to stress. These are thickness ($h$), the Young's modulus ($E$), and the Poisson's ratio ($\nu$) \parencite{Bhaskar2021Plates, Kandaz2021FiniteMicroplates}. While real materials exhibit much more complex behaviours requiring additional parameters, these parameters capture the essential physics of how thin plates distribute and respond to forces in the linear-elastic regime \parencite{Bhaskar2021Plates}. 

The thickness ($h$) of a material represents its physical dimension perpendicular to its primary surface. In structural analysis, thickness plays a fundamental role in load distribution and structural integrity \parencite{Huda2022MechanicalMaterials}. When a material experiences a force, its thickness determines how that force disperses through the material's volume \parencite{Huda2022MechanicalMaterials}. Thicker elements, therefore, generally demonstrate enhanced load-bearing capacity because they provide more material volume through which forces can be distributed, thereby reducing the concentration of stress at any given point. Analogously, a community with stronger institutions, more robust governance, or denser social capital could “distribute” conflict stress more effectively, reducing the concentration of harm at any one point.

Young's modulus ($E$) serves as an important measure of a material's stiffness, quantifying its resistance to deformation under applied forces. This property is defined mathematically as the ratio of stress ($\sigma$) to strain ($\varepsilon$), where stress is force per unit area ($\frac{F}{A}$), and strain is the change in length per original length ($\frac{\Delta\\L}{L_0}$) \parencite{Huda2022MechanicalMaterials}. In practical terms, a higher Young's Modulus indicates that a material strongly resists deformation under loading conditions (e.g., steel), while a lower value suggests that the material will experience more substantial displacement under the same load (e.g. rubber). In social terms, a high $E$ might correspond to a region that is less sensitive to the same "shock"—perhaps due to economic resources, effective leadership, or social norms that mitigate conflict’s immediate disruption. A lower $E$ means even modest events can trigger substantial displacement or chaos.

Poisson's Ratio ($\nu$) describes the relationship between a material's deformation in different directions. Specifically, it is defined as the ratio of lateral strain to axial strain, expressing how much a material contracts perpendicular to the direction of stretching \parencite{Huda2022MechanicalMaterials}. When a material experiences an axial force (pulling or pushing along one direction), it typically responds with a corresponding change in its perpendicular direction. Practically, when a material is stretched longitudinally, it generally becomes narrower in its cross-section. In social contexts, this could represent how impacts propagate between different locations or communities. For instance, a high Poisson's ratio might characterise well-connected regions where stress in one area rapidly spreads to affect neighbouring areas through existing social networks, supply chains, or institutional partnerships. Conversely, a low Poisson's ratio might represent more isolated communities where impacts remain relatively localised due to limited connectivity or weak inter-community ties. This parameter could, therefore, capture how the social fabric's interconnectedness influences the spatial spread of conflict events. 

It is important to recognise that this linear-elastic approximation, while useful, represents a deliberate simplification. Real communities, like real materials, might show more complex responses. They may not return to their original state, similar to how a metal sheet remains bent after excessive force. Their stress response might also change over time, comparable to how a rubber band loses elasticity with repeated stretching. Most critically, they might suddenly fail when stress exceeds critical thresholds, like a plate that cracks when bent too far. Modelling these more complex behaviours would require additional parameters and more sophisticated mathematical frameworks. For instance, capturing permanent changes would require defining yield strength—the point at which a community begin to show lasting transformation rather than resilient recovery. Understanding how repeated conflicts alter the "social fabric" itself would need hardening parameters that describe how material properties evolve under stress. Predicting system breakdown would demand fracture toughness parameters characterising resistance to complete institutional collapse. 

However, starting with the linear-elastic approach and its three core parameters provides several important advantages. First, it offers a less complex way to mathematically model key dynamics identified in the literature, particularly the propagation of effects through space and the interaction of multiple stressors. Second, the parameters have clear physical meanings that can guide interpretation in social contexts, making the framework more accessible to practitioners. Third, this foundation can be systematically extended to include more complex behaviours in the future, allowing the gradual incorporation of additional social science insights as the framework develops.


\section{Finite Element Analysis Framework}

To translate these physical concepts into practical analysis tools, the framework employs the \ac{FEM} based on the Kirchhoff-Love plate theory. This classic plate theory provides the mathematical framework for analysing how thin elastic plates deform when subjected to load in a transversal direction, i.e. perpendicular to the plate's surface \parencite{Larson2013TheApplications}.

The fundamental hypotheses characterising Kirchhoff-Love plate theory are: 

\begin{enumerate}
    \item Straight lines normal to the plate's mid-surface remain straight and normal to that surface after deformation \parencite{Larson2013TheApplications}. This assumption neglects shear and twisting deformation effects, meaning that the plate deforms primarily through bending \parencite{Kandaz2021FiniteMicroplates}. 
    \item The plate's thickness remains constant during deformation \parencite{Larson2013TheApplications}. This assumption, combined with the first, means that membrane effects (in-plane deformation) and through-thickness changes are neglected \parencite{Kandaz2021FiniteMicroplates}.
\end{enumerate}

Because the theory assumes that the plate is very thin compared to its other dimensions, it is sufficient to describe the entire deformation by tracking just the deflection to its middle surface \parencite{Larson2013TheApplications}. This simplifies the mathematical complexity while retaining the essential physics of how impacts spread. Under these assumptions, the displacement field ($u$) representing the primary impact simplifies to three components: a vertical displacement ($w$) and two rotations ($\theta_x$, $\theta_y$) that are directly related to the slope of the vertical displacement \parencite{Kandaz2021FiniteMicroplates}. 

The governing differential equation under these assumptions takes the form \parencite{Bhaskar2021Plates}: 
\begin{equation} \label{eq:kirchhoff-love-plate-equation}
    D\nabla\\^4w = q 
\end{equation}
where $D$ represent the plate's bending stiffness, $w$ denotes the displacement and $q$ represents the applied load. The operator $\nabla ^4$ is called the biharmonic operator \parencite{Bhaskar2021Plates}. It represents taking derivatives with respect to both spatial coordinates (x and y) twice. 

The plate's bending stiffness $D$ is given by \parencite{Bhaskar2021Plates, Larson2013TheApplications}:
\begin{equation} \label{eq:bending-stiffness}
    D = \frac{Eh^3}{12\cdot(1-\nu^2)} 
\end{equation}
Here, $D$ combines the three material parameters already described above—the thickness ($h$), Young's modulus ($E$), and Poisson's ratio ($\nu$)—into a single measure of the plate's resistance to bending. The cubic dependence on thickness ($h^3$) makes this parameter particularly influential. Young's modulus ($E$) appears linearly, representing how strongly the material resits deformation, while Poisson's ratio ($\nu$) moderates the coupling between different directions of bending. The practical influence of these parameters will be tested in the following sections. 

\begin{figure}[H]
    \centering
    \includegraphics[width=0.75\linewidth]{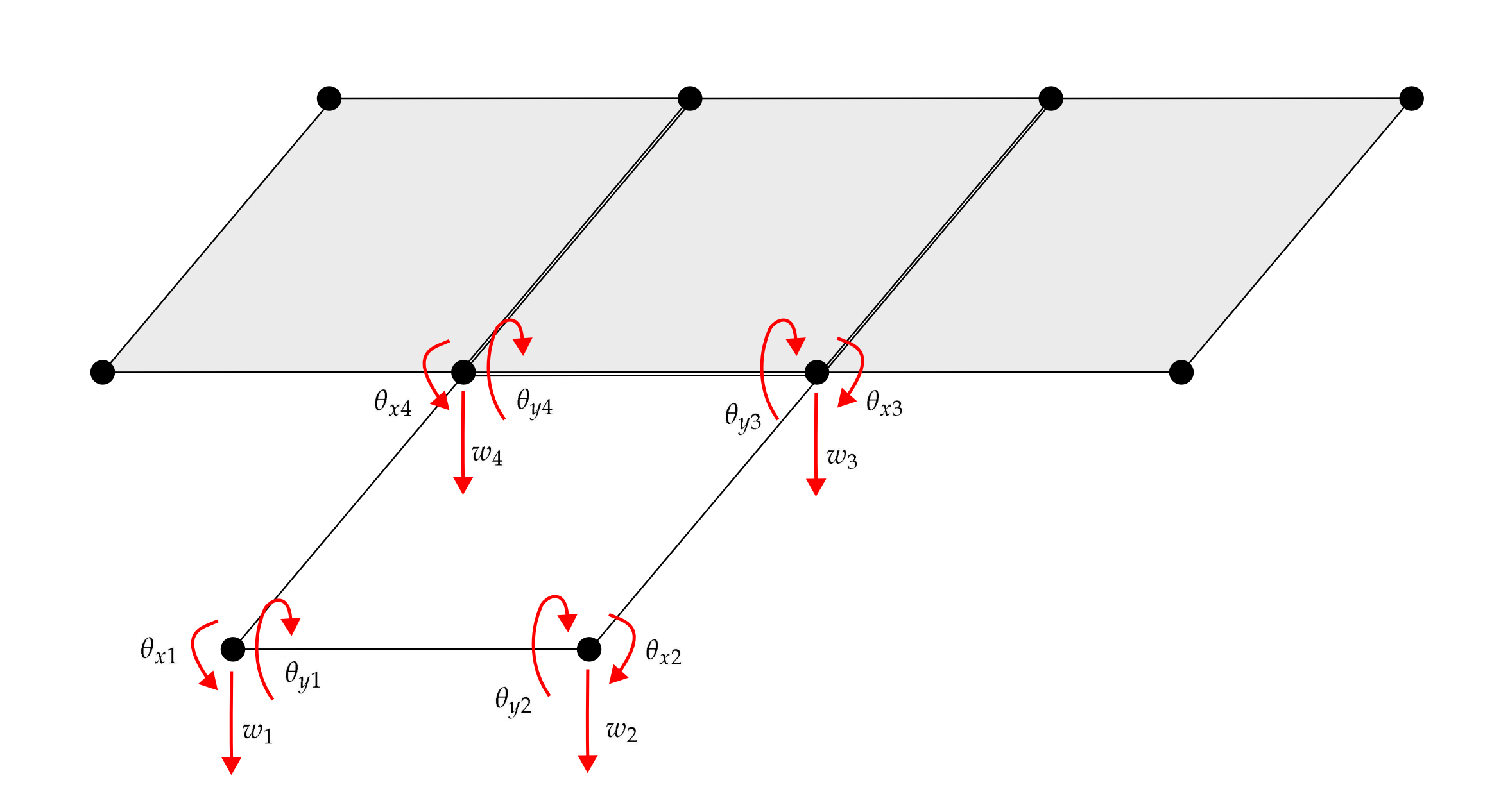}
    \caption[Finite element mesh representation]{Finite element mesh representation showing a four-node quadrilateral element with three degrees of freedom per node: vertical displacement ($w$) and rotations about the x- and y-axes ($\theta_x$, $\theta_y$). These nodal \ac{DOFs} determine the element's deformation behaviour under applied loads \parencite{Kandaz2021FiniteMicroplates}.}
    \label{fig:mesh-modal-DOFs}
\end{figure}

To solve the governing differential equation numerically, the domain (the plate) is discretised into a mesh of small elements \parencite{Kandaz2021FiniteMicroplates}. For plate bending analysis, each node in this mesh has three \ac{DOFs}, as visualised in Figure \ref{fig:mesh-modal-DOFs}, and as already described above for the displacement field components: the vertical displacement $w$ and the two rotations $\theta_x$ and $\theta_y$ \parencite{Kandaz2021FiniteMicroplates}. 

This discretisation transforms the continuous problem into a system of equations that can be solved computationally. However, as the plate is not indefinite, the boundary conditions at the edges need to be specified as well. Boundary conditions constrain all or some \ac{DOFs} of the nodes at the edges. These nodes can, therefore, no longer be displaced and/or rotated, which influences the displacement and stress propagation to some extent. Because these constraints do not stem from the material properties themselves, it is important to examine these influences through tests in the following sections to avoid distorting the observed impacts unintentionally. By systematically assembling and solving the resulting equations, the displacement field for each node is obtained. In a social context, this displacement could be interpreted as a measure of how strongly that location is being impacted. The suitability of displacement as an impact measure, however, is further tested and explored in the next chapter.


\section{Python-based Finite Element Analysis Implementation}

To develop these theoretical insights into a practical tool for conflict analysis, a custom Python-based \ac{FEA} framework was developed rather than relying on commercial \ac{FEA} software. This decision was driven by several key requirements specific to conflict analysis: 

\begin{enumerate}
    \item \textbf{Direct integration with geospatial data}: The implementation needed to handle geospatial data formats efficiently, allowing direct import of social indicators from \ac{GIS} sources (i.e. \texttt{.tif} files) to map them to material properties. This enables easier data integration into the analysis. 
    \item \textbf{Flexible data processing}: Conflict event data, typically available in a tabular format like CSV, needed to be easily incorporated as loading conditions. The custom implementation allows direct processing of such data sources to define force configurations. 
    \item \textbf{Modular design}: The framework needed to be highly adaptable for various mapping approaches between social indicators and material properties. The implementation in a Jupyter Notebook with Python facilitates easy modification of these relationships without changing the core \ac{FEA} implementation. 
\end{enumerate}

\begin{figure}[H]
    \centering
    \includegraphics[width=0.6\linewidth]{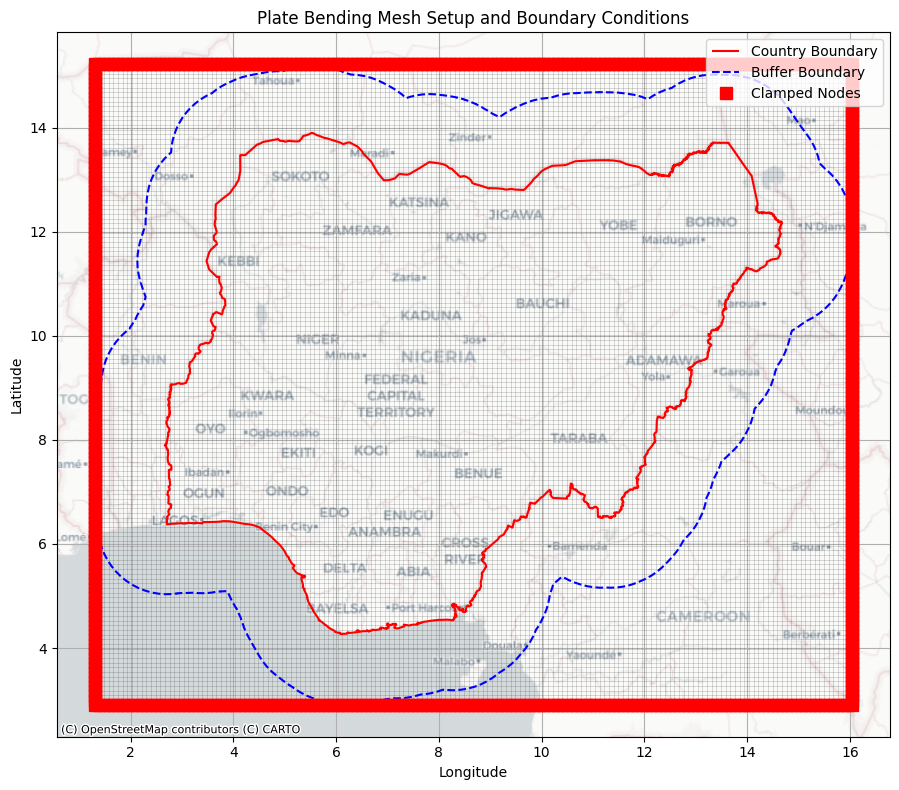}
    \caption[Computational domain setup for finite element analysis]{Computational domain setup for \ac{FEA} of Nigeria, showing the national boundary (red solid line), extended buffer zone boundary (blue dashed line), and applied clamped boundary conditions (red bars). The mesh grid overlay illustrates the spatial discretisation used for the finite element implementation.}
    \label{fig:fea-plate-buffer}
\end{figure}

The implementation centres around a custom \texttt{FEAMesh2D} class that generates the mesh, including geographic coordinate handling, manages element-wise material properties, enforces boundary conditions (either "clamped", which constrains all \ac{DOFs} or "simply supported", which only constrains the vertical displacement $w$), solves the plate bending equations, and visualises the results within a geographic context. It can also define a buffer zone around the main analysis region (Figure \ref{fig:fea-plate-buffer}), preventing boundary effects from artificially constraining or skewing results near the edges of the analysis area.

Such an open-source, Python-based framework also aligns well with the calls in social science for transparent, reproducible workflows. Additionally, it also enables easier future expansions, such as integrating advanced \ac{FEA} methods (e.g., machine-learning-assisted solvers) if the linear-elastic model is to be extended for more complex "plastic" or time-evolving scenarios \parencite{Maurizi2022PredictingNetworks}. For this thesis, however, the linear-elastic approach provides a clear, interpretable starting point to see how repeated conflict loads accumulate, how they might overlap, and how local socio-economic vulnerabilities can effectively be translated into "material" properties. 

By adopting the mathematics of plate bending, it becomes possible to model how conflict events, treated as forces, interact with underlying vulnerabilities and resilience characteristics. The next chapter details how preliminary tests of this framework confirm that simple changes in "thickness" or "elasticity" can significantly alter the "impact" of conflicts.


\chapter{Implementation Testing \& Behavioural Analysis}

As established in the previous chapter, validating the behaviour of the custom implementation and exploring potential limitations is essential before developing the framework for translating social and conflict characteristics. This chapter presents systematic tests of various plate parameters and force configurations possible within the Python implementation, examining how different combinations affect displacement patterns and stress distributions. 

The analysis measured three key outputs: displacement magnitudes, the affected area\footnote{The "affected area" measures the percentage of all mesh nodes whose displacement exceeds 10\% of the maximum displacement.}, and the von Mises stress\footnote{Von Mises stress is a commonly used stress measure in structural analysis. It combines the multi-axis stress state into a single scalar value, providing a straightforward way to compare the stress in a material to its yield stress \parencite{Huda2022MechanicalMaterials}.}. Tests were conducted using a plate model approximating Nigeria's dimensions with a 150 km buffer zone, measuring roughly 1620 km by 1370 km. The model employs a uniform 10 km mesh resolution with clamped boundary conditions, except where specifically boundary conditions effects were examined. Figure \ref{fig:fea-plate-buffer} illustrates this basic plate configuration. 

The testing methodology follows a systematic progression: 

\begin{enumerate}
    \item Uniform plate property tests examining the effects of varying thickness $h$, Young's modulus $E$, and Poisson's ratio $\nu$.
    \item Heterogenous plate property tests investigating the impact of spatial variations in material parameters. 
    \item Boundary condition tests analysing edge effects and buffer requirements. 
    \item Force property tests exploring magnitude, distribution, radius, and interaction effects. 
\end{enumerate}

This comprehensive testing ensured the implementation behaved as theoretically expected and identified any constraints or considerations necessary for the subsequent social conflict translation framework.


\section{Plate Property \& Boundary Condition Tests}

The investigation of the model behaviour began with a systematic examination of plate properties, progressing from uniform configurations to spatially varying parameters and boundary effects. This approach establishes a fundamental understanding of the relationships between material parameters and plate response characteristics.

\subsection{Uniform Plate Properties - Tests}

Initial testing focused on understanding how individual material parameters—thickness $h$, Young's modulus $E$, and Poisson's ratio $\nu$—influence plate behaviour across physically meaningful ranges. Each parameter was varied independently while applying a single point force of 1.0e9 N at the plate's centre (longitude 8.67° and latitude 9.06°). During each test series, non-investigated parameters were held at the following baseline values: 

\begin{itemize}
    \item Thickness Effects: $E$ = 5e9 Pa,  $\nu$ = 0.3
    \item Young's Modulus Effects: $h$ = 2,000 m,  $\nu$ = 0.3
    \item Poisson's Ratio Effects: $h$ = 2,000 m, $E$ = 5e9 Pa
\end{itemize}

\subsubsection{Thickness Effects}

\begin{figure}[H]
    \centering
    \includegraphics[width=0.60\linewidth]{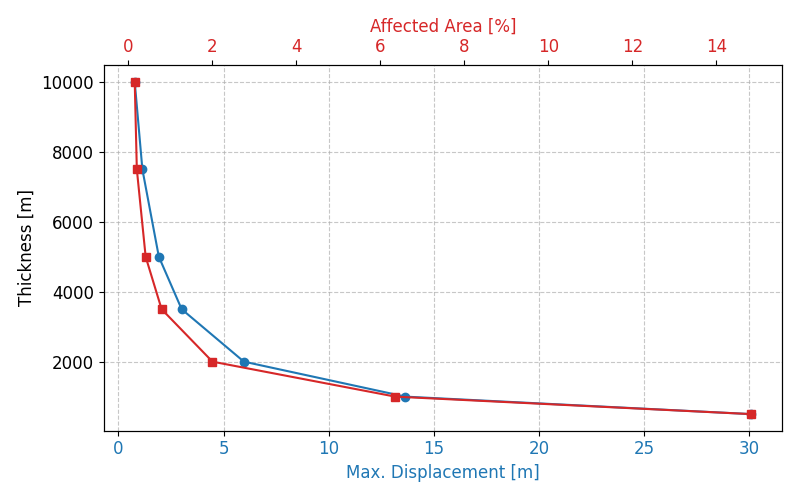}
    \caption[Effect of plate thickness on maximum displacement and affected area]{Effect of plate thickness on maximum displacement and affected area. The dual-axis plot demonstrates the inverse relationship between thickness and both response measures. The blue curve tracks maximum displacement (m) against the bottom axis, while the red curve shows the percentage of affected area against the top axis.}
    \label{fig:plate_material_properties_thickness__displacement-affected-area}
\end{figure}

\begin{figure}[H]
    \centering
    \includegraphics[width=0.60\linewidth]{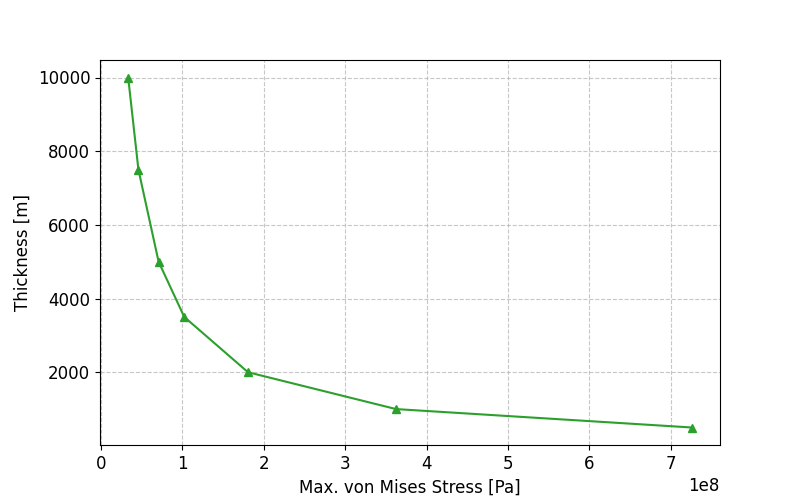}
    \caption[Relationship between plate thickness and maximum von Mises stress]{Relationship between plate thickness and maximum von Mises stress. The curve illustrates the non-linear decay of the maximum von Mises stress with increasing thickness.}
    \label{fig:plate_material_properties_thickness__von-Mises-stress}
\end{figure}

The analysis of the thickness variations revealed a pronounced inverse relationship between plate thickness and deformation characteristics. This relationship directly reflects the cubic relationship ($h^3$) in the bending stiffness equation (Equation \ref{eq:bending-stiffness}). When increasing the thickness from 500 m to 10,000 m, the maximum displacement decreased by nearly two orders of magnitude, from 30.095 m to 0.7738 m (Figure \ref{fig:plate_material_properties_thickness__displacement-affected-area}). Simultaneously, the affected area contracted from 14.83 \% to 0.16 \% of the total plate surface (Figure \ref{fig:plate_material_properties_thickness__displacement-affected-area}), while the maximum von Mises stress decreased from 7.2598e+08 Pa to 3.3336e+07 Pa (Figure \ref{fig:plate_material_properties_thickness__von-Mises-stress}). 

Figure \ref{fig:plate_material_properties_thickness__displacement-affected-area} illustrates this relationship between thickness and displacement magnitude. The curve exhibits clear non-linear behaviour, with the steepest response changes occurring at lower thickness values. This non-linearity could be used to suggest that communities with already low baseline resilience (represented by low thickness) might experience disproportionately larger impacts from the same conflict event compared to more robust regions. 


Examination of the displacement and stress patterns across different thickness values (Figures \ref{fig:plate_material_properties_thickness__displacement-affected-area} and  \ref{fig:plate_material_properties_thickness__von-Mises-stress}) also reveal three distinct response regimes. In the low-thickness regime ($h < 1,000$ m), even minor variations in thickness produce substantial changes in all response measures. The intermediate regime (2,000 m $< h <$ 5,000 m) exhibits more moderate sensitivity, where response changes remain noticeable but less dramatic. Above 5,000 m, the high-thickness regime demonstrates minimal additional changes in response characteristics, suggesting a potential saturation point in the relationship between structural robustness and impact resistance. 

Table \ref{tab:plate-material-properties-thickness} presents the complete quantitative results of the thickness variation testing. The systematic reduction in all three measured parameters—displacement, the affected area, and von Mises stress—demonstrates the consistency of thickness effects across different measures of the plate response. This comprehensive influence of thickness on plate behaviour suggests it could serve as a primary parameter for representing fundamental community resilience in the social fabric analogy. 

\begin{table}[H]
    \centering
    \begin{tabular}{>{\raggedright\arraybackslash}p{100pt}|>{\raggedright\arraybackslash}p{100pt}|>{\raggedright\arraybackslash}p{100pt}|>{\raggedright\arraybackslash}p{100pt}} \hline
         \textbf{Thickness $h$ [m]}&  \textbf{Max. Displacement [m]}&  \textbf{Affected Area [\%]}& \textbf{Max. von Mises Stress [Pa]}\\  \hline \hline
         500&  3.0095e+01&  14.83& 7.2598e+08\\ \hline 
         1,000&  1.3642e+01&  6.36& 3.6257e+08\\ \hline 
         2,000&  5.9667e+00&  2.01& 1.8059e+08\\ \hline 
         3,500&  3.0062e+00&  0.80& 1.0227e+08\\ \hline
         5,000& 1.9211e+00& 0.42&7.0716e+07\\\hline
         7,500& 1.1377e+00& 0.21&4.5917e+07\\\hline
         10,000& 7.7380e-01& 0.16&3.3336e+07\\\hline
    \end{tabular}
    \caption[Effect of plate thickness variation on displacement characteristics and internal stress]{Effect of plate thickness variation on displacement characteristics and internal stress. These results demonstrate a systematic reduction in maximum displacement, affected area, and von Mises stress with increasing thickness.}
    \label{tab:plate-material-properties-thickness}
\end{table}

\begin{figure}[H]
\centering
\begin{tabular}{cc}
\includegraphics[width=0.45\textwidth]{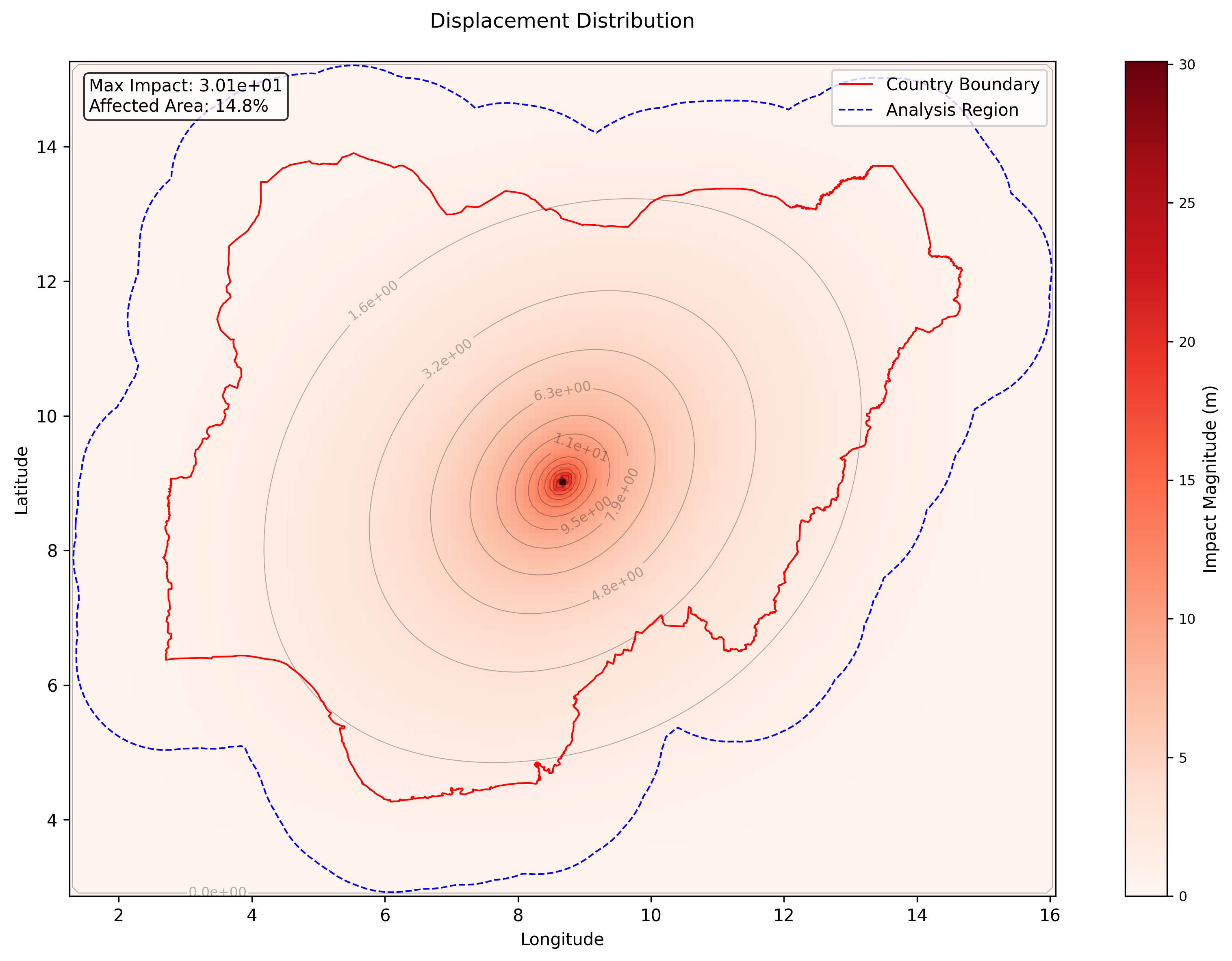}& \includegraphics[width=0.45\textwidth]{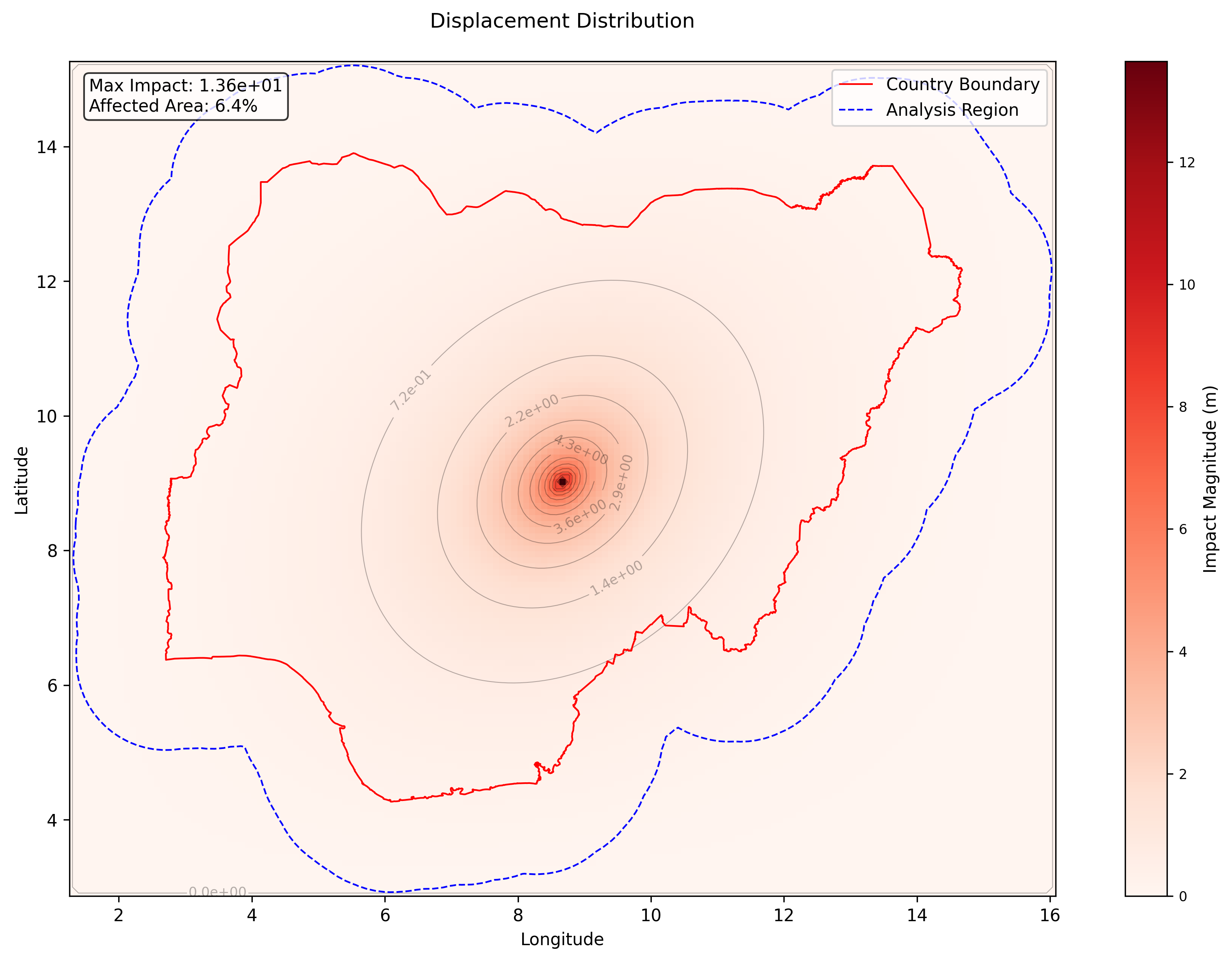}\\
(a) & (b) \\[6pt]
\includegraphics[width=0.45\textwidth]{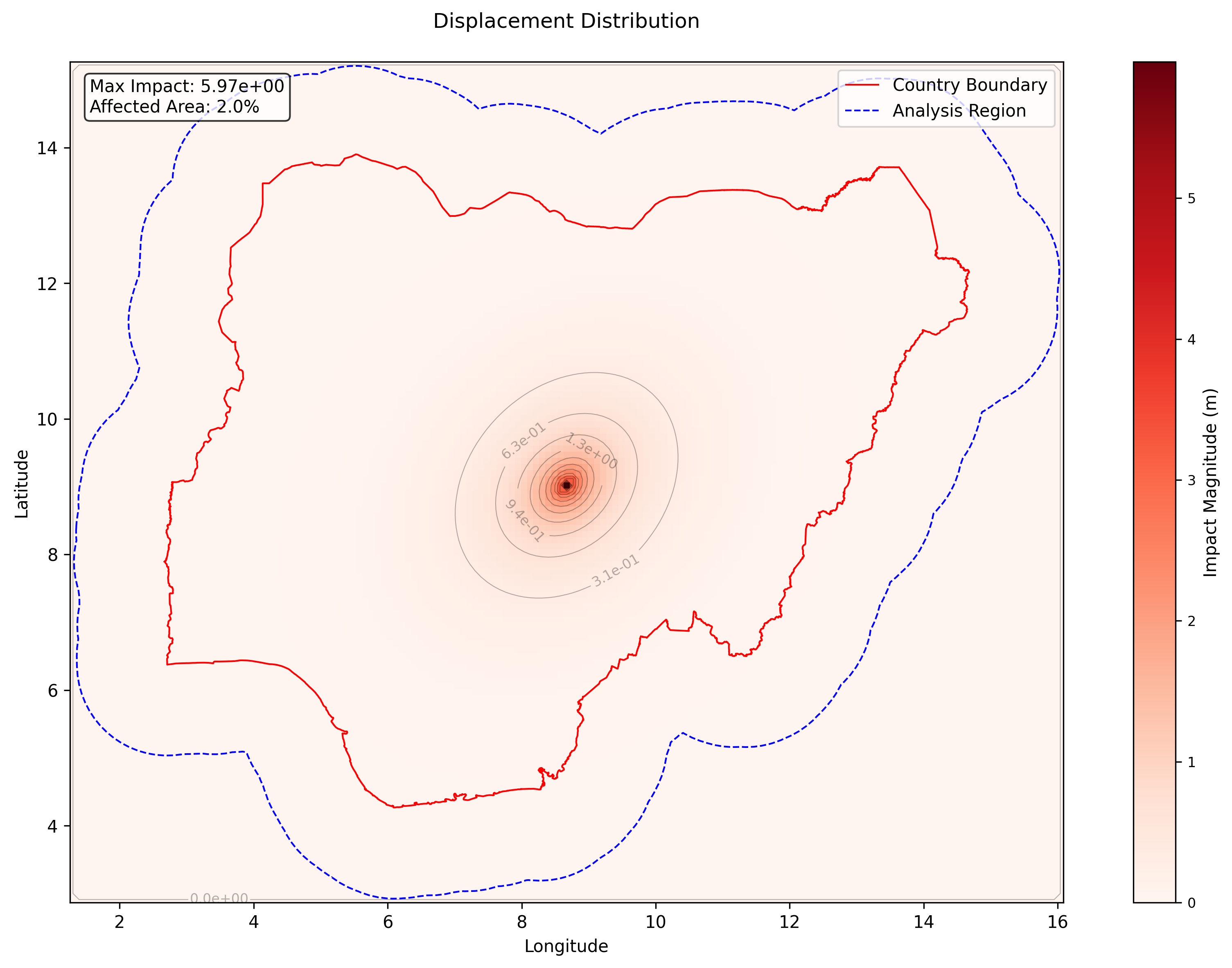}& \includegraphics[width=0.45\textwidth]{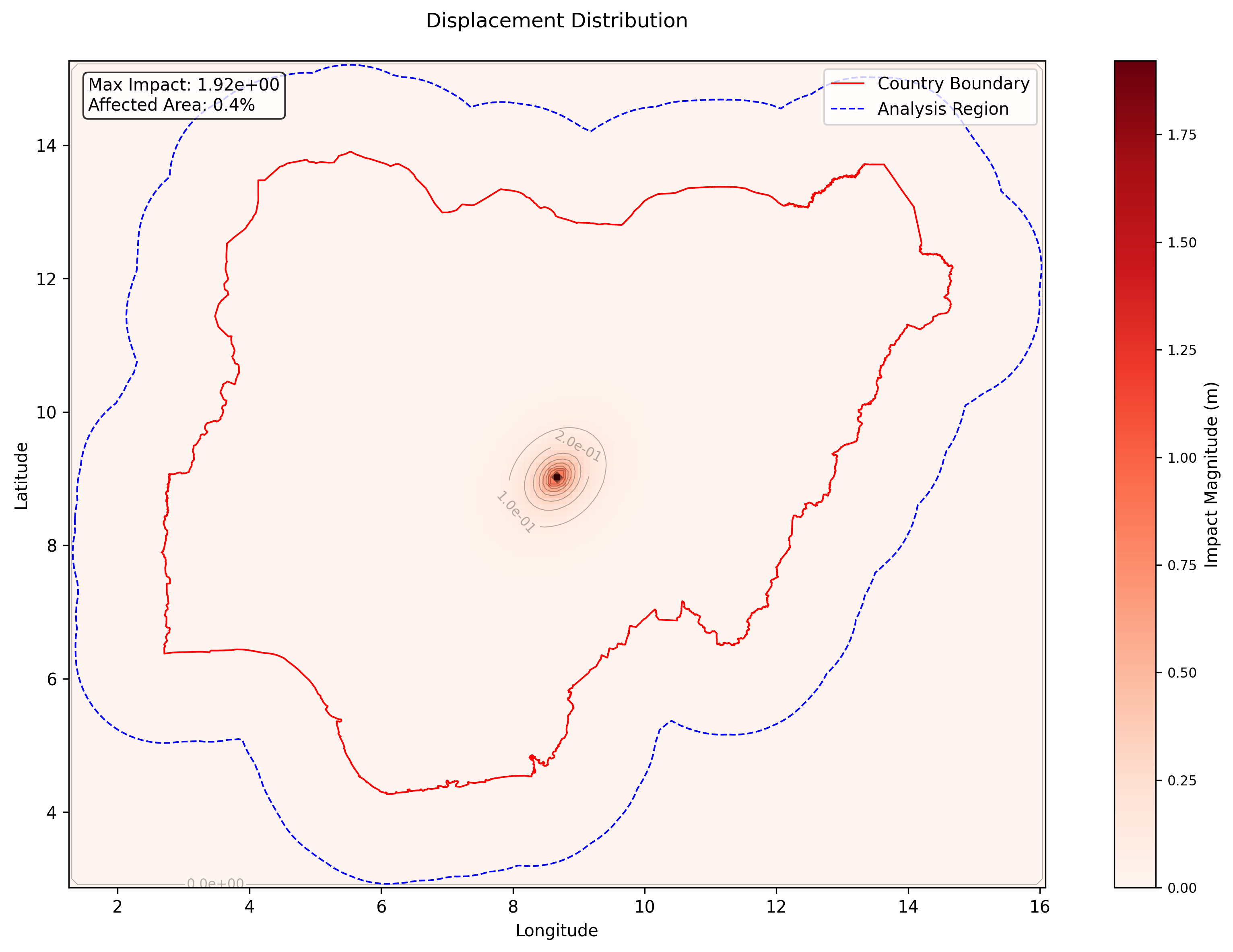}\\
(c) & (d) \\\end{tabular}
\caption[2D displacement field visualisations for increasing plate thickness]{2D displacement field visualisations for increasing plate thickness: \textbf{(a)} 500 m, \textbf{(b)} 1,000 m, \textbf{(c)} 2,000 m, and \textbf{(d)} 5,000 m. The colour scale represents displacement magnitude, with more intensive reds indicating larger displacements.}
\label{fig:plate_material_properties_thickness__displacement-visualisation}
\end{figure}

The spatial extent of these thickness effects can clearly be observed in Figure \ref{fig:plate_material_properties_thickness__displacement-visualisation}, which presents a series of 2D displacement visualisations across four representative thickness values. At 500 m thickness (Figure \ref{fig:plate_material_properties_thickness__displacement-visualisation} \textbf{(a)}), the displacement field shows extensive spread from the central force application point, with significant deformation extending across a large portion of the analysis region. As the thickness increases to 2,000 m (Figure \ref{fig:plate_material_properties_thickness__displacement-visualisation} \textbf{(c)}), the displacement field becomes notably more concentrated, with steeper gradients between the peak displacement and surrounding areas. By 5,000 m thickness (Figure \ref{fig:plate_material_properties_thickness__displacement-visualisation} \textbf{(d)}), the displacement field is highly localised, demonstrating the plate's enhanced ability to resist deformation at greater thicknesses.  

These visualisations provide important insights beyond the numerical data, revealing how thickness influences not only the magnitude but also the spatial pattern of displacement. The progression from diffuse to concentrated displacement fields has important implications for modelling social systems, for instance, when communities with greater baseline resilience (represented by higher thickness) experience less severe impacts from similar events while also better containing their spread. 

\subsubsection{Young's Modulus Effects}

The analysis of Young's modulus variations revealed a distinct inverse proportional relationship with displacement magnitude while maintaining consistent patterns in stress distribution and affected area. As documented in Table \ref{tab:plate-material-properties-youngs-modulus}, increasing Young's modulus from 1.0e7 Pa to 1.0e11 Pa produces systematic reductions in maximum displacement from 2.9833e+03 m to 2.9833e-01 m, following an exact order-of-magnitude correspondence. Notably, both the affected area and maximum von Mises stress remain constant at 2.01 \% and 1.8059e+08, respectively, across all tested modulus values.

\begin{table}[H]
    \centering
    \begin{tabular}{>{\raggedright\arraybackslash}p{100pt}|>{\raggedright\arraybackslash}p{100pt}|>{\raggedright\arraybackslash}p{100pt}|>{\raggedright\arraybackslash}p{100pt}} \hline 
        \textbf{Young's Modulus $E$ [Pa]}&  \textbf{Max. Displacement [m]}&  \textbf{Affected Area [\%]}&  \textbf{Max. von Mises Stress [Pa]}\\ \hline \hline
         1.0e7&  2.9833e+03&  2.01& 1.8059e+08\\ \hline 
         1.0e8&  2.9833e+02&  2.01& 1.8059e+08\\ \hline 
         1.0e9&  2.9833e+01&  2.01& 1.8059e+08\\ \hline
 1.0e10& 2.9833e+00& 2.01&1.8059e+08\\\hline
 1.0e11& 2.9833e-01& 2.01&1.8059e+08\\\hline
    \end{tabular}
    \caption[Effects of Young's modulus variation on plate response characteristics]{Effects of Young's modulus variation on plate response characteristics. The results demonstrate perfect inverse scaling of maximum displacement while maintaining a constant affected area and von Mises stress.}
    \label{tab:plate-material-properties-youngs-modulus}
\end{table}

\begin{figure}[H]
    \centering
    \includegraphics[width=0.75\linewidth]{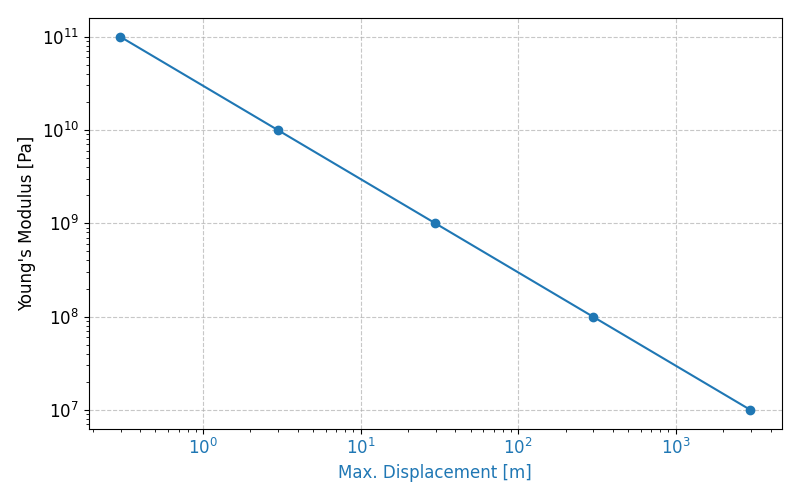}
    \caption[Relationship between Young's modulus and the maximum displacement]{Relationship between Young's modulus and the maximum displacement plotted on logarithmic scales.}
    \label{fig:plate_material_properties_youngs-modulus__displacement}
\end{figure}

Figure \ref{fig:plate_material_properties_youngs-modulus__displacement} illustrates this relationship on a logarithmic scale, revealing a perfect linear relationship between Young's modulus and the maximum displacement. This linear behaviour in log space confirms the direct inverse proportionality between these parameters, where each order of magnitude in Young's modulus produces a corresponding order of magnitude decrease in displacement. The consistent slope of -1 in this log plot provides strong validation that the implementation correctly captures the theoretical relationship between $E$ and the bending stiffness $D$ in Equation \ref{eq:bending-stiffness}. 

The constant affected area and von Mises stress across different Young's modulus values have important implications for modelling social systems. This behaviour suggests that Young's modulus acts purely as a scaling factor, modifying the magnitude of the response without altering its spatial distribution or internal stress patterns. In the context of the social fabric framework, this property makes Young's modulus particularly suitable for representing factors that uniformly enhance or diminish a community's response to conflict events without fundamentally changing how those impacts propagate through the social system.

\subsubsection{Poisson's Ratio Effects}

The analysis of Poisson's ratio variations revealed more subtle and complex relationships compared to the other material parameters. As shown in Table \ref{tab:plate-material-properties-poissons-ratio}, increasing Poisson's ratio from 0.1 to 0.45 produces moderate but consistent change across all response measures. The maximum displacement increases linearly from 5.1491 m to 6.5208 m, while the affected area decreases from 2.37 \% to 1.72 \%, and the maximum von Mises stress shows a slight decline from 1.8074e+08 Pa to 1.8041e+08 Pa. The latter two measures exhibit particularly interesting behaviour in their influence on the localisation of the plate's response.  

\begin{table}[H]
    \centering
    \begin{tabular}{>{\raggedright\arraybackslash}p{100pt}|>{\raggedright\arraybackslash}p{100pt}|>{\raggedright\arraybackslash}p{100pt}|>{\raggedright\arraybackslash}p{100pt}} \hline 
         \textbf{Poisson's Ratio $\nu$}&\textbf{Max. Displacement [m]}& \textbf{Affected Area [\%]}& \textbf{Max. von Mises Stress [Pa]}\\ \hline \hline
         0.1&  5.1491e+00&  2.37& 1.8074e+08\\ \hline 
         0.2&  5.5667e+00&  2.18& 1.8067e+08\\ \hline 
         0.3&  5.9667e+00&  2.01& 1.8059e+08\\ \hline 
         0.4&  6.3435e+00&  1.84& 1.8048e+08\\ \hline
 0.45& 6.5208e+00& 1.72&1.8041e+08\\\hline
    \end{tabular}
    \caption[Impact of Poisson's ratio variation on plate response characteristics]{Impact of Poisson's ratio variation on plate response characteristics. The results demonstrate moderate but consistent changes across all measures, with increasing ratios leading to larger peak displacements but more concentrated impact zones.}
    \label{tab:plate-material-properties-poissons-ratio}
\end{table}

This behaviour stems from Poisson's ratio role in describing the coupling between deformations in perpendicular directions. When a plate undergoes bending, Poisson's ratio determines how strains in one direction induce complementary strains in perpendicular directions \parencite{Huda2022MechanicalMaterials}. Specifically, when a force creates downward displacement (bending) along one axis, a higher Poisson ratio leads to greater curvature development in the perpendicular direction. This cross-directional coupling means that the plate's response becomes more "coupled", and deformation in any direction increasingly influences deformation in perpendicular directions as Poisson's ratio increases. 

This coupling appears mathematically in the bending stiffness (Equation \ref{eq:bending-stiffness}) through the $(1-\nu^2)$ term in the denominator. As Poisson's ratio increases, this term decreases, effectively reducing the overall bending stiffness. The reduced stiffness allows greater displacement for the same applied force, explaining the observed increase in maximum displacement with higher Poisson's ratios. Simultaneously, the stronger coupling between directions tends to concentrate the deformation pattern, resulting in the observed decrease in the affected area. 

The upper limit, which must not be reached, is set to 0.5, as higher Poisson's ratios would imply that a material's volume decreases under tension (or increases under compression), which is physically impossible for linear-elastic materials \parencite{Brnada2019AStresses}. Given that Poisson's ratio appears as a squared term in the bending stiffness equation, a natural lower limit of 0 emerges, creating a narrow but physically meaningful range for this parameter used for linear-elastic plates. 

\begin{figure}[H]
    \centering
    \includegraphics[width=0.6\linewidth]{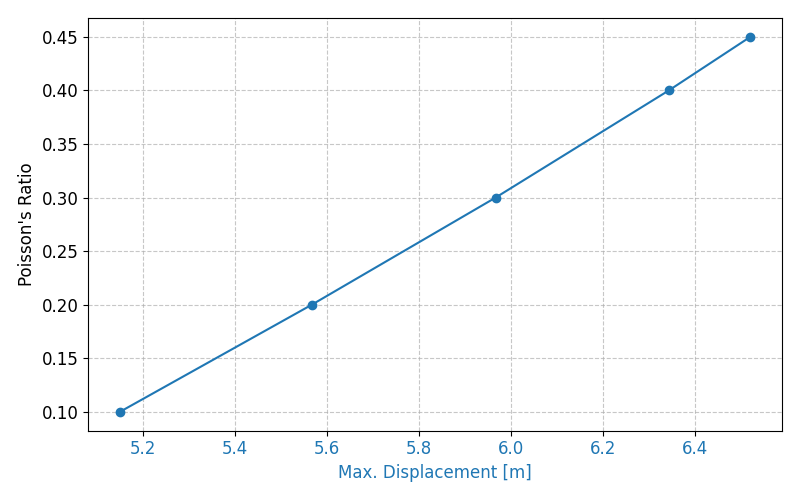}
    \caption[Linear relationship between Poisson's ratio and maximum displacement]{Linear relationship between Poisson's ratio and maximum displacement, demonstrating a consistent increase in peak response with higher Poisson's ratios.}
    \label{fig:plate_material_properties_poissons-ratio__displacement}
\end{figure}

\begin{figure}[H]
    \centering
    \includegraphics[width=0.6\linewidth]{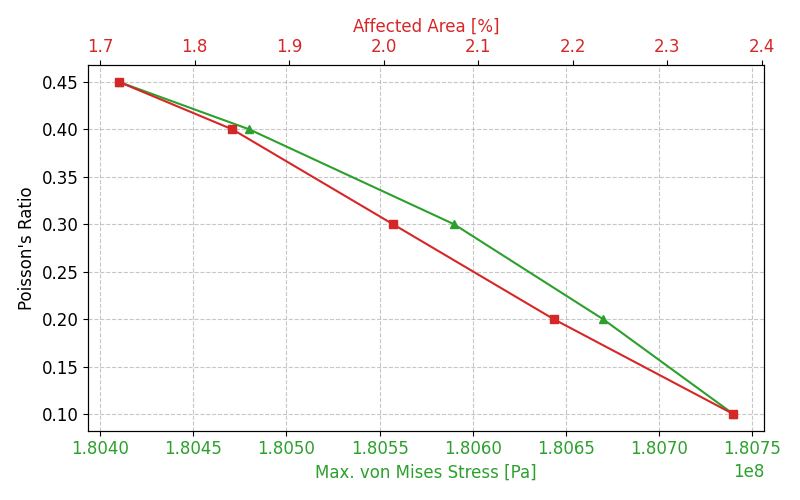}
    \caption[Dual effects of Poisson's ratio on the maximum von Mises stress and the affected area]{Dual effects of Poisson's ratio on the maximum von Mises stress (green) and the affected area (red), showing inverse relationships that indicate increasing localisation of impact with higher Poisson's ratios.}
    \label{fig:plate_material_properties_poissons-ratio__von-Mises-stress-affected-area}
\end{figure}

Figure \ref{fig:plate_material_properties_poissons-ratio__displacement} illustrates the linear relationship between Poisson's ratio and the maximum displacement within this constrained range. The complementary relationships between Poisson's ratio, the affected area, and von Mises stress are shown in Figure \ref{fig:plate_material_properties_poissons-ratio__von-Mises-stress-affected-area}, where the dual-axis plot reveals how higher Poisson's ratios tend to concentrate deformation effects while reducing the overall affected area. 

These constrained but complex effects make Poisson's ratio particularly suitable for fine-tuning impact propagation patterns in social fabric modelling, especially in representing how interconnected communities might concentrate or distribute impacts. However, its limited range and more subtle influence suggest it should serve as a secondary parameter rather than a primary driver of system behaviour.

\subsection{Varying Plate Properties - Tests}

\begin{figure}[H]
    \centering
    \includegraphics[width=1.0\linewidth]{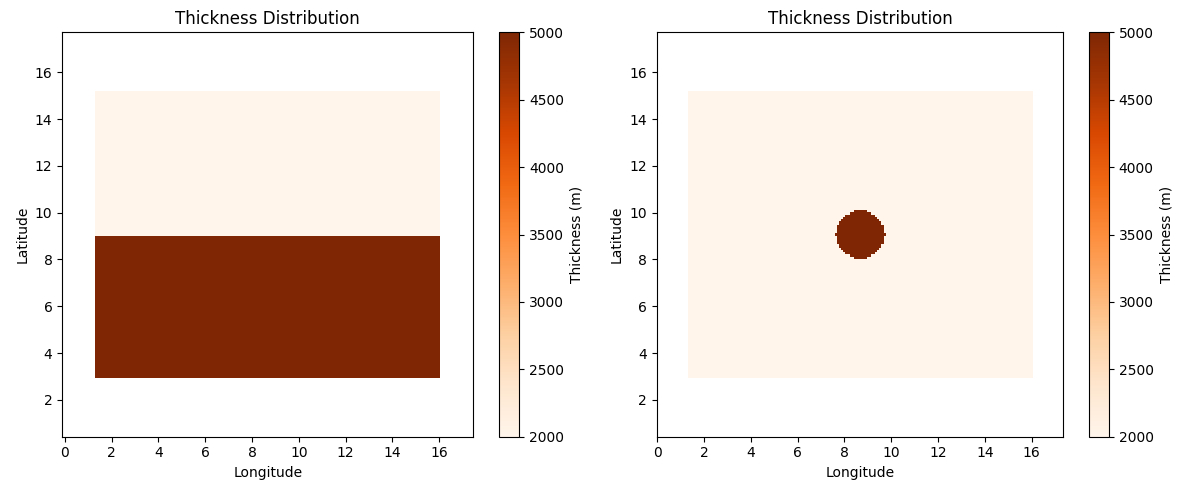}
    \caption[Visualisation of the spatial property variation patterns used in testing]{Visualisation of the spatial property variation patterns used in testing: (left) north-south division pattern showing distinct material properties for northern and southern regions, (right) centre zone pattern showing a circular region of distinct properties within a surrounding area of different characteristics.}
    \label{fig:varying-plate-properties-example}
\end{figure}

Following the analysis of uniform properties, testing progressed to examine the effects of spatially varying material parameters. This is important for understanding how the model represents heterogeneous social structures, where resilience and vulnerability factors may vary significantly across regions. Two spatial patterns, in particular, were tested as illustrated in Figure \ref{fig:varying-plate-properties-example}: a north-south division creating distinct zones with different properties and a central zone configuration examining the interaction between small regions of distinct properties with larger surrounding areas. 

\subsubsection{North-South Division Patterns}

The first spatial variation tests divided the plate into northern and southern halves with distinct materials. Three separate tests were conducted, varying thickness, Young's modulus, and Poisson's ratio independently while keeping other parameters at constant baseline values ($E$ = 5e9 Pa, $\nu$ = 0.3, $h$ = 2000 m). To investigate the effects of forces within these heterogeneous zones, a force of $F$ = 1e9 N with constant distribution over an area with a radius of $r$ = 0.5°, was subsequently applied to the northern zone (8.67°, 9.56°), the boundary separating these two zones (8.67°, 9.06°), and the southern zone (8.67°, 8.56°).  

\begin{figure}[H]
    \centering
    \begin{tabular}{cc}
         \includegraphics[width=0.45\textwidth]{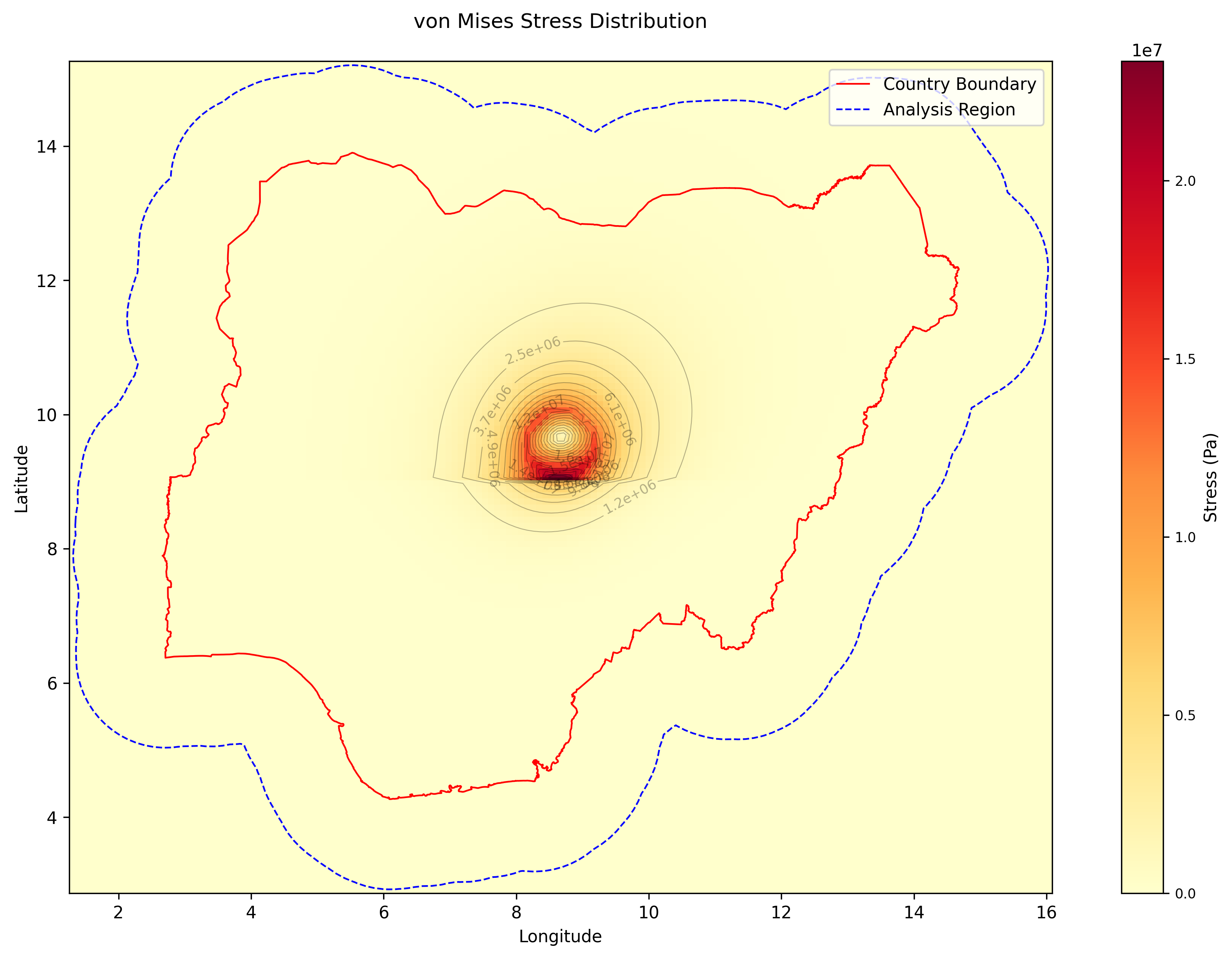}&  \includegraphics[width=0.45\textwidth]{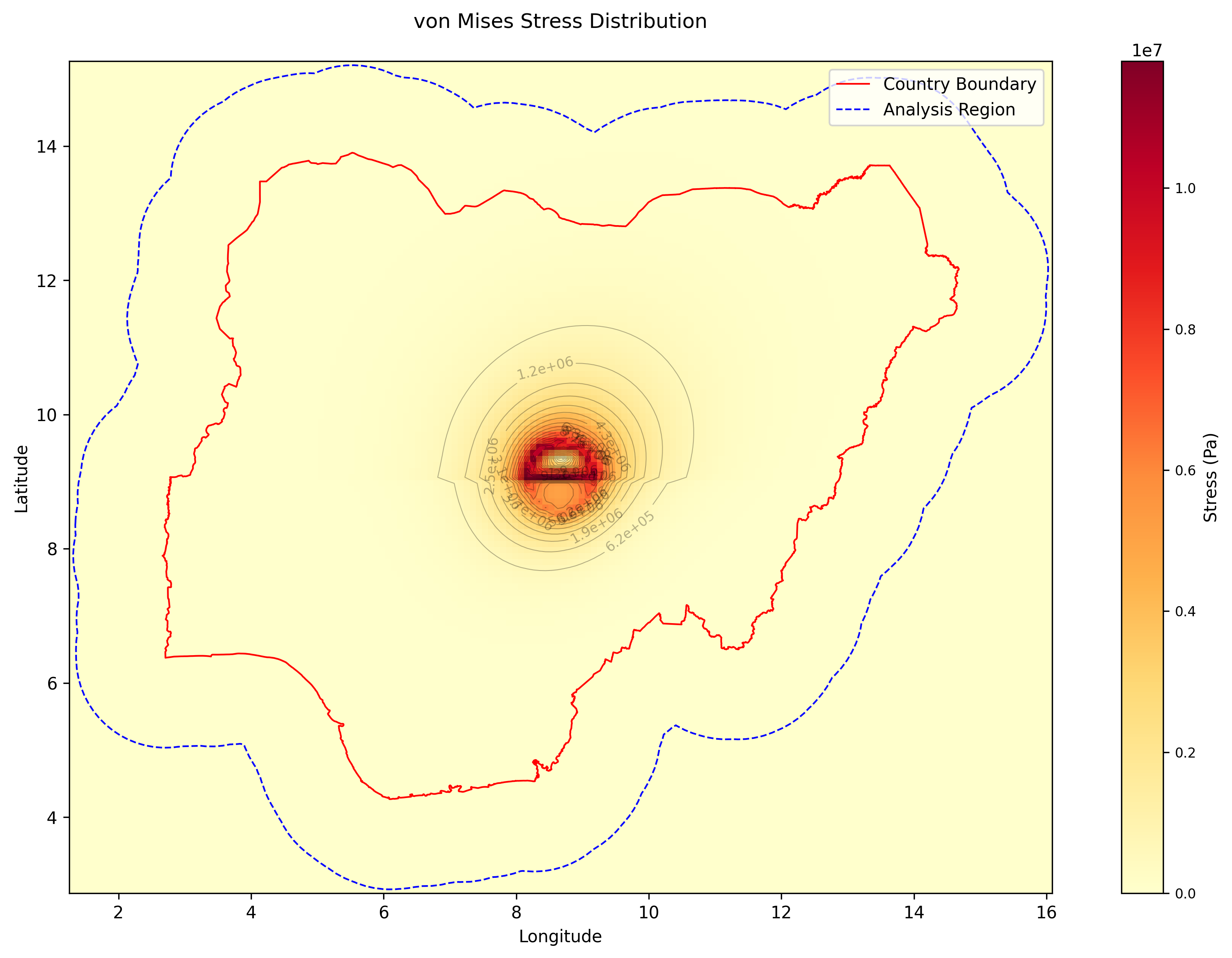}\\
         (a)&(b)\\
 \includegraphics[width=0.45\textwidth]{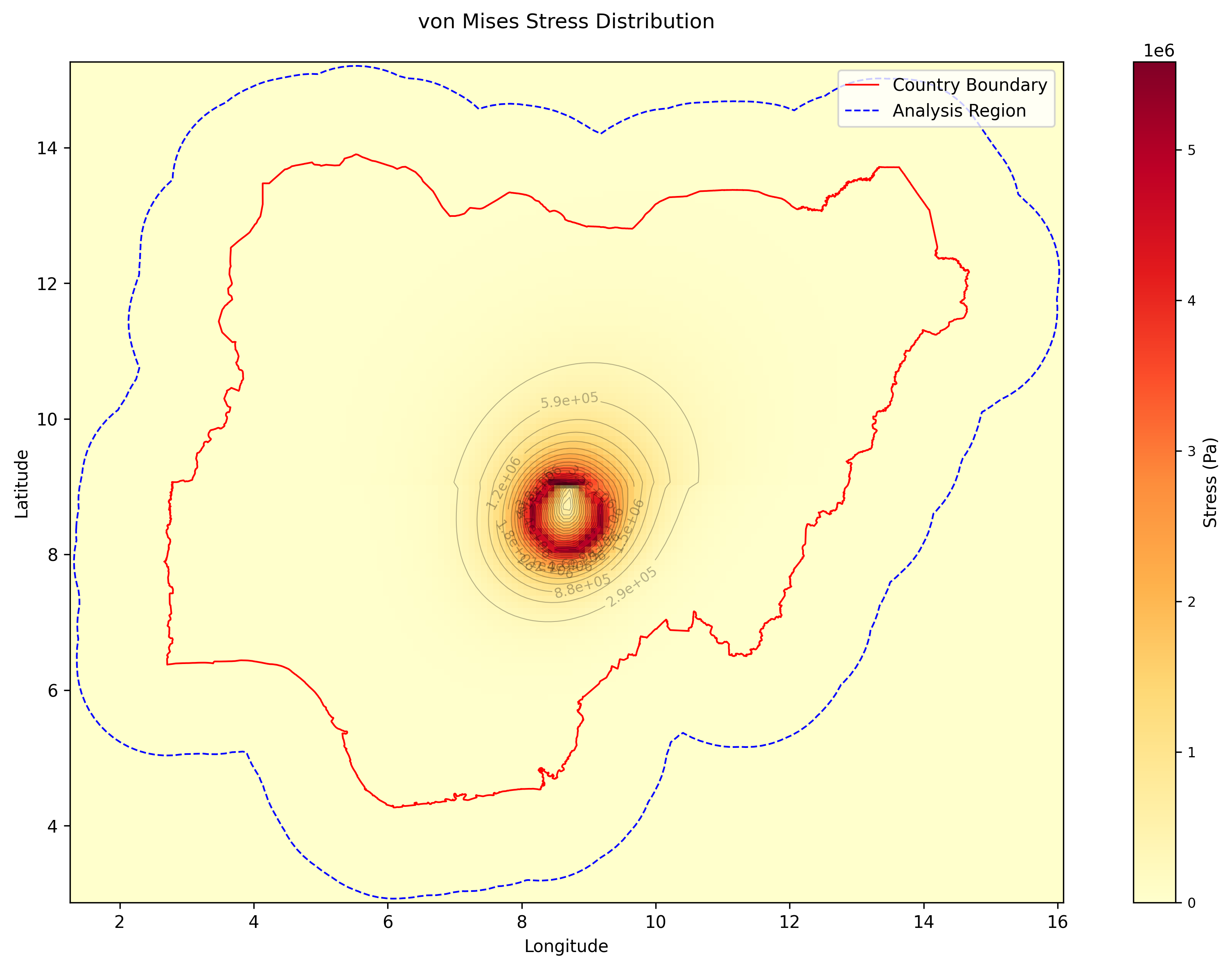}& \\
 (c) & \\
    \end{tabular}
    \caption[Von Mises stress distributions for thickness-divided plate configuration]{Von Mises stress distributions for different thicknesses of the northern (2,000 m) and southern (5,000 m) halves of the plate: \textbf{(a)} force applied in the northern region, showing broader stress propagation in the thinner material, \textbf{(b)} force applied at the boundary, revealing asymmetric stress distribution across the thickness transition, \textbf{(c)} force applied in the southern region, demonstrating a more localised stress pattern in the thicker material. The stress patterns reveal distinct interaction effects at property boundaries.}
    \label{fig:varying-properties-ns-divisions-thickness-visualisations}
\end{figure}

The thickness variation test comparing regions of $h$ = 2,000 m (north) and $h$ = 5,000 m (south), revealed complex interaction effects at zone boundaries. Figure \ref{fig:varying-properties-ns-divisions-thickness-visualisations} illustrates these effects through von Mises stress distributions for the three force application scenarios. When force is applied to the northern regions with lower thickness (Figure \ref{fig:varying-properties-ns-divisions-thickness-visualisations} \textbf{(a)}), the stress pattern shows broader spread and higher intensity compared to application in the thicker southern region (Figure \ref{fig:varying-properties-ns-divisions-thickness-visualisations} \textbf{(c)}). Most notably, force application at the boundary between these regions (Figure \ref{fig:varying-properties-ns-divisions-thickness-visualisations} \textbf{(b)}) produces a distinctly asymmetric stress pattern, with greater stress propagation into the thinner northern region.

The boundary behaviours can also be observed in the quantitative results (Table \ref{tab:varying-properties-ns-divisions}), where force application at the thickness transition zone produces a maximum displacement of 0.92857 m, which is significantly lower than the maximum displacements observed for either individual thickness (5.9667 m for $h$ = 2,000 m and 1.9211 m for $h$ = 5,000 m). The stress visualisations (Figure \ref{fig:varying-properties-ns-divisions-thickness-visualisations}) suggest that this reduction in displacements could stem from the thicker region providing additional support at the boundary, effectively creating a stiffening effect that isn't present in uniform configurations. 

\begin{table}[H]
    \centering
    \begin{tabular}{>{\raggedright\arraybackslash}p{70pt}|>{\raggedright\arraybackslash}p{85pt}|>{\raggedright\arraybackslash}p{85pt}|>{\raggedright\arraybackslash}p{85pt}|>{\raggedright\arraybackslash}p{85pt}} \hline
        \textbf{Varying Property}&\textbf{Location (long, lat) [°]}&\textbf{Max. Displacement [m]}&\textbf{Affected Area [\%]}&\textbf{Max. von Mises Stress [Pa]}\\\hline \hline
         $h$&(8.67, 9.56)& 1.5089e+00& 4.41&2.3354e+07\\ \cline{2-5}
          North 2,000 m&(8.67, 9.06)& 9.2857e-01& 3.81&1.1796e+07\\ \cline{2-5}
 South 5,000 m& (8.67, 8.56)& 4.3163e-01& 3.89&5.5832e+06\\ \hline
         $E$&(8.67, 9.56)& 6.7365e+01& 3.72&2.8068e+07\\ \cline{2-5}
 North 1e8 Pa& (8.67, 9.06)& 2.9740e+01& 2.56&3.1787e+07\\ \cline{2-5}
          South 5e9 Pa&(8.67, 8.56)& 2.8858e+00& 8.60&1.9809e+07\\ \hline
         $\nu$&(8.67, 9.56)& 1.8314e+00& 8.02&1.6484e+07\\ \cline{2-5}
 North 0.2& (8.67, 9.06)& 1.8758e+00& 7.71&1.6908e+07\\ \cline{2-5}
          South 0.4&(8.67, 8.56)& 1.9357e+00& 7.37&1.7024e+07\\ \hline
    \end{tabular}
    \caption[Response characteristics for north-south divided plate configurations under varying material properties]{Response characteristics for north-south divided plate configurations under varying material properties. The results suggest complex interaction effects, particularly at boundaries between regions of different properties.}
    \label{tab:varying-properties-ns-divisions}
\end{table}

Young's modulus variation testing ($E$ = 1e8 Pa north, $E$ = 5e9 Pa south) demonstrated similar interaction effects that deviated from the uniform property behaviour. The uniform tests demonstrated perfect inverse displacement scaling with $E$, suggesting a 50-fold displacement increase as $E$ decreases 50-fold. However, in the divided configuration, the northern region ($E$ = 1e8 Pa) shows a maximum displacement of 67.365 m, while the southern region ($E$ = 5e9 Pa) shows 2.8858 m—a ratio of approximately 23.3 rather than the expected 50. This deviation from the linear scaling production again suggests a significant interaction effect between regions of different properties. 

Poisson's ration variations produced results more closely aligned with uniform testing behaviour, primarily influencing the distribution pattern of deformations rather than their magnitude. This aligns with earlier findings about Poisson's ratio's role in determining how impacts propagate through the material.

These findings, particularly the visual evidence of stress pattern modification at property transition, have important implications for modelling social systems. They suggest that sharp transitions in social characteristics might create unexpected behaviour patterns that cannot be predicted from uniform property analysis alone. The observed boundary effects could represent either artificial computational artefacts that need to be considered when interpreting results or potentially meaningful representations of how abrupt changes in social conditions influence conflict impact patterns.

\subsubsection{Centre Zone Patterns}

The second spatial variation test examined interactions between a small circular region (1° radius) with distinct properties embedded within a larger surrounding area. Two complementary configurations were tested: a "strong" centre surrounded by a "weaker" region and, conversely, a "weak" centre within a "stronger" surrounding area. For both scenarios, a force is applied sequentially to the centre of the small zone (8.67°, 9.06°), the transition zone between the two areas (9.67°, 9.06°), and a point in the larger outer area (10.67°, 9.06°) with a magnitude of $F$ = 1e9 N constantly distributed over an area with a radius of $r$ = 0.5°.

The configurations for both the "strong" and "weak" areas were established according to insights from the observed uniform behaviour so far:

\begin{itemize}
    \item "Strong" configuration (high bending stiffness): $E$ = 5e9 Pa, $\nu$ = 0.4, $h$ = 5,000 m
    \item "Weak" configuration (low bending stiffness): $E$ = 1e8 Pa, $\nu$ = 0.2, $h$ = 2,000 m
\end{itemize}

The centre zone test results shown in Table \ref{tab:varying-properties-center-zone-patterns} indicate significant localisation effects that exceed simple property-based predictions. In the strong centre configuration, a force applied to the centre produces relatively small maximum displacements of 0.40832 m, while the same force on the outer region causes substantially larger displacements of 80.132 m. This behaviour inverted in the weak centre configuration, with a maximum displacement of 61.137 m at the centre compared to 0.38067 m at the outer region.

 \begin{table}[H]
    \centering
    \begin{tabular}{>{\raggedright\arraybackslash}p{70pt}|>{\raggedright\arraybackslash}p{85pt}|>{\raggedright\arraybackslash}p{85pt}|>{\raggedright\arraybackslash}p{85pt}|>{\raggedright\arraybackslash}p{85pt}} \hline
        \textbf{Varying Property}&\textbf{Location (long, lat) [°]}&\textbf{Max. Displacement [m]}&\textbf{Affected Area [\%]}&\textbf{Max. von Mises Stress [Pa]}\\\hline \hline
         &(8.67, 9.06)& 4.0832e-01& 7.62&4.4737e+06\\ \cline{2-5}
          strong centre &(9.67, 9.06)& 2.1871e+01& 2.97&2.4459e+07\\ \cline{2-5}
 & (10.67, 9.06)& 8.0132e+01& 5.72&1.9774e+07\\ \hline
         &(8.67, 9.06)& 6.1137e+01& 1.36&1.6609e+07\\ \cline{2-5}
 weak centre& (9.67, 9.06)& 2.0944e+01& 0.90&1.6422e+07\\ \cline{2-5}
          &(10.67, 9.06)& 3.8067e-01& 2.48&4.6919e+06\\ \hline
    \end{tabular}
    \caption[Response characteristics for centre zone configurations under varying force locations]{Response characteristics for centre zone configurations under varying force locations.}
    \label{tab:varying-properties-center-zone-patterns}
\end{table}

This contrast in stress propagation patterns is further reflected in the affected area measurements in Table \ref{tab:varying-properties-center-zone-patterns}. The strong centre shows varying affected areas (7.62 \% for central loading, reducing to 2.97 \% at the transition zone, then increasing to 5.72 \% in the outer region), suggesting an interaction effect between the strong centre and weak surroundings. The weak centre configuration, despite showing higher peak displacements, consistently affects smaller areas  (1.36 \%, 0.90 \%, and 2.48 \%, respectively), indicating that the stronger surrounding material helps contain the overall spread of effects even as it allows larger local deformations. 

\begin{figure}[H]
\centering
\begin{tabular}{cc}
\includegraphics[width=0.45\textwidth]{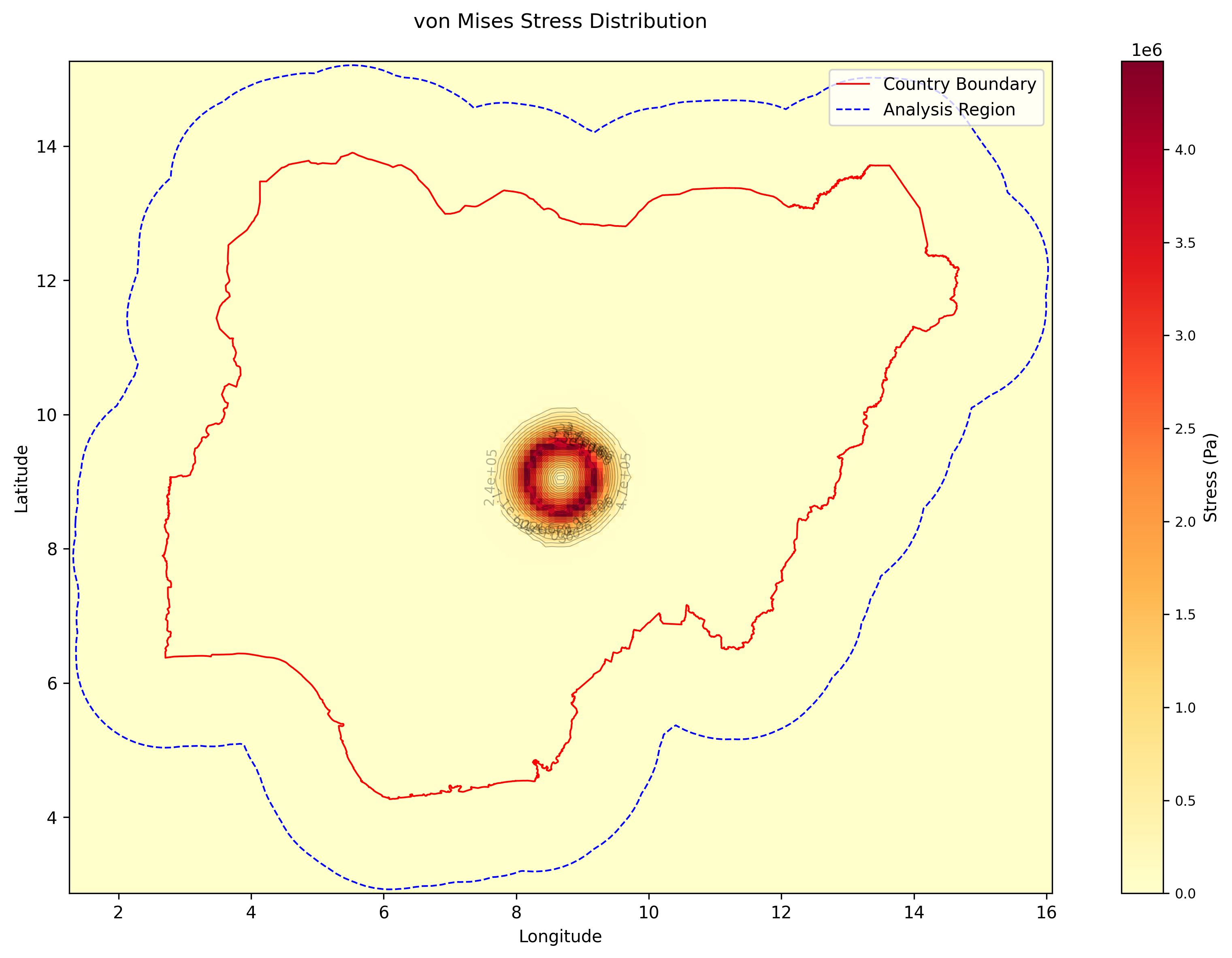}& \includegraphics[width=0.45\textwidth]{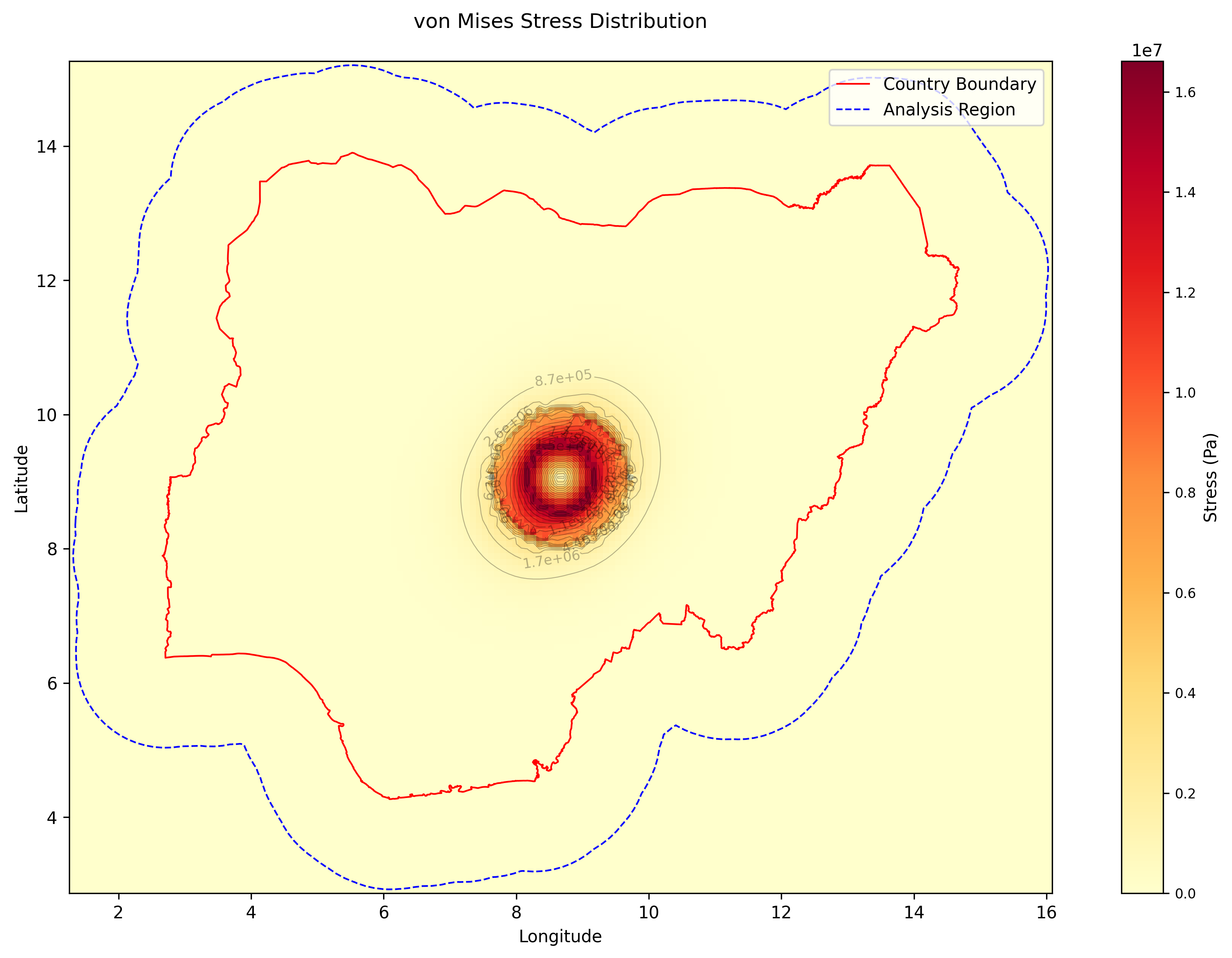}\\  
 (a)&(b)\\\end{tabular}
\caption[Von Mises stress patterns for contrasting centre zone configurations with central force application]{Von Mises stress patterns for contrasting centre zone configurations with central force application: \textbf{(a)} strong centre configuration showing localised stress containment, \textbf{(b)} weak centre configuration exhibiting broader stress distribution despite stronger surroundings.}
\label{fig:varying-properties-centre-zone-patterns-visualisations}
\end{figure}

The von Mises stress distributions, visualised in Figure \ref{fig:varying-properties-centre-zone-patterns-visualisations}, provide further insights into how different centre zone configurations influence stress propagation. The strong centre configuration (Figure \ref{fig:varying-properties-centre-zone-patterns-visualisations} \textbf{(a)}) demonstrates strong stress containment, with the stress pattern showing sharp localisation and minimal spread beyond the central region. In contrast, the weak centre configuration (Figure \ref{fig:varying-properties-centre-zone-patterns-visualisations} \textbf{(b)}) exhibits broader stress distribution, with effects propagating well beyond the central zone despite the surrounding stronger material.

These results reveal that local property variations create complex interaction effects extending beyond simple superposition behaviours. For the social fabric modelling, this suggests that localised resilience characteristics (represented here by the strong centre) might effectively contain and absorb impacts, while localised vulnerabilities could lead to broader impact distribution even when surrounded by more resilient regions. 

\subsection{Boundary Condition Tests}

Since the mesh is secured at the outer edges, these constraints may affect the displacement and stress distributions within the plate. While a 150 km buffer zone around the area of interest has been used in previous tests to minimise these effects, systematic testing of force application near boundaries is necessary to quantify any remaining influences. 

\begin{figure}[H]
    \centering
    \includegraphics[width=0.55\linewidth]{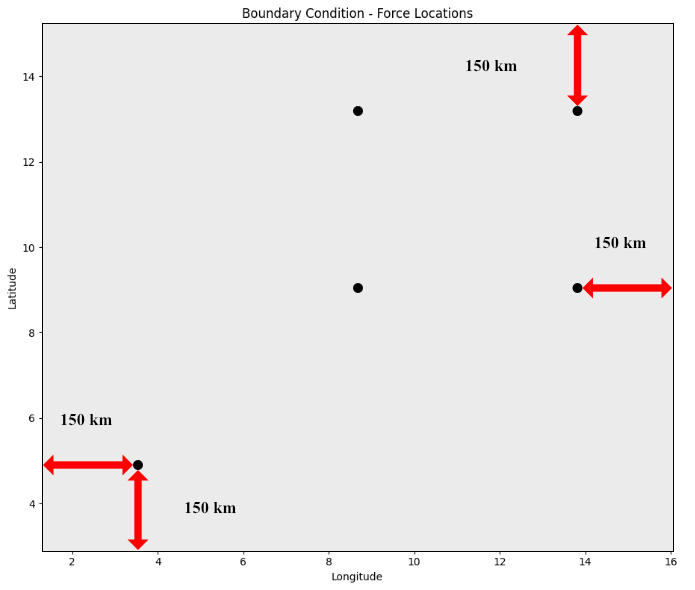}
    \caption[Test configuration for boundary condition analysis showing force application locations]{Test configuration for boundary condition analysis showing force application locations (black dots) positioned 150 km from plate edges and corners. Red arrows indicate the distance from these boundaries.}
    \label{fig:boundary-condition-locations}
\end{figure}

For all boundary condition tests, consistent parameters were used: Young's modulus $E$ = 5e9 Pa, Poisson's ratio $\nu$ = 0.3, and thickness $h$ = 2000 m. A force of $F$ = 1e9 N was applied with linear distribution over a radius $r$ = 0.5°. The plate edges were constrained using "clamped-all" boundary conditions, which completely restrict the movement of nodes at the boundaries by constraining both displacement and rotation. 

As illustrated in Figure \ref{fig:boundary-condition-locations}, force applications were tested at various locations near the plate edges and corners, all maintaining a 150 km distance from the boundaries. The test points include locations near the northern edge (8.67°, 13.20°), eastern edge (13.81°, 9.06°), northeastern corner (13.81°, 13.20°), and southwestern corner (3.53°, 4.92°). These locations were compared against a reference force application at the plate's centre (8.67°, 9.06°).

The results, documented in Table \ref{tab:location-dependency-clamped-all}, reveal varying degrees of boundary influence across different response measures. For the maximum displacement and the maximum von Mises stress, the deviations from the centre-applied force remain relatively modest, with relative errors between 1.30 \% - 2.44 \% and 0.04 \% - 0.90 \%, respectively. However, the affected areas show more substantial variations, with relative errors ranging from 9.81 \% to 19.00 \%, indicating that boundary proximity more strongly influences the spatial extent of impact than its magnitude.

\begin{table}[H]
    \centering
    \begin{tabular}{>{\raggedright\arraybackslash}p{130pt}|>{\raggedright\arraybackslash}p{100pt}|>{\raggedright\arraybackslash}p{100pt}|>{\raggedright\arraybackslash}p{100pt}} \hline
    \textbf{Location (long, lat) [°]}&\textbf{Max. Displacement [m]}&\textbf{Affected Area [\%]}&\textbf{Max. von Mises Stress [Pa]}\\ \hline  \hline
 \textit{(8.67, 9.06), centre}& \textit{2.2990e+00}& \textit{6.42}&\textit{1.9376e+07} \\\hline
         (8.67, 13.20), 150 km from the northern edge&  2.2669e+00 
(1.40 \%)&  5.56
(13.40 \%)& 1.9550e+07
(0.90 \%)\\ \hline 
         (13.81, 9.06), 150 km from the eastern edge&  2.2691e+00
(1.30 \%)&  5.79
(9.81 \%)& 1.9400e+07
(0.12 \%)\\\hline
 (13.81, 13.20), northeastern corner - 150 km from both edges& 2.2428e+00
(2.44 \%)& 5.22
(18.69 \%)&1.9369e+07
(0.04 \%)\\\hline
 (3.53, 4.92), southwestern corner - 150 km from both edges& 2.2467e+00
(2.27 \%)& 5.20
(19.00 \%)&1.9423e+07
(0.24 \%)\\\hline
    \end{tabular}
    \caption[Effect of force location on plate response with 150 km buffer and clamped boundary conditions]{Effect of force location on plate response with 150 km buffer and clamped boundary conditions. The results show comparisons between the central force application (reference case) and forces applied near edges and corners. Values in parentheses indicate relative errors compared to the reference case, demonstrating increasing boundary influence near corners.}
    \label{tab:location-dependency-clamped-all}
\end{table}

Changing the boundary conditions from "clamped-all" to "simply-supported-all," which only restricts vertical displacement at the edges, yields identical test results as presented in Table \ref{tab:location-dependency-clamped-all}. Although differences can be observed in very small plates (1 m x 1 m), the large size of this plate effectively mitigates any discrepancies between different boundary condition types.

To investigate these boundary effects further, additional buffer distances were examined to explore the impact of the clamped boundaries on maximum displacement and maximum von Mises stress\footnote{Unlike the earlier tests, the affected area is omitted since it does not offer a significant comparison metric for plates of varying sizes, as it depends on the total number of nodes, which also increases with size.}. The same test force was applied at the southwestern corner (3.53°, 4.92°), progressively increasing the distance to the outer edges. This choice stemmed from the previous tests, which revealed relatively high errors in all measures (Table \ref{tab:location-dependency-clamped-all}). To assess the error, the central force was also evaluated at these varying buffer radii. The results presented in Table \ref{tab:location-dependency-comparison}, demonstrate that large buffers progressively reduce boundary influences, with relative errors in maximum displacement increasing from 2.27 \% at 150 km to 0.41 \% at 500 km.

\begin{table}[H]
    \centering
    \begin{tabular}{>{\raggedright\arraybackslash}p{130pt}|>{\raggedright\arraybackslash}p{130pt}|>{\raggedright\arraybackslash}p{130pt}} \hline
        \textbf{Buffer [m]} &\textbf{Max. Displacement [m]}&\textbf{Max. von Mises Stress [Pa]}\\\hline \hline
         150,000  & 2.2990e+00 (reference)&1.9376e+07 (reference)\\ \cline{2-3}
          & 2.2467e+00 (2.27 \%)&1.9423e+07 (0.24 \%)\\ \hline
         300,000  & 2.6279e+00 (reference)&2.0275e+07 (reference)\\ \cline{2-3}
          & 2.6048e+00 (0.88 \%)&2.0320e+07 (0.22 \%)\\ \hline
         500,000  & 3.0212e+00 (reference)&2.1256e+07 (reference)\\ \cline{2-3}
          & 3.0335e+00 (0.41 \%)&2.1455e+07 (0.94 \%)\\ \hline
    \end{tabular}
    \caption[Effect of buffer size on boundary condition influences]{Effect of buffer size on boundary condition influences. For each buffer size, the reference row shows results for the force applied at the centre (8.67°, 9.06°), while subsequent rows show results for the force applied at the southwestern corner  (3.53°, 4.92°). Relative errors (shown in parentheses) demonstrate the decreasing boundary influence with larger buffers.}
    \label{tab:location-dependency-comparison}
\end{table}

\begin{figure}[H]
\centering
\begin{tabular}{cc}
\includegraphics[width=0.45\textwidth]{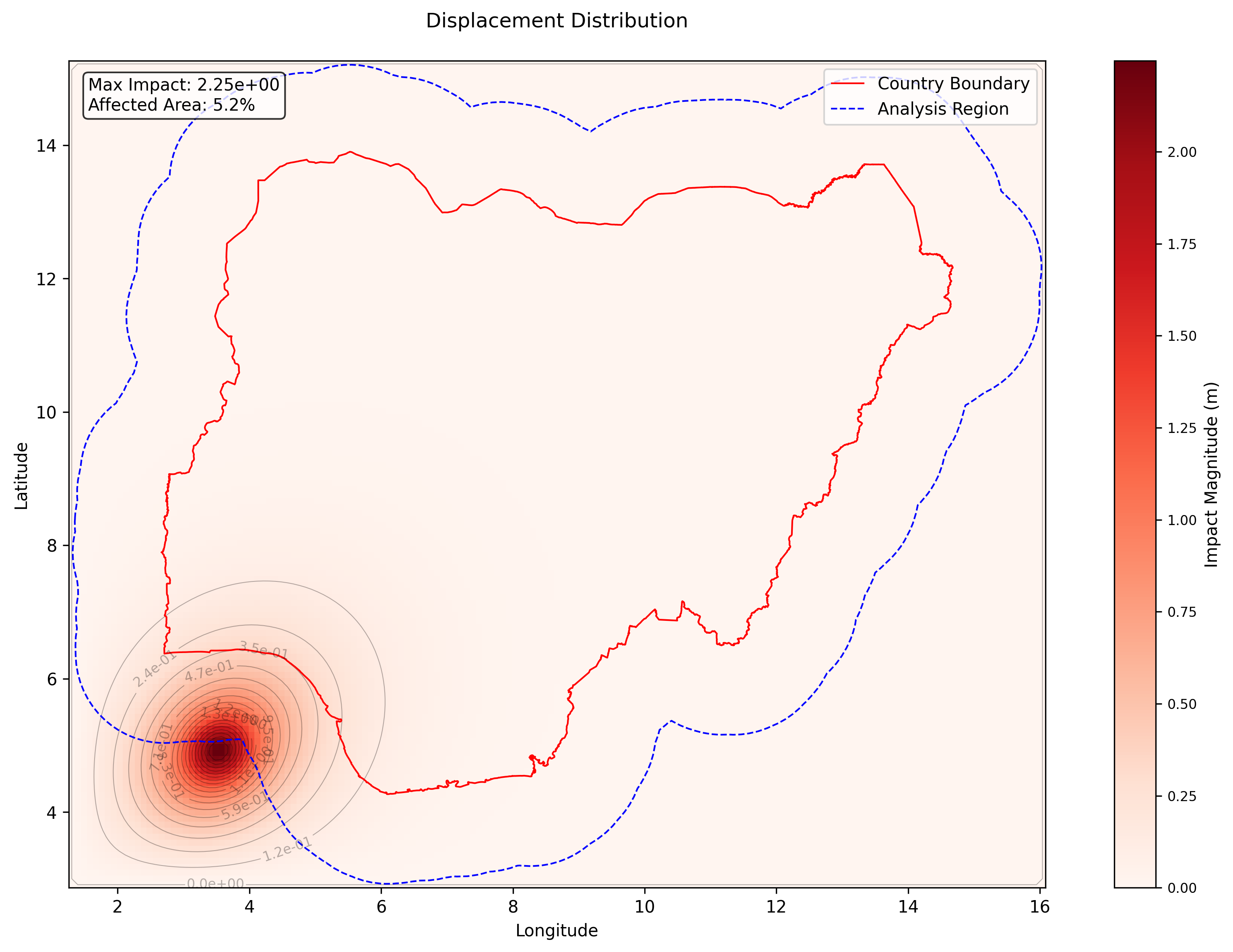}& \includegraphics[width=0.45\textwidth]{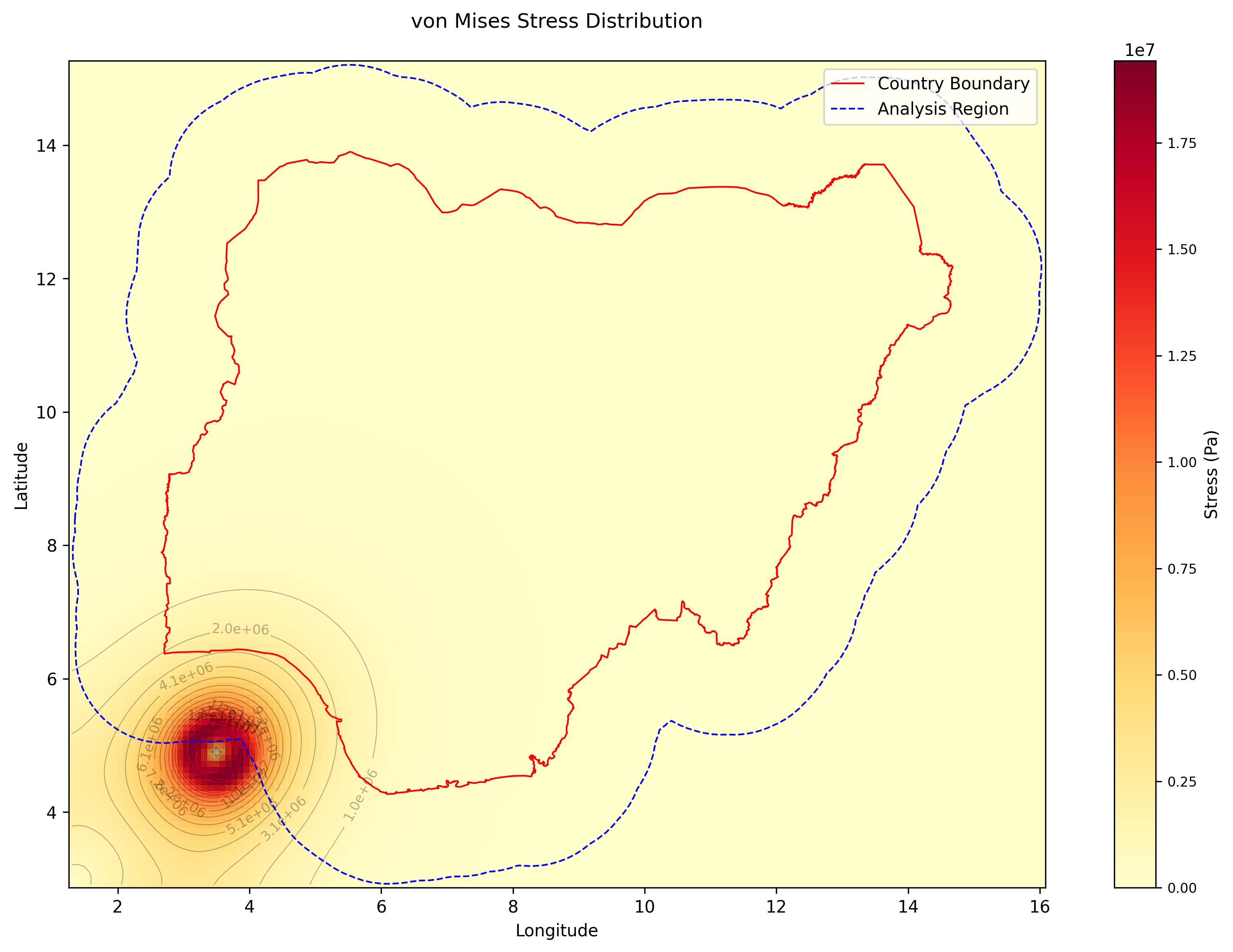}\\
(a) & (b) \\[6pt]
\includegraphics[width=0.45\textwidth]{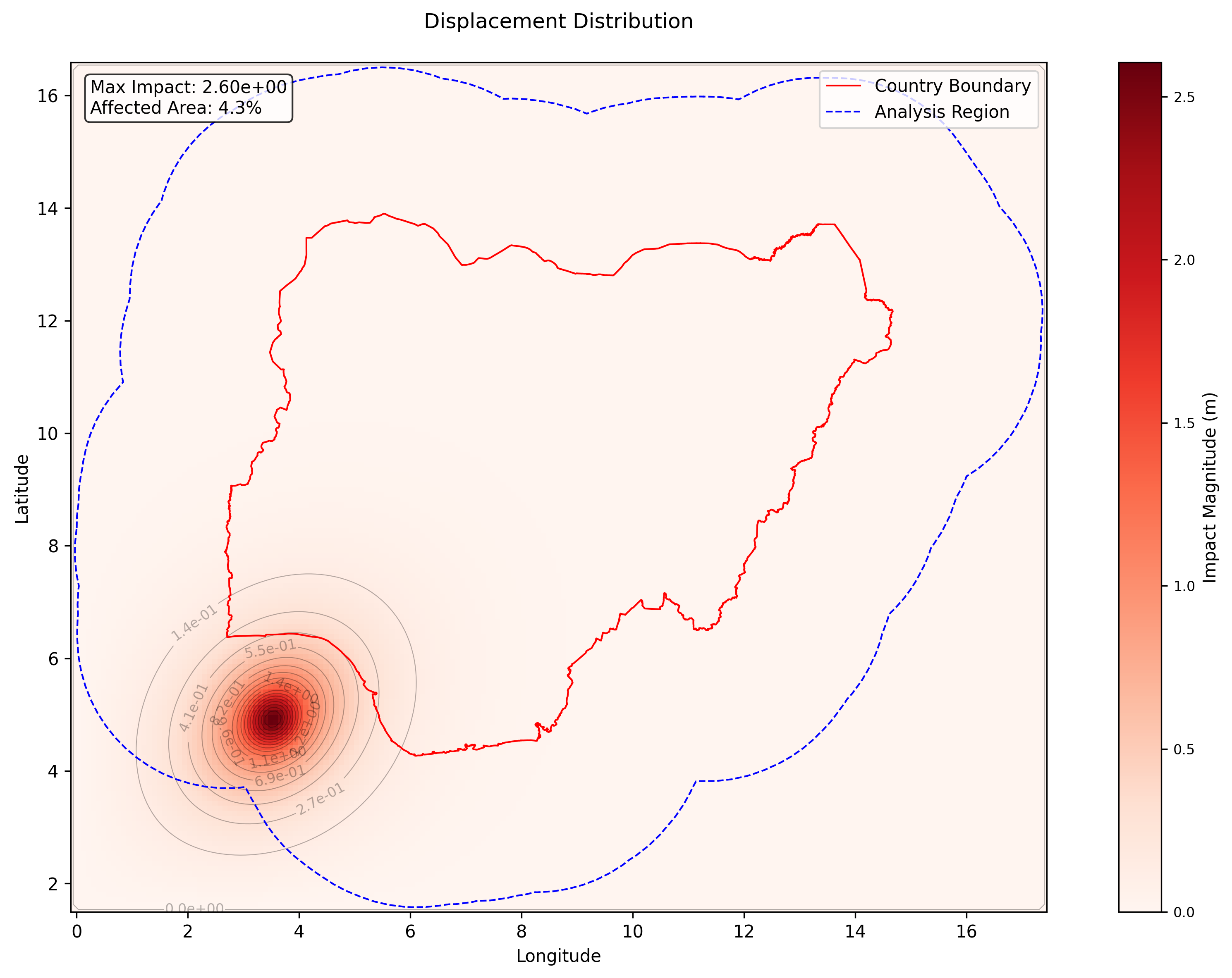}& \includegraphics[width=0.45\textwidth]{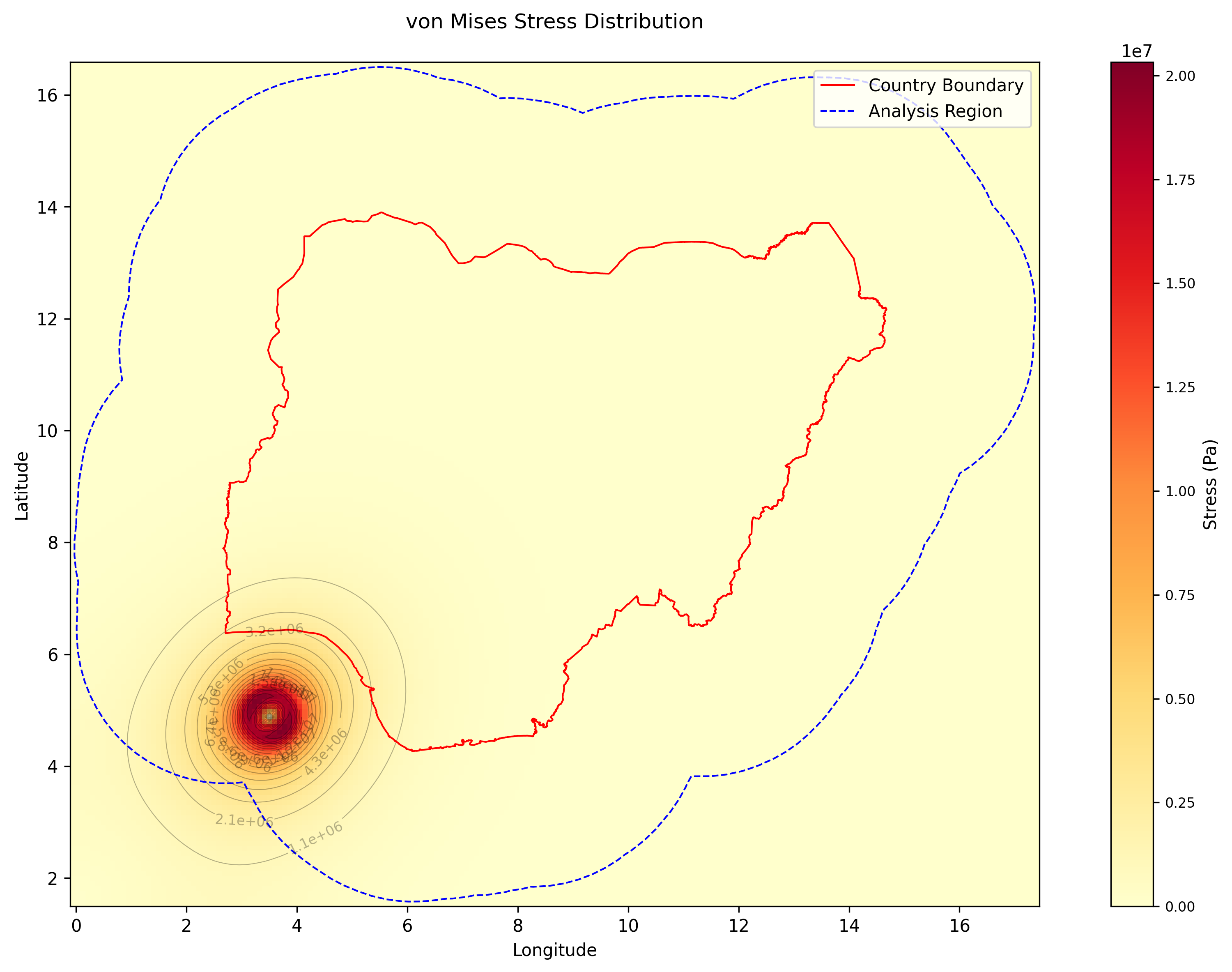}\\
(c) & (d) \\
 \includegraphics[width=0.45\textwidth]{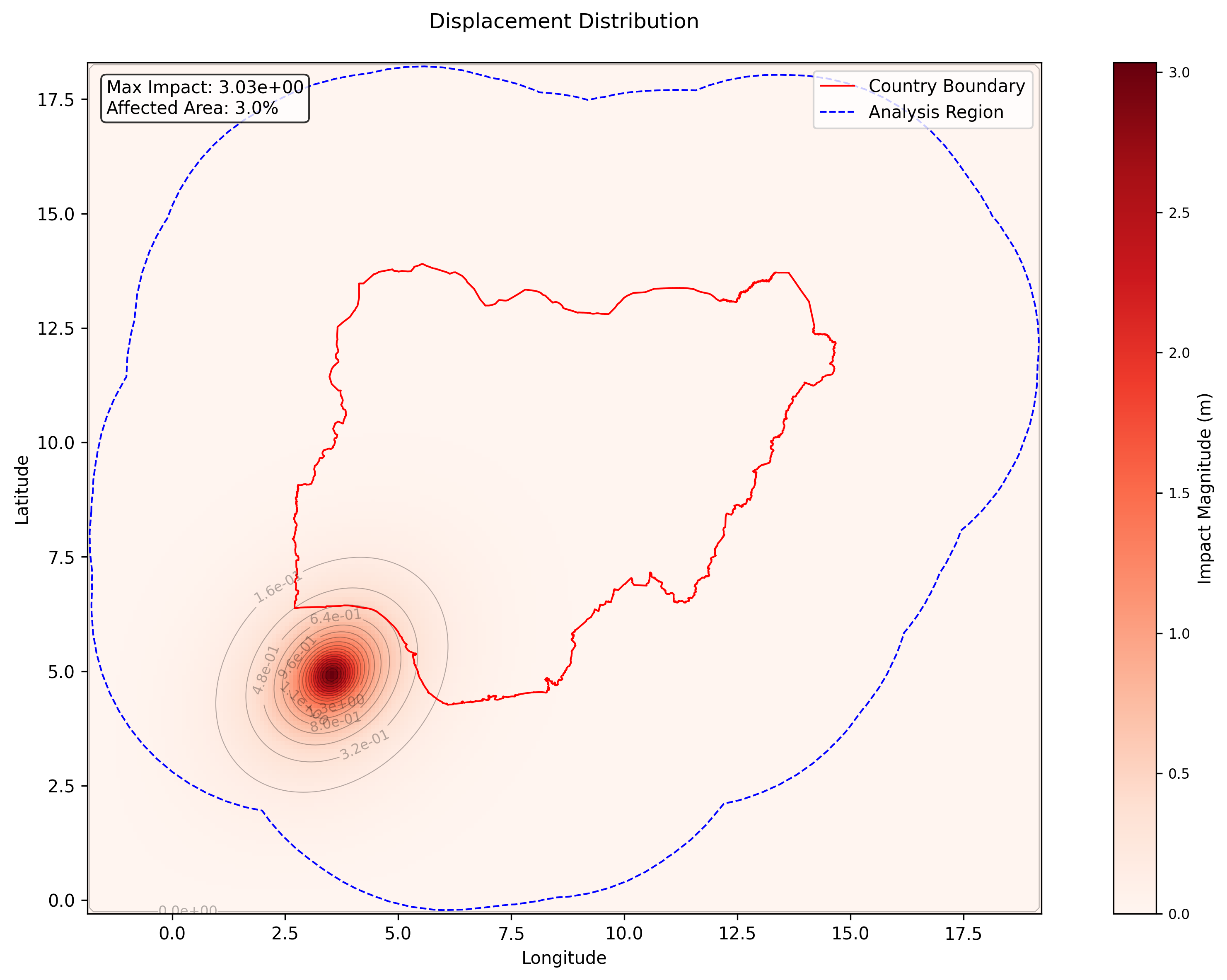}& \includegraphics[width=0.45\textwidth]{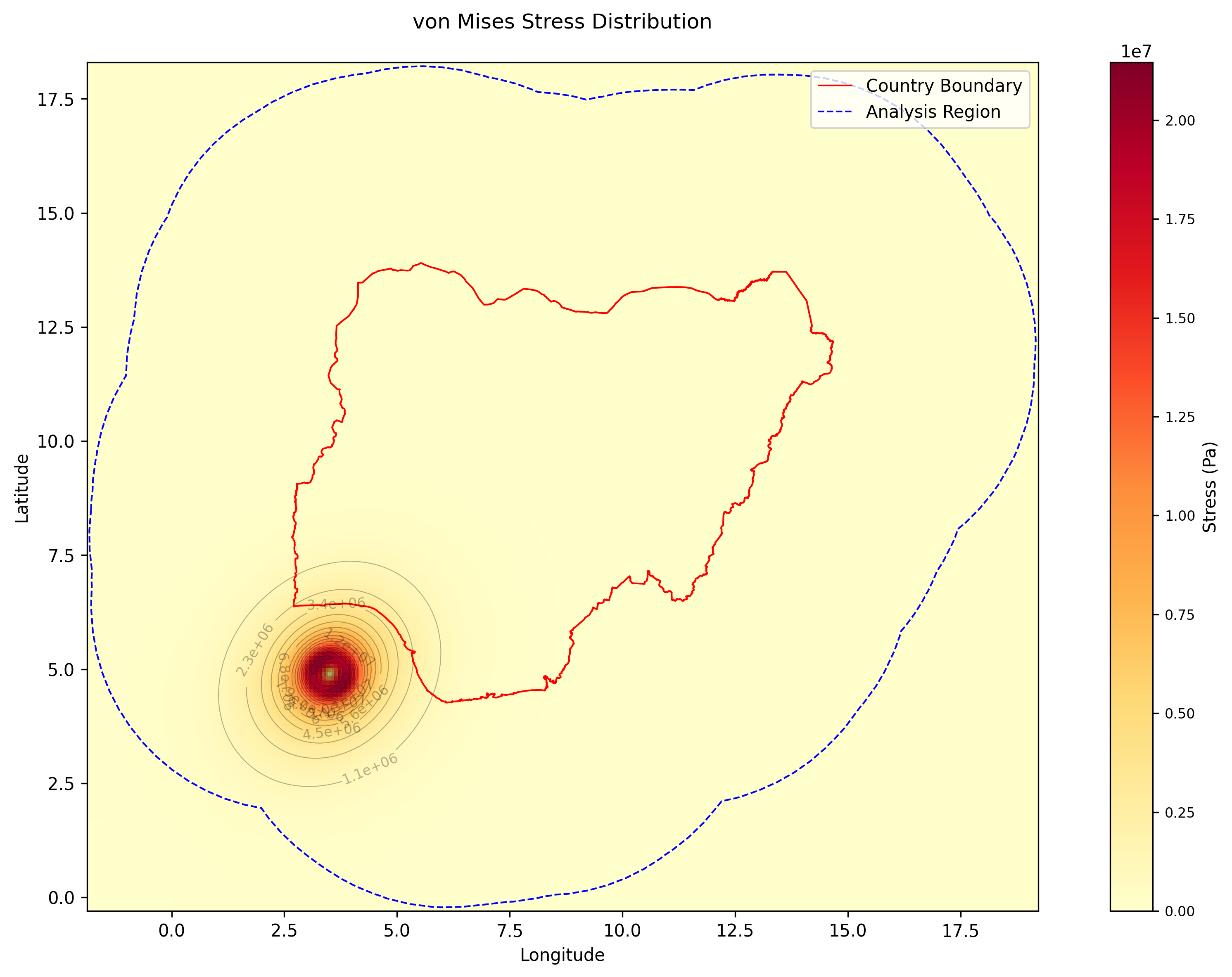}\\
 (e) & (f)\\\end{tabular}
\caption[Comparison of displacement and von Mises stress distributions for increasing buffer sizes]{Comparison of displacement (left) and von Mises stress (right) distributions for increasing buffer sizes: \textbf{(a)}, \textbf{(b)} 150 km buffer, \textbf{(c)}, \textbf{(d)} 300 km buffer, and \textbf{(e)}, \textbf{(f)} 500 km buffer. }
\label{fig:location-dependency-comparison-visualisation}
\end{figure}

The visual evidence for buffer effectiveness is particularly clear in Figure \ref{fig:location-dependency-comparison-visualisation}, which compares the displacement and von Mises stress distributions across different buffer sizes. At 150 km (Figure \ref{fig:location-dependency-comparison-visualisation} \textbf{(a)}, \textbf{(b)}), some asymmetry is visible in the response pattern near boundaries. The 300 km buffer (Figure \ref{fig:location-dependency-comparison-visualisation} \textbf{(c)}, \textbf{(d)}), shows improved symmetry, while the 500 km buffer (Figure \ref{fig:location-dependency-comparison-visualisation} \textbf{(e)}, \textbf{(f)}), produces nearly perfect radial symmetry in both displacement and stress patterns, suggesting effective isolation from boundary influence. 

The comprehensive analysis of boundary effects leads to several important conclusions for the framework's implementation. While a 150 km provides reasonable accuracy for tests applications where forces can deliberately be applied at the centre, for practical applications where conflict events may occur anywhere within the region of interest, larger buffer of 300-500 km become essential to ensure minimal artificial influence from boundary conditions. The visualisations of displacement and stress pattern provide clear evidence that a 300 km buffer substantially improves the symmetry of the distributions, while 500 km approach ideal behaviour with nearly perfect radial distributions. 

This requirement for substantial buffer zones also has important implications for computational efficiency, as larger buffers increase the total mesh size and consequently the computational resources needed. However, the improved accuracy and reliability of the results justify this computational overhead, particularly when modelling social systems where artificial boundary effects could lead to misinterpretation of conflict impact patterns. Therefore a minimum of 300 km buffer should be maintained at all times for actual analysis scenarios. Furthermore, these buffer requirements should be considered fundamental implementation parameters rather than variable choices, as they ensure the physical authenticity of the model's predictions. 

\section{Force Property Tests}

After establishing the basic behaviour of the plate (the "social fabric") through material property and boundary condition testing, the analysis proceeded with examining the force characteristics and their influences on the plate's response. Similar to the plate test, variations in the individual force parameters (magnitude $F$, distribution pattern, and radius $r$) were analysed first, followed by more complex scenarios involving multiple interacting forces.

\subsection{Single Force Tests}

Initial testing focused on understanding how individual force parameters affect the plate response. For all single-force tests, baseline plate properties were kept at $E$ = 5e9 Pa, $\nu$ = 0.3, and $h$ = 2000 m, with force application at the plate's centre (8.67°, 9.06°). 

The following static force parameters were used for the respective tests:

\begin{enumerate}
    \item Magnitude Effects: single point force without distribution
    \item Distribution Effects: force $F$ = 1e9 N, radius $r$ = 0.5°
    \item Radius Effects: force $F$ = 1e9 N, constant distribution
\end{enumerate}

\subsubsection{Magnitude Effects}

The analysis of the force magnitude variations revealed strictly linear relationships with the maximum displacement and the maximum von Mises stress. As shown in Table \ref{tab:force-properties-magnitude}, varying the force magnitude from 1.0e5 N to 1.0e11 N produced proportional changes in maximum displacement and the maximum con Mises stress, while the affected area remained constant at 2.01 \%. This linear scaling confirms operation within the linear elastic regime of the model.

The affected area remaining constant across all magnitude variations has important implications for modelling conflict impacts. It suggests that force magnitude alone cannot alter the spatial extent of the impact. This requires either a change in the force distribution pattern or the radius when only modelling the conflict event. For the social fabric modelling, this means while event intensity (represented by force magnitude) directly scales the severity of the impact, the spatial reach of that impact depends on other factors, such as the event's spatial characteristics or the community's inherent resilience/vulnerability characteristics. 

\begin{table}[H]
    \centering
    \begin{tabular}{>{\raggedright\arraybackslash}p{100pt}|>{\raggedright\arraybackslash}p{100pt}|>{\raggedright\arraybackslash}p{100pt}|>{\raggedright\arraybackslash}p{100pt}} \hline 
         \textbf{Magnitude $F$ [N]}&\textbf{Max. Displacement [m]}&\textbf{Affected Area [\%]}&\textbf{Max. von Mises Stress [Pa]}\\ \hline \hline
         1.0e5&  5.9667e-04&  2.01& 1.8059e+04\\ \hline 
         1.0e7&  5.9667e-02&  2.01& 1.8059e+06\\ \hline 
         1.0e9&  5.9667e+00&  2.01& 1.8059e+08\\ \hline 
         1.0e11&  5.9667e+02&  2.01& 1.8059e+10\\\hline
    \end{tabular}
    \caption[Force magnitude scaling effects]{Force magnitude scaling effects show linear relationships with displacement and stress while maintaining a constant affected area. The results demonstrate perfect proportionality across six orders of magnitude, confirming the linear-elastic behaviour of the model.}
    \label{tab:force-properties-magnitude}
\end{table}

\subsubsection{Distribution Pattern Effects}

\begin{figure}[H]
    \centering
    \includegraphics[width=1.0\linewidth]{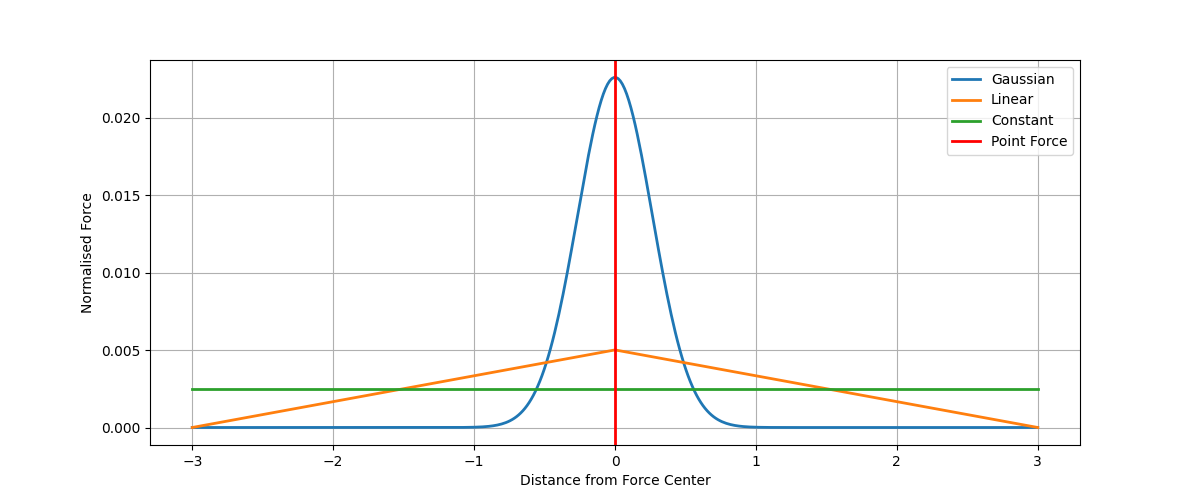}
    \caption[Comparison of possible force distribution patterns]{Comparison of possible force distribution patterns showing normalised force intensity versus distance from the application centre. The point force (red) represents localised loading, while Gaussian (blue), linear (orange), and constant (green) distributions spread the same total force over the specified radius with different intensity profiles.}
    \label{fig:force-distribution-patterns}
\end{figure}

While adjusting the magnitude of the force alone won't change the affected area, it is also possible to not only apply the force to a single point but to an area with a specified distribution. The developed \ac{FEA} approach allows the selection of four different distribution patterns: constant, linear, Gaussian and a point force. As illustrated in Figure \ref{fig:force-distribution-patterns}, these patterns represent distinctly different ways of applying the same total force over a specified area.

A point force represents the most concentrated application, with the entire force magnitude acting on a single node. The Gaussian distribution creates a bell-curved pattern with high central concentration that rapidly diminishes with distance. The linear distribution provides a gradual decline from the centre to the edge, while the constant distribution spreads the force uniformly across the specified radius. 

The testing results, shown in Table \ref{tab:single-forces-distribution}, demonstrate how these distribution patterns produce significantly different response characteristics despite maintaining the same total force magnitude ($F$ = 1e9 N) and application radius ($r$ = 0.5°). The constant distribution, which spreads the force most evenly, produces the lowest maximum displacement (1.8892 m)  and maximum von Mises stress (1.6354e+07 Pa) but affects the largest area (7.75 \%). Conversely, the Gaussian distribution generates the highest maximum displacement (4.7425 m) and maximum von Mises stress (1.2364e+08 Pa) while affecting the smallest area (2.77 \%), reflecting its concentrated loading. The linear distribution shows intermediate behaviour, with a maximum displacement of 2.2990 m and an affected area of 6.42 \%, providing a middle ground between localisation and spread. 

\begin{figure}[H]
\centering
\begin{tabular}{cc}
\includegraphics[width=0.45\textwidth]{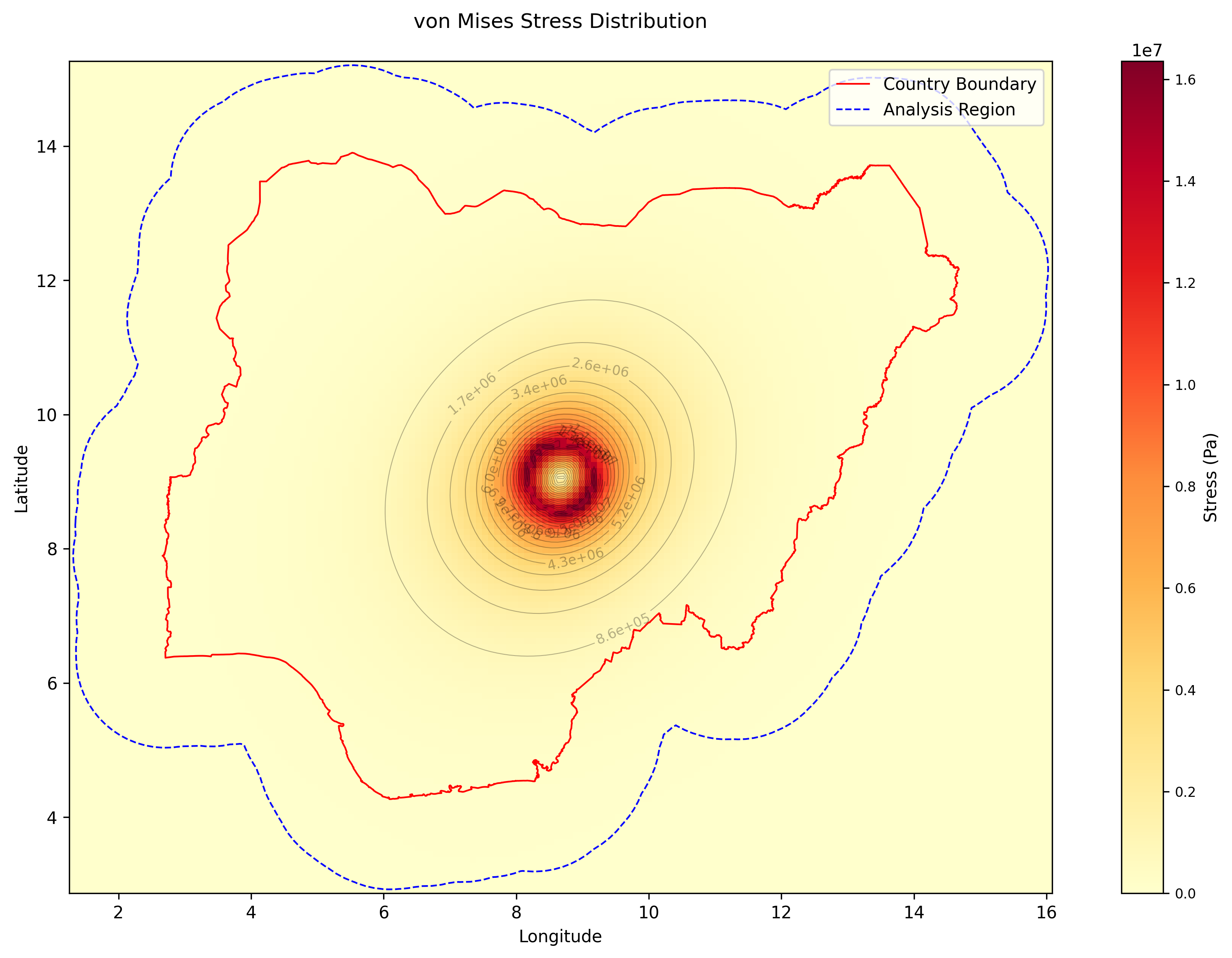}& \includegraphics[width=0.45\textwidth]{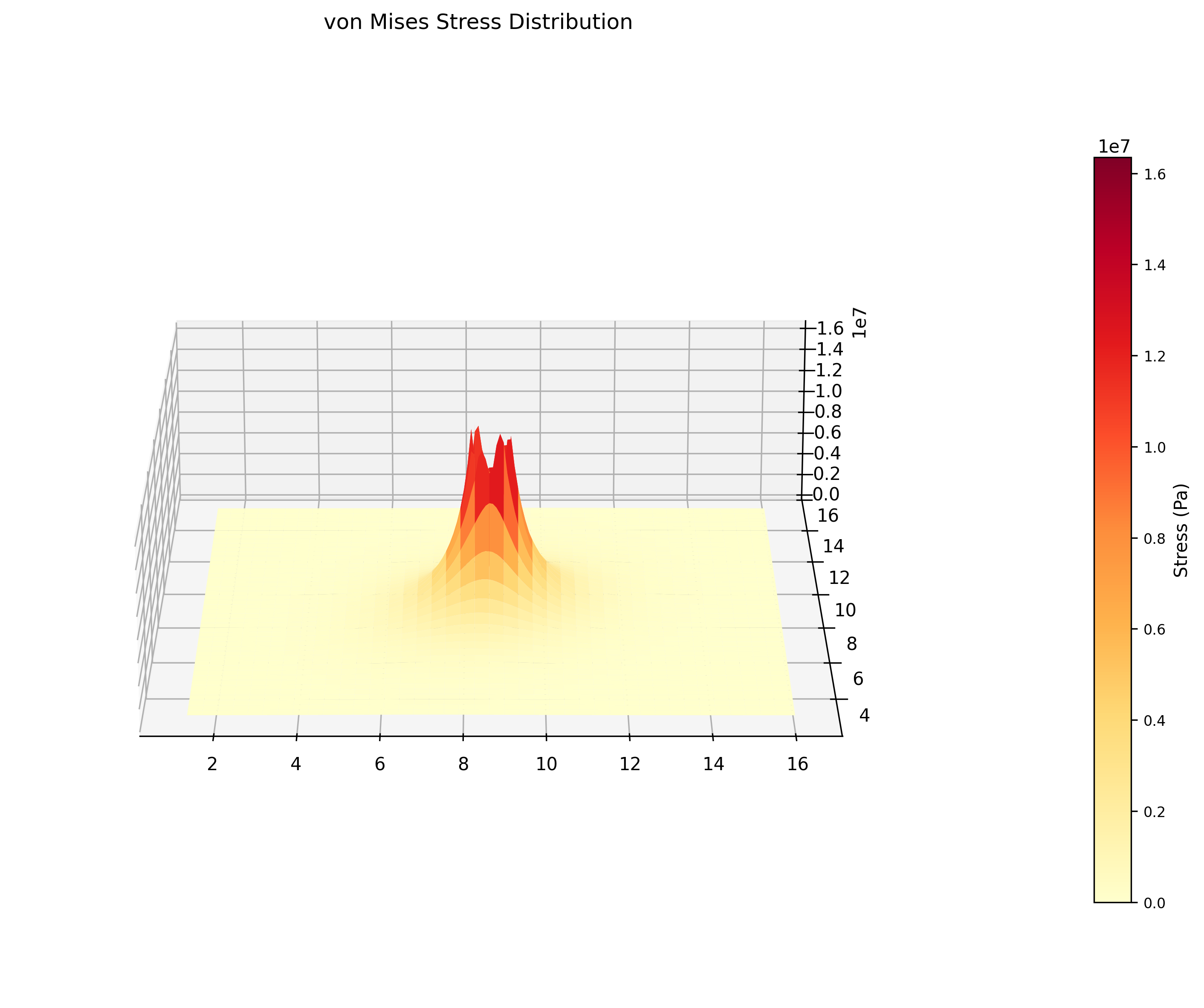} \\
(a) & (b) \\[6pt]
\includegraphics[width=0.45\textwidth]{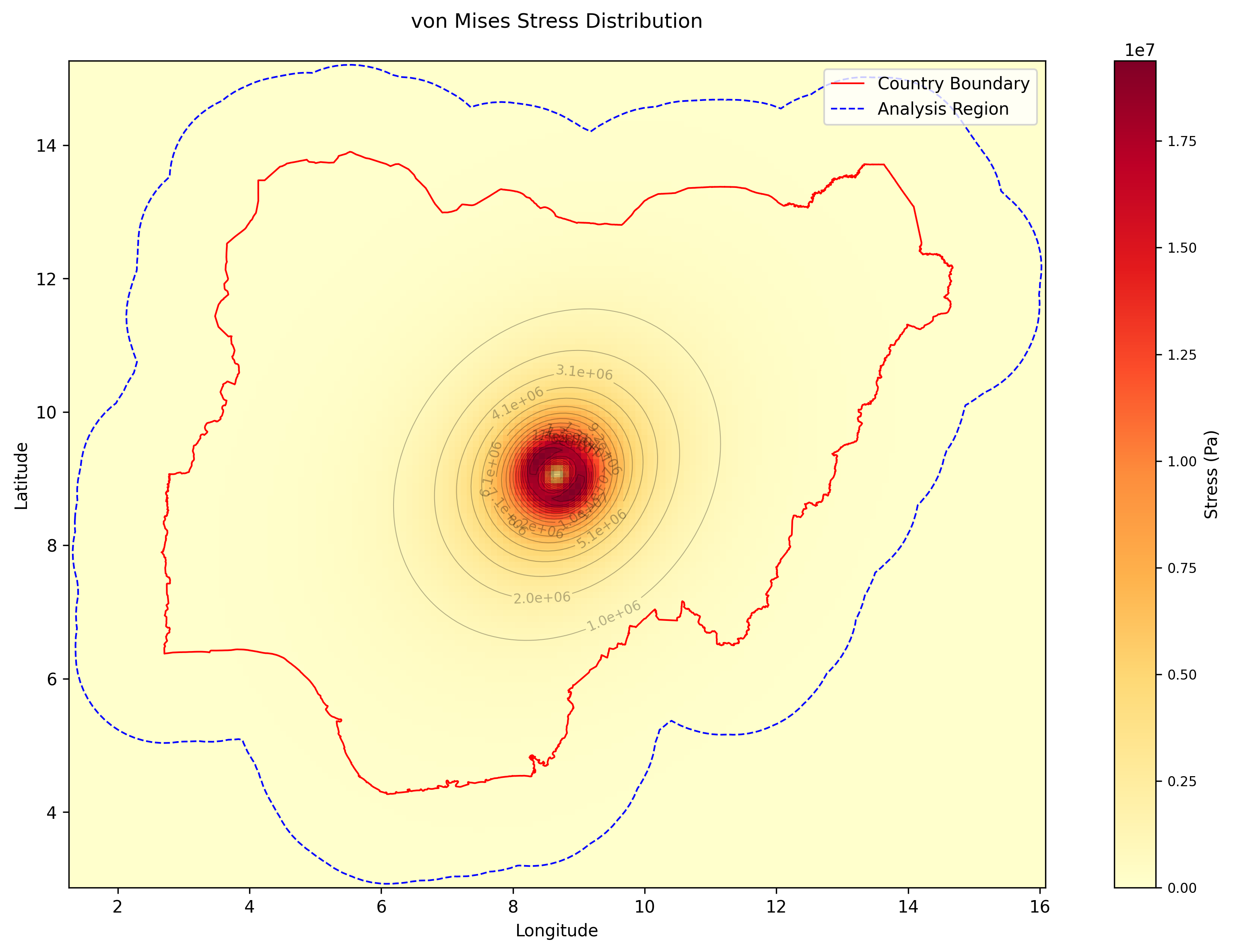} & \includegraphics[width=0.45\textwidth]{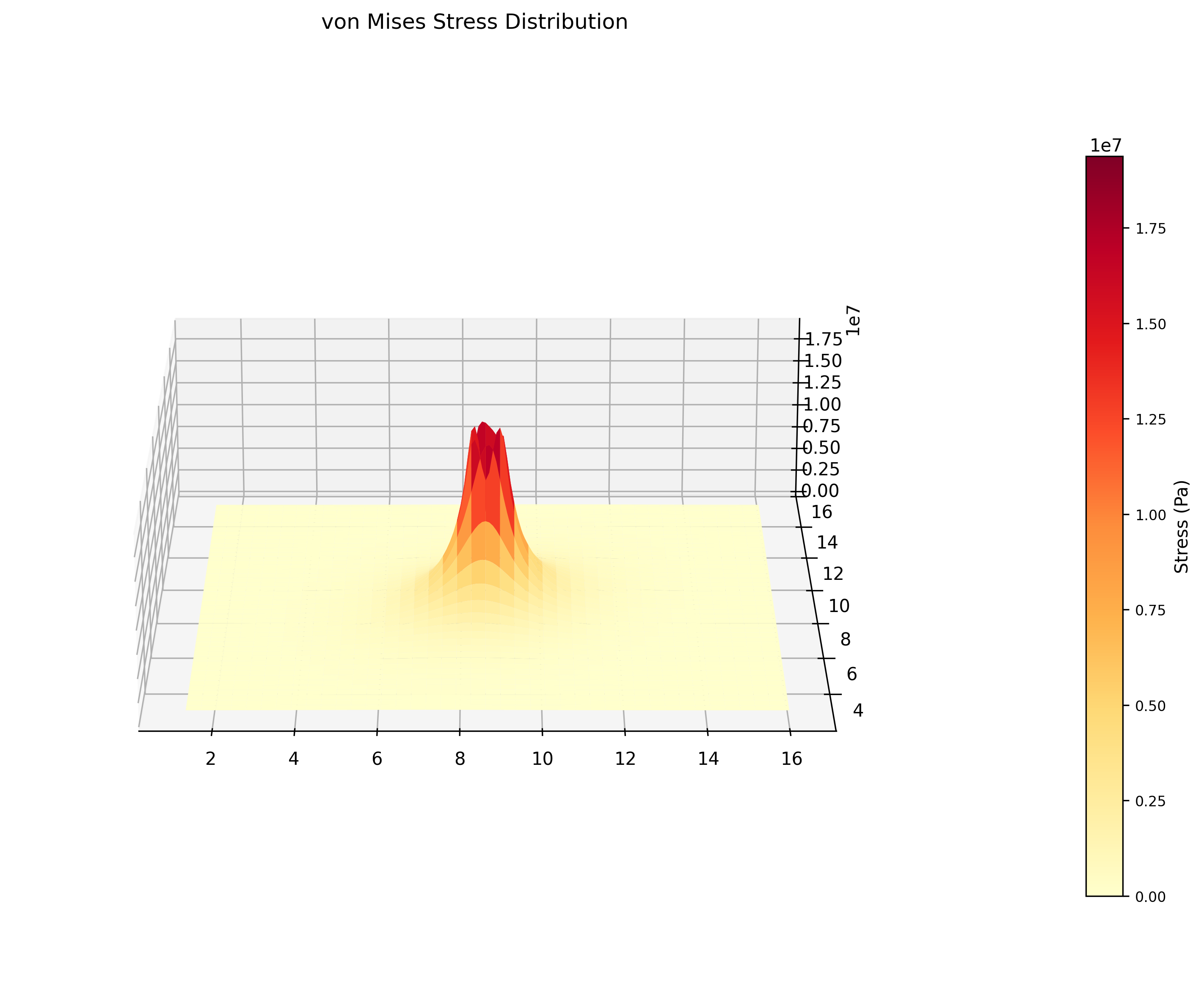} \\
(c) & (d) \\
 \includegraphics[width=0.45\textwidth]{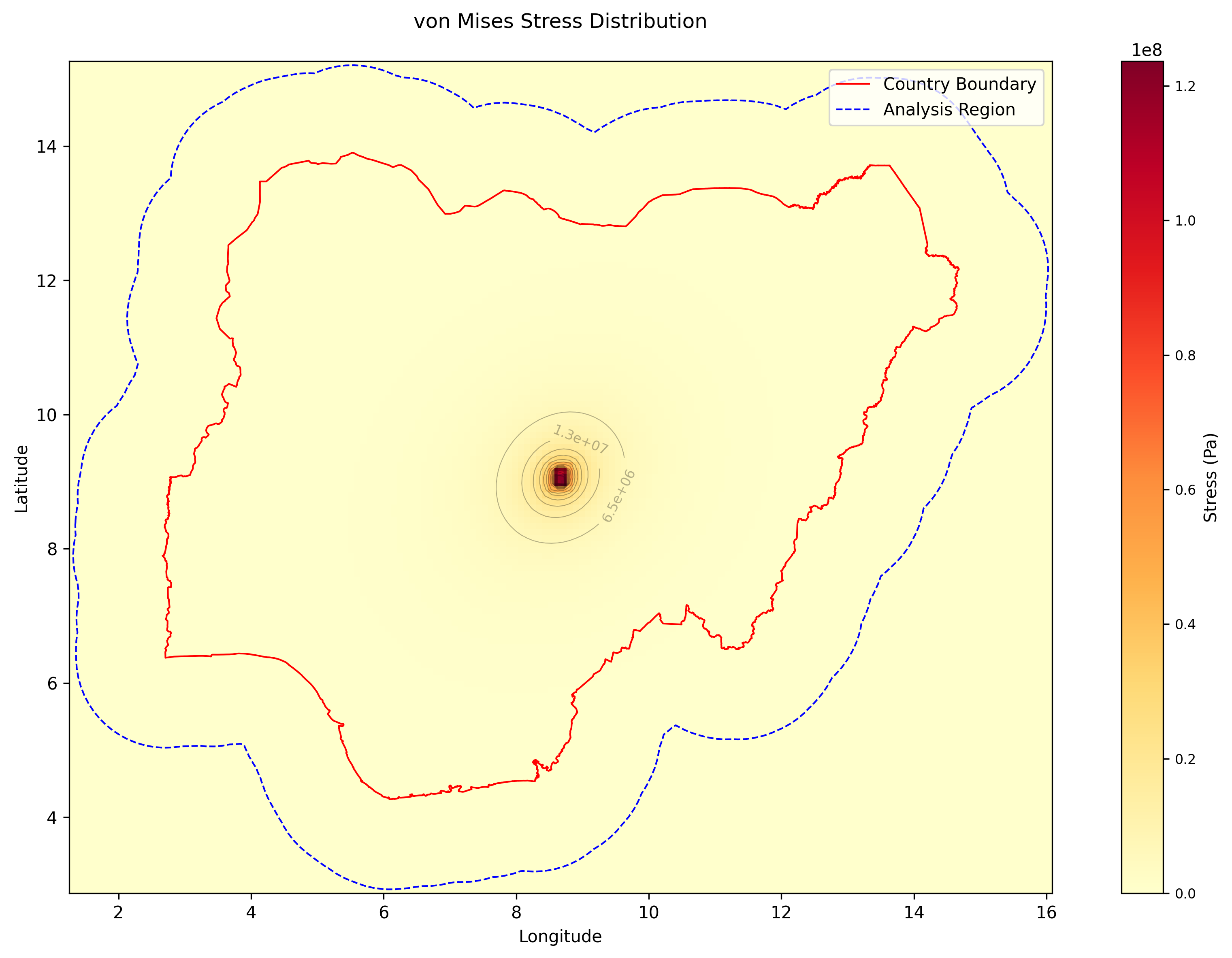} & \includegraphics[width=0.45\textwidth]{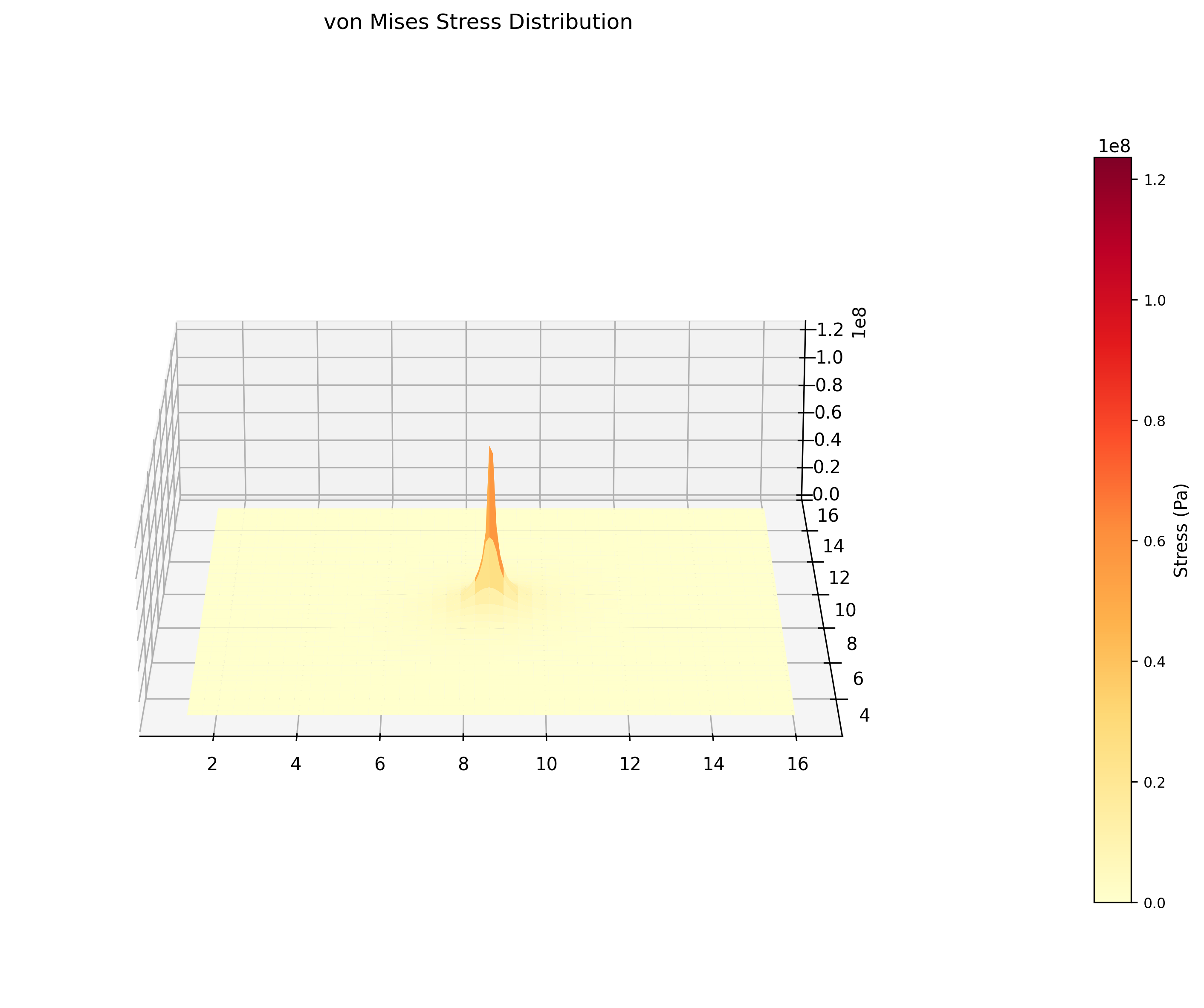} \\
 (e) & (f)\\\end{tabular}
\caption[Comparison of von Mises stress distributions for different force application patterns]{Comparison of von Mises stress distributions for different force application patterns using both 2D contour plots (left column) and 3D surface visualisations (right column): \textbf{(a)}, \textbf{(b)} constant distribution, \textbf{(c)}, \textbf{(d)} linear distribution, and \textbf{(e)}, \textbf{(f)} Gaussian distribution.}
\label{fig:force-distribution__von-Mises-stress-visualisation}
\end{figure}

\begin{table}[H]
    \centering
    \begin{tabular}{>{\raggedright\arraybackslash}p{100pt}|>{\raggedright\arraybackslash}p{100pt}|>{\raggedright\arraybackslash}p{100pt}|>{\raggedright\arraybackslash}p{100pt}} \hline
         \textbf{Distribution}&\textbf{Max. Displacement [m]}&\textbf{Affected Area [\%]}&\textbf{Max. von Mises Stress [Pa]}\\ \hline \hline
 constant& 1.8892e+00& 7.75&1.6354e+07\\\hline
         linear&  2.2990e+00&  6.42& 1.9376e+07\\ \hline 
         gaussian&  4.7425e+00&  2.77& 1.2364e+08\\\hline
    \end{tabular}
    \caption[Effect of force distribution patterns on plate response characteristics]{Effect of force distribution patterns on plate response characteristics. The results show distinct relationships between distribution type and response measures.}
    \label{tab:single-forces-distribution}
\end{table}

The varying impact patterns are visualised in Figure \ref{fig:force-distribution__von-Mises-stress-visualisation}, which shows both 2D and 3D representations of the von Mises stress distributions. The constant distribution (Figure \ref{fig:force-distribution__von-Mises-stress-visualisation} \textbf{(a)}, \textbf{(b)}) shows a broad stress pattern with notably low stress at the centre of the application area. The linear distribution  (Figure \ref{fig:force-distribution__von-Mises-stress-visualisation} \textbf{(c)}, \textbf{(d)}) produces a more graduated stress distribution with moderate central concentration, while the Gaussian distribution  (Figure \ref{fig:force-distribution__von-Mises-stress-visualisation} \textbf{(e}, \textbf{(f)}) creates highly localised stress peak with sharp gradients. 

The visualisation also highlights an important limitation of using von Mises stress as an impact measure in the social fabric framework. While mathematically correct, the stress patterns—especially the low central stress in constant distributions—do not align intuitively with how social impacts might manifest. For instance, the centre of a conflict zone would likely experience significant impacts, contrary to what the stress distribution suggests. This observation suggests that displacement might serve as a more appropriate metric for quantifying social impact. However, this will be further explored in the following sections.

\subsubsection{Radius Effects}

In addition to a specific distribution, the radius of the area to which the force should be applied must be provided as well. The testing examined radius variations from 0.05° to 2.0° while maintaining a constant force magnitude ($F$ = 1e9 N) and using the constant distribution pattern. The results shown in Table \ref{tab:force-properties-radius} demonstrate a clear transition from localised to distributed deformation patterns as the radius increases. 

\begin{table}[H]
    \centering
    \begin{tabular}{>{\raggedright\arraybackslash}p{100pt}|>{\raggedright\arraybackslash}p{100pt}|>{\raggedright\arraybackslash}p{100pt}|>{\raggedright\arraybackslash}p{100pt}} \hline
         \textbf{Radius $r$ [°]}&\textbf{Max. Displacement [m]}&\textbf{Affected Area [\%]}&\textbf{Max. von Mises Stress [Pa]}\\ \hline \hline 
 0.05& 4.5909e+00& 2.92&1.1923e+08\\\hline 
         0.1&  4.1327e+00&  3.32& 9.9294e+07\\ \hline 
         0.5&  1.8892e+00&  7.75& 1.6354e+07\\ \hline 
         1&  1.1549e+00&  12.18& 6.8586e+06\\ \hline 
         2&  5.6180e-01&  22.80& 2.3738e+06\\\hline
    \end{tabular}
    \caption[Effect of the force radius on displacement characteristics and stress distribution]{Effect of the force radius on displacement characteristics and stress distribution.}
    \label{tab:force-properties-radius}
\end{table}

\begin{figure}[H]
    \centering
    \includegraphics[width=0.6\linewidth]{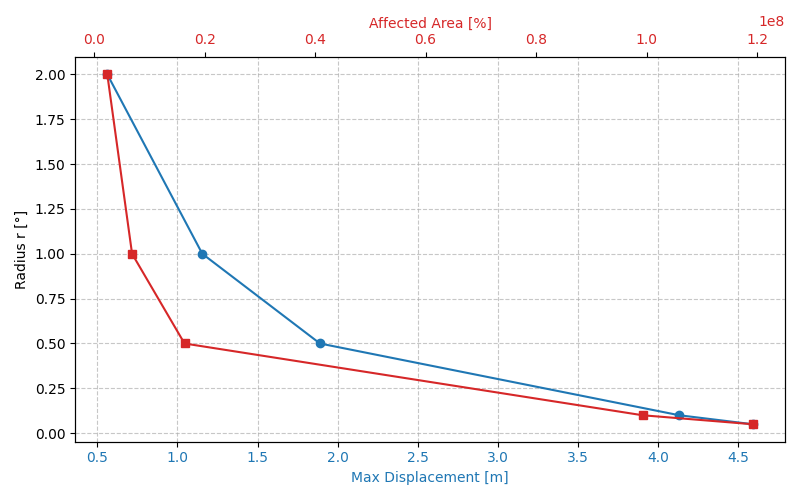}
    \caption[Relationship between force radius, the maximum displacement and the affected area]{Relationship between force radius and plate response measures: maximum displacement (blue curve, bottom axis) and affected area percentage (red curve, top axis).}
    \label{fig:force_properties_radius__displacement-affected-area}
\end{figure}

As illustrated in Figure \ref{fig:force_properties_radius__displacement-affected-area}, the maximum displacement and affected areas show inverse relationships with increasing radius. The maximum displacement exhibits approximately exponential decay, decreasing from 4.5909 m at $r$ = 0.05° to 0.5618 m at $r$ = 2.0°. Simultaneously, the affected area expands significantly from 2.92 \% to 22.80 \%, indicating a broader but less intense impact distribution. 

\begin{figure}[H]
    \centering
    \includegraphics[width=0.6\linewidth]{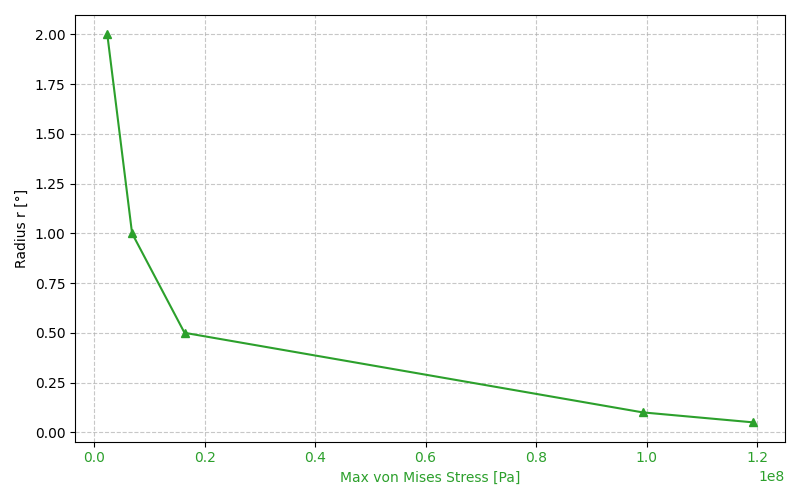}
    \caption[Effect of the force radius on the maximum von Mises stress]{Effect of the force radius on the maximum von Mises stress.}
    \label{fig:force_properties_radius__von-Mises-stress}
\end{figure}

The maximum von Mises stress, shown in Figure \ref{fig:force_properties_radius__von-Mises-stress}, follows an approximately logarithmic reduction, decreasing from 1.1923e+08 Pa to 2.3738e+06 Pa as the radius increases. This behaviour aligns with the expected stress mechanics, where distributing a constant force over larger areas necessarily reduces local stress intensity. 

For the social fabric framework, these results indicate that the radius parameter allows for control over the balance between impact intensity and spatial reach, possibly representing different types of violence ranging from localised incidents to more extensive regional conflicts.

\subsection{Force Interaction Tests}

After analysing the characteristics of individual forces, multiple forces applied simultaneously to the same plate were tested to investigate potential interactions. All tests maintained consistent plate parameters: Young's modulus $E$ = 5e9, Poisson's ratio $\nu$ = 0.3, and thickness $h$ = 2000 m. Two distinct test configurations were established: 

\begin{enumerate}
    \item Two-Force Magnitude Effects: Forces applied with constant distribution, radius $r$ = 0.5°
    \item Two-Force Distance Effects: Forces applied with constant distribution, radius $r$ = 0.5°, magnitude $F$ = 5e8 N
\end{enumerate}

\subsubsection{Two-Force Magnitude Effects} 

The initial interaction tests examined the interaction between two forces with various magnitudes. Understanding these interactions is important for recognising potential superposition effects. Previously, a single force of $F$ = 1e9 N was applied to the plate. These tests involve applying twice the initial force, with two forces of $F$ = 1e9 N, half the initial force, with two forces of $F$ = 1e8, and a quarter of the initial force, with two forces of $F$ = 2.5e8 N. The distance remained 1° for all test scenarios. Different force ratios, 10:1, 2:1, and 1:2, were also tested. 

As documented in Table \ref{tab:single-forces-distribution}, a single force with a constant distribution of 1e9 N creates a maximum displacement of 1.8892 m, a maximum von Mises stress of 1.6354e7 Pa and affects 7.75 \% of the total area. As documented in Table \ref{tab:two-force-magnitude-effects}, introducing an additional equal force raises the maximum displacement to 2.7044 m (approximately a 1.43 times increase), the maximum von Mises stress to 1.9501e+07 Pa (around a 1.19 times increase), and the affected area to 10.63 \% (about a 1.37 times increase). This indicates that the forces do not simply combine, which would have doubled these measures, but rather interact and partially reinforce one another. 

\begin{table}[H]
    \centering
   \begin{tabular}{>{\raggedright\arraybackslash}p{80pt}|>{\raggedright\arraybackslash}p{80pt}|>{\raggedright\arraybackslash}p{80pt}|>{\raggedright\arraybackslash}p{80pt}|>{\raggedright\arraybackslash}p{80pt}} \hline
         \textbf{Force 1 [N], (8.67°, 9.06°)}&\textbf{Force 2 [N], (9.67°, 9.06°)}&\textbf{Max. Displacement [m]}&\textbf{Affected Area [\%]}&\textbf{Max. von Mises Stress [Pa]}\\ \hline  \hline 
 1e9& 1e9& 2.7044e+00& 10.63&1.9501e+07\\ \hline 
 5e8& 5e8& 1.3522e+00& 10.63&9.7504e+06\\\hline 
         2.5e8&  2.5e8&  6.7610e-01&  10.63&4.8752e+06\\\hline \hline 
 1e9& 1e8& 1.9563e+00& 8.27&1.6566e+07\\\hline
 1e9& 5e8& 2.2479e+00& 9.75&1.7811e+07\\\hline
 1e9& 2e9& 4.4876e+00&  9.79&3.5851e+07\\\hline
    \end{tabular}
    \caption[Two-force interaction effects showing response characteristics for various magnitude combinations]{Two-force interaction effects showing response characteristics for various magnitude combinations at 1° distance. Values demonstrate partial reinforcement between forces rather than simple addition of individual effects.}
    \label{tab:two-force-magnitude-effects}
\end{table}

Comparing the equal force pairs in Table \ref{tab:two-force-magnitude-effects} shows that halving both forces results in a perfect linear scaling, which likewise halves the maximum displacement and maximum von Mises stress. This behaviour mirrors the magnitude effect observed with a single force. Similarly, the affected area remains unchanged at 10.63 \%. Examination of unequal force combinations reveals that the stronger force usually dictates the response pattern, whereas the weaker force continues to influence overall behaviours. This relationship appears to intensify as the ratio between the forces increases. 

\begin{figure}[H]
\centering
\begin{tabular}{cc}
\includegraphics[width=0.45\textwidth]{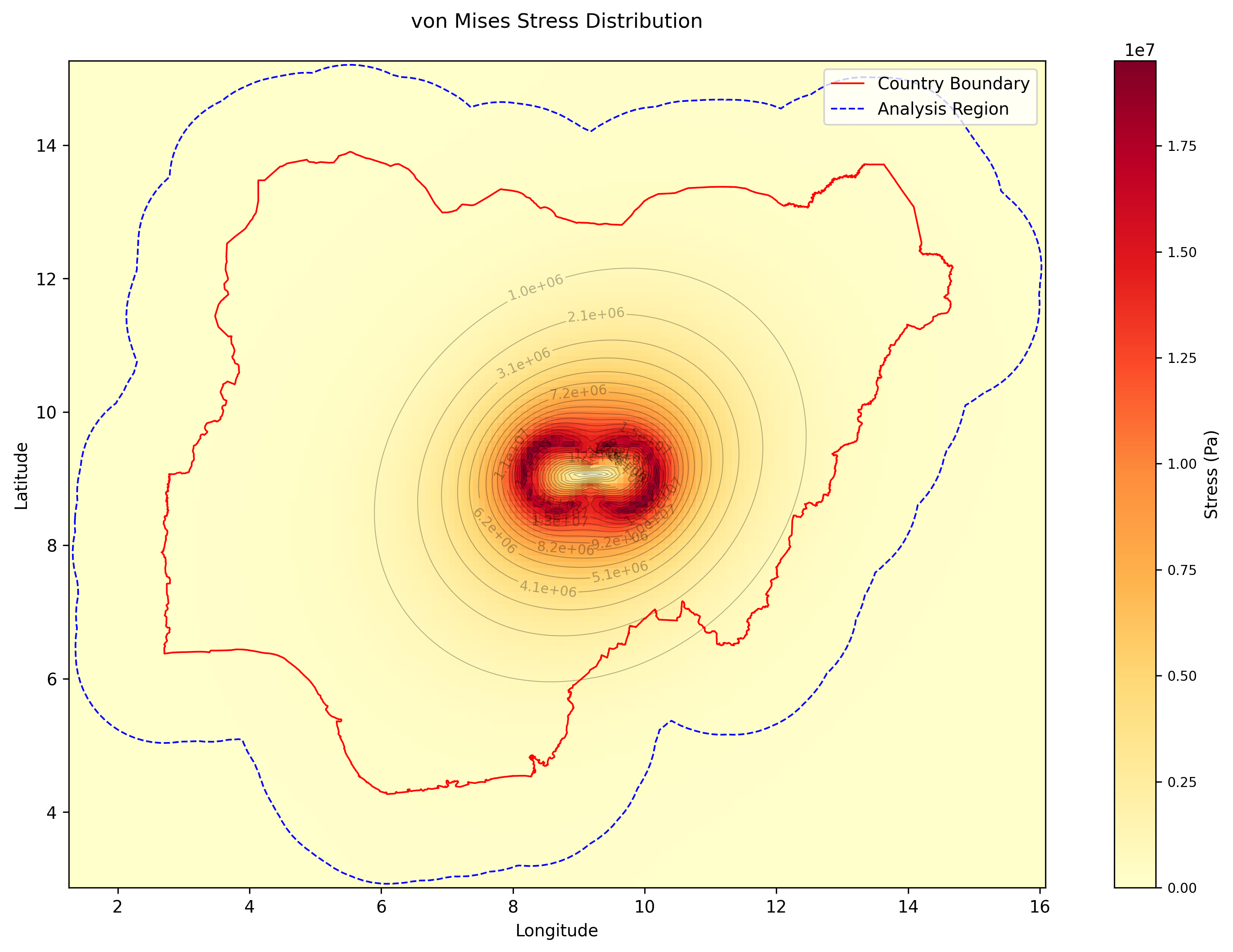}& \includegraphics[width=0.45\textwidth]{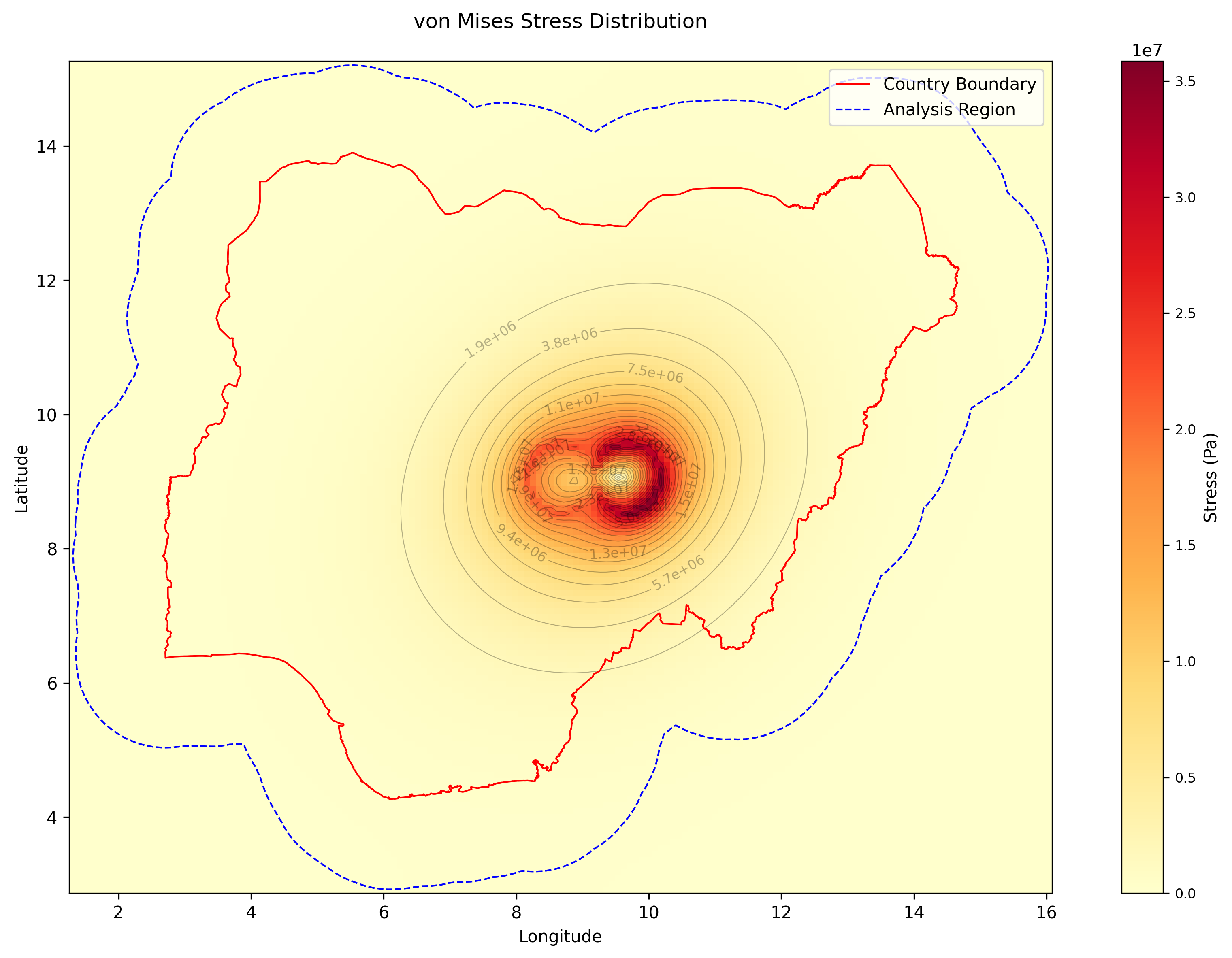} \\  
 (a)&(b)\\\end{tabular}
\caption[Von Mises stress distribution for two-force interactions at 1° distance]{Von Mises stress distribution for two-force interactions at 1° distance: \textbf{(a)} equal forces of 1e9 N at (8.67°, 9.06°) and 1e9 N at (9.67°, 9.06°), showing a symmetric stress pattern with reduced stress in the intermediate region, \textbf{(b)} unequal forces of 1e9 N at (8.67°, 9.06°) and 2e9 N at (9.67°, 9.06°), that show a stress field by the stronger force.}
\label{fig:two-force-magnitude-effects-visualisations}
\end{figure}

However, the von Mises stress distributions for different force combinations, shown in Figure \ref{fig:two-force-magnitude-effects-visualisations}, highlight significant limitations in using physical stress as a measure of social impact again. As observed in the earlier distribution tests, the stress levels are notably low in regions where social stress should be significantly high, particularly at the centres of the force application. Furthermore, beyond the limitations of using von Mises stress to measure social stress at the event's centre, the interaction visualisations illustrate a translation issue for the areas between concurrent events. 

When examining equal forces (Figure \ref{fig:two-force-magnitude-effects-visualisations} \textbf{(a)}), the intermediary area appears "suspended" between the events, therefore showing reduced stress levels. This pattern becomes even more pronounced with unequal forces (Figure Figure \ref{fig:two-force-magnitude-effects-visualisations} \textbf{(b)}), where the stronger force ($F$ = 2e9 N) dominates the stress field. In contrast to this physically correct behaviour, social systems would likely experience intensified stress in areas affected by multiple overlapping conflict events. 

These findings suggest that while the physical model captures important aspects of force interaction, care must be taken in interpreting von Mises stress patterns, particularly in cases of multiple overlapping events. This confirms the earlier assumption, suggesting that the displacement field could provide a more reliable metric for analysing compound effects, as well as for more widespread impacts within the social fabric model. 

\subsubsection{Two-Force Distance Effects}

The second set of interaction tests examined how separation distance influences the combined effects of two forces ($F$ = 5e8 N each). When two forces are perfectly aligned (distance = 0°), their combined effect equals that of one force of the same combined magnitude ($F$ = 1e9 N). As documented in Table \ref{tab:two-force-distance-effects}, increasing separation between forces produces systematic change in the response characteristics. The maximum displacement decreases from 1.8892 m at 0° separation to 0.9820 m at 3° separation, while the affected area expands from 7.75 \% to 15.69 \%.

\begin{table}[H]
    \centering
   \begin{tabular}{>{\raggedright\arraybackslash}p{65pt}|>{\raggedright\arraybackslash}p{65pt}|>{\raggedright\arraybackslash}p{65pt}|>{\raggedright\arraybackslash}p{65pt}|>{\raggedright\arraybackslash}p{65pt}|>{\raggedright\arraybackslash}p{65pt}} \hline
         \textbf{Force 1 (long, lat) [°]}&\textbf{Force 2 (long, lat) [°]}&\textbf{Distance [°]}&\textbf{Max. Displacement [m]}&\textbf{Affected Area [\%]}&\textbf{Max. von Mises Stress [Pa]}\\ \hline \hline 
 (8.67, 9.06)& (8.67, 9.06)& 0& 1.8892e+00& 7.75&1.6354e+07\\ \hline  \hline 
  (8.67, 9.06)& (8.92, 9.06)&0.25& 1.8547e+00& 7.89&1.5200e+07\\ \hline 
  (8.67, 9.06)& (9.17, 9.06)&0.5& 1.7742e+00& 8.24&1.3852e+07\\\hline 
         (8.67, 9.06)&  (9.67, 9.06)&1.0&  1.3522e+00&  10.63&9.7504e+06\\\hline
 (8.67, 9.06)& (10.67, 9.06)&2.0& 1.0518e+00& 13.80&8.6514e+06\\\hline
 (8.67, 9.06)& (11.67, 9.06)&3.0& 9.8200e-01& 15.69&8.3420e+06\\\hline
    \end{tabular}
    \caption[Two-force distance effects showing systematic changes in the response measures]{Two-force distance effects showing systematic changes in the response measures with increasing force separation.}
    \label{tab:two-force-distance-effects}
\end{table}

\begin{figure}[H]
\centering
\begin{tabular}{cc}
\includegraphics[width=0.45\textwidth]{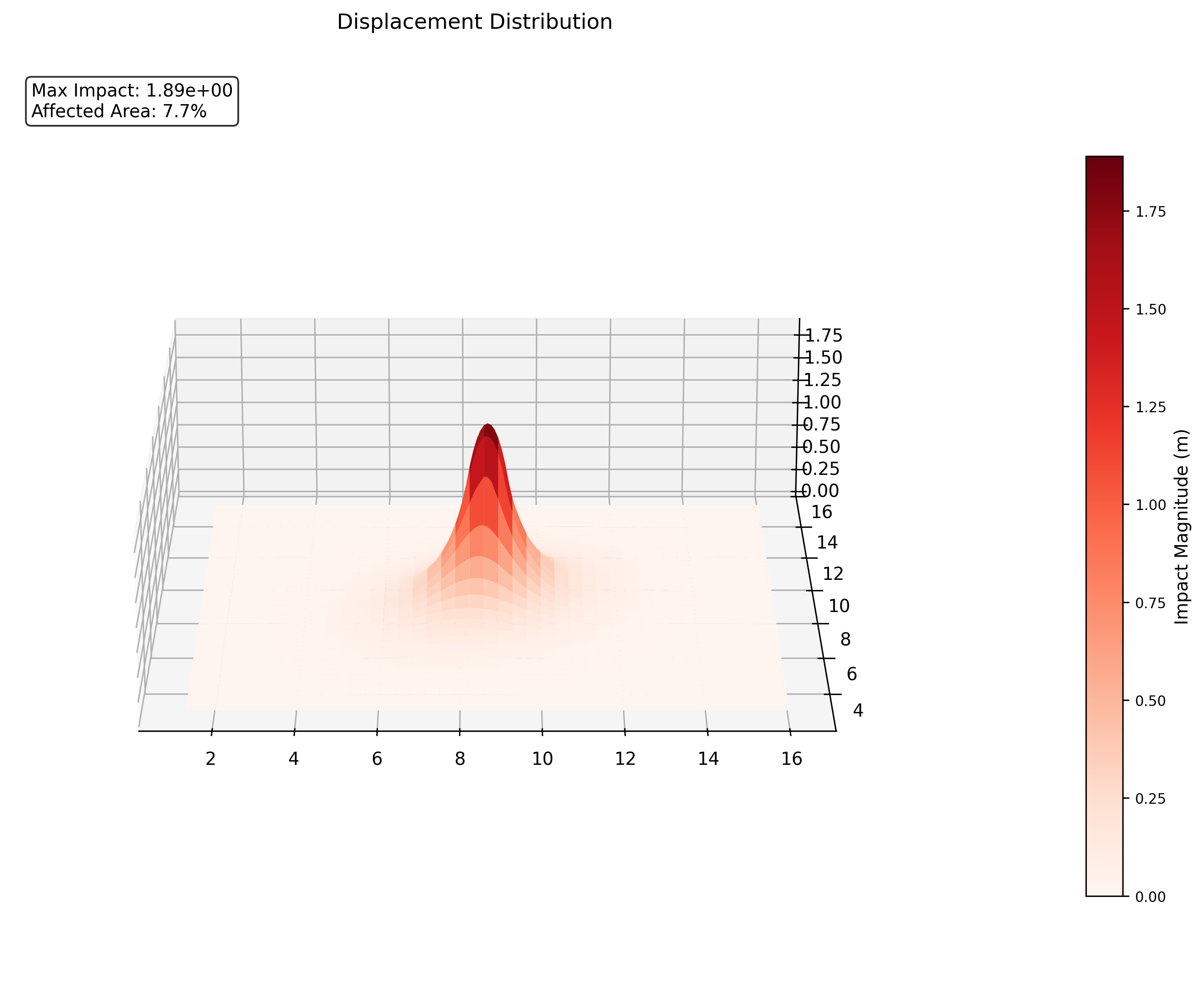}& \includegraphics[width=0.45\textwidth]{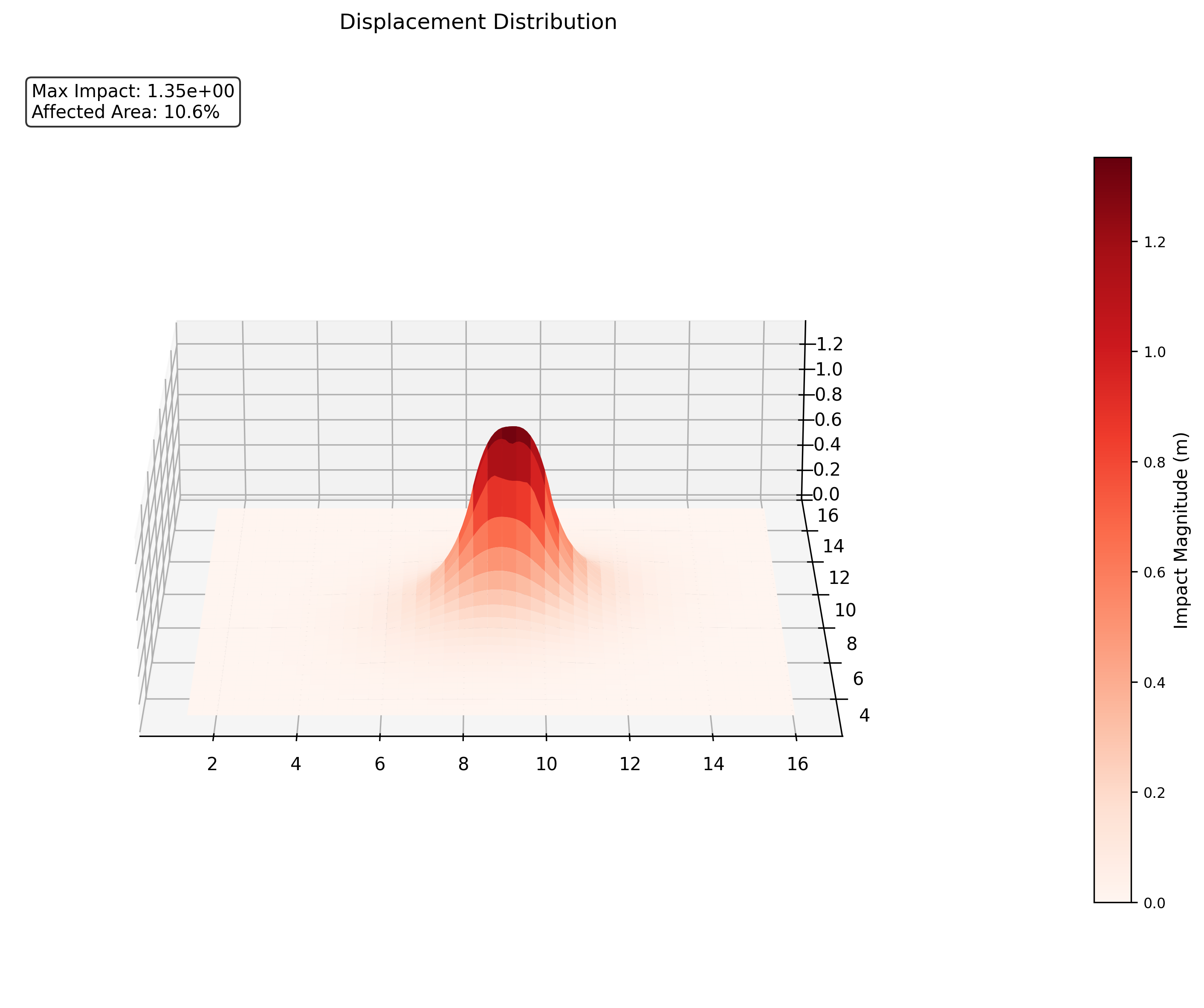}\\  
 (a)&(b)\\
 \includegraphics[width=0.45\textwidth]{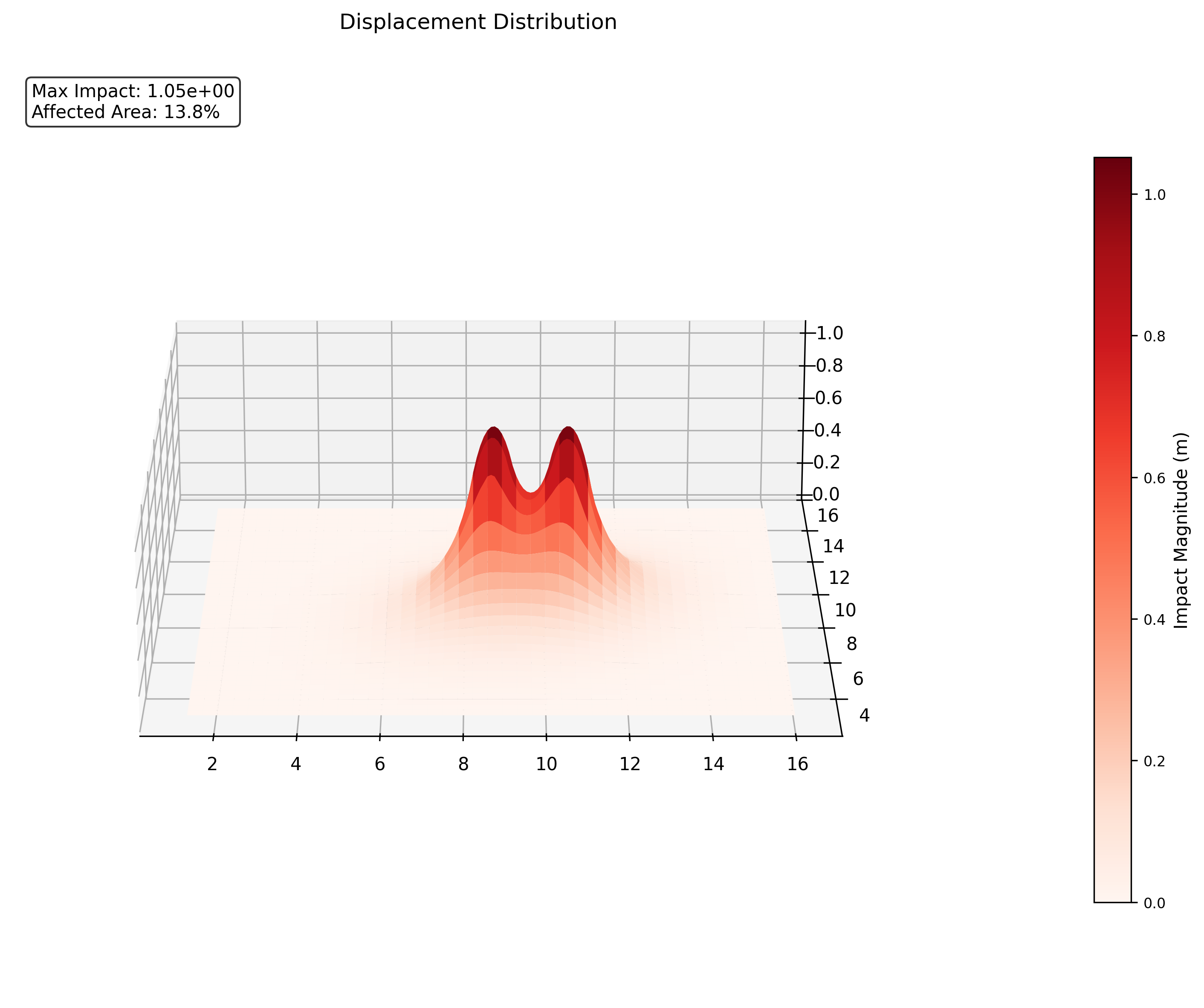}&\includegraphics[width=0.45\textwidth]{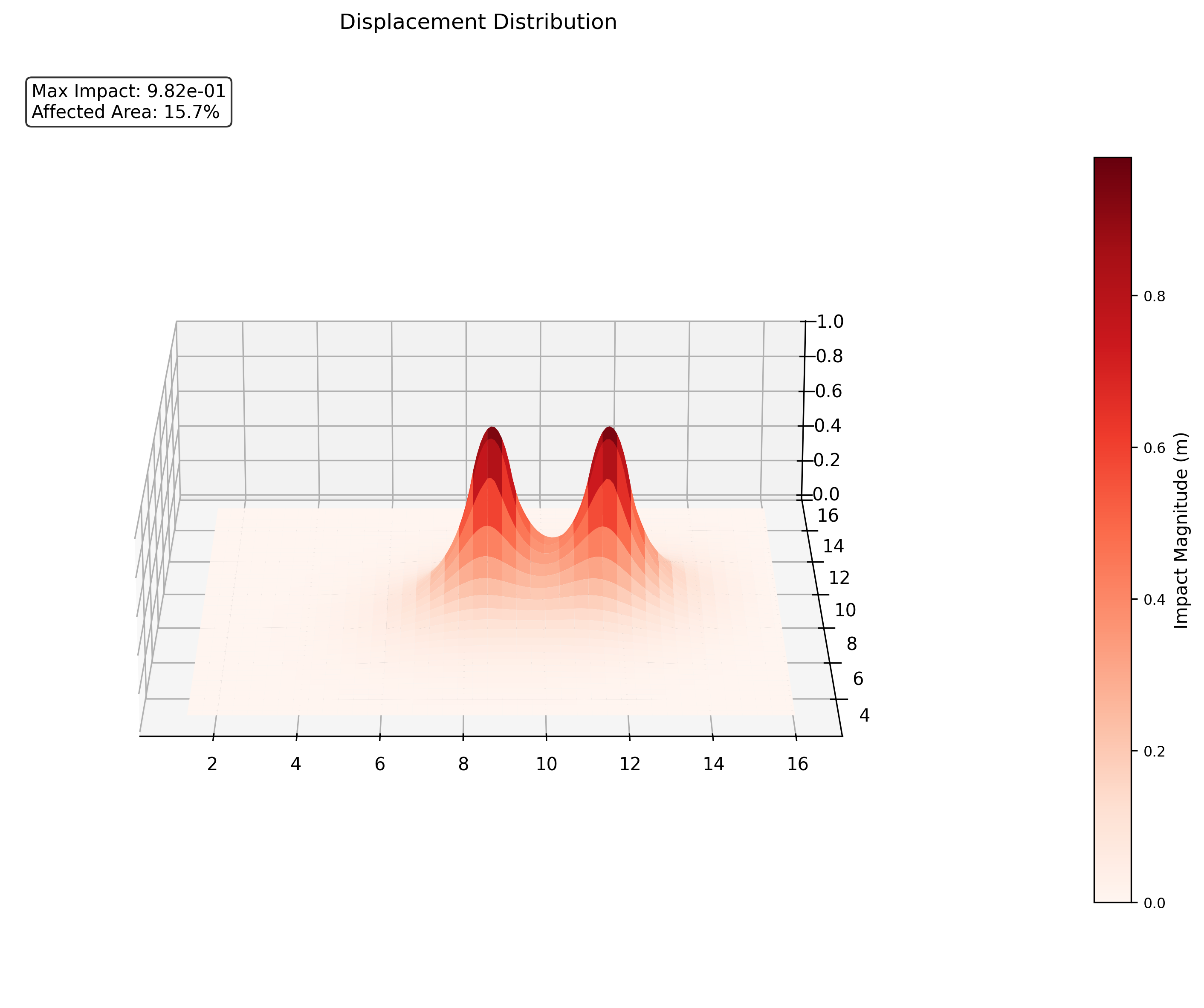}\\
 (c)&(d)\\\end{tabular}
\caption[3D visualisation of the displacement fields for increasing force separation]{3D visualisation of the displacement fields for increasing force separation: \textbf{(a)} 0° apart, showing complete force combination, \textbf{(b)} 1.0° apart, \textbf{(c)} 2.0° apart, \textbf{(d)} 3.0° apart.}
\label{fig:two-force-distance-effects-visualisations}
\end{figure}

The 3D displacement visualisations in Figure \ref{fig:two-force-distance-effects-visualisations} provide additional insight into how force interaction patterns evolve with distance. At 0° separation (Figure \ref{fig:two-force-distance-effects-visualisations} \textbf{(a)}), the forces combine to create a single intense peak. As the separation increases to 1° (Figure \ref{fig:two-force-distance-effects-visualisations} \textbf{(b)}), the response begins to split into distinct peaks while maintaining significant interaction. Further separation to 2° and 3° (Figure \ref{fig:two-force-distance-effects-visualisations} \textbf{(c)}, \textbf{(d)}) produces increasingly independent displacement fields, though some interaction effects persist even at these large distances. 

For the social fabric framework, this implies that when conflict events occur in close proximity, they produce intensified but localised effects. As spatial separation increases, the individual events maintain their distinct character while affecting a broader total area. This behaviour aligns with expectations from social systems, where concentrated conflicts cause intense localised effects, whereas conflicts affecting an entire area have broader yet less intense impacts. However, the same issue regarding von Mises stress as an indicator of social stress becomes evident here, too. The displacement, on the other hand, appears to show the more appropriate superposition behaviour to be used as a social impact measure.


\section{Summary of Findings of the Model Behaviour Tests}

The analysis of the model's behaviour reveals two fundamental findings that validate the framework's capability to model and measure impacts. First, tests involving forces distributed over an area or overlapping forces revealed significant limitations in using von Mises stress. The stress measure shows low values in the centre of areas where force is applied and significantly lower stress when events overlap. While this accurately represents physical material stress, it fails to represent social impacts in these scenarios properly. As confirmed through the literature analysis, social systems would be expected to show intensified impacts at the centre of a large conflict event or when multiple events overlap. In contrast, displacement demonstrates behaviour that aligns with these expectations, showing appropriate superposition and spatial distribution patterns.

Second, through systematic boundary condition testing, it was confirmed that displacement measurements can be made independent of edge effects by using an adequate buffer zone of over 300 km. This validation is crucial as it ensures that the computational implementation can produce reliable results unaffected by artificial boundaries. Together, these findings establish displacement as the verified measure for analysing impact propagation within the framework.

Beyond these fundamental validations, the analysis also reveals clear relationships between model parameters and system response, which will inform the subsequent translation of social and conflict indicators into model properties. The Table \ref{tab:model-behaviour-summary} below summarises these observed behaviours and their potential implications for social and conflict behaviour.

These systematic relationships between physical parameters and plate response provide a robust foundation for translating social and conflict characteristics within the social fabric framework. The clear behavioural pattern and discussed measurement approach enable the development of precise mapping strategies that preserve both mathematical validity and social meaning, as detailed in the following chapter. 

\begin{table}[H]
    \begin{tabular}
    {>{\raggedright\arraybackslash}p{100pt}>{\raggedright\arraybackslash}p{170pt}>{\raggedright\arraybackslash}p{170pt}} \hline
         \textbf{Parameter}&\textbf{Observed Effects}&\textbf{Potential Social/Conflict Implications}\\ \hline \hline 
 \textbf{Thickness $h$}&Increasing $h$ exponentially reduces displacement, affected area, and stress; below \~ 1000 m, even slight changes of $h$ lead to large shifts in the plate's behaviour; above \~ 5000 m, changes become minimal with increasing $h$.&Baseline structural "resilience" or "stability"; thinner implies greater vulnerability and greater impact spread; thicker implies more robustness.\\ \hline 
         \textbf{Young's Modulus $E$}&Higher $E$ reduces displacement proportionally without altering the stress distribution and the affected area. &  Acts as "sensitivity" to a violent event. High $E$ mean less deformation for the same intensity, implying stiffness or resistance to conflict. \\  \hline 
         \textbf{Poisson's Ratio $\nu$}&Overall small changes; higher $\nu$ slightly increases max. displacement and decreases affected area and stress; influences the localisation of the plate's response.&  Tunes how stresses spread or localise; adjusting $\nu$ can represent whether social tensions concentrate or disperse in response to conflict.\\  \hline \hline 
         \textbf{Force Magnitude $F$}&Scales displacement and stress linearly, leaving the affected area unchanged. &  Direct parameter for conflict intensity; larger force leads to more severe societal strain without changing how far the effects spread. \\  \hline 
         \textbf{Force Distribution (constant, linear, Gaussian)}&Gaussian distribution concentrates impact with highest central displacement; constant distribution shows broader impact zone with lower peak values; linear distribution shows intermediate behaviour.& Reflects how conflicts can be localised or widespread; from a focused "hotspot" (Gaussian) to a more diffused unrest (constant).\\ \hline 
        \textbf{Force Radius $r$}&Larger radius increases the affected area but reduces the max. displacement and stress; force gets more distributed.&Controls "reach" of the event's impact; a larger radius implies more widespread but less intense local effects; diffuse social upheaval vs. tightly focused incidents. \\ \hline 
        \textbf{Multiple Forces \& Interactions}&Combined impacts depend on distance and magnitudes; displacement shows appropriate superposition behaviour.&Multiple conflict events interact in complex ways; proximity and intensity influence overall societal impact; effects can compound; \\ \hline 
    \end{tabular}
    \caption[Summary of findings or the implementation testing and behavioural analysis]{Summary of findings or the implementation testing and behavioural analysis}
    \label{tab:model-behaviour-summary}
\end{table}


\chapter{Translation Framework for Social \& Conflict Characteristics}

This chapter develops a systematic framework for translating social and conflict characteristics into material properties and forces for a physics-informed model. Rather than prescribing specific indicators or fixed mappings, this framework aims to provide domain experts with the flexibility to select relevant indicators while ensuring physically meaningful parameter combinations. 

The framework builds directly on the validation testing results from the previous chapter, which established clear relationships between physical parameters and system response. These relationships provide the mathematical foundation for modelling how different social characteristics influence conflict impact and propagation. Most importantly, the testing confirmed that displacement patterns can reliably capture both localised impact and broader spillover effects while maintaining independence from computational boundary conditions when properly implemented.


\section{Translating Social Characteristics to Material Properties}

Drawing on the insights from the literature review, particularly regarding the multifaceted nature of conflict impacts and the interplay between vulnerability and resilience, this framework established distinct roles for each physical parameter that align with their mathematical relationships in the bending stiffness equation (Equation \ref{eq:bending-stiffness}). 

Thickness ($h$) serves as the primary parameter representing a community's fundamental capacity to absorb shocks and impacts. Its cubic relationship in the bending stiffness equation creates distinct response regimes that mirror how communities with different levels of vulnerability respond to shocks (Figure \ref{fig:plate_material_properties_thickness__displacement-affected-area}). Just as a thin plate shows significant deformation under even small forces, communities with low 'thickness' exhibit high vulnerability, where even minor disruptions can trigger cascading failures. This behaviour aligns with findings from conflict literature about how fragile states often experience severe deterioration of health services and social systems from relatively limited violence \parencite{Bendavid2021TheChildren, Norris2007CommunityReadiness, Tapsoba2023TheConflict}. 

The middle range provides proportional responses characteristic of moderately resilient or vulnerable communities. In this regime, communities maintain basic functionality while showing predictable strain under stress, similar to how conflict-affected regions might maintain essential services but experience graduate degradation as violence intensifies \parencite{Amberg2023ExaminingAnalysis, Wagner2018ArmedAnalysis}. The high thickness range represents robust communities where additional strengthening yields diminishing returns, matching observations about how strong institutional capacity and infrastructure create baseline resilience that helps communities withstand significant disruption \parencite{Cutter2008ADisasters, Cutter2014TheResilience, Norris2007CommunityReadiness}. 

Young's modulus ($E$) controls the magnitude of displacement responses without altering its distribution pattern, creating a simple linear scaling relationship as demonstrated in the implementation testing (Figure \ref{fig:plate_material_properties_youngs-modulus__displacement}). This behaviour makes it ideal for representing socioeconomic factors that modify how strongly a community reacts to conflict stress. For instance, testing showed that halving $E$ doubles displacement magnitude while preserving the spatial pattern effects, which is similar to how economic resources might buffer impact intensity without changing the fundamental way a community responds \parencite{Cutter2008ADisasters, Norris2007CommunityReadiness}.

Poisson's ratio ($\nu$) influences both displacement magnitude and spatial distribution through its role in coupling deformation in different directions. As explained in the previous section, it can vary between 0 and $<$0.5. Through its influence on spatial distribution, it is most suitable for representing connectivity factors that determine how impacts propagate through the social fabric. Higher values reflect a stronger connection between neighbouring areas, leading to a broader impact distribution, similar to how well-connected communities might share both resources and stress through social networks, institutional partnerships, and physical infrastructure \parencite{Cappelli2024LocalAfrica, Hanze2024WhenDisasters}. This aligns with findings from the literature review about how conflict effects often propagate through existing social and economic networks, creating spillover effects well beyond the immediate impact zone \parencite{Cappelli2024LocalAfrica, Schutte2011DiffusionWars, Tapsoba2023TheConflict}. 

However, it is important to note that the effects that can be modelled with Poisson's ratio are very limited. This is because its mathematical role in the bending stiffness equation (Equation \ref{eq:bending-stiffness}) is constrained through the $(1-\nu^2)$ term in the denominator. With $\nu$ also limited to the range [0, 0.5), this term can only vary between 1 and 0.75, providing much less leverage than the cubic effect of the thickness or the linear scaling of Young's modulus. Additionally, while Poisson's ratio does influence how deformation spreads between different directions, this coupling effect is secondary to the overall magnitude of deformation controlled by the other parameters. This makes Poisson's ratio most suitable for fine-tuning the spatial distribution of effects rather than controlling their fundamental character or magnitude.

\subsection{Indicator Selection \& Classification}

To ensure flexibility, it needs to be possible to select data from various categories to characterise community resilience and vulnerability. This should also enable using existing vulnerability and resilience frameworks such as \ac{DROP} from \textcite{Cutter2008ADisasters} and \ac{BRIC} from \textcite{Cutter2014TheResilience} to build on existing research and methods.  

Infrastructure-related indicators—including critical facilities, transportation networks, and lifeline systems \parencite{Cutter2008ADisasters}—form a foundation for thickness parameters due to their role in providing basic absorption capacity. These may include both the physical presence of facilities and their operational characteristics, such as capacity and robustness \parencite{Norris2007CommunityReadiness}. Environmental factors that create additional climate-related vulnerabilities may influence this foundational capacity as well, reflecting the compound nature of vulnerability identified in conflict zones \parencite{Cappelli2024LocalAfrica, Hanze2024WhenDisasters}. 

Institutional indicators can represent another component in determining the baseline resilience and, therefore, the thickness. These can encompass governance structures, emergency response systems, and mitigation planning capabilities \parencite{Cutter2008ADisasters}. This institutional strength may act as a prerequisite for other forms of resilience, as strong governance systems can enhance the effectiveness of physical infrastructure, while weak institutions may undermine even substantial physical resources \parencite{Caso2023The2018}. This aligns with observations from conflict literature that institutional breakdown often precedes and amplifies other vulnerabilities \parencite{Cappelli2024LocalAfrica, Caso2023The2018, Norris2007CommunityReadiness, Sinha2023WaterApproach}. 

Socioeconomic indicators are mapped to Young's modulus due to their linear scaling behaviour demonstrated in the implementation testing. These could include economic diversity, employment rates, income distribution, social cohesion measures, and community resource levels \parencite{Cappelli2024LocalAfrica, Cutter2008ADisasters}. 

Connectivity indicators are particularly well-suited for mapping to Poisson's ratio, given its role in determining how impacts propagate between regions. These indicators can encompass both physical and social networks, including transportation linkages, communication infrastructure, supply chains, as well as social networks, institutional partnerships, and cultural ties \parencite{Cutter2008ADisasters, Cutter2014TheResilience}.

\subsection{Parameter Mapping Methodology}

The translation process itself involves three fundamental steps. First, social indicators must be normalised to comparable scales, accounting for their natural ranges and distributions. Second, these normalised values must be translated into factors that can either have an increasing or decreasing effect on the physical parameter through appropriate response functions that capture how social characteristics may influence community resilience or vulnerability. Finally, multiple indicators affecting the same parameter must be combined in ways that preserve both physical meaning and social reality.

For each parameter, the framework establishes a neutral baseline representing typical or average conditions (e.g., based on the middle ranges identified in implementation testing). Variations from this baseline are then computed based on indicator values, with careful attention to maintaining physically valid parameter ranges. This approach ensures that the resulting parameter fields produce meaningful displacement patterns that can be interpreted in terms of social impact.

\subsubsection{1. Normalisation \& Response Functions}

To enable consistent mapping, each social indicator is first normalised to a range of [-1,1] according to:

\begin{equation} \label{eq:indicator-normalisation}
    x_{normalized} = \frac{x_{actual} - x_{mid}}{x_{max} - x_{mid}}
\end{equation}

where $x_{mid}$ represents neutral conditions. The normalised indicator values represent different cases. Pure vulnerability indicators occupy the negative domain [-1,0], pure resilience indicators occupy the positive domain [0,1], and indicators capture both vulnerability and resilience effects spanning the full range [-1,1].

To allow various mappings of the vulnerability and resilience ranges for indicators and to establish a uniform process for purely vulnerability or resilience indicators, the positive and negative ranges will be handled independently, employing only absolute values. To create a diminishing or increasing impact on the respective parameter, their signs will be altered to correspond with the direction of their specific effect. The choice between these functions should reflect how different social indicators influence community characteristics.

\paragraph{\textbf{Option 1: Linear Response}}\hfill \break
 
The linear response function (Equation \ref{eq:linear-response}) creates directly proportional modifications to material properties. For instance, \textcite{Cutter2014TheResilience} mentions the number of public schools per 10,000 as an indicator of school restoration potential and as a direct resilience concept. Therefore, each incremental change in the number of schools would create a proportional change in community resilience. 

\begin{figure}[H]
\centering
\begin{tabular}{cc}
\includegraphics[width=0.45\textwidth]{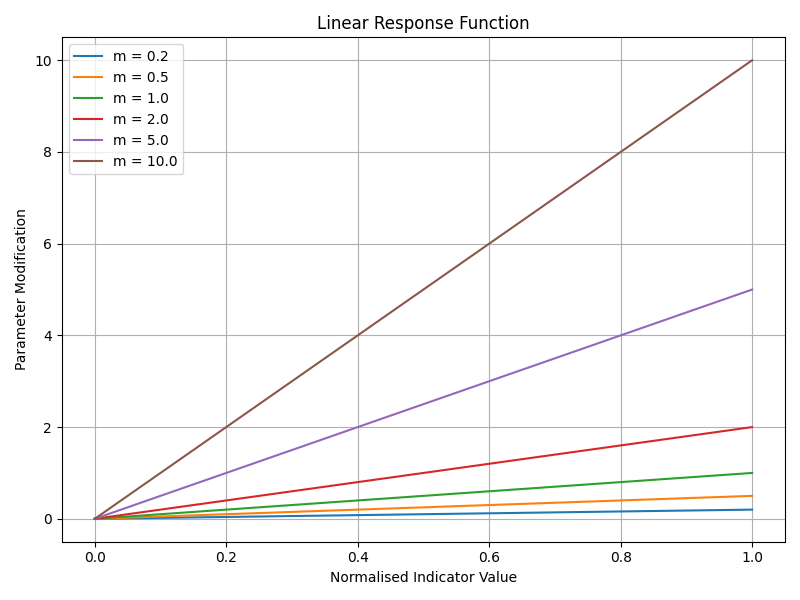}& \includegraphics[width=0.45\textwidth]{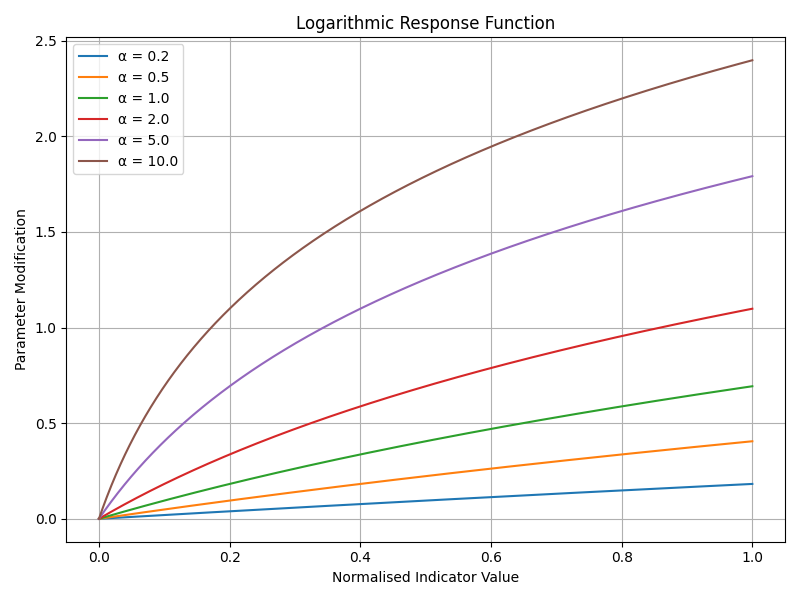}\\
(a) & (b) \\[6pt]
\includegraphics[width=0.45\textwidth]{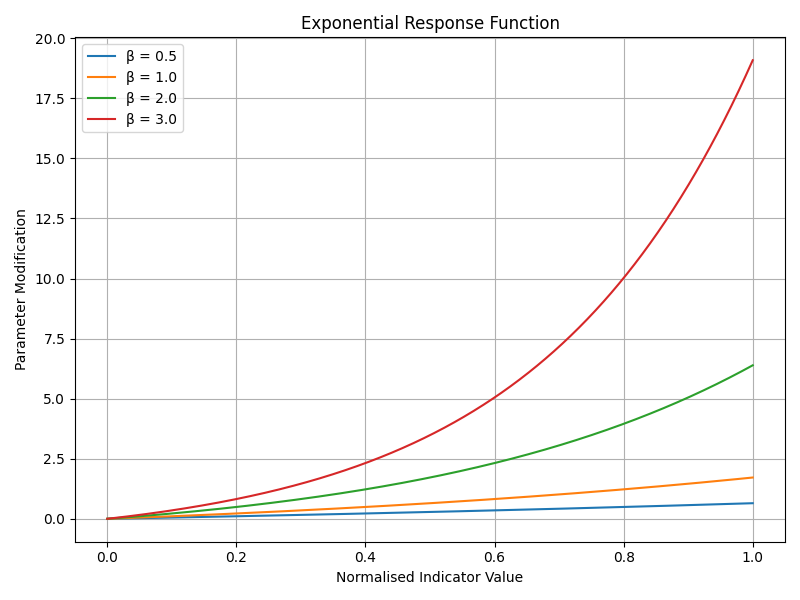}& \includegraphics[width=0.45\textwidth]{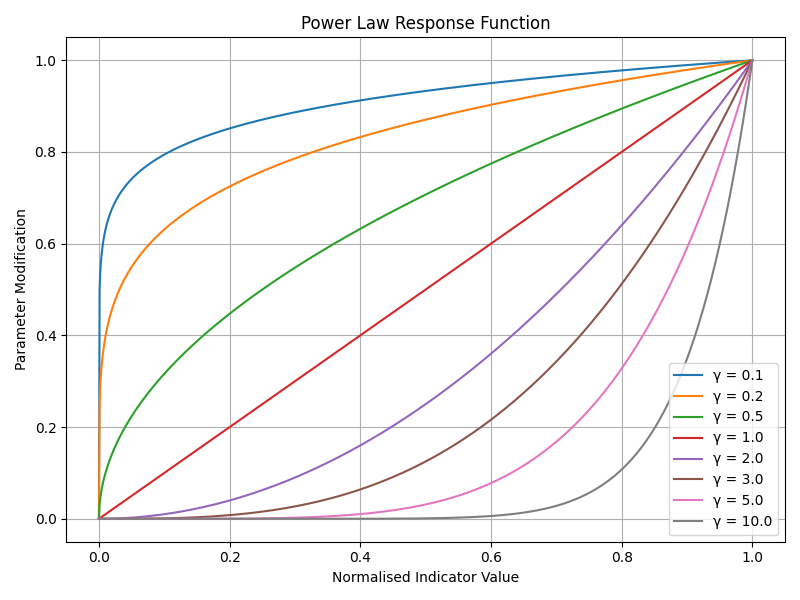}\\
(c) & (d) \\
 \includegraphics[width=0.45\textwidth]{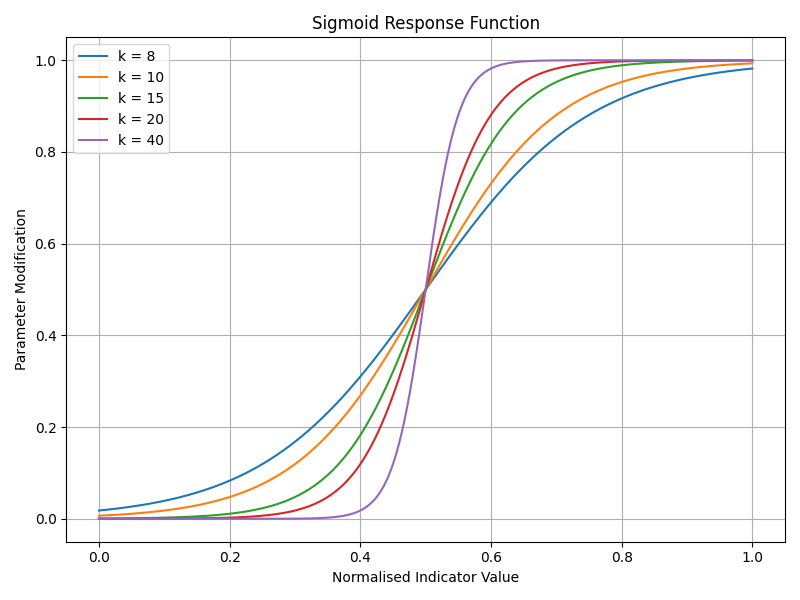}& \\
 (e) & \\\end{tabular}
\caption[Visualisations of response functions for translating social indicators, showing the effect of varying influence parameters]{Visualisations of response functions for translating social indicators, showing the effect of varying influence parameters: \textbf{(a)} linear response with slope $m$, \textbf{(b)} logarithmic response with scaling $\alpha$, \textbf{(c)} exponential response with growth rate $\beta$, \textbf{(d)} power-law response with exponent $\gamma$) and \textbf{(e)} sigmoid response with steepness $k$. Each function maps normalised indicator values to different response patterns.}
\label{fig:response-function-visualisations}
\end{figure}

When mapping this infrastructure indicator to thickness, for example, communities would show direct scaling between a number of schools and their baseline capacity to absorb impacts. The slope $m_r$ controls how strongly these infrastructure changes positively modify the plate property ($m_v$ the modification in a negative way) (Figure \ref{fig:response-function-visualisations} \textbf{(a)}). While probably most indicators reach a limit at some point, the linear response function might provide a good baseline for various cases, especially when there is no clear evidence for more complex relationships. Furthermore, the linear response function can also be defined with different slopes for different regimes to capture varying sensitivity in different ranges. This could be suitable for indicators where there is empirical evidence for distinct response regimes. 

\begin{equation}\label{eq:linear-response}
    f(x) = 
    \begin{cases}
        -1 \cdot m_vx & \text{ if } x < 0 \text{ (vulnerability)}, \text{with }  m_v > 0\\
        +1 \cdot m_rx & \text{ if } x \geq 0 \text{ (resilience)}, \text{with }  m_r > 0
    \end{cases}
\end{equation}

\paragraph{\textbf{Option 2: Logarithmic Response}}\hfill \break

The logarithmic function (Equation \ref{eq:logarithmic-response}) creates strong initial effects that diminish as the indicator value increases in magnitude. This could be the case, for example, for institutional indicators like institutional knowledge about disaster preparedness or mitigation, as mentioned by \textcite{Cutter2014TheResilience}. When mapping this institutional capacity to thickness, for instance, initial implementations of disaster preparedness training might create significant improvements in a community's baseline resilience. However, once these foundational programmes are established, additional expansions might yield progressively smaller enhancements. The parameters $\alpha_v$ and $\alpha_r$ control how quickly these diminishing returns set in, and could be derived from observed saturation points in social characteristics (Figure \ref{fig:response-function-visualisations} \textbf{(b)}). 

\begin{equation}\label{eq:logarithmic-response}
    f(x) = 
    \begin{cases}
        -1 \cdot \ln(1 + \alpha_v|x|) & \text{ if } x < 0 \text{ (vulnerability)}, \text{with }  \alpha_v > 0 \\
        +1 \cdot \ln(1 + \alpha_r|x|) & \text{ if } x \geq 0 \text{ (resilience)}, \text{with } \alpha_r > 0
    \end{cases}
\end{equation}

\paragraph{\textbf{Option 3: Exponential Response}}\hfill \break

To capture indicators where effects intensity progressively as values become more extreme, an exponential function (Equation \ref{eq:exponential-response}) could be used. This pattern would be particularly relevant for indicators that might have minimal impact on near-neutral conditions but create rapidly growing effects as they deviate further. This could be the case, for example, for patterns in jurisdictional coordination, where an increasing number of governments or jurisdictions create rapidly accelerating complexity and potential for coordination breakdown \parencite{Cutter2014TheResilience}. When mapping this to Poisson's ratio, for instance, small initial increases in jurisdictional complexity may have limited effects, but continued fragmentation leads to accelerating weakness in the community's ability to coordinate responses and distribute stress. The parameters $\beta_v$ and $\beta_r$ control this acceleration rate (Figure \ref{fig:response-function-visualisations} \textbf{(c)}). As the exponential behaviour does not level off, it must be used cautiously, as its strong non-linearity can create parameter modifications that overwhelm other indicators' effects.

\begin{equation}\label{eq:exponential-response}
    f(x) = 
    \begin{cases}
        -1 \cdot (e^{\beta_v|x|} - 1) & \text{ if } x < 0 \text{ (vulnerability)}, \text{with }  \beta_v > 0\\
        +1 \cdot (e^{\beta_r|x|} - 1) & \text{ if } x \geq 0 \text{ (resilience)}, \text{with }  \beta_r > 0
    \end{cases}
\end{equation}

\paragraph{\textbf{Option 4: Power Law Response}}\hfill \break

The power law response function (Equation \ref{eq:power-law-response}) capture behaviour patterns that are similar to, but more flexible than, the logarithmic and exponential functions. With the exponents $\gamma_v$ and $\gamma_r$ controlling the fundamental behaviour, it can model two distinct patterns (Figure \ref{fig:response-function-visualisations} \textbf{(d)}). For $\gamma < 1$, it produces a pattern similar to the logarithmic function, but with even stronger initial effects that level off more dramatically. This could be suitable for indicators where even small improvements from very poor conditions create substantial gains, but additional improvements yield strongly diminishing returns. Conversely, with $\gamma > 1$, it creates a pattern similar to the exponential function but with more pronounced late-state acceleration. This could capture cases, where an indicator needs to reach a considerable level before signifiant effects occur, but then produces even more dramatic acceleration than an exponential function. The flexibility of the power law function makes it particularly useful when the logarithmic or exponential functions almost, but don't quite, capture an indicator's behaviour pattern. 

\begin{equation}\label{eq:power-law-response}
    f(x) = 
    \begin{cases}
        -1 \cdot (|x|^{\gamma_v}) & \text{ if } x < 0 \text{ (vulnerability)}, \text{with }  \gamma_v > 0 \\
        +1 \cdot (|x|^{\gamma_r}) & \text{ if } x \geq 0 \text{ (resilience)}, \text{with }  \gamma_r > 0
    \end{cases}
\end{equation}

Since the input $x$ is already normalised between -1 and 1, the power law response also maintains these bounds. This ensure consistent scaling behaviour when combining multiple indicators, without the risk of dominating behaviours like that of the exponential response function. The choice of $\gamma$ value determines only the shape of the response curve between these fixed endpoints.

\paragraph{\textbf{Option 5: Sigmoid Response}}\hfill \break

The sigmoid function (Equation \ref{eq:sigmoid-response}) could represent cases where indicator exhibit distinct threshold behaviour with relatively stable states at both extremes and a concentrated transition zone in between. This creates an S-shaped response curve where initial changes in an indicator produce minimal effects, followed by a zone of transition, before again stabilising with minimal additional effects. The parameters $k$ and $x_0$ provide control over this behaviour pattern (Figure \ref{fig:response-function-visualisations} \textbf{(e)}). The transition point $x_0$ determines where the changing points occurs, while the steepness parameter $k$ controls how abruptly this transition takes place. A larger $k$ creates a sharper threshold effects, while a smaller $k$ produces a more gradual transition between states. This could make the sigmoid function particularly useful for indicators where communities demonstrate resistance up to a certain threshold, then undergo rapid change before stabilising at a new level. 

\begin{equation}\label{eq:sigmoid-response}
    f(x) = 
    \begin{cases}
        -1 \cdot (\frac{1}{1 + e^{-(k_v(|x| - x_{0_v}))}}) & \text{ if } x < 0 \text{ (vulnerability)}, \text{with }  k_v > 0 \\
        +1 \cdot (\frac{1}{1 + e^{-(k_r(|x| - x_{0_r}))}})) & \text{ if } x \geq 0 \text{ (resilience)}, \text{with }  k_r > 0
    \end{cases}
\end{equation}

To guide the selection of appropriate response functions for different indicators, Table \ref{tab:response-function-selelction} provides an overview of suggestions for when to use each function type based on the indicator's characteristics and expected behaviour.


\subsubsection{2. Dependencies \& Interactions}

Before combining individual indicator effects into final material properties, it is important to recognise and account for potential dependencies between social indicators. While indicators that show a high degree of collinearity with other indicators could be dropped upfront \parencite{Cutter2014TheResilience}, other dependencies might occur at different levels, each requiring specific consideration in the translation framework.

While dependencies can capture complex relationships between indicators, they should be used sparingly and only when strongly justified, as each additional dependency increases the model's complexity and reduces transparency.

\paragraph{\textbf{Option 1: Threshold Dependency}}\hfill \break

\begin{figure}[H]
    \centering
    \includegraphics[width=0.75\linewidth]{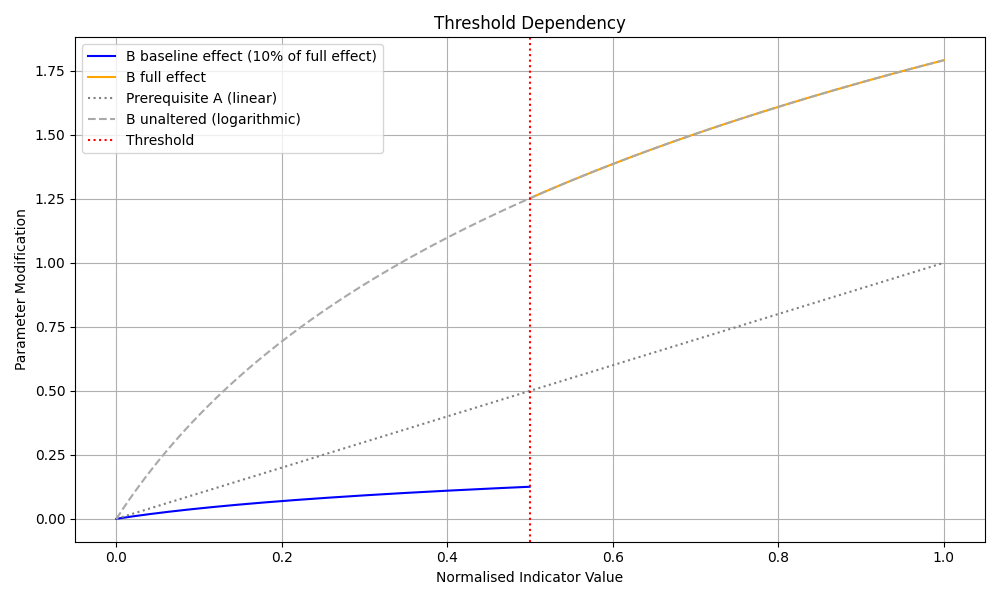}
    \caption[Threshold dependency option for response functions]{Threshold dependency: The effectiveness of indicator $B$ depends on indicator $A$ reaching a defined threshold (vertical red dashed line). When $A$ is below this threshold, $B$ operates at a baseline level (here, 10 \% of its full potential). The plot shows $B$'s unaltered logarithmic response (grey dashed line), prerequisite $A$'s linear response (dotted line), and $B$'s actual response both at baseline (blue) and full effect (orange).}
    \label{fig:threshold-dependency}
\end{figure}

The most fundamental are critical dependencies, where one indicator's function might require another to be present at a minimum level. As described by \textcite{Sinha2023WaterApproach,} for example, water infrastructure systems may completely fail without electrical power for pumps, regardless of the water system's own robustness. Similarly, \textcite{Norris2007CommunityReadiness} notes how emergency response capabilities can depend on communication infrastructure so that even very good emergency services can become ineffective without basic communication capability. These critical dependencies could be captured through a threshold-based scaling function (Equation \ref{eq:threshold-dependency}).

\begin{equation}\label{eq:threshold-dependency}
    f_{effective_B}(x_A, x_B) = 
    \begin{cases}
        base_B \cdot f_B(x_B) & \text{if } f_A(x_A) < threshold \\
        f_B(x_B) & \text{if } f_A(x_A) \geq threshold
    \end{cases}
\end{equation}

where indicator $B$ depends on indicator $A$. The threshold determines the minimum level of the prerequisite indicator needed for the dependent indicator to function effectively. Both indicators $A$ and $B$ can use any of the previously defined functions. For example, in Figure \ref{fig:threshold-dependency}, the indicator $B$ follows a logarithmic response, while the indicator $A$ uses a linear response. When $f_A(x_A)$ falls below the threshold, $B$ influence reduces to its baseline response, which is a fraction of its full response defined by the $base_B$ parameter. Once the prerequisite indicator $A$ exceeds the threshold, $B$ returns to its full influence. 

It is also possible to reverse this behaviour, so the effect of dependent indicator $B$ will follow the defined influence until a certain threshold is met by indicator $A$ after which $B$ reduces to the $base_B$ percentage of its full effect.

\begin{table}[H]
    \centering
    \begin{tabular}{>{\raggedright\arraybackslash}p{100pt}|>{\raggedright\arraybackslash}p{110pt}|>{\raggedright\arraybackslash}p{110pt}|>{\raggedright\arraybackslash}p{110pt}} \hline
         \textbf{Response Function}& \textbf{Suggested Use} & \textbf{Potential Indicators}&\textbf{Key Considerations}\\ \hline \hline
         \textbf{Linear}& 
         \begin{itemize}[nosep] 
            \item Direct proportional relationships
            \item Clear, consistent scaling effects 
            \item Uniform sensitivity across range
        \end{itemize}& \begin{itemize}[nosep]
            \item Number of schools
            \item Hospital beds per capita
            \item Emergency response units
        \end{itemize}&\begin{itemize}[nosep]
            \item Most transparent
            \item Good baseline choice when relationship unclear
            \item Can combine multiple slopes for different ranges
        \end{itemize}\\ \hline
         \textbf{Logarithmic}&  
         \begin{itemize}[nosep]
            \item Diminishing returns
            \item Initial improvements most crucial
            \item Resource saturation effects
        \end{itemize}& \begin{itemize}[nosep]
            \item Basic healthcare access
            \item Disaster preparedness training
            \item Communication infrastructure
        \end{itemize}&\begin{itemize}[nosep]
            \item Strong initial effects
            \item Rapidly plateaus
            \item Good for foundational capabilities
        \end{itemize}\\ \hline
         \textbf{Exponential}&
         \begin{itemize}[nosep]
            \item Accelerating effects
            \item Compound growth dynamics
            \item Rapid deterioration scenarios
        \end{itemize}& \begin{itemize}[nosep]
            \item Jurisdictional complexity
            \item Coordination breakdown
            \item System cascade failures
        \end{itemize}&\begin{itemize}[nosep]
            \item Use sparingly
            \item Can dominate other effects
            \item Risk of oversensitivity
        \end{itemize}\\ \hline
         \textbf{Power Law}& 
         \begin{itemize}[nosep]
            \item Need flexibility between early/late effects
            \item Complex scaling relationships
            \item Variable sensitivity requirements
        \end{itemize}& \begin{itemize}[nosep]
            \item Institutional capacity
            \item Social network density
            \item Resource distribution
        \end{itemize}&\begin{itemize}[nosep]
            \item Most versatile
            \item Can mimic both log and exponential
            \item Requires careful parameter selection
        \end{itemize}\\ \hline
          \textbf{Sigmoid}& 
         \begin{itemize}[nosep]
            \item Clear threshold behaviour
            \item Distinct state transitions
            \item Resistance then rapid change
        \end{itemize}& \begin{itemize}[nosep]
            \item Critical mass effects
            \item Social tipping points
            \item Institutional breakdown
        \end{itemize}&\begin{itemize}[nosep]
            \item Good for binary-like transitions
            \item Requires threshold knowledge
        \end{itemize}\\ \hline
    \end{tabular}
    \caption[Overview of response function selection criteria]{Overview of response function selection criteria, including potential usage scenarios, example applications, and important considerations for each function type.}
    \label{tab:response-function-selelction}
\end{table}

\paragraph{\textbf{Option 2: Amplifying/Dampening Dependency}}\hfill \break

\begin{figure}[H]
    \centering
    \includegraphics[width=0.8\linewidth]{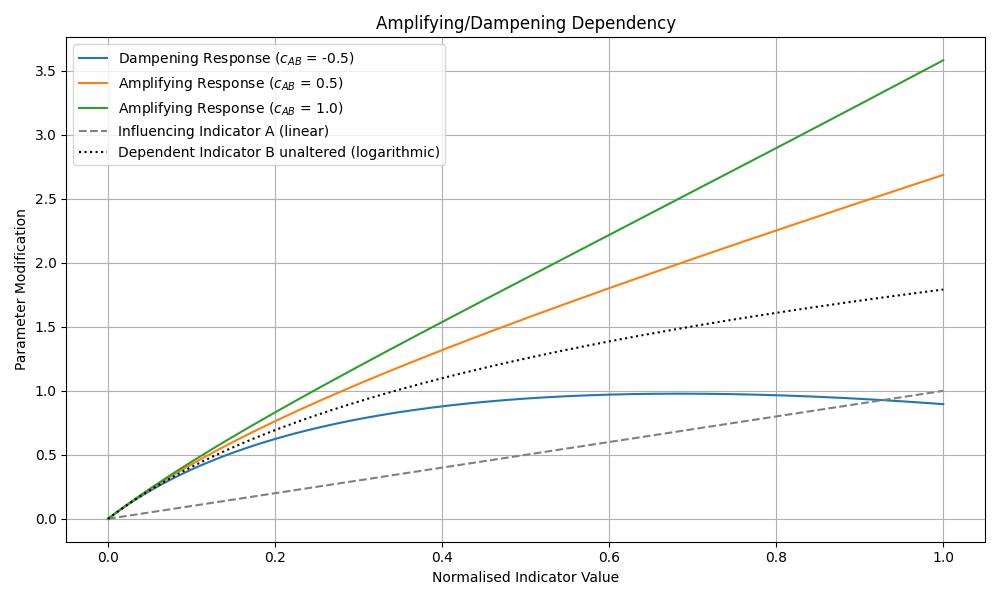}
    \caption[Amplifying and dampening dependency option for response functions]{Amplifying and dampening dependency: Visualisation of how indicator $A$ modifies indicator $B$'s response through different values of $c_{AB}$. Positive values of $c_{AB}$ amplify $B$'s response, while negative values dampen it relative to the unaltered baseline. The plot shows indicator $A$'s linear response (dashed line), $B$'s unmodified logarithmic response (dotted line), and $B$'s modified responses under different $c_{AB}$ values.}
    \label{fig:amplify-dampen-dependency}
\end{figure}

Another form of dependency could occur with amplifying or dampening effects, where one indicator's impact depends on the state of another. For instance, \textcite{Norris2007CommunityReadiness} indicates a relationship between institutional capacity and resource deployment, where strong institutions can amplify the benefits of available resources, while weak governance may fail to effectively even substantial resources. These interactions can be captured through a multiplicative modification of response functions (Equation \ref{eq:amplify-dampen-dependency}).  

\begin{equation}\label{eq:amplify-dampen-dependency}
    f_{compound_B}(x_A, x_B) = f_B(x_B) \cdot (1 + f_A(x_A) \cdot c_{AB})
\end{equation}

where $c_{AB}$ represents the strength of the effect indicator $A$ has on indicator $B$ . This allows both amplifying effects ($c_{AB} > 0$) where indicator $A$ enhances indicator $B$ impact and dampening effects ($c_{AB} < 0$) where indicator $A$ dampens indicator $B$ impact (Figure \ref{fig:amplify-dampen-dependency}). 

\begin{figure}[H]
    \centering
    \includegraphics[width=0.8\linewidth]{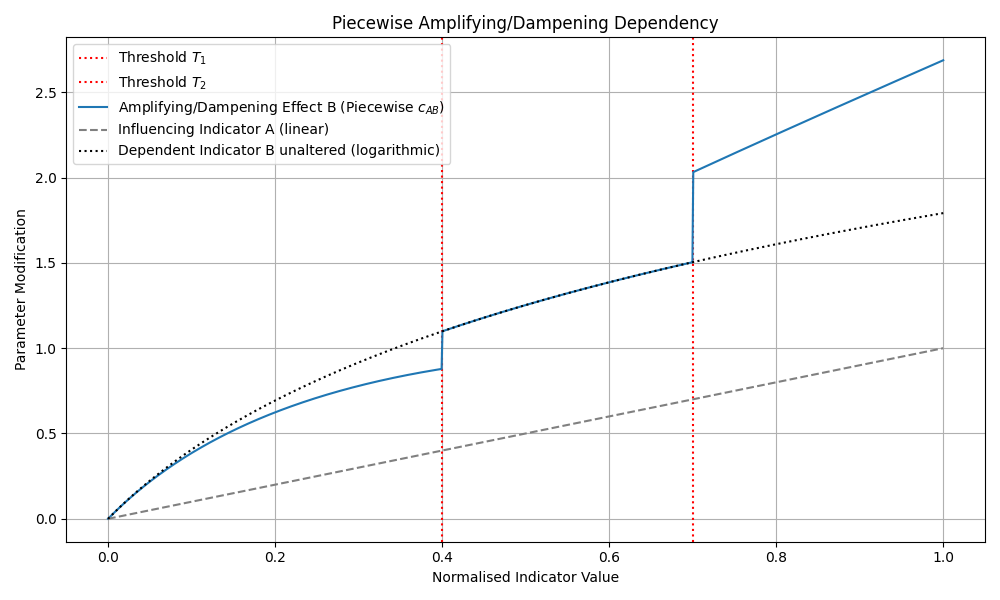}
    \caption[Piecewise amplifying/dampening dependency option for response functions]{Piecewise amplifying/dampening dependency: The influence of $A$ on $B$ shifts across defined ranges, with thresholds $T_1$ and $T_2$ marking transition points between varying $c_{AB}$ effects.}
    \label{fig:amplify-dampen-dependency-ranges}
\end{figure}

When indicator $A$ should only have an amplifying or dampening effect below or above certain thresholds ($T$), $c_{AB}$ can also be defined for certain ranges of $A$ (Equation \ref{eq:amplify-dampen-dependency-thresholds}). This allows, for example, indicator $A$ to have a dampening effect on the indicator $B$ until threshold $T_1$ is reached, then no effect between the thresholds $T_1$ and $T_2$, and finally, an amplifying effect beyond $T_2$ (Figure \ref{fig:amplify-dampen-dependency-ranges}). 

\begin{equation}\label{eq:amplify-dampen-dependency-thresholds}
    c_{AB}(x_A) = 
    \begin{cases}
         c_{AB_{low}} & \text{if } f_A(x_A) < T_1 \\
         c_{AB} & \text{if } T_1 \leq f_A(x_A) < T_2 \\
        c_{AB_{high}} & \text{if } f_A(x_A) \geq T_2
    \end{cases}
\end{equation}

\subsubsection{3. Parameter Combination}

When multiple social indicators influence the same physical parameters, their effects must be combined in a way that preserves both mathematical validity and social meaning. This combination must handle both resilience and vulnerability indicators while maintaining physically meaningful ranges and ensuring symmetric scaling behaviour around the baseline. 

\paragraph{\textbf{Exponential Mapping Combination for the Thickness and Young's modulus}}\hfill \break

For the physical parameter $p$ (thickness $h$, or Young's modulus $E$), indicator effects are combined through an exponential mapping (Equation \ref{eq:parameter-combination}).

\begin{equation}\label{eq:parameter-combination}
    p_{combined} = p_{base} \cdot \exp\left(\sum_{i=1}^n w_i f_i(x_i)\right), \text{ with } \sum_{i=1}^n w_i = 1
\end{equation}

where $p_{base}$ represents the neutral parameter value, $f_i(x_i)$ are the individual response function and $w_i$ are importance weights that sum to 1. For resilience indicators, $f_i(x_i)$ provides positive values representing enhancement, while for vulnerability indicators, $f_i(x_i)$ provides negative values representing reduction.  

This exponential mapping ensures perfect symmetry in scaling behaviour. For example, if a response function gives $g(x) = \ln(2) \approx 0.693$, as a resilience indicator, it will scale the parameter by $\exp(0.693) \approx 2.0$ (doubling), while as a vulnerability indicator, it will scale by $\exp(-0.693) \approx 0.5$ (halving). This symmetric behaviour is important for maintaining balanced effects between resilience and vulnerability indicators once they are translated into effects using the response functions.  

\begin{figure}[H]
    \centering
    \includegraphics[width=0.8\linewidth]{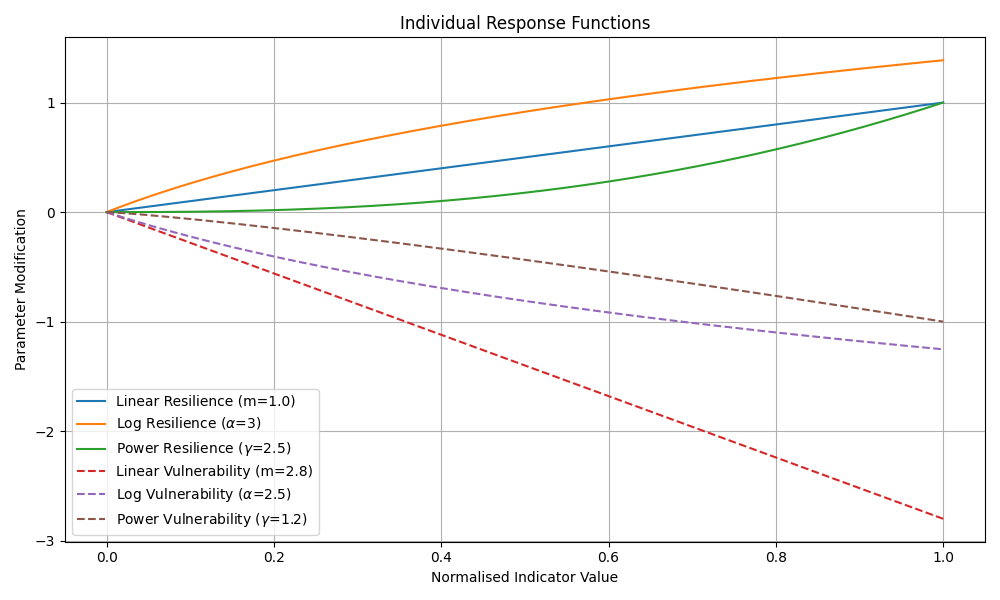}
    \caption[Comparison of individual response functions with selected parameter values - exponential combination mapping]{Comparison of individual response functions with selected parameter values, illustrating distinct resilience and vulnerability dynamics using linear, logarithmic, and power-law response functions.}
    \label{fig:parameter-combination-h-E_individual}
\end{figure}

The exponential mapping provides several key advantages for parameter combination. It guarantees that parameter values always remain positive as $\exp(\sum f_i \cdot w_i )$ is strictly greater than zero regardless of the input. This matches the physical requirement that the thickness and Young's modulus cannot be negative or zero. Additionally, it provides natural bounds on vulnerability effects. While resilience indicators can enhance parameters significantly if needed, vulnerability indicators create asymptotic reduction approaching but never reaching zero. 

\begin{figure}[H]
    \centering
    \includegraphics[width=0.85\linewidth]{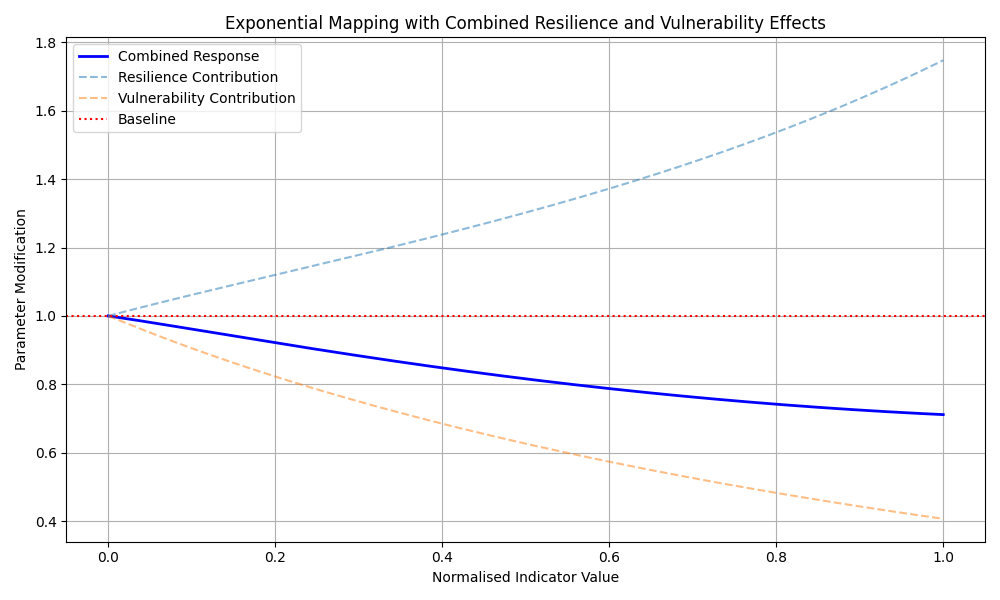}
    \caption[Combined resilience and vulnerability indicators effects - exponential combination mapping]{Combined resilience and vulnerability indicator effects, as shown in Figure \ref{fig:parameter-combination-h-E_individual}, and the resulting total combined response for the final parameter modification. The following weights ($w_i$) were used: 0.2 for linear resilience and vulnerability and 0.15 for logarithmic and power law resilience and vulnerability.}
    \label{fig:parameter-combination-h-E_combined}
\end{figure}

An example of different individual response functions (Figure \ref{fig:parameter-combination-h-E_individual}) being combined into the final modification that will be multiplied with the  $p_{base}$ baseline value can be seen in Figure \ref{fig:parameter-combination-h-E_combined}. 

\paragraph{\textbf{Sigmoid Mapping Combination for Poisson's ratio}}\hfill \break

For the Poisson's ratio, however, a different approach is necessary due to its strict physical bounds. Unlike thickness or Young's modulus, which can technically increase without bound, Poisson's ratio must remain between 0 and $<$0.5, as explained before. Therefore, the combination for Poisson's ratio uses a sigmoid-based mapping (Equation \ref{eq:poisson-ratio-combination})

\begin{equation}\label{eq:poisson-ratio-combination}
    \nu_{combined} = \frac{0.5}{1 + \exp(k \cdot \sum_{i=1}^n f_i(x_i)\cdot w_i)}, \text{ with } \sum_{i=1}^n w_i = 1
\end{equation}

Since the weighted sum ($\sum_{i=1}^n f_i(x_i)\cdot w_i)$) ranges between -1 and 1 when response functions are appropriately designed, the scaling parameter $k$ can be chosen to provide optimal sensitivity in this range. A value of $k = 2.5$ ensures that inputs of $\pm 1$ map to approximately 0.04 and 0.46 for the Poisson's ratio. In contrast to the exponential mapping used for the other parameters, the sigmoid mapping already maps to the full available range, and no multiplication with a baseline value is necessary. 

Another important difference is that resilience indicators will decrease Poisson's ratio while vulnerability indicators will increase it. To ensure a consistent workflow, the normalisation and mapping using the response functions will stay the same, and the relationship is implemented through Equation \ref{eq:poisson-ratio-combination}. The positive sign before the scaling parameter $k$ ensures that vulnerability indicators (negative $f_i$) increase $\nu$ while resilience indicators (positive $f_i$) decrease it. This behaviour aligns with the physical meaning of Poisson's ratio, where higher values indicate stronger coupling and greater spread of impacts, while lower values lead to more localised impacts.

The effectiveness of this mapping depends primarily on choosing appropriate response functions for indicators affecting the Poisson's ratio. The power law response function, the sigmoid response function and the logarithmic response functions with $\alpha < 1.7$ naturally compress their output to this range while maintaining a good sensitivity to changes. 

For other functions, when an indicator shows exponential behaviour or extreme responses, the sigmoid mapping will saturate toward 0 and 0.5, which might cause unwanted restrictions when choosing the suggested $k$ value above. In these cases, where different sensitivity is needed for such extreme effects, the parameter $k$ can be adjusted to either lower values for more gradual transitions or higher values for sharper changes around zero.

\begin{figure}[H]
    \centering
    \includegraphics[width=0.85\linewidth]{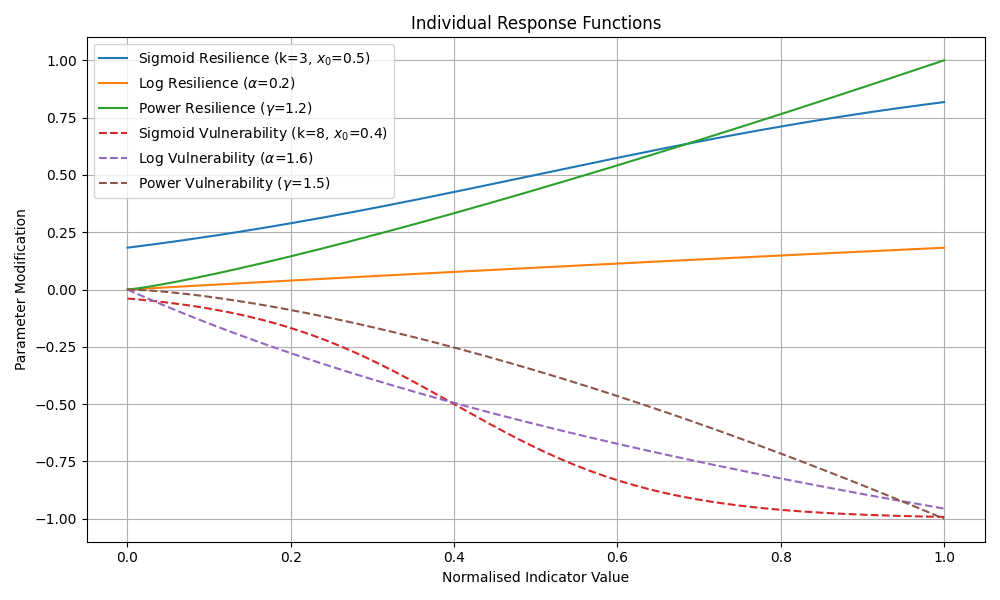}
    \caption[Individual response functions for resilience and vulnerability effects - sigmoid combination mapping]{Individual response functions for resilience and vulnerability effects, using sigmoid, logarithmic, and power-law response functions with specified parameter values. These functions illustrate how different behaviours are mapped to Poisson's ratio.}
    \label{fig:parameter-combination-poisson-ratio_individual}
\end{figure}

An example of different individual response functions (Figure \ref{fig:parameter-combination-poisson-ratio_individual}) being combined into the final Poisson's ratio can be seen in Figure \ref{fig:parameter-combination-h-E_combined}. To avoid any extreme responses while using $k = 2.5$, only sigmoid response functions, power law functions and logarithmic functions with $\alpha < 1.7$ were used.

\begin{figure}[H]
    \centering
    \includegraphics[width=0.85\linewidth]{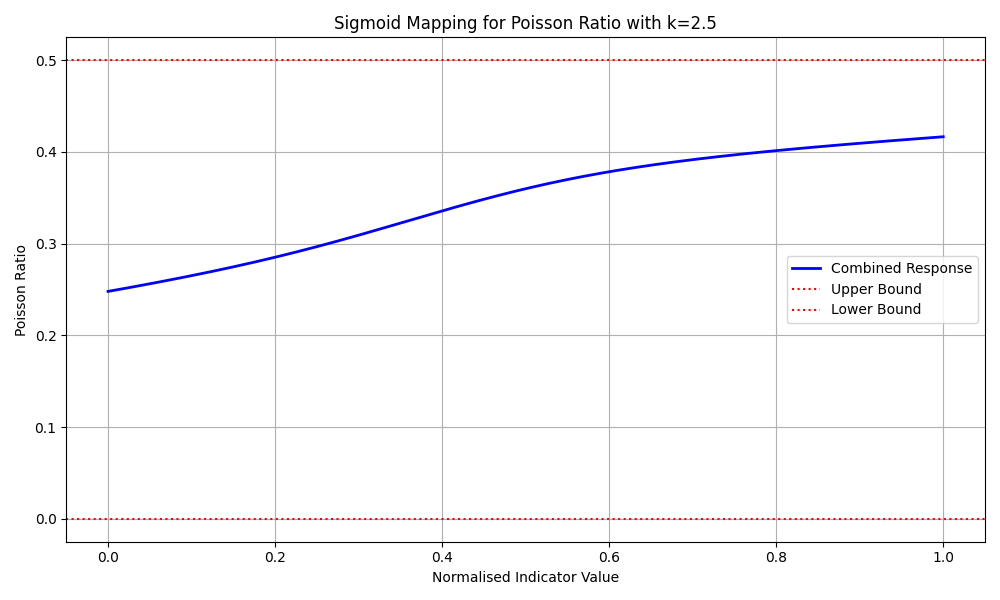}
    \caption[Combined resilience and vulnerability indicator effect - sigmoid combination mapping]{Combined resilience and vulnerability indicator effects, as shown in Figure \ref{fig:parameter-combination-poisson-ratio_individual}, and the resulting total combined response for the final Poisson’s ratio. The following weights ($w_i$) were used: 0.1 for sigmoid resilience and 0.3 for sigmoid vulnerability, 0.05 for logarithmic resilience and 0.3 for logarithmic vulnerability, and 0.05 for power law resilience and 0.2 for power law vulnerability.}
    \label{fig:parameter-combination-poisson-ratio_combined}
\end{figure}

For both approaches, the weighting factors $w_i$ provide control over each indicator's relative importance and can be calibrated through empirical evidence or expert assessment. Through careful selection of response functions, this framework provides a robust and customisable foundation for translating social indicators into material properties, capturing a range of enhancement and reduction effects while preserving mathematical validity and physical constraints. 

While material properties discussed above capture the underlying social fabric's characteristics, the following section describes how conflict events are translated into forces acting upon this fabric. The interaction between these components—material properties and applied forces—ultimately determines tie impacts and how they propagate through communities. This separation allows independent calibration of social vulnerability/resilience factors and conflict event characteristics while maintaining their interconnected effects in the final analysis.


\section{Translating Conflict Characteristics to Force Properties}

Translating conflict events into physical forces requires consideration of how different event characteristics influence both immediate impact and broader societal effects. While various conflict datasets exist, they typically share fundamental attributes that characterise violent events: intensity measures (e.g., number of fatalities), events classifications, and temporal-spatial coordinates \parencite{Tapsoba2023TheConflict}.

\ac{ACLED} provides a comprehensive example of such datasets, offering detailed event categorisation along with fatality counts, targeting information, and precise geographic coordinates \parencite{Raleigh2023PoliticalChoices}. It will be, therefore, used as an example and referenced throughout the next paragraphs and sections. As shown in Table \ref{tab:acled-event-types}, events are categorised into six primary types: Battles, Protests, Riots, Explosions/Remote violence, Violence against civilians and Strategic developments, with further refinement into specific sub-event types \parencite{ACLED2024ACLED2024}.

\begin{table}[H]
    \begin{tabular}{>{\raggedright\arraybackslash}p{100pt}|>{\raggedright\arraybackslash}p{170pt}|>{\raggedright\arraybackslash}p{170pt}}
    \hline
    \textbf{Event type} & \textbf{Sub-event type} & \textbf{Disorder type} \\ \hline\hline
    Battles & Government regains territory & Political violence \\ \cline{2-3}
            & Non-state actor overtakes territory & Political violence \\ \cline{2-3}
            & Armed clash & Political violence \\ \hline
    Protests & Excessive force against protesters & Political violence; Demonstrations \\ \cline{2-3}
             & Protest with intervention & Political violence; Demonstrations \\ \cline{2-3}
             & Peaceful protest & Demonstrations \\ \hline
    Riots    & Violent demonstration & Demonstrations \\ \cline{2-3}
             & Mob violence & Demonstrations \\ \hline
    Explosions/Remote violence & Chemical weapon & Political violence \\ \cline{2-3}
                                & Air/drone strike & Political violence \\ \cline{2-3}
                                & Suicide bomb & Political violence \\ \cline{2-3}
                                & Shelling/artillery/missile attack & Political violence \\ \cline{2-3}
                                & Remote explosive/landmine/IED & Political violence \\ \cline{2-3}
                                & Grenade & Political violence \\ \hline
    Violence against civilians  & Sexual violence & Political violence \\ \cline{2-3}
                                & Attack & Political violence \\ \cline{2-3}
                                & Abduction/forced disappearance & Political violence \\ \hline
    Strategic developments & Agreement & Strategic developments \\ \cline{2-3}
                           & Arrests & Strategic developments \\ \cline{2-3}
                           & Change to group/activity & Strategic developments \\ \cline{2-3}
                           & Disrupted weapons use & Strategic developments \\ \cline{2-3}
                           & Headquarters or base established & Strategic developments \\ \cline{2-3}
                           & Looting/property destruction & Strategic developments \\ \cline{2-3}
                           & Non-violent transfer of territory & Strategic developments \\ \cline{2-3}
                           & Other & Strategic developments \\ \hline
    \end{tabular}
    \caption[ACLED conflict event and sub-event types]{\ac{ACLED} Event types from the Armed Conflict Location \& Event Data CODEBOOK \parencite{ACLED2024ACLED2024}} 
    \label{tab:acled-event-types}
\end{table}

For each event, \ac{ACLED} records the number of reported fatalities, providing a quantitative measure that can be used to capture different levels of intensity \parencite{Tapsoba2023TheConflict}. Importantly, these fatality counts are recorded without minimum thresholds so that an event does not need to reach a certain number of casualties to be included in the dataset \parencite{ACLED2024ACLED2024}. When reports differ or provide vague estimates, \ac{ACLED}  uses the most conservative figures, making these counts a reliable lower bound for event intensity \parencite{ACLED2024ACLED2024}. \ac{ACLED} also flags whether civilians were the main or only targets of violence through its civilian targeting indication.

The spatial attributes of events are captured through geographic coordinates along with associated precision codes ranging from 1 to 3 \parencite{ACLED2024ACLED2024}. A precision code of 1 indicates exact location information, such as a specific town \parencite{ACLED2024ACLED2024}. Code 2 signals nearby locations (e.g., near a town), while code 3 is used for larger regions where only approximate locations are known \parencite{ACLED2024ACLED2024}. This spatial uncertainty has direct implications for how force parameters should be applied, with events with lower precision codes needing either to be excluded or warrant broader force distribution to reflect uncertainty. Given these precision limitations, force application on spatial scales finer than about 5-10 km risks, implying greater location certainty that the underlying data supports and should, therefore, be avoided when using this dataset.

Building on the implementation testing results, this framework maps conflict characteristics to three key force parameters: magnitude ($F$), distribution pattern, and radius ($r$). The magnitude determines impact intensity, showing linear scaling with displacement as demonstrated in the model testing. The distribution pattern shapes how effects concentrate or disperse from the event location. The radius parameter controls the geographic extent of the impact, with testing showing clear relationships between radius size and both maximum displacement and the affected area. 

An important challenge in conflict analysis involves representing the temporal dimension of violent events. While dynamic changes in the social fabric properties—such as deteriorating infrastructure or weakening institutions—require more sophisticated time-dependent material models beyond the scope and foundational framework of this thesis, the analysis must still account for multiple events occurring at different times. As described in the literature review, existing approaches offer either aggregate events over time windows, losing temporal granularity, or analyse each in isolation, missing potential compounding effects. This framework addresses this limitation by incorporating temporal decay functions that modify force parameters based on time elapsed since each event. This allows for a static "snapshot" analysis that captures the diminishing but persistent effects of past events alongside more recent occurrences. 

For instance, a major violent event from several months ago might still contribute significant stress to the social fabric, though less intensely than when it occurred, while simultaneously, a recent smaller event adds its acute effects to this underlying strain \parencite{Mittermaier2024TheV1.1}. This superposition behaviour demonstrated in the implementation testing supports this approach, showing how multiple forces can interact to produce compound effects that reflect the reality of overlapping conflict impacts. This temporal scaling must be carefully calibrated to reflect how different types of events fade or persist in their societal impact. The following sections detail the specific mapping and mathematical formulations for each parameter.

\subsection{Force Magnitude Mapping}

The force magnitude must capture both the immediate intensity of conflict events as well as their diminishing impact over time. For any event $e$ at the time $t$,  the total force magnitude $F_e(t)$ can be expressed through Equation \ref{eq:force-magnitude}.

\begin{equation}\label{eq:force-magnitude}
    F_e(t) = F_{base}(e) \cdot I(e) \cdot T(\Delta t), \text{ with } \Delta t = t - t_e
\end{equation}

This multiplicative formulation allows independent scaling of the base event type magnitude $F_{base}(e)$ through event-specific intensity characteristics $I(e)$ and temporal decay $T(\Delta t)$. The implementation testing demonstrated that force magnitudes scale linearly with displacement while maintaining consistent distribution patterns across multiple orders of magnitude, ensuring that relative differences between event types remain meaningful regardless of the absolute magnitude chosen.

\subsubsection{1. Base Event Type Magnitude}

For the default plate parameter used in the implementation testing with a thickness of 2000 m, a force of 1e9 N produces meaningful displacement patterns of around 6 m that can be clearly visualised and analysed. Around this value, the relative magnitudes between event types can establish a hierarchy of event impacts based on their characteristics and intensity levels (e.g., establishing Explosions/Remote violence with $F_{base}(\text{Explosions/Remote violence}) =$ 1e9 N as 100 \% and then ranking all other relevant event types as a percentage of that). However, it is important to note that an intensity scaling will be applied later, so values should be chosen without accounting for casualties or whether civilians were impacted or not. 

While event types like battles are usually directly associated with conflict and can directly be mapped through $F_{base}(e)$ with additional scaling, protests, particularly peaceful protests, might not immediately appear as "conflict" events. However, they could still serve as important indicators of societal tension that may escalate into violent conflict but might require additional consideration of the local political context. As \textcite{Mittermaier2024TheV1.1} notes, protest events carry fundamentally different meanings across political systems, as they could represent normal democratic expression in one country and could signal serious societal upheaval in another.  

To account for this variation, democracy ratings like the \ac{V-Dem} liberal democracy index \parencite{Gerring2024V-DemV14} or similar measures should be incorporated when translating protest events into forces \parencite{Mittermaier2024TheV1.1}. Following \textcite{Mittermaier2024TheV1.1}'s methodology, the impact of protests is weighted by the inverse of the democracy index. This approach prevents inflation of protest significance in countries where protests and demonstrations are an integral part of the political system while appropriately scaling up their importance in contexts where public protests face significant restrictions \parencite{Mittermaier2024TheV1.1}.

The adjusted base force magnitude for protest events can be expressed through Equation \ref{eq:demonstration-adjustment}.

\begin{equation}\label{eq:demonstration-adjustment}
    F_{base_{adjusted}}(\text{Protests}) = F_{base}(\text{Protests})\cdot (1 - V_{dem})
\end{equation}

This contextual scaling is particularly important given \ac{ACLED}'s comprehensive recording of protest events, regardless of size or peaceful nature. Rather than excluding peaceful protests entirely, this approach allows them to contribute appropriately weighted stress to the social fabric model, reflecting their role as potential precursors to more serious conflict.

\subsubsection{2. Event-Specific Intensity Scaling}

The event intensity scaling $I(e)$ incorporates the number of fatalities and characteristics if civilians were targeted or the main target through Equation \ref{eq:intensity-scaling}. 

\begin{equation}\label{eq:intensity-scaling}
    I(e) = S(n_f) \cdot (1 + \gamma_c \cdot C_t)
\end{equation}

Here, $S(n_f)$ represents the fatality-based scaling function while $C_t \in {0, 1}$ indicates whether civilians were the primary target. The parameter $\gamma_c$ controls the additional impact of this civilian targeting factor. This scaling adjustment may be necessary to account for the increased impact of conflict on civilians \parencite{Idler2024ConflictViolence}, as well as the significantly higher indirect deaths, which constitute at least 75 \% of total mortality in 11 to 15 armed conflicts, according to \textcite{Bendavid2021TheChildren}. While not explicitly mentioned by \textcite{ACLED2024ACLED2024}, indirect deaths occurring with a time delay may not be recorded, as they might not clearly be associated with a specific event. This effect can be accounted for by adjusting the parameter $\gamma_c$. For example, a value of 0.75 for $\gamma_c$ , would represent a 75 \% increase in impact magnitude for events where civilians were targeted. 

The fatality scaling function $S(n_f)$ require some additional consideration. Direct linear scaling $S(n_f) = 1 + n_f$ would suggest that each additional death creates the same marginal increase in social impact. However, this might not always be intended and might not capture realistic scaling in response to violence. While a single violent death might lead to significant social trauma and disruption, an increase in deaths may not have a proportionate effect on social impact. 

Conversely, pure logarithmic scaling $S(n_f) = 1 + log(1 + n_f)$, as also used by \textcite{Mittermaier2024TheV1.1}, might better capture this marginal impact but potentially risks understating meaningful differences between larger events. Therefore, this framework suggests a combined approach that enables a more flexible and modifiable scaling option by integrating both approaches in the scaling Equation \ref{eq:fatality-scaling-function}.

\begin{equation} \label{eq:fatality-scaling-function}
    S(n_f) = \alpha_c \cdot (1 + n_f) + (1-\alpha_c) \cdot (1 + log(1+ n_f))
\end{equation}

The weighting variable $\alpha_c \in [0, 1]$ determines the balance between linear and logarithmic scaling of fatality impacts. One possible option would be to derive $\alpha_c$ also from \ac{ACLED} event types. 

This event type-based determination of $\alpha_c$ recognises that fatalities in different forms of violent events may characteristically have different patterns of societal impact, as it was already assumed for the baseline magnitudes. So, for instance, fatalities from explosions or battles might scale more linearly as each additional death might also represent a proportional increase in destruction and disruption of infrastructure and services. In contrast, fatalities during protests might show more logarithmic scaling, where initial deaths cause substantial societal shock, but additional casualties do not proportionally increase the event's broader destabilising effect.

\subsubsection{3. Temporal Decay}

The temporal decay function $T(\Delta t)$ models how event impacts diminish over time while still allowing past events to maintain persistent influence in the analysis. This addresses a key limitation of some existing approaches that aggregate events over specific time windows or analyse them in isolation. For any event, the days that passed since its occurrence $\Delta t = t - t_e$ modify its force magnitude through an exponential decay function. While an exponential decay function alone provides a mathematically convenient way to model diminishing influence, its asymptotic approach to zero rather than reaching it completely may not accurately reflect reality. There likely exists a point beyond which an event no longer meaningfully influences current societal conditions. Therefore, the modified decay function (Equation \ref{eq:temporal-decay}) cuts off events once they fall to 1 \% of their original magnitude. 

\begin{equation}\label{eq:temporal-decay}
        T(\Delta t) = 
        \begin{cases}
            \exp(-\lambda \cdot \Delta t) & \text{if } \Delta t < \frac{\ln(100)}{\lambda} \\
            0
        \end{cases}
\end{equation}

The exponential form ensures events maintain a certain influence over time, reflecting that conflict events can persist well beyond their immediate aftermath \parencite{Mittermaier2024TheV1.1, Verwimp2019TheConflict}. The decay rate $\lambda$ determines how quickly this influence diminishes, with higher values indicating faster decay. The half-life of impact (time until magnitude reduces by 50 \%) can be calculated as $\ln(2)/\lambda$, providing an intuitive way to calibrate the parameter. 

Following the same principle of differentiating event types as used for base magnitude, decay rates could be systemically assigned based on how different forms of violent events create lasting impacts. Events involving infrastructure damage or territorial change might warrant lower $\lambda$ values as their effects persist through altered physical and institutional landscapes. Events primarily creating social disruption might decay faster while still maintaining meaningful medium-term influence. For example, given $\lambda = 0.003$ per day, the half-life would be approximately 231 days ($\ln(2)/0.003$), and the complete cut-off would occur when $\Delta t = \ln(100)/0.003 \approx 1,535$ days, or approximately in 4.2 years.

\subsection{Force Distribution Selection}

The distribution pattern determines how an event's impact concentrates or disperses from its recorded location. The implementation testing demonstrated four distinct options: point force (no distribution), Gaussian, linear, and constant distributions. A point force represents the most concentrated application, creating the highest local displacement as the entire force magnitude acts on a single point. The other distribute the same total force magnitude over an area according to their characteristic profiles, necessarily reducing the maximum displacement but affecting broader regions. Gaussian distributions produce relatively high central displacement with a small affected area, linear distributions show intermediate behaviour, while constant distributions create the lowest maximum displacement but affect the largest area.

The most appropriate distribution pattern can be determined primarily at the sub-event type level in the case of \ac{ACLED}'s dataset, as these seem to capture different forms of potential spread in greater detail. For instance, within the Explosions/Remote violence category, air strikes and suicide bombs might create distinctly different spatial impact patterns than artillery attacks despite falling under the same primary event type. Further modification could also be based on specific contextual information or more detailed event information, as, for example, provided in \ac{ACLED}'s notes field.

\subsubsection{Option 1: Point Force}

Point forces might be appropriate for highly localised events that affect specific individuals or very small groups, such as targeted violence against a single person. However, they should be used carefully, as location precision limitations of the conflict data mean that assuming perfectly localised point impacts might imply greater spatial precision than the data supports.

\subsubsection{Option 2: Gaussian}
 
Gaussian distributions provide an appropriate pattern for focused events while still capturing some spread of impacts. This might particularly apply to precision strikes (e.g., air/drone strikes, targeted bombings) and deliberately targeted violence, where impacts concentrate strongly at a recorded point before rapidly diminishing. The Gaussian shape allows, therefore, for appropriately high central displacement while acknowledging spatial spread. However, it is important to note that spreading should not account for the indirect and spillover effects, as these will be determined through the interplay between the force and the fabric.

\subsubsection{Option 3: Linear}

Linear distributions create a moderate central peak that gradually declines with distance, making them more suitable for events with a clear epicentre but also substantial direct effects in surrounding areas. This might apply particularly to battles, where while the most intense fighting may concentrate in specific locations, surrounding communities might experience significant direct impacts like displacement and restricted movement as well.

\subsubsection{Option 4: Constant}

A constant distribution is most suitable for events that create widespread but relatively uniform effects across a large region. For example, protests might exert societal pressure that affects entire areas in a relatively uniform manner. Another reason to use a constant distribution might be spatial uncertainty, which may arise from limitations in data or the nature of the event itself. For instance, while an attack occurs at a specific location, protests and demonstrations might be dynamic, even though they are recorded at a fixed point.

\subsection{Force Radius Determination}

The radius parameter $r$ determines the geographic extent over which an event's impact is spread in combination with the selected distribution. Implementation testing demonstrated clear relationships between radius size and impact characteristics as well. Larger radii increase the affected areas but reduce the maximum displacement by distributing the force over broader regions. Similar to the force magnitude, the total radius $r_e(t)$ for any event $e$ at the time $t$, can be expressed through Equation \ref{eq:force-radius}.

\begin{equation}\label{eq:force-radius}
    r_e(t) = r_{base}(e) \cdot E(\Delta t), \text{ with } \Delta t = t - t_e
\end{equation}

\subsubsection{1. Base Radius}

In this equation $r_{base}(e)$ represent the base radius, which can also be determined by event characteristics and their specific general or empirically validated spatial expansion. For explosions and remote violence, base radii could correspond with typical weapon ranges or blast radii, while protests and riots might use radii reflecting typical gathering patterns. While research like \textcite{Wagner2018ArmedAnalysis} revealed impacts up to 100 km ($\approx$ 50 km radius) from conflict sites, it is important to note that only the direct observable spatial extent should be expressed through this parameter, not potential spillovers or indirect impacts.

\subsubsection{2. Temporal Expansion}

$E(\Delta t)$, defined in Equation \ref{eq:radius-expansion}, allows to model an increasing impact zone over time to account for the spread of delayed indirect effects, news distribution and potential spreading of fear \parencite{Tapsoba2023TheConflict}. 

\begin{equation}\label{eq:radius-expansion}
    E(\Delta t) = 1 + \beta_c \cdot (1 - \exp(-\mu \cdot \Delta t))
\end{equation}

It also uses an exponential function, however, with an increasing instead of a decreasing influence. $\beta_c$ controls the maximum radius expansion, which sets the limit of the maximum extension as a percentage of the initial radius (e.g., $\beta_c = 1.0$ would mean a 100 \% increase or doubling of the initial $r_{base}(e)$ radius. $\mu$ determines the expansion rate and can be calibrated with the same term as the temporal decay parameter $\lambda$ using $\ln(2)/\mu$. However, here, it marks the time for the expansion to reach half of its maximum values. This expansion rate could also be set based on event types, with highly dynamic events potentially showing faster expansion than more localised incidents. The approximate time after which the radius reached nearly its full expansion can also be calculated using $4\ln(10)/\mu$.


\section{Summary of the Translation Framework}

\begin{figure}[H]
    \centering
    \includegraphics[width=0.82\linewidth]{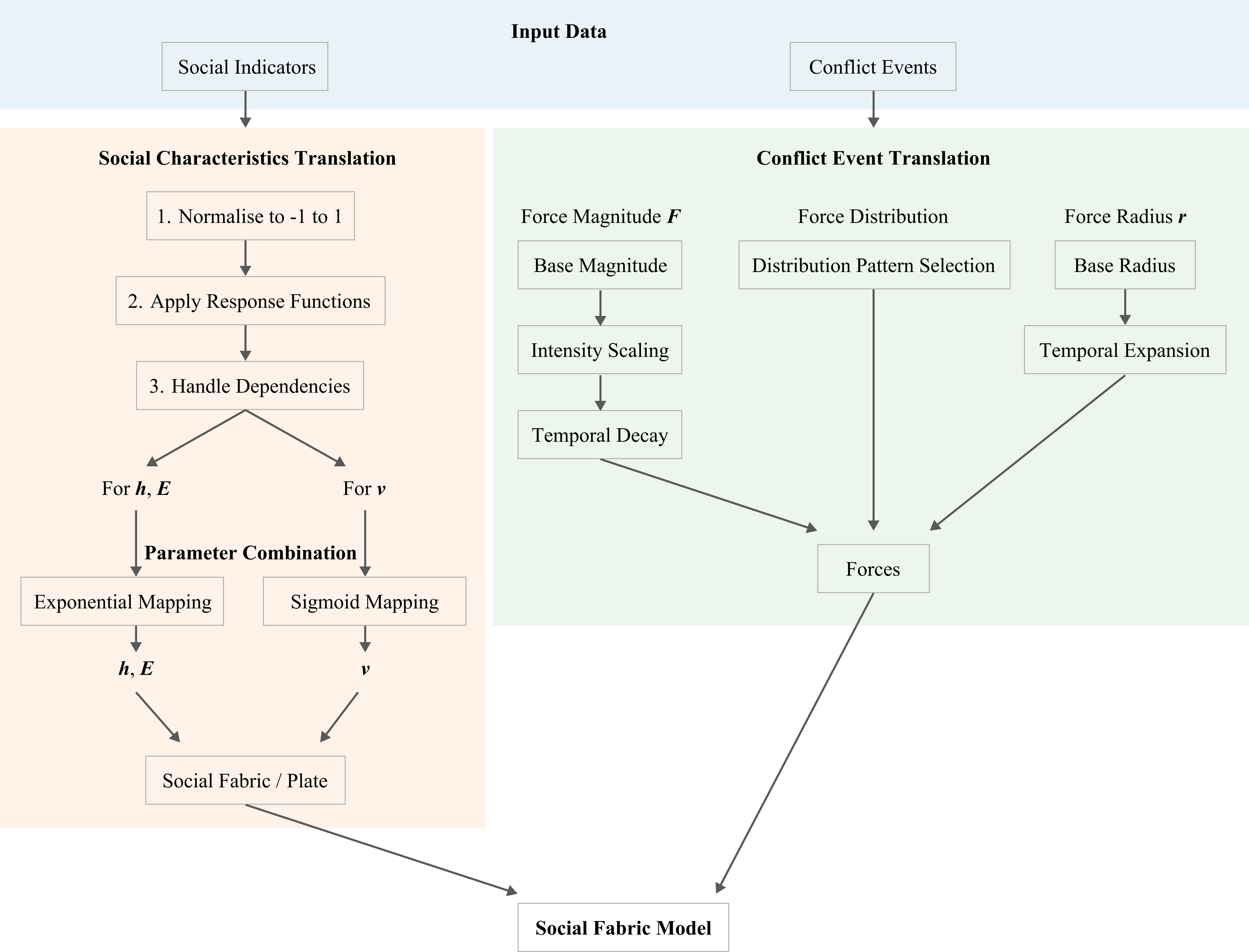}
    \caption[Visualisation of the complete translation workflow for social indicators and conflict events]{Visualisation of the complete translation workflow for social indicators and conflict events.}
    \label{fig:translation-framework-workflow-visualisation}
\end{figure}

This framework provides a systematic approach for translating social and conflict characteristics into physical parameters that can then be analysed with the proposed physics-based "social fabric" model. It establishes a flexible foundation that allows domain experts to determine specific mappings and parameters while ensuring meaningful combinations. The translation process handles two main types of inputs: social characteristics that define the material properties of the modelled fabric and conflict events that are transformed into forces acting upon this fabric. Both translations follow structured approaches that preserve physical validity while capturing a variety of social and conflict dynamics. A full overview of these processes can be seen in Figure \ref{fig:translation-framework-workflow-visualisation}. The following chapter demonstrates its practical application through a proof-of-concept of Nigeria, showing how social indicators and conflict events can be mapped to create a comprehensive analysis of conflict impact. 

\chapter{Proof-of-Concept Application}

This chapter presents a proof-of-concept application demonstrating how the translation framework can be implemented in practice. Before diving into the technical details, it is valuable to consider how this framework might transform traditional conflict analysis approaches. When analysing conflict impacts in a region like Nigeria during 2018, conventional methods often rely on separate analyses of conflict events and socioeconomic factors. Research might plot event locations, create impact buffers, and qualitatively assess how these interact with local conditions—a process that can miss crucial interaction effects. 

The physics-based approach enables an integrated analysis approach instead. It begins by constructing a "social fabric" that represents the region's underlying characteristics. Infrastructure quality and institutional strength, for example, become "material properties" that determine how well different areas absorb impacts. Economic resources might influence immediate response capabilities, while social connectivity shapes how effects propagate through communities. This foundation then interacts with conflict events, which are transformed into forces with varying intensities and distributions. Major battles or explosions may create stronger forces, while different forms of violence generate distinct impact patterns. As the forces interact with the social fabric, regions with stronger underlying properties show great resistance to impacts, while areas with compounding vulnerabilities or conflicts may experience amplified stress. 

Before proceeding with the implementation, several important considerations must be addressed. While not demonstrated in this proof-of-concept application, the framework requires careful preprocessing of the input data, including an analysis of correlations between indicators to avoid double-counting effects. In cases with numerous potential indicators, dimension reduction techniques may be necessary to identify the most influential factors while maintaining model interpretability. Furthermore, the spatial and temporal resolution of different datasets should be harmonised to ensure consistency.

It is critical to emphasise that the results presented in this chapter serve solely illustrative purposes and should not be used for actual conflict analysis. The datasets or types of data used were selected based on their frequent mention in conflict analysis literature and their public availability. Whereas they demonstrate the framework's capabilities, proper implementation would require rigorous validation against empirical data, expert calibration of parameters, and careful consideration of the local context. While all data are processed and clipped to Nigeria's geographic extent, no interpolation or extrapolation is performed to compensate for temporal differences between datasets and possible collinearities are neglected. This limitation means that indicators from different years are treated as concurrent, which would require careful consideration in actual analysis. For conflict events, georeferenced data from \ac{ACLED} is used to demonstrate force parameter mapping.


\section{Social Indicator Mapping - Demonstration}

Following the indicator selection and classification descriptions in the previous chapter, social indicators were assigned to physical parameters based on their behavioural characteristics. Infrastructure, environmental, and demographic indicators were mapped to thickness, reflecting their role in determining baseline resilience and vulnerability. Economic measures were translated to Young's modulus to capture stress response capacity, while connectivity metrics informed Poisson's ratio to model impact propagation patterns. The following sections detail this translation process. A comprehensive summary table of all social indicators and their mapping parameters can be found in Appendix \ref{app:social-indicator-mapping-parameters}.

\subsection{Thickness Mapping}

The thickness parameter fundamentally determines how the modelled social fabric responds to and absorbs impacts. Four indicators are selected to illustrate this baseline resilience: critical infrastructure, environmental stress, health systems and demographic dependency.

\subsubsection{Critical Infrastructure Spatial Index}

\begin{figure}[H]
    \centering
    \includegraphics[width=1\linewidth]{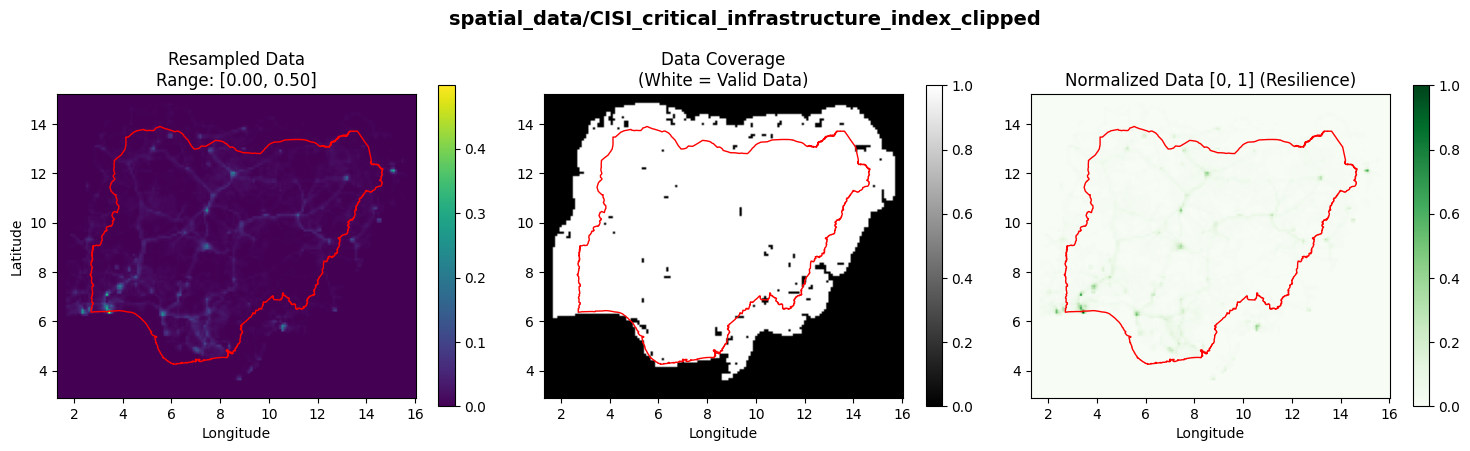}
    \caption[Critical Infrastructure Spatial Index - indicator data]{\ac{CISI} data: (left) resampled data, (centre) data coverage showing valid data in white, and (right) normalised data scaled between 0 and 1 representing resilience.}
    \label{fig:cisi-mapping}
\end{figure}

The \ac{CISI} from \textcite{Nirandjan2022AInfrastructure} provides a composite measure of infrastructure density and distribution. This aligns with findings from \textcite{Cutter2008ADisasters, Norris2007CommunityReadiness} about how critical physical infrastructure creates fundamental absorption capacity against shocks. The index ranges from 0 to 1, where 1 represents the highest critical infrastructure density (Figure \ref{fig:cisi-mapping}). 

Assuming that a higher density of critical infrastructure has an increasingly positive effect on resilience, this indicator will be mapped only to the resilience range. To model a substantial increasing effect in the beginning with diminishing improvement when approaching the highest density values, the power law response function is selected with $\gamma_r = 0.5$. 

\subsubsection{Standard Precipitation Index}

The \ac{SPI} from \textcite{Funk2014AMonitoring} capture environmental stress through rainfall patterns. This reflects findings from \textcite{Cappelli2024LocalAfrica, Hanze2024WhenDisasters, Stojetz2024ShockingNigeria} about how environmental conditions create compound vulnerabilities in conflict zones. The \ac{SPI} ranges from -2 (extreme drought) to +2 (extreme wet), with 0 representing normal conditions. 

\begin{figure}[H]
    \centering
    \includegraphics[width=1\linewidth]{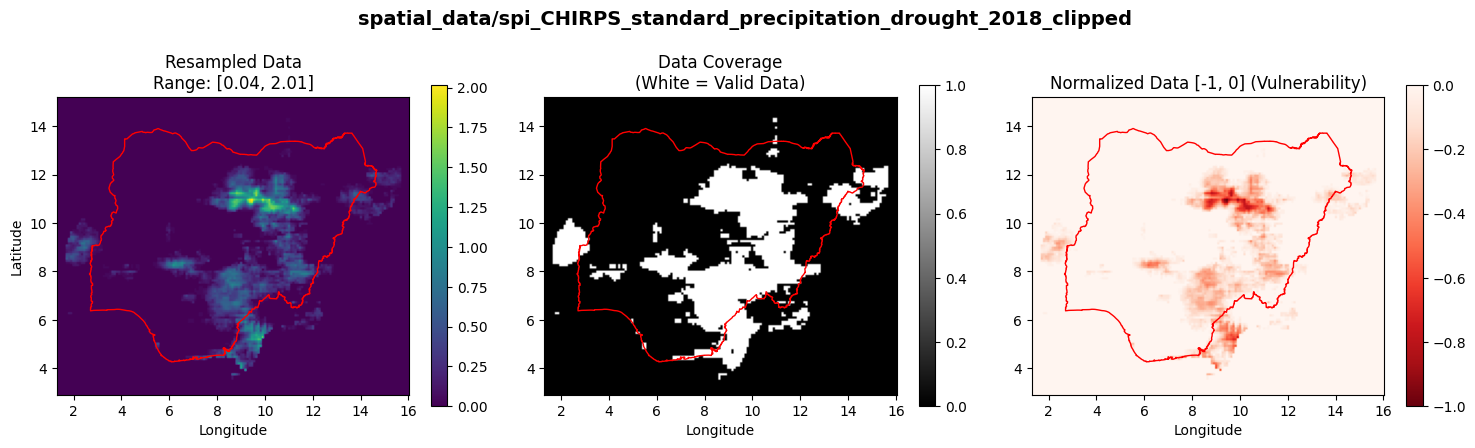}
    \caption[Standard Participation Index - drought - indicator data]{\ac{SPI} drought data: (left) resampled data, (centre) data coverage showing valid data in white, and (right) normalised data scaled between 0 and -1 representing vulnerability.}
    \label{fig:SPI-drought-mapping}
\end{figure}

\begin{figure}[H]
    \centering
    \includegraphics[width=1\linewidth]{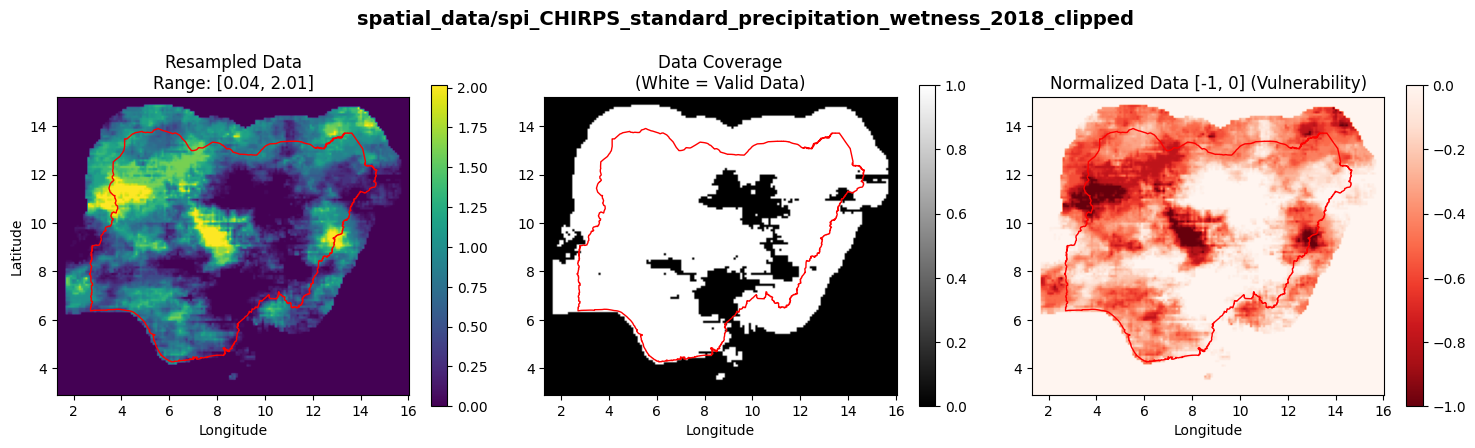}
    \caption[Standard Participation Index - wetness - indicator data]{\ac{SPI} wetness data: (left) resampled data, (centre) data coverage showing valid data in white, and (right) normalised data scaled between 0 and -1 representing vulnerability.}
    \label{fig:SPI-wetness-mapping}
\end{figure}

The original data was split during the preprocessing of the files, allowing for the mapping of both regimes—wetness and dryness—as vulnerability indicators. Therefore, the standard normalisation can be applied (Figure \ref{fig:SPI-drought-mapping}, and Figure \ref{fig:SPI-wetness-mapping}). The power law response function is used for both indicator regimes, with $\gamma_v = 1.2$ based on the assumption that moderate rainfall or drought might be manageable, but extreme conditions trigger a rapid increase in vulnerability. 


\subsubsection{Health Infrastructure Density}

This health infrastructure indicator is derived from the \ac{CISI} by \textcite{Nirandjan2022AInfrastructure} and focuses on the spatial distribution and density of healthcare facilities. Because it is also part of the \ac{CISI}, there is inevitably some overlap with other indicators. However, as previously noted, this correlation is disregarded for simplicity. As highlighted by \textcite{Bendavid2021TheChildren, Kadir2018TheChildren}, healthcare infrastructure is crucial in determining how communities respond to conflict-related injuries, diseases, and long-term health impacts. 

To demonstrate dependency behaviour, it is assumed that the effectiveness of health infrastructure depends on its accessibility. For that, the walking-only travel time to healthcare facilities dataset from \textcite{Weiss2020GlobalFacilities} is used, which measures the access in minutes. If this walking-time indicator reaches values above 0.03, equivalent to approximately 30 minutes of walking after normalisation, the health infrastructure's effectiveness reduces to 40 \% of its potential impact, reflecting how even well-equipped facilities might provide limited benefit if they cannot be reached in time.

\begin{figure}[H]
    \centering
    \includegraphics[width=1\linewidth]{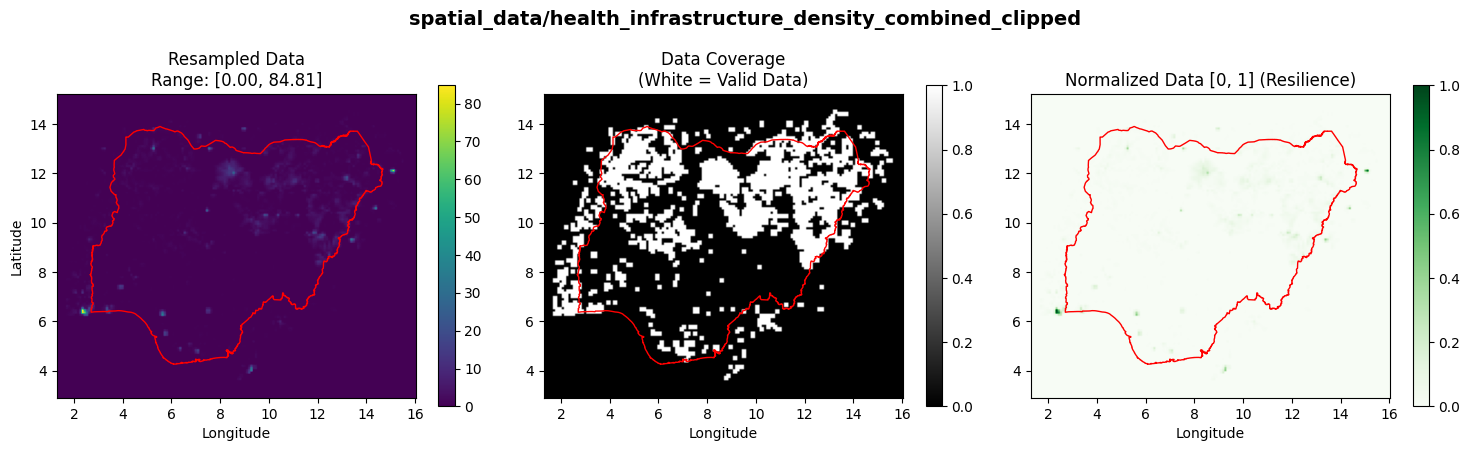}
    \caption[Healthcare infrastructure density - indicator data]{Healthcare infrastructure density data: (left) resampled data, (centre) data coverage showing valid data in white, and (right) normalised data scaled between 0 and 1 representing resilience.}
    \label{fig:healthcare-infrastructure}
\end{figure}

\begin{figure}[H]
    \centering
    \includegraphics[width=1\linewidth]{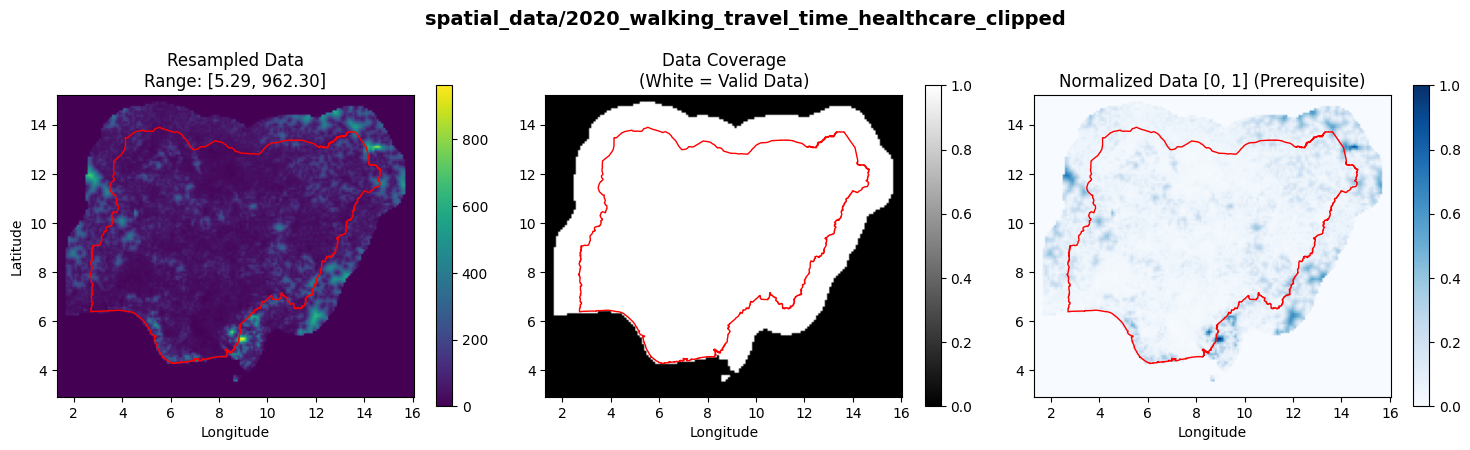}
    \caption[Walking-only travel time to healthcare facilities - indicator data]{Walking-only travel time to healthcare facilities data: (left) resampled data, (centre) data coverage showing valid data in white, and (right) normalised data scaled between 0 and 1 representing the prerequisite for the healthcare infrastructure resilience effect.}
    \label{fig:healthcare-travel-time}
\end{figure}

Both indicators are normalised to a range [0, 1], while the health indicator is treated as a resilience influence and the travel time solely as a prerequisite (Figure \ref{fig:healthcare-infrastructure}, and Figure \ref{fig:healthcare-travel-time}). For the health infrastructure index, the logarithmic response function is selected with $\alpha_r = 2.5$ capturing how initial healthcare capacity creates substantial resilience, but additional facilities show diminishing returns once basic coverage is achieved. The prerequisite walking-only travel time to healthcare facilities is modelled as a linear effect with $m_p = 1.0$.


\subsubsection{Dependency Ratio}

The dependency ratio from \textcite{WorldPop2016WorldPopAfrica} measures the proportion of the population aged 0-14 and over 65 relative to the working-age population. This demographic indicator was included under the assumption that it reflects the heightened vulnerabilities of these age groups, who may face additional challenges during evacuation and might require greater medical support in crisis situations \parencite{Amberg2023ExaminingAnalysis, Bendavid2021TheChildren, Wagner2018ArmedAnalysis}. 

\begin{figure}[H]
    \centering
    \includegraphics[width=1\linewidth]{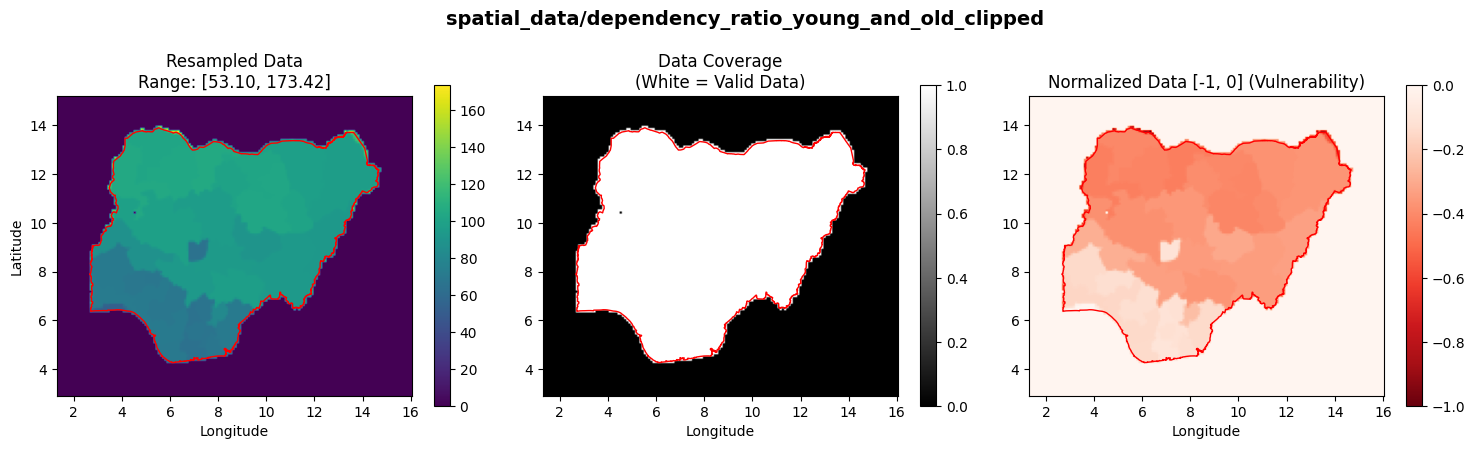}
    \caption[Population ages 0-14 and over 65, relative to the working-age population - indicator data]{Population ages 0-14 and over 65, relative to the working-age population percentage: (left) resampled data, (centre) data coverage showing valid data in white, and (right) normalised data scaled between 0 and -1 representing vulnerability.}
    \label{fig:dependency-ratio}
\end{figure}

Assuming that these demographic dependencies create accelerating strain on community resources and possibly hinder certain response measures, an exponential function is selected with $\beta_v = 0.8$. This captures how communities might manage moderate dependency ratios but face rapidly mounting challenges as the proportion of the dependent population grows (Figure \ref{fig:dependency-ratio}). 

\subsubsection{Parameter Combination - Thickness}

Using the exponential mapping, which will automatically be used for the thickness effects combination in the Python implementation, these four indicators are combined to determine the final thickness distribution. The weighting factors are based on assumptions of their importance in conflict resilience, as identified in the literature review. 

Healthcare infrastructure (as resilience) is assigned the highest weight 0.3 due to the assumption that it has the greatest influence on baseline conditions with respect to potential conflict impacts \parencite{Cutter2008ADisasters, Norris2007CommunityReadiness}. Environmental stresses of drought and wetness (as vulnerabilities) are assigned weights of 0.2 each to account for the increased pre-existing vulnerabilities caused by climate-related conditions \parencite{Cappelli2024LocalAfrica, Caso2023The2018}. General critical infrastructure (as resilience) and demographic dependencies (as vulnerability) were assigned lower weights (0.15 each) subsequently.   

\begin{equation}\label{eq:combined-thickness}
    h_{combined} = h_{base} \cdot \exp(0.3 \cdot f_{health} + 0.2 \cdot f_{spi-drought} + 0.2 \cdot f_{spi-wetness} + 0.15 \cdot f_{csis} + 0.15 \cdot f_{dependency})
\end{equation}

A baseline thickness of $h_{base} = 2,500$ m is selected based on the implementation testing results, which showed this value provides good sensitivity to both enhancement and reduction effects for force in the range magnitude 1e9 N. 

\begin{figure}[H]
    \centering
    \includegraphics[width=0.8\linewidth]{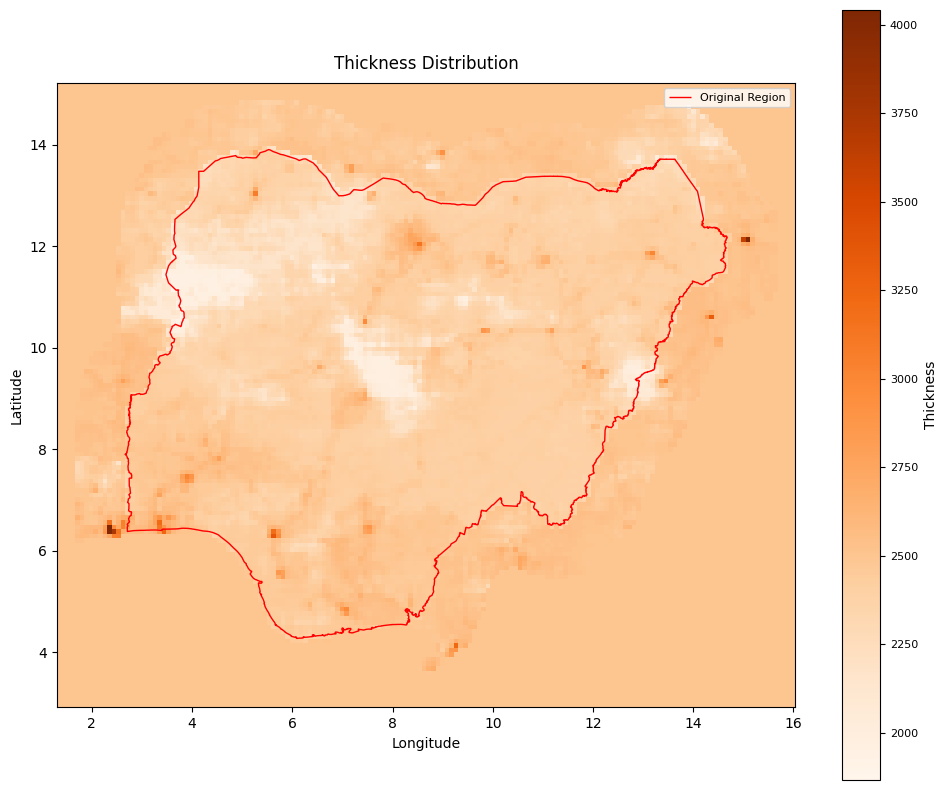}
    \caption[Combined thickness distribution across Nigeria]{Spatial distribution of the combined thickness $h$, derived from the weighted effects of various indicators. Values range from approximately 2,000 m (lower resilience) to 4,000 m (higher resilience).}
    \label{fig:combined-thickness-distribution}
\end{figure}

The final combined thickness distribution (Figure \ref{fig:combined-thickness-distribution}) shows notable variations between approximately 2,000 m and 4,000 m across the country. Areas with stronger healthcare infrastructure and lower environmental stress exhibit higher thickness values, particularly visible in parts of the southwest. In contrast, regions with higher dependency ratios and environmental vulnerabilities show reduced thickness, indicating lower baseline resilience to conflict impacts.

\subsection{Young's Modulus Mapping}

Young's modulus controls the magnitude of displacement response without altering distribution patterns by making it more elastic or stiffer. Three indicators are selected to illustrate different economic dimensions altering the response capabilities: \ac{GDP}, poverty rates, and childhood poverty.

\subsubsection{Gross Domestic Product}

The gridded \ac{GDP} \ac{PPP} data from \textcite{Kummu2018Gridded1990-2015} provides spatially disaggregated economic output measures. As \textcite{Cappelli2024LocalAfrica, Cutter2008ADisasters, Norris2007CommunityReadiness} indicate, economic resources determine a community's capacity to resist and recover from impacts. 

\begin{figure}[H]
    \centering
    \includegraphics[width=1\linewidth]{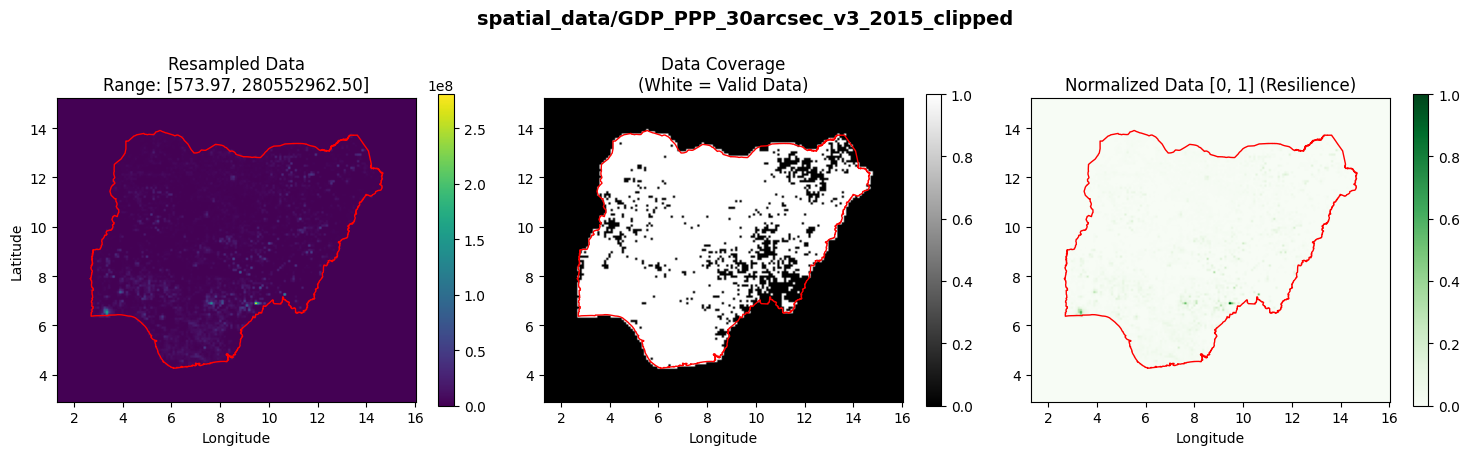}
    \caption[Gross Domestic Product Purchasing Power Parity - indicator data]{\ac{GDP} \ac{PPP}:  (left) resampled data, (centre) data coverage showing valid data in white, and (right) normalised data scaled between 0 and 1 representing resilience.}
    \label{fig:gdp-ppp}
\end{figure}

The \ac{GDP} values are normalised to [0, 1], treating it purely as a resilience indicator (Figure \ref{fig:gdp-ppp}). Assuming that economic resources show linear relationships with response capabilities, the linear response function is selected with $m_r = 1.0$. Therefore, additional economic resources create proportional increases in the resilience of the social fabric in this case. 

\subsubsection{Poverty Rate}

The poverty rate from \textcite{Tatem2013WorldPopPoverty} measures the proportion of people living on less than \$2 per day. This captures an important vulnerability dimension, as \textcite{Kaila2023TheNigeria, Le2022TheCountries, Norris2007CommunityReadiness} note that poor households have few coping strategies and often experience the largest setbacks. 

\begin{figure}[H]
    \centering
    \includegraphics[width=1\linewidth]{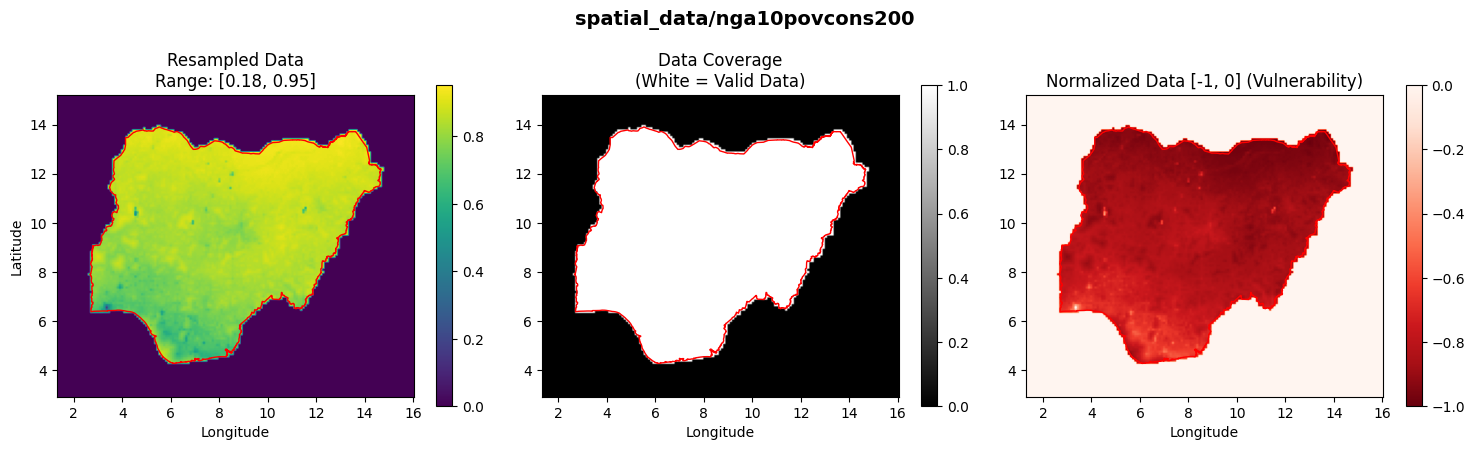}
    \caption[Proportion of people living on less than \$2 per day - indicator data]{Proportion of people living on less than \$2 per day: (left) resampled data, (centre) data coverage showing valid data in white, and (right) normalised data scaled between 0 and -1 representing vulnerability.}
    \label{fig:poverty-rates}
\end{figure}

With proportions ranging from 0 to 1, the indicator is treated as a pure vulnerability measure, with higher proportions indicating greater vulnerability (Figure \ref{fig:poverty-rates}). The power law response function with $\gamma_v = 1.5$ is selected, assuming that the vulnerability effects accelerate as poverty becomes more prevalent. 

\subsubsection{Childhood Poverty}

The proportion of children aged 12-23 months born to the poorest households from \textcite{Utazi2023ACountries} provides an additional poverty measure focused on children. As \textcite{Le2022TheCountries} specifically indicates, children from relatively poor families face especially heightened risks during conflicts. 

\begin{figure}[H]
    \centering
    \includegraphics[width=1\linewidth]{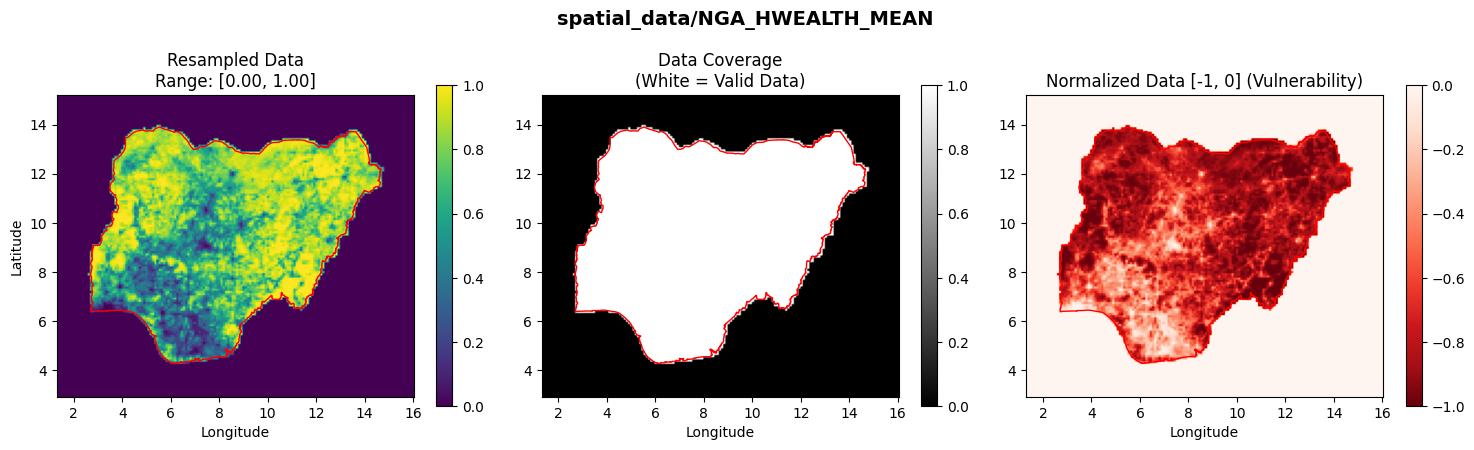}
    \caption[Proportion of children aged 12-23 months born to the poorest households - indicator data]{Proportion of children aged 12-23 months born to the poorest households: (left) resampled data, (centre) data coverage showing valid data in white, and (right) normalised data scaled between 0 and -1 representing vulnerability.}
    \label{fig:childhood-poverty}
\end{figure}

Like general poverty rates, this indicator is treated as a pure vulnerability measure (Figure \ref{fig:childhood-poverty}). The power law response function with $\gamma_v = 0.5$ creates strong initial vulnerability effects from the beginning. This assumes that even modest levels of child poverty significantly compromise community resilience, as children represent a particularly culinary population during conflicts \parencite{Le2022TheCountries}. 

\subsubsection{Parameter Combination - Young's Modulus}

Using the exponential mapping for Young's modulus effects combination, the three economic indicators are combined based on their roles as either resilience or vulnerability factors. \ac{GDP} (as resilience) is assigned the highest weight of 0.6, assuming that economic resources represent the primary capacity for immediate stress response \parencite{Cappelli2024LocalAfrica}. The general poverty rate (as vulnerability) and childhood poverty (as vulnerability) are both assigned weights of 0.2 each due to their roles in reflecting household-level response capabilities \parencite{Kaila2023TheNigeria, Le2022TheCountries, Norris2007CommunityReadiness}.  

\begin{equation}\label{eq:combined-youngs-modulus}
    E_{combined} = E_{base} \cdot \exp(0.6 \cdot f_{gdp} + 0.2 \cdot f_{pov} + 0.2 \cdot f_{child-pov})
\end{equation}

A baseline Young's modulus of $E_{base} = 5e9$ Pa is selected based on implementation testing results, which demonstrated good sensitivity to both enhancement and reduction at this value. 

\begin{figure}[H]
    \centering
    \includegraphics[width=0.8\linewidth]{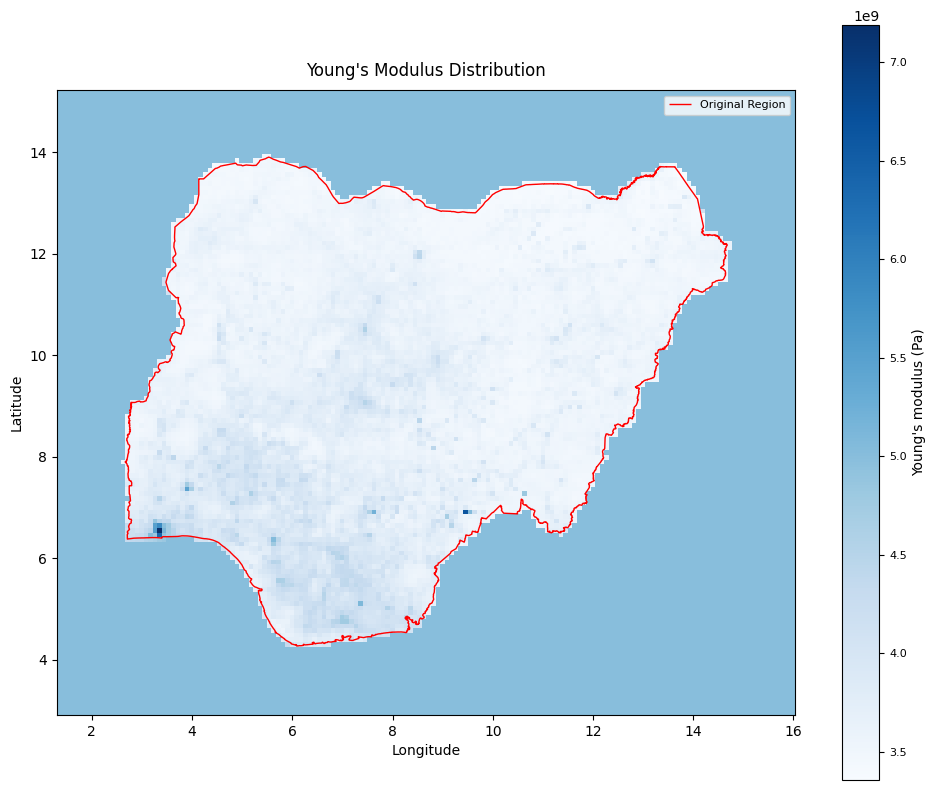}
    \caption[Combined Young's modulus distribution across Nigeria]{Spatial distribution of Young's modulus $E$, derived from the weighted effects of various indicators. Values range from approximately 3.5e9 Pa (higher sensitivity) to 7.0e9 Pa (lower sensitivity).}
    \label{fig:combined-youngs-modulus-distribution}
\end{figure}

The final combined Young's modulus distribution (Figure \ref{fig:combined-youngs-modulus-distribution}) ranges from approximately 3.5e9 Pa to 7.0e9 Pa, with clear spatial heterogeneity. Urban centres and areas of higher economic activity displace elevated values, suggesting an enhanced ability to resist deformation under stress. The influence of poverty indicators is particularly visible in rural regions, where lower Young's modulus values indicate potentially higher sensitivity to conflict impacts. 

\subsection{Poisson's Ratio Mapping}

Poisson's ratio influences how impacts propagate, making it suitable for connectivity indicators that determine distribution and spread patterns, either of information, fear or resources. Two indicators are selected to illustrate different aspects of this connectivity: population density and road infrastructure

\subsubsection{Population Density}

The \ac{UN}-adjusted population density data from \textcite{WorldPop2020WorldPopAdjusted} captures the spatial distribution of people. Assuming that fear and conflict effects are spreading through social networks and community ties, a denser populated area might experience faster effect propagation than less densely populated areas \parencite{Tapsoba2023TheConflict}. 

\begin{figure}[H]
     \centering
     \includegraphics[width=1\linewidth]{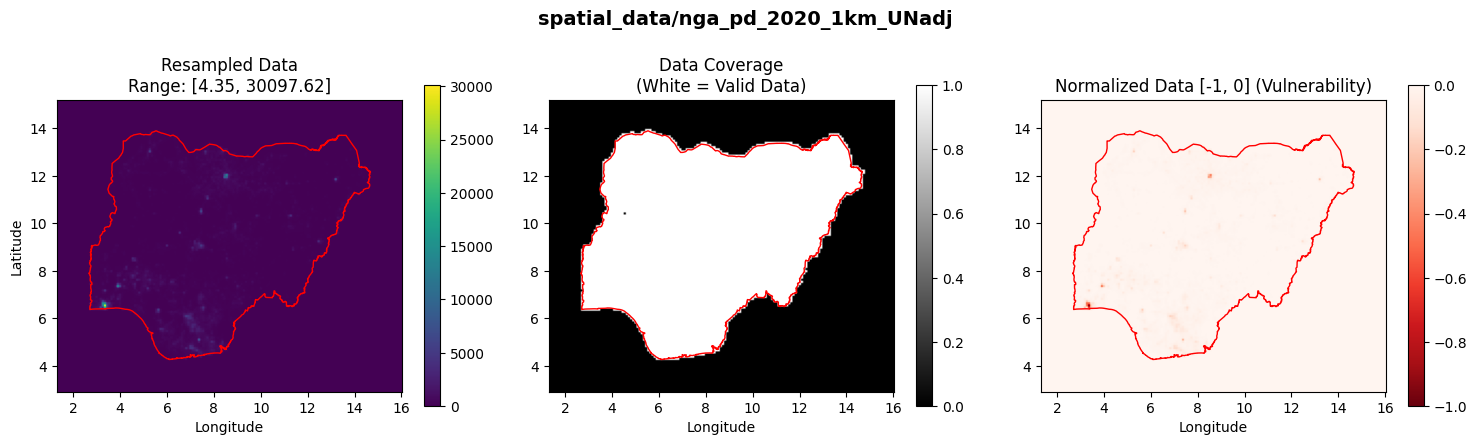}
     \caption[United Nations-adjusted population density - indicator data]{\ac{UN}-adjusted population density data: (left) resampled data, (centre) data coverage showing valid data in white, and (right) normalised data scaled between 0 and -1 representing vulnerability.}
     \label{fig:population-density}
 \end{figure}

Population density values are normalised and treated as a vulnerability factor (Figure \ref{fig:population-density}). The linear response function with $m_v$ = 2 is selected, assuming that increasing population density created proportionally stronger pathways for impact propagation through social networks. 

\subsubsection{Road Density}

Road density from \textcite{Meijer2018GlobalInfrastructure} measures the spatial concentration of all road types. This physical connectivity indicator is used assuming that transportation networks can enhance community resilience by facilitating resource distribution and maintaining critical connections during conflicts. 

\begin{figure}[H]
    \centering
    \includegraphics[width=1\linewidth]{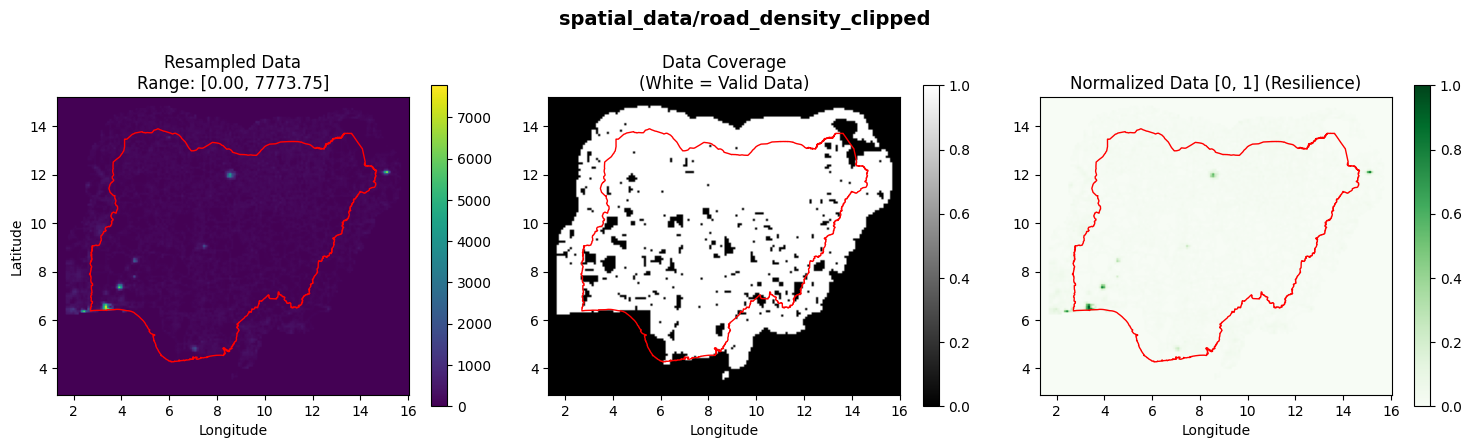}
    \caption[Road density of all road types - indicator data]{Road density of all road types: (left) resampled data, (centre) data coverage showing valid data in white, and (right) normalised data scaled between 0 and 1 representing resilience.}
    \label{fig:road-density}
\end{figure}

The density values are normalised and affect the resilience (Figure \ref{fig:road-density}). The logarithmic response function with  $\alpha_r = 2$ is chosen under the premise that initial road connection might substantially enhance resource distribution capabilities, while additional roads yield diminishing returns in terms of further resilience improvements. 

\subsubsection{Parameter Combination - Poisson's Ratio}

The two connectivity indicators are combined using the sigmoid-based mapping approach. Population density (as vulnerability) receives a weight of 0.6, assuming a stronger role because of its representation of social connectivity and the spread of fear \parencite{Tapsoba2023TheConflict}. Road infrastructure (as resilience) is weighted at 0.4, capturing the physical connectivity that might be necessary for resource distribution and emergency responses. 

\begin{equation}\label{eq:combined-poissons-ratio}
    \nu_{combined} = \frac{0.5}{1 + \exp(2.5 \cdot (0.6 \cdot f_{pop} + 0.4 \cdot f_{road}))}
\end{equation}

The scaling parameter was set to $k = 2.5$, to ensure appropriate distribution over approximately the full range of the Poisson's ratio of [0, 0.5). 

\begin{figure}[H]
    \centering
    \includegraphics[width=0.8\linewidth]{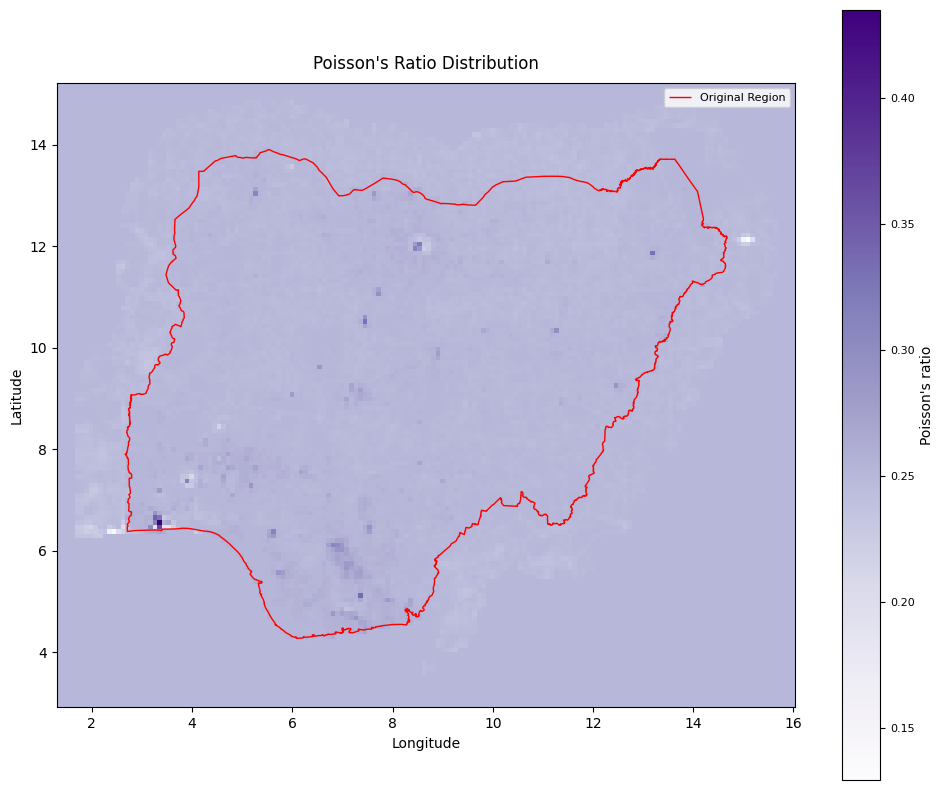}
    \caption[Combined Poisson's ratio distribution across Nigeria]{Spatial distribution of Poisson's ratio $\nu$, derived from the weighted effects of various indicators. Values range from approximately 0.15 (more localised impact propagation) to 0.4 (stronger impact coupling).}
    \label{fig:combined-poissons-ratio-distribution}
\end{figure}

The final combined Poisson's ratio distribution (Figure \ref{fig:combined-poissons-ratio-distribution}) varies between approximately 0.15 and 0.4, capturing the interplay between population density and road infrastructure. Higher values in densely populated areas suggest increased potential for impact propagation, while better-connected but less populated regions show lower values, indicating more localised response patterns. This distribution particularly highlights the urban-rural divide in terms of impact propagation potential \parencite{Amberg2023ExaminingAnalysis, Bendavid2021TheChildren}. 

\section{Conflict Event Mapping - Demonstration}

To demonstrate the translation of conflict events into forces, this proof-of-concept uses \ac{ACLED} data for Nigeria from 2018. This year is selected as it represents approximately the average of all the years covered by the social indicator datasets. 

As indicated in the previous section \ac{ACLED} provides a spatial precision code ranging from 1 to 3 for each event, where 1 indicates that the exact location is known, 2 signals nearby locations, and 3 represent larger regions \parencite{ACLED2024ACLED2024}. Given the importance of spatial precision for force application, only events with precision code 1 are included in this demonstration. For Nigeria in 2018, \ac{ACLED} recorded 2,050 events in total. Of these, 1,307 events (63.76 \%) had precision code 1, while 687 events (33.51 \%) had precision code 2, and 56 events (2.73 \%) had precision code 3.

Strategic developments, accounting for 117 events (5.71 \%), are excluded from the analysis as they capture primarily contextual information, with events, such as peace agreements, arrests, or base establishments, representing shifts in political dynamics rather than direct sources of societal stress that could be meaningfully translated into physical forces \parencite{ACLED2024ACLED2024}. The distribution of events and sub-events in the remaining data is shown in Appendix \ref{app:acled-data-distribution-analysis}.

\subsection{Force Magnitude Mapping}

The complete force magnitude parameters are summarised in Table \ref{tab:force-magnitude-mapping}. Similar to the social indicator mappings, these parameter choices are primarily based on assumptions and insights from the literature review rather than empirical validation for demonstration purposes. 

\begin{table}[H]
    \centering
   \begin{tabular}{>{\raggedright\arraybackslash}p{170pt}|>{\raggedright\arraybackslash}p{100pt}|>{\raggedright\arraybackslash}p{80pt}|>{\raggedright\arraybackslash}p{80pt}} \hline
         \textbf{Event Type}&\textbf{$F_{base}$  [N]}&\textbf{$\alpha_c$}&\textbf{$\lambda$}\\ \hline \hline
 \textbf{Explosions/Remote violence}& 1.0e9 N (reference)& 0.4& 0.0064\\ \hline
  \textbf{Battles}& 8.0e8 N (80 \%)&0.4& 0.0064\\ \hline 
  \textbf{Violence against civilians}& 6.0e8 N (60 \%)&0.2& 0.0128\\\hline 
         \textbf{Riots}&  3.0e8 N (30 \%)&0.3&  0.0256\\\hline
 \textbf{Protests}& 1.0e8 N (10 \%)*&0.3& 0.0256\\ \hline
    \end{tabular}
    \caption[Force magnitude parameters for different conflict event types]{Force magnitude parameters for different conflict types. Base magnitudes ($F_{base}$) are shown as absolute values and percentages of the reference event type (Explosions/Remote violence). The fatality scaling balance ($\alpha_c$) determines the mix of linear and logarithmic fatality effects, while the temporal decay rate ($\lambda$) controls how quickly event impacts diminish. *The protest base magnitude is further adjusted by the \ac{V-Dem} Liberal Democracy Index, where the final magnitude is calculated as $F_{base} \cdot (1-V_{dem})$ to account for varying political contexts.}
    \label{tab:force-magnitude-mapping}
\end{table}

\subsubsection{Base Event Type Magnitude Mapping}

Explosions/Remote violence serves as the reference category ($F_{base}(\text{Explosions/Remote violence}) = 1.0e9$ N) due to its concentrated destructive power and sophisticated weaponry \parencite{HeidelbergInstituteforInternationalConflictResearchHIIK2024Conflict2023}. Battles are assumed to exert 80 \% of this reference magnitude ($F_{base}(\text{Battles}) = 8.0e8$ N), as they combine multiple weapon types with coordinated personnel deployment but might create more dispersed impacts. 

Violence against civilians is assigned 60 \% of the reference magnitude ($F_{base}(\text{Violence against civilians}) = 6.0e8$ N). While potentially employing less sophisticated weapons, these events can create severe societal impacts through deliberate civilian targeting \parencite{ACLED2024ACLED2024}. Riots receive 30 \% ($F_{base}(\text{Riots}) = 3.0e8$ N), based on the assumption that they involve lighter weaponry but still have the potential for rapid escalation and property destruction \parencite{ACLED2024ACLED2024}. 

Protests are assumed to start at 10 \% before political context scaling through the \ac{V-Dem} index, following \textcite{Mittermaier2024TheV1.1}'s methodology.  In the case of Nigeria in 2018, the Liberal Democracy Index is 0.4, which results in a $(1 - 0.4) = 0.6$ adjustment-factor for protest events \parencite{Gerring2024V-DemV14}. While protesters themselves are peaceful, they may face violence from other actors \parencite{ACLED2024ACLED2024}.

\subsubsection{Event-Specific Intensity Scaling Mapping}

The intensity scaling combines fatality counts and civilian targeting. Setting the civilian targeting multiplier to $\gamma_c = 0.75$ reflects the findings of \textcite{Bendavid2021TheChildren} that indirect civilian casualties often exceed direct fatalities by over 75 \%.

The fatality scaling function uses a mix of linear and logarithmic effects governed by $\alpha_c$. Explosions/Remote violence and battles are assumed to have slight logarithmic-dominated scaling with $\alpha_c = 0.4$, as they might also represent a proportional increase in destruction. To capture the societal shocks and the assumption that an increase in deaths might not be proportionate to it, violence against civilians is assigned $\alpha = 0.2$ and riots and protests, which might be slightly more balanced $\alpha = 0.3$ \parencite{Mittermaier2024TheV1.1}.

\subsubsection{Temporal Decay Mapping}

The temporal decay function models how event impacts diminish over time. The cut-off can—the time when the event impact has decreased to only 1 \% of its original impact—be determined with $\ln(100)/\lambda$. Battles and explosions/remote violence are assumed to have an impact up until 2 years after the event happened ($\lambda = 0.0064$), reflecting potential sustained infrastructure damage and displacement effects. The impact of violence against civilians is set to reach up to 1 year ($\lambda = 0.0128$), balancing acute harm with persistent community trauma. Riots and protests are assumed to show a decreasing impact for up to 6 months ($\lambda = 0.0256$), reflecting their potentially shorter-term direct effects while acknowledging that underlying tensions can still persist.

\subsection{Force Distribution Mapping}

The complete force distribution selection for each sub-event type is summarised in Table \ref{tab:force-distribution-mapping}. The distribution pattern determines how an event's direct impact propagates from its recorded location. The implementation testing demonstrated four distinct patterns, all of which are employed in this demonstration given the restriction to precision code 1 events.

\begin{table}[H]
    \begin{tabular}{>{\raggedright\arraybackslash}p{120pt}|>{\raggedright\arraybackslash}p{160pt}|>{\raggedright\arraybackslash}p{160pt}} \hline
    \textbf{Event type} & \textbf{Sub-event type} & \textbf{Distribution}\\ \hline\hline
    Battles & Government regains territory & Linear\\ \cline{2-3}
            & Non-state actor overtakes territory & Linear\\ \cline{2-3}
            & Armed clash & Linear\\ \hline
    Protests & Excessive force against protesters & Constant\\ \cline{2-3}
             & Protest with intervention & Constant\\ \cline{2-3}
             & Peaceful protest & Constant\\ \hline
    Riots    & Violent demonstration & Constant\\ \cline{2-3}
             & Mob violence & Constant\\ \hline
    Explosions/Remote violence & Chemical weapon & Linear\\ \cline{2-3}
                                & Air/drone strike & Gaussian\\ \cline{2-3}
                                & Suicide bomb & Gaussian\\ \cline{2-3}
                                & Shelling/artillery/missile attack & Linear\\ \cline{2-3}
                                & Remote explosive/landmine/IED & Gaussian\\ \cline{2-3}
                                & Grenade & Gaussian\\ \hline
    Violence against civilians  & Sexual violence & Point Force\\ \cline{2-3}
                                & Attack & Point Force\\ \cline{2-3}
                                & Abduction/forced disappearance & Point Force\\ \hline
    \end{tabular}
    \caption[Force distribution patterns for conflict sub-event types]{Force distribution patterns are assigned to different \ac{ACLED} sub-event types. Point forces represent highly localised impacts, Gaussian distribution captures focused events with limited spread, linear distributions model gradual impact decline, and constant distributions represent uniform effects across an area.} 
    \label{tab:force-distribution-mapping}
\end{table}

Point forces create the highest local concentration by applying the entire force magnitude to a single point. Given the high spatial precision requirement, this pattern might be appropriate for highly targeted violence that affects specific individuals or very small groups. Within the violence against civilians category, sexual violence, targeted attacks, and abductions can be assumed to demonstrate such precise targeting, making them suitable for point force application. 

Gaussian distributions are assumed for precision military strikes and focused explosive events. Air/drone strikes, suicide bombs, remote explosives/landmines, and grenades might create intense impacts that concentrate strongly at their recorded points before rapidly diminishing. This pattern there, therefore, captures a high central displacement with limited but acknowledged spatial spread of the event itself while leaving broader indirect effects to emerge through interaction with the social fabric. 

Linear distributions are used for events that create graduate effects from a clear epicentre. All battle sub-events are modelled with linear patterns, assuming that fighting expands outward, creating zones of gradually decreasing intensity. Chemical weapons and artillery/missile attacks similarly warrant linear distributions, supposing their effects also spread through space, affecting broader areas with declining intensity from their release or impact points. 

Constant distributions are assigned to all protest and riot events, assuming that these events create uniform societal pressure across affected areas. While precise coordinates might have been recorded, their actual effects might spread more broadly without a specific point of impact. This also acknowledges the dynamic nature of these kinds of events, as they might even move through an area rather than being static at the recorded location.

\subsection{Force Radius Mapping}

The radius determines the geographic extent of an event's impact in conjunction with its distribution pattern. While studies like \textcite{Wagner2018ArmedAnalysis} found conflict effects up to 100 km from event sites, these often include indirect and spillover impacts. As this approach leaves the broader impact to emerge through the interaction with the social fabric, the radii need to be much smaller, only accounting for direct effects. The complete force radius parameters mapped to event types are summarised in Table \ref{tab:force-radius-mapping}.

\subsubsection{Base Radius Mapping}

Battles receive a base radius of 25 km, assuming that direct combat operations affect an area about the size of a large urban district or several connected settlements. This radius also accounts for immediate military movements, deployment of forces, and direct restrictions on civilian movement in the combat zone \parencite{ACLED2024ACLED2024}. 

Explosions/Remote violence events are assigned 10 km radii, based on the assumption that immediate weapon effects and direct infrastructure damage occur within this zone. This creates a reasonable area for direct impacts while leaving longer-range effects to emerge through the social fabric properties. 

Violence against civilians employs point forces requiring no radius parameter, assuming highly localised direct impacts on specific targets. While the psychological and social impacts of such violence might spread widely, the direct event itself is assumed to remain concentrated. 

Protests and riots are assigned 5 km radii, assuming immediate public disruptions spanning central urban districts and their immediate surroundings while also accounting for the dynamics of crowd movements over a larger area.

\subsubsection{Temporal Expansion Mapping}

To account for the possible spatial expansion of the impact zone,  each event type (apart from purely point-based incidents) is assigned a fraction of the base radius by which it can expand $\beta_c$ and a rate at which the impact zone approaches the expanded limit $\mu$. It is important to note that these parameters are intended solely to capture the direct evolution of an event's impact zone. Broader repercussions are expected to arise from how these forces interact with the social fabric rather than being specific in the expansion parameters themselves. 

In this demonstration, military events, such as battles and explosions/remote violence, are given a moderate expansion fraction ($\beta_c = 0.5$) alongside a slower rate expansion rate ($\mu = 0.0512$), implying roughly six months until their direct impact reaches its maximum size. This assumes that combat zones might expand gradually as forces manoeuvre, new positions are established, and civilian movement patterns adjust over time in response to ongoing hostilities.  

Civil unrest events, like protests and riots, employ a lower expansion parameter ($\beta_c = 0.1$) but a faster rate so that the maximum expansion is reached in approximately 30 days ($\mu = 0.3070$). This assumes that although the overall additional spread might remain relatively limited, it emerges quickly as crowds mobilise and the public responds. 

Violence against civilians, on the other hand, is applied as a point-based force with no associated radius and, therefore, expansion, resembling one-off incidents.

\begin{table}[H]
    \centering
   \begin{tabular}{>{\raggedright\arraybackslash}p{170pt}|>{\raggedright\arraybackslash}p{100pt}|>{\raggedright\arraybackslash}p{80pt}|>{\raggedright\arraybackslash}p{80pt}} \hline
         \textbf{Event Type}&\textbf{$r_{base}$ [m]}&\textbf{$\beta_c$}&\textbf{$\mu$}\\ \hline \hline 
 \textbf{Explosions/Remote violence}& 10,000& 0.5& 0.0512\\ \hline
  \textbf{Battles}& 25,000&0.5& 0.0512\\ \hline 
  \textbf{Violence against civilians}& -&-& -\\\hline 
         \textbf{Riots}&  5,000&0.1&  0.3070\\\hline
 \textbf{Protests}& 5,000&0.1& 0.3070\\ \hline
    \end{tabular}
    \caption[Force radius parameters for different event types]{Force radius parameters for different event types. Base radius ($r_{base}$) defines the initial spread extent of impact in meter, with battles covering the largest area and protests/riots the smallest. Violence against civilians uses point forces without radius parameters (-). The expansion fraction ($\beta_c$) and rate ($\mu$) control how impact zones grow over time.}
    \label{tab:force-radius-mapping}
\end{table}

\subsection{Conflict Event Mapping Results}

\begin{figure}[H]
\centering
\begin{tabular}{cc}
\includegraphics[width=0.4\textwidth]{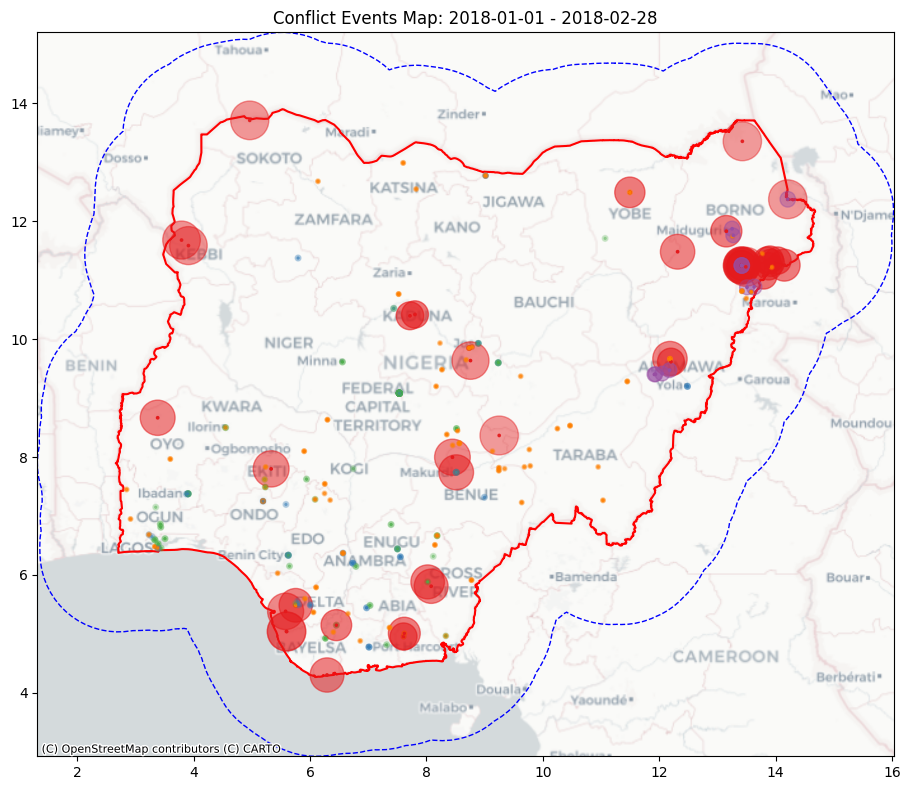}& \includegraphics[width=0.4\textwidth]{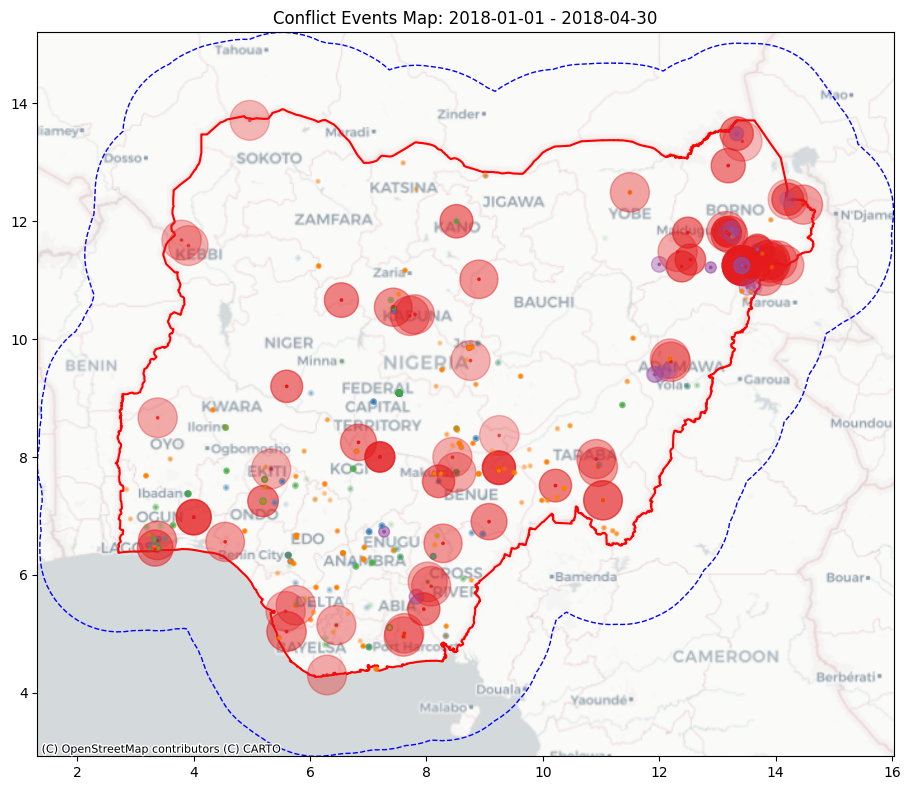}\\
(a) & (b) \\[6pt]
\includegraphics[width=0.4\textwidth]{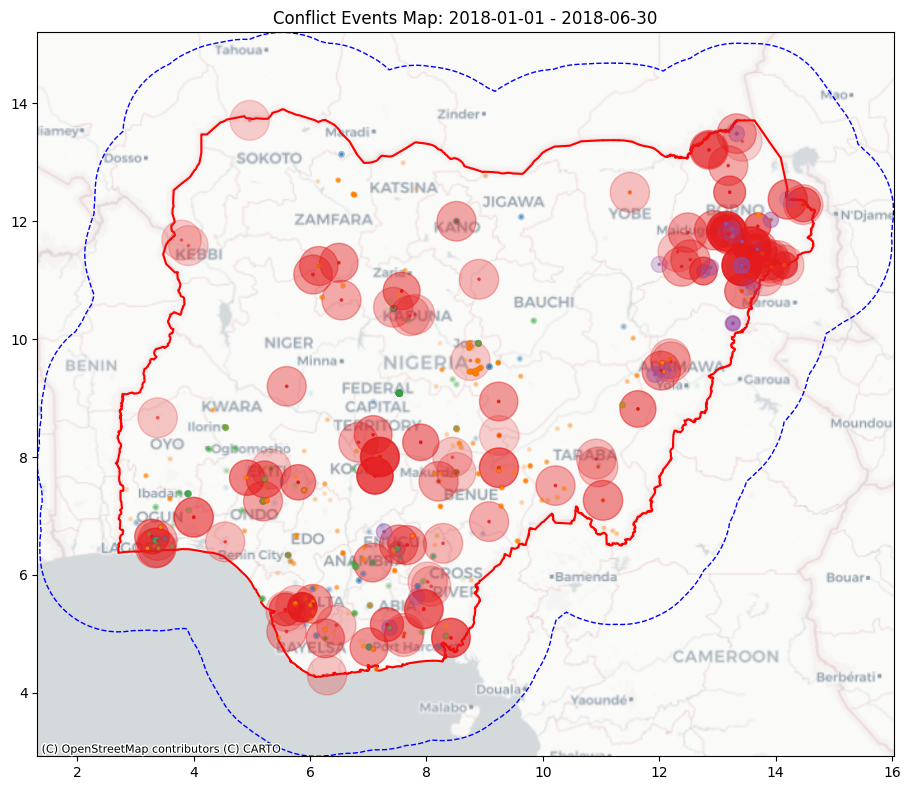}& \includegraphics[width=0.4\textwidth]{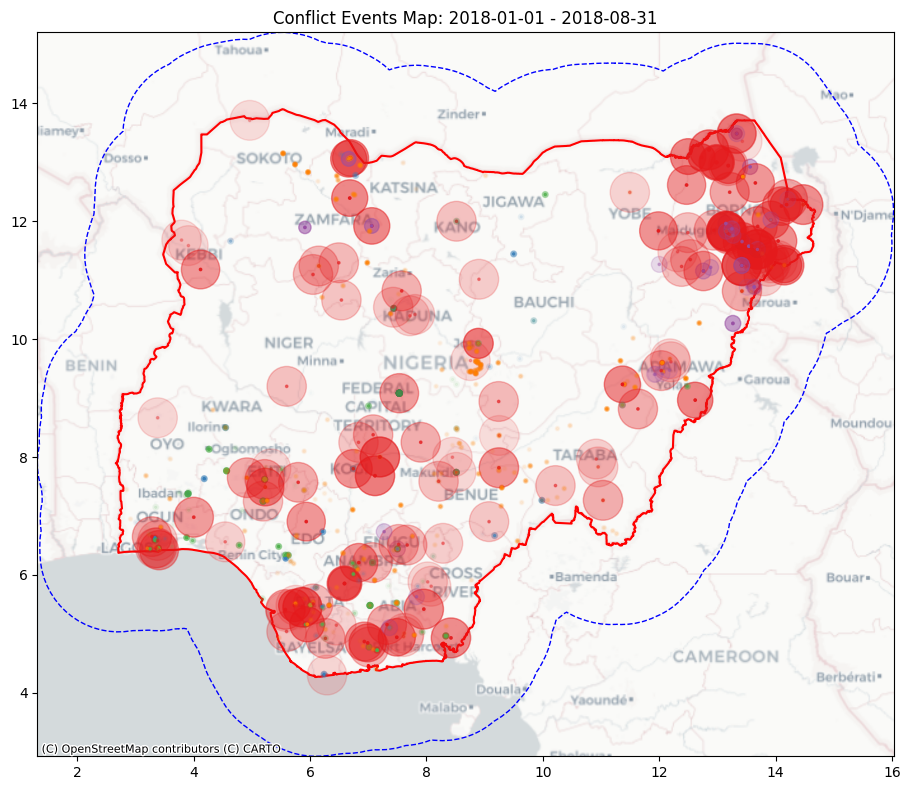}\\
(c) & (d) \\
 \includegraphics[width=0.4\textwidth]{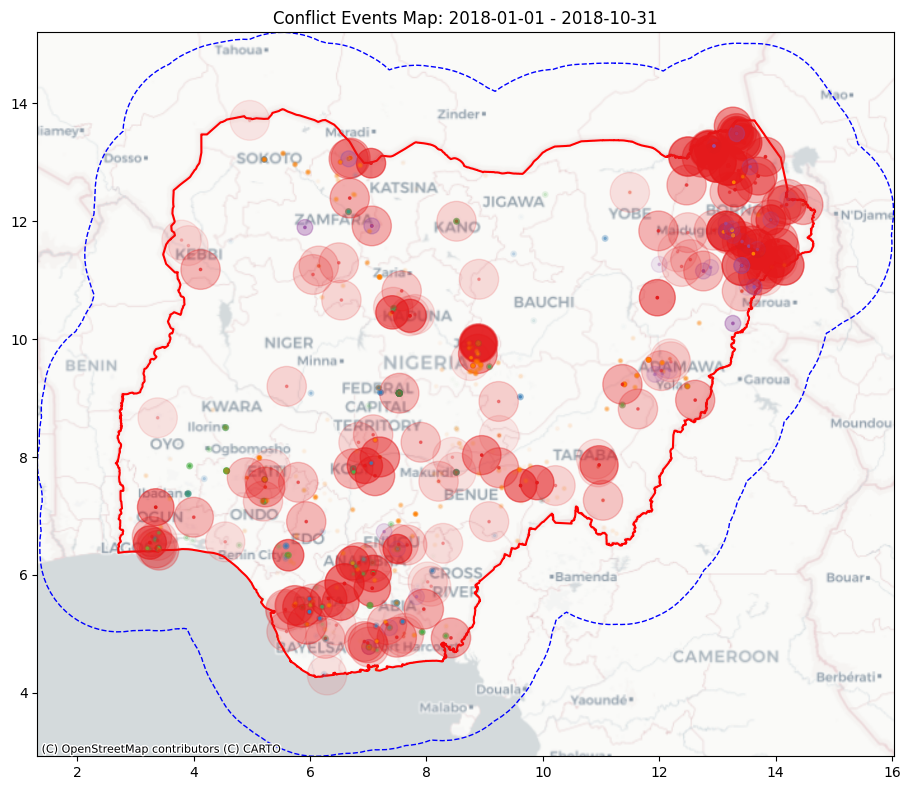}& \includegraphics[width=0.4\textwidth]{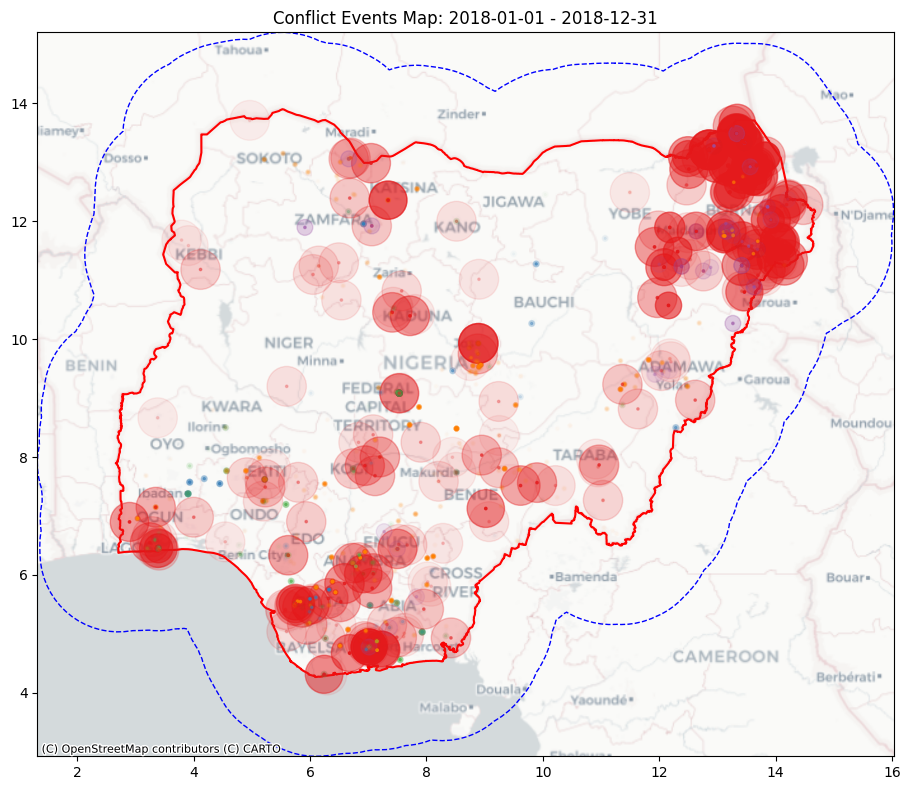}\\
 (e) & (f)\\\end{tabular}
\caption[Conflict events across Nigeria in 2018, shown in bimonthly increments.]{Conflict events for Nigeria in 2018, displayed in six cumulative snapshots labelled \textbf{(a)} through \textbf{(f)} in bimonthly increments. The circles are colour-coded by event type: battles in red, explosions/remote violence in purple, violence against civilians in orange, riots in blue, and protests in green. Circle transparency indicates temporal decay, with older events appearing more translucent. The red boundary marks the national border, while the dotted blue boundary denotes the buffer region.}
\label{fig:conflict-events-mapping-visualisation}
\end{figure}

Figure \ref{fig:conflict-events-mapping-visualisation} shows a bimonthly sequence of snapshots from January to December 2018, visualising the forces and some of their properties (distributions are not visually represented) generated by the conflict-event mappings. The magnitude of these forces, including their decay over time, is represented by the intensity of the colours. The radii of the circles reflect the resulting zones of influence, including their potential expansion over time, as determined by the mapping specifications. Each event type is colour-coded for clarity: battles appear in red, explosions/remote violence in purple, violence against civilians in orange, riots in blue, and demonstrations in green. 

The temporal evolution of events is particularly evident in Yobe State, in the northeastern part close to the border, where a battle-type event early in 2018 clearly demonstrates the dynamics of the temporal mapping. Following its occurrence, the event's radius visibly expands between snapshots Figure \ref{fig:conflict-events-mapping-visualisation} \textbf{(a)} and Figure \ref{fig:conflict-events-mapping-visualisation} \textbf{(b)}, indicating a growing zone of influence. At the same time, the circle becomes increasingly translucent in subsequent maps, illustrating the gradual decay of the event's impact over time.

When multiple events or their remaining influence overlap, it is also visible how the colours get stronger in the overlapping areas. While this doesn't follow any mathematical combination at this point, it already visualised how overlapping events will result in cumulative effects later. In contrast to buffer-based approaches these circular areas do not indicate the final areas of impact just yet. They should just just resemble the direct effects—observable in empirical data, e.g., from areas of destruction from satellite imagery—rather than accounting for any indirect impacts.

Overall, these visualisations demonstrate how raw conflict data can be transformed into spatially explicit force representations with meaningful variations in intensity, extent, and duration. As forces as well as the social fabric they are applied to are defined now, they can be combined using the described \ac{FEM} and analysed as illustrated in the next section.

\section{Model Implementation - Demonstration}

Before the interaction between mapped social characteristics and conflict forces can be analysed, a couple of final implementation steps must be completed to generate the visual outputs. These steps build directly on the theoretical framework and the implementation testing described in the previous chapters. 

The first step involves generating an appropriate computational mesh using the \texttt{FEAMesh2D} of the custom Python implementation. For this demonstration, the mesh is set to a rectangle that contains Nigeria's national borders with a 350 km buffer zone. This buffer size was chosen based on the findings from the implementation testing chapter, which demonstrated that buffer zones exceeding 300 km effectively counteract boundary condition influences on displacement measures. The mesh resolution is set to 10 km per element as it was used in the implementation testing, proven to provide sufficient detail while maintaining reasonable computation times. 

\begin{figure}[H]
    \centering
    \includegraphics[width=1\linewidth]{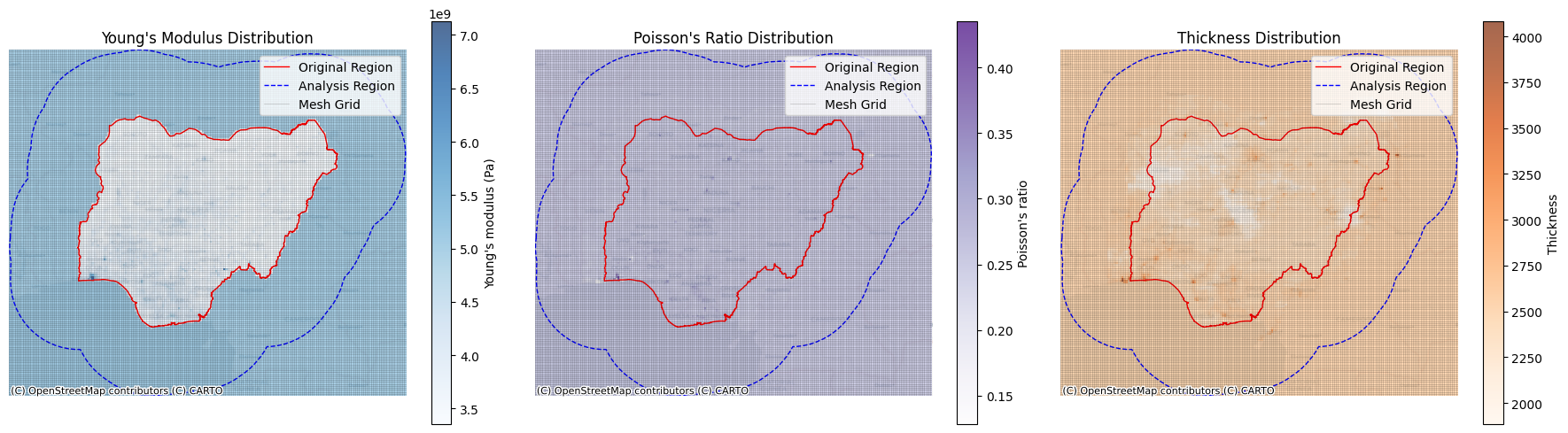}
    \caption[Physical parameter distribution for FEA implementation]{Final physical parameter distributions prepared for \ac{FEA} implementation: Young's modulus $E$, Poisson's ratio $\nu$, and thickness $h$ (from the left to the right). The mesh grid overlay illustrates the 10 km spatial discretisation used for the finite element implementation.}
    \label{fig:FEA-parameter-mesh}
\end{figure}

The material properties derived from the social indicator mapping are then assigned to this base mesh, representing the physical plate (Figure \ref{fig:FEA-parameter-mesh}). The spatially varying thickness, Young's modulus, and Poisson's ratio distribution, visualised in earlier Figures \ref{eq:combined-thickness}, \ref{eq:combined-youngs-modulus}, and \ref{eq:combined-poissons-ratio}, are therefore creating a computational representation of the social fabric with locally specific response characteristic.   

The force representations of conflict events, shown in Figure \ref{fig:conflict-events-mapping-visualisation}, are then applied to this materially heterogeneous plate. The \ac{FEM} implementation solves the governing differential equations (Equation \ref{eq:kirchhoff-love-plate-equation}) to determine how these forces create displacements throughout the social fabric. These displacement patterns, which emerge from the interaction between mapped social characteristics and conflict force, can then be visualised either as 2D contour plots or 3D surface deformations.


\section{Proof-of-Concept Results \& Visualisations} 

This section will showcase and describe the results of the implementation above. However, they are still solely presented as a proof-of-concept rather than a validated analysis at this point. It must also be noted that only events from 2018 are used here, while in principle, battles from 2017 would still have partial influence as their decay spans roughly two years. However, these were excluded for simplicity in this demonstration. Furthermore, as this approach requires spatially distributed social indicators, which are sometimes clipped directly to the Nigerian border, baseline values had to replace missing data outside these zones. This omission might be particularly relevant in the northeast, where the international border reduces the availability of cross-border indicators. 

Conceptually, the result shows the net effect of every event that occurred during 2018, each decaying to reflect its remaining impact by December 31. This means that all events' magnitudes are computed for that date according to their temporal rules, and each force is then applied simultaneously to the finite-element mesh encoding thickness, Young's modulus and Poisson's ratio, resulting in the combined displacements. Extension of this framework could perform genuine time-stepping analyses, though that would require additional assumptions or data about how the social fabric itself might evolve under repeated stress. 

\begin{figure}[H]
    \centering
    \includegraphics[width=1.00\linewidth]{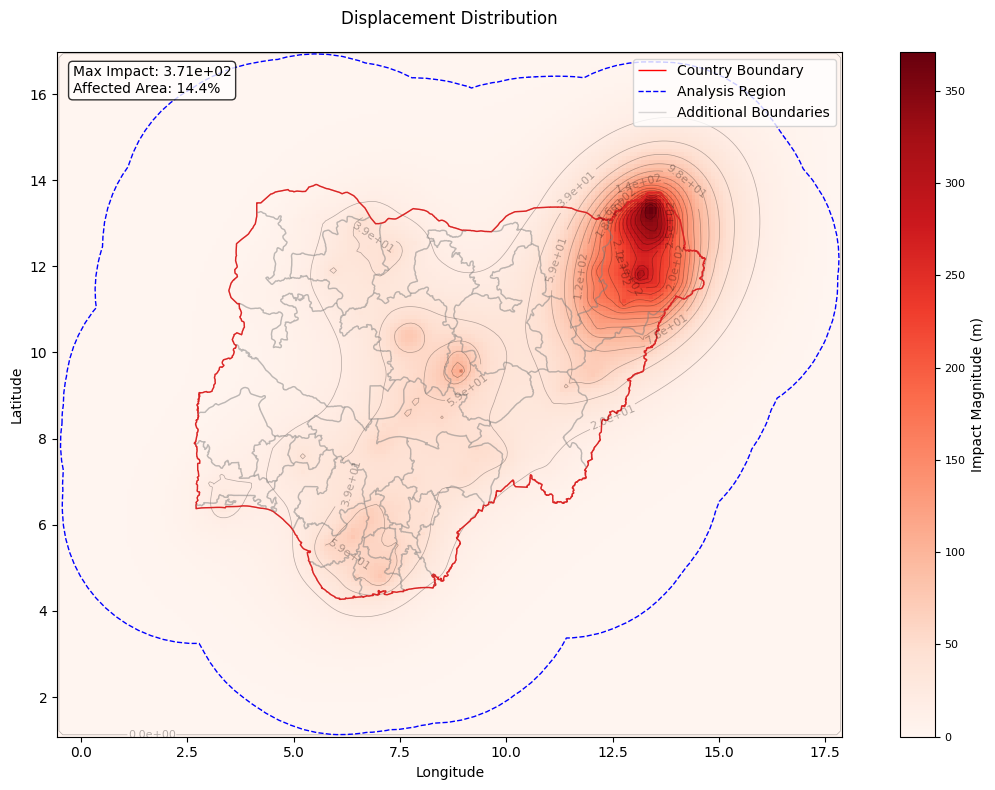}
    \caption[2D contour displacement visualisation]{2D contour displacement visualisation showcasing the proof-of-concept results for conflict events in Nigeria in 2018.}
    \label{fig:2d-displacement-result}
\end{figure}

Figure \ref{fig:2d-displacement-result} displays the result as a two-dimensional displacement field. It reveals a maximum displacement of 371 m, with 14.4 \% of the total area showing significant impact, defined as displacements exceeding 10 \% of the maximum displacement as established in the implementation testing chapter. Larger regions of minimal displacement occur in the southwest, where relatively high resilience and fewer devastating conflict events coincide. In contrast, regions of intense overlapping conflict events, in combination with relatively high vulnerability indicators, especially in the northeast, display very high displacements.

\begin{figure}[H]
    \centering
    \includegraphics[width=1\linewidth]{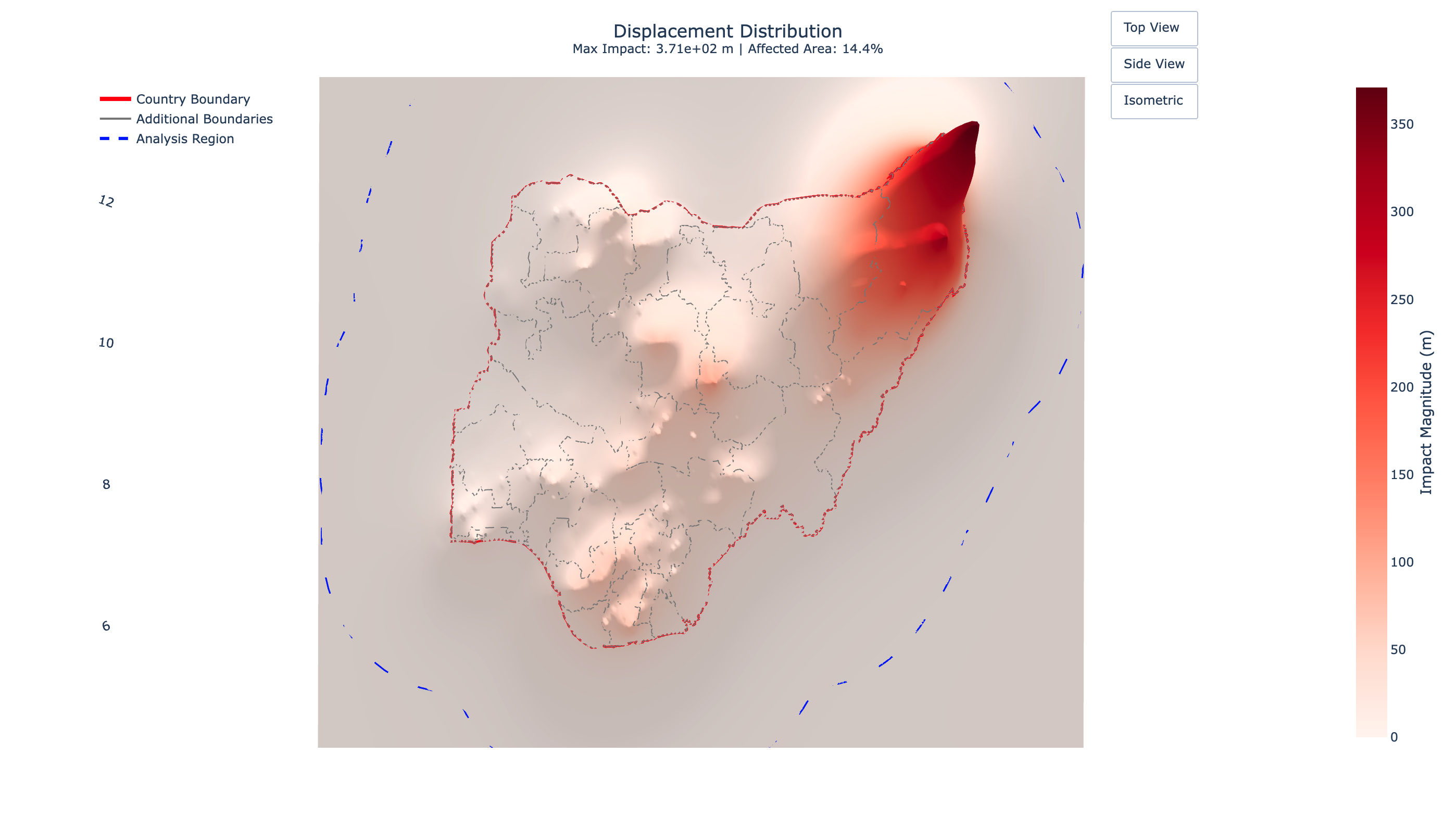}
    \caption[Top view of the 3D displacement visualisation]{Top view of the 3D displacement visualisation showcasing the proof-of-concept results for conflict events in Nigeria in 2018.}
    \label{fig:3d-displacement-result-TOP}
\end{figure}

\begin{figure}[H]
    \centering
    \includegraphics[width=1\linewidth]{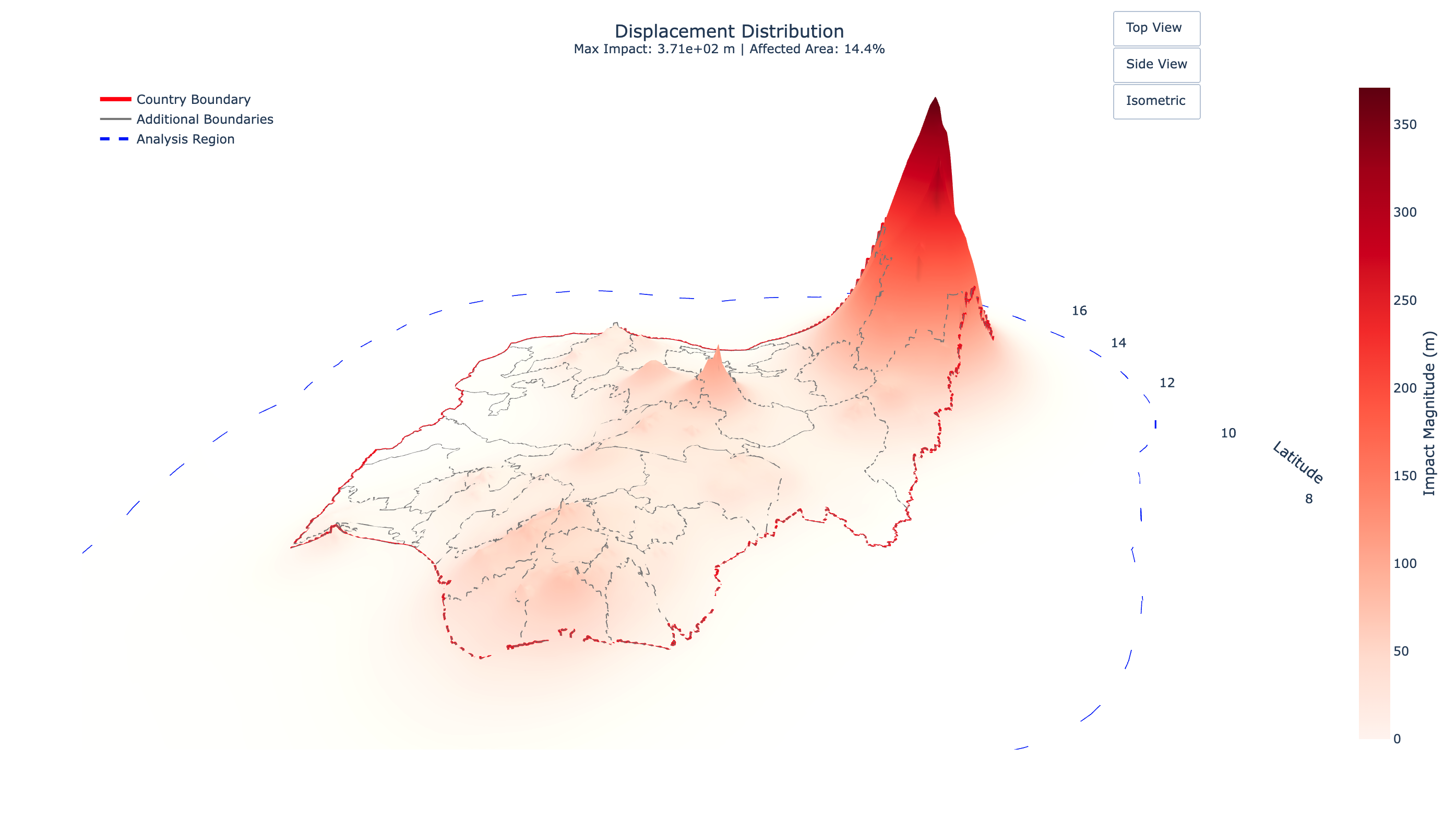}
    \caption[Southeast view of the 3D displacement visualisation]{Southeast view of the 3D displacement visualisation showcasing the proof-of-concept results for conflict events in Nigeria in 2018.}
    \label{fig:3d-displacement-result-ISO-SE}
\end{figure}

\begin{figure}[H]
    \centering
    \includegraphics[width=1\linewidth]{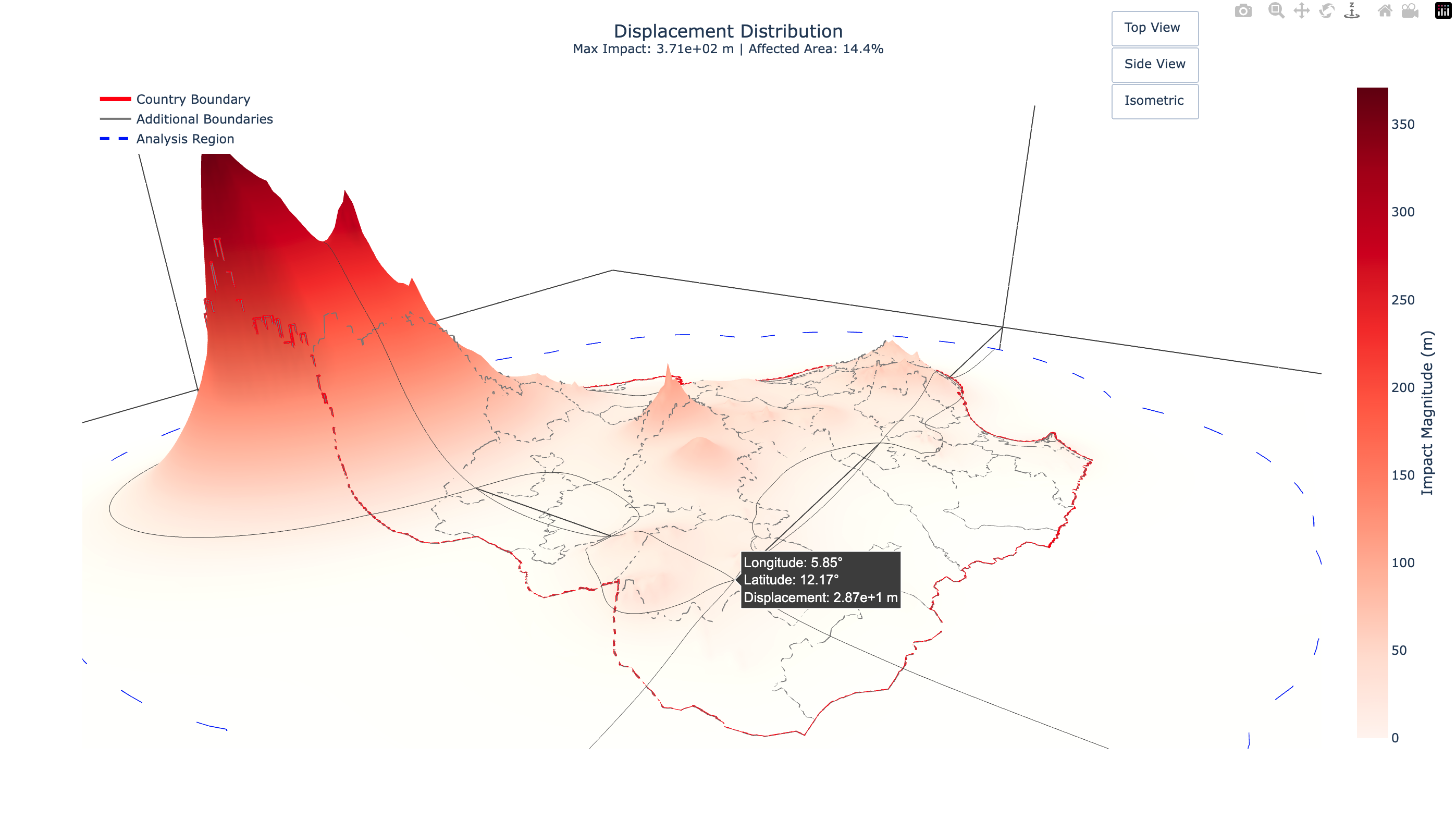}
    \caption[Northwest view of the 3D displacement visualisation]{Northwest view of the 3D displacement visualisation showcasing the proof-of-concept results for conflict events in Nigeria in 2018.}
    \label{fig:3d-displacement-result-ISO-NW}
\end{figure}

Figures \ref{fig:3d-displacement-result-TOP}, \ref{fig:3d-displacement-result-ISO-SE}, and \ref{fig:3d-displacement-result-ISO-NW} provide a three-dimensional view of the same results with the vertical axis showing displacement and the horizontal axes showing latitude and longitude. Larger peaks on the surface correspond to locations experiencing more concentrated impacts. 

Small adjustments in the model parameter can substantially alter these final displacement patterns. Decreasing the temporal decay rate, for instance, would prolong an event's impact, consequently raising the cumulative impact in areas with repeated or protracted violence as well. On the opposite, accelerating decay would decrease how strongly older events remain in the system. Similarly, higher baseline parameters would significantly improve the modelled resilience and, therefore, decrease the impact of violent events as well. Recalibrating the weights assigned to certain social indicators could also shift which places experience the largest peaks. For example, if child poverty is deemed particularly crucial for susceptibility, increasing its weight in Young's modulus mapping concentrates higher displacements in areas with many vulnerable children. 

Regarding possible interpretations, these absolute displacements shouldn't be treated as absolute but rather as comparative measures of the intensity of conflict impacts. Observing that one region's maximum displacement is twice or three times that of another can be understood as the first being correspondingly more "impacted". Another option would also be to define thresholds based on empirical data or domain expert evaluations, indicating at what level a region experiences serious humanitarian disasters or social collapses. The framework could also be used for temporal or scenario comparisons. Re-solving the model for a hypothetical intervention, such as strengthened health infrastructure or improved roads, would reduce the local "social fabric's" susceptibility and produce smaller peak displacements for the same event intensities. Repeating the analysis for different years could also reveal whether conflict patterns are worsening in specific locations or where certain forms of targeted interventions effectively mitigate repeated shocks.  

In summary, the aggregated 2018 displacement field demonstrates the framework's capacity to combine event-level conflict data and spatially heterogeneous social indicators into a layer that can be used to assess the potential impact these events exert on the respective regions. Several prominent hotspots reflect the confluence of intense, recurrent events with locally weaker baseline conditions, while other areas either show only scattered peaks or minimal displacements due to better baseline resilience or fewer incidents. Overlaps in time and space produce distinct compound effects, aligning with studies that highlight how indirect harm and cumulative shocks can far exceed the damage from discrete events alone. No single displacement value can be read as an immediate predictor of collapse, but relative comparisons and threshold-based mapping could highlight which zones warrant more in-depth scrutiny or urgent resources. Although the outputs in this proof-of-concept are strictly illustrative, they confirm the feasibility of a physics-inspired approach to capture the temporal, spatial, and socio-economic dimensions of conflict impact in a single, interpretable surface. 

\chapter{Discussion, Implications \& Conclusion}

\section{Key Insights}

The results presented in this thesis suggest that a physics-based "social fabric" analogy can enrich the theoretical and empirical understanding of how conflict impacts unfold and accumulate in space and time. As described in the literature review, conflict research often highlights spillovers, protracted violence, and indirect harms that extend well beyond direct destruction and casualties \parencite{Cappelli2024LocalAfrica, Schutte2011DiffusionWars, Tapsoba2023TheConflict}. By conceptualising social indicators as "material properties", this approach demonstrates how minor and independent events can combine into pronounced impact peaks if the local "fabric" is not resilient enough or too vulnerable to absorb repeated shocks. By systematically combining conflict event data with socio-economic indicators, the framework underscores that conflict dynamics are not purely discrete but emerge from many interlinked factors. 

This perspective also aligns with the literature review findings, highlighting the interactions between conflict severity and the multi-dimensional vulnerabilities of affected populations \parencite{Norris2007CommunityReadiness, Slone2021ChildrensSupport}. However, it also advances this concept by offering a framework that integrates these interactions into a single "displacement" measure while still allowing separate modelling of the "impact" and the "impacted".

This physics-based approach, therefore, provides an alternative, particularly to studies and methods that rely on fixed zones of influence only determined by the conflict events themselves. It highlights that impacted areas are a product of multiple factors like pre-existing conditions and different event intensities and their interaction in spatial or temporal overlays. The gradients of the resulting displacement fields allow contagion and diffusion processes to be visible. 

It, therefore, provides a theoretical bridge between discrete event analysis and continuous impact processes, between local vulnerabilities and regional conflict dynamics, and between quantitative measurement and qualitative understanding of conflict impacts. The framework encourages viewing conflict not as a series of isolated incidents but as an ongoing interaction between external stresses and community characteristics within an evolving "social fabric".


\section{Comparison with Existing Approaches}

While actual analysis results and differences cannot be properly compared due to the lack of verification of the results of the proof-of-concept application, some differences, possible advantages, or persistent limitations can still be highlighted. Whereas some spatial conflict analyses rely on uniform radii or kernel density for event locations, the presented model visualises how far and strongly violence impacts a specific location based on each location's pre-existing conditions and capacities. This shifts away from simply drawing fixed-distance buffers by letting local vulnerabilities or resilience directly shape the radius of an event's influence. 

Furthermore, unlike purely regression-based or econometric methods, which treat socio-economic data as explanatory factors predicting conflict occurrence, the physics-based analogy transforms these variables into material properties that simulate "mechanical" interaction with each violent "force". At the same time, it still depends on reliable georeferenced event data and, as of this first version, raster-based socio-economic indicators, a limitation shared by most macro-level frameworks.

In contrast to fine-grained household surveys or qualitative investigations—which are essential for understanding micro-level impacts—this approach targets broad spatial coverage, potentially even complementing those local perspectives by highlighting where repeated events might escalate stress in areas of high vulnerability. With that, it also echoes the strengths of multi-indicator and composite indices of producing broader measures that can also be compared and tracked across regions while allowing for separate modelling of the "impact" (the conflict events) and the "impacted" (the social fabric).


\section{Strengths, Limitations, and Ethical Considerations}
 
One of the key strengths of this approach is its ability to unify diverse indicators—ranging from poverty rates to health infrastructure density—into explicit, continuous measures of how conflict interacts with local conditions. By choosing specific indicators and weighting them to capture specific vulnerabilities or resilience factors, the model can also better account for the impact on most vulnerable groups, such as children, who tend to be disproportionally affected but can be overlooked when focusing on broader population averages. Moreover, by using a continuum mechanics model instead of discrete simulation, the framework can produce smooth gradients that approximate real-world spillovers and cumulative effects, offering an alternative to abrupt administrative boundary or polygon-based analyses. 

This model also provides outlooks for further theoretical exploration and conversation around conflict escalation by suggesting an actual mechanism for how repeated impacts accumulate on a fragile "fabric" until it reaches a tipping point. Separating "impact" from the "impacted" facilitates cross-regional comparisons on a larger scale, as long as underlying data layers are harmonised or adjusted appropriately. In addition, its transparent physical metaphor—which has even been used in other conflict and resilience/vulnerability contexts before—can help with interpretation among non-specialists and bridge academic research with policy or humanitarian applications. Importantly, it draws upon well-established computational methods from physics and material science, ensuring mathematical rigour without requiring researchers to reinvent fundamental numerical methods. This makes it also possible to benefit from rapid developments of new \ac{FEA} solvers or processes, like using \ac{AI} for improved computational efficiency or for modelling more complex composite materials \parencite{Maurizi2022PredictingNetworks}.

Despite these benefits, several constraints remain. First, the social fabric itself is assumed static, overlooking how protracted conflicts may destroy infrastructure, force migration, or degrade institutions over time. A single snapshot inevitably risks mismatches with rapidly changing conditions on the ground. Second, conflict event data, especially in active or remote zones, may be incomplete or biased by selective reporting, leading to distorted force distributions and possibly giving false impressions of safe corridors or heavily impacted hotspots. Third, determining thresholds for "significant displacements" remains somewhat subjective: users of the model might define 10 \% or 20 \% of the maximum displacement as "critical", but no universally accepted metric ties these relative measures to concrete humanitarian thresholds (e.g. famine risk, forced migration trigger). 

This also leads to critical ethical considerations. Although the thesis' demonstration is still preliminary, misuse may emerge and should already be considered in further developments and research. Labelling certain areas as "fragile" or "high-risk" could cause marginalisation if such labels influence resource distribution or security planning without local consultation. Furthermore, the perceived reliability and accuracy of a physics-based model, drawn from disciplines renowned for their predictive accuracy, might be tempting to treat outputs as definitive forecasts rather than heuristics measures. Unchecked, this could potentially result in misguided interventions or neglect of areas where hidden vulnerabilities or most vulnerable groups do not register in the available data.


\section{Future Research \& Extensions}
 
Further research should address validation as well as both model and data limitations. To calibrate or test the displacement outputs, future studies could compare them against observed infrastructure damage, forced displacement records, or micro-level survey data. Pilot studies in well-documented-conflict zones—where high-resolution data exists alongside robust humanitarian assessments—would be ideal to see if peaks in the mode align with actual humanitarian crises or displacement surges. Identifying suitable control variables, such as historical frequencies of child malnutrition or post-conflict rebuilding measures, could further clarify how accurately the model can capture cumulative fragility. 

As mentioned earlier, it would also be valuable to incorporate time-evolving social-fabric parameters to reflect how communities might degrade or rebuild in response to ongoing conflict and how the vulnerabilities and resilience factors change. While this thesis focuses on linear-elastic bending as an initial approach, more complex plastic or fracture mechanics could also model sudden "breaks" or irreversible damages. Such extensions may also explore the reverse process of reconstruction or "healing", which, although not typically a feature of real materials, can represent how development interventions, policy changes, or humanitarian aid might restore local capacity over time.


\section{Implications for Practice}

In real-world applications, different stakeholders might adopt this modelling framework to identify hotspots, test "what-if" scenarios, or integrate conflict stress analyses into early warning systems. At present, selecting parameter values for social indicators and conflict event mapping requires careful, expert judgment, especially given that each socio-economic layer shapes the final displacement field. While the proof-of-concept applications in the previous chapter mapped all indicators first before running the final model to explore and show the different capabilities and options for mappings, the recommended practice would be to adopt an iterative process. This way, default values grounded in literature or prior analyses could provide a starting point from which the resulting stress maps can be compared to known patterns of historical conflict severity. From there, parameters can be adjusted and fine-tuned to improve alignment. 

Data requirements are also important to consider. The model needs consistent, high-resolution (preferably aligned with the chosen resolution of the analysis mesh) socio-economic indicators in a raster format, accurate geolocations of conflict events, and a meaningful classification for event types. As the approach scales up to broader regions or cross-country analyses, differences in data quality could create uneven reliability across the displacement surface. Cross-country standardisation would require harmonising event definitions, intensities, and baseline reference scales for social indicators. 

Once implemented carefully, however, the method could be integrated into existing humanitarian or development frameworks—for example, by automatically updating the conflict "forces" in near real-time, thus enabling a rolling snapshot of areas accumulating dangerous levels of stress. Furthermore, if combined with other approaches and proper verification that considers potential ethical distortions, the displacement field could also feed into more elaborate early warning systems, further refining crisis detection and resource allocation.


\section{Conclusion}

This thesis set out to develop and demonstrate a physics-based "social fabric" approach capable of exploring how conflict events and local socio-economic conditions interact in a single, cohesive framework. Specifically, it pursued three objectives: (1) to formulate a conceptual modelling approach that translates social indicators into "material properties" and conflict events into "forces" within a unified structure, (2) to showcase a proof-of-concept illustrating how overlapping events and vulnerabilities compound over space and time, and (3) to establish the foundations for subsequent validation and collaboration across disciplines. 

Regarding the first objective, the thesis introduced a physics-based analogy in which resilience and vulnerability factors determine the physical properties of a "plate", while conflict events serve as external loads acting upon this "social fabric". By systematically merging conflict data and socio-economic indicators, the model demonstrates how apparently discrete violent events can overlap and create cumulative stress peaks in regions lacking sufficient absorptive capacity or with persistent vulnerabilities. 

For the second objective, a proof-of-concept application was implemented to highlight how this model might operate in practice. Using example data on conflict events and selected socio-economic variables, the results, in combination with the thorough implementation testing, were able to show how repeated and protracted violence can cause amplified impacts, especially in areas suffering from high vulnerabilities and low resilience factors. While no formal validation was undertaken, these simulations affirmed that the approach could indeed represent indirect or spillover effects in a single displacement surface—an advantage over simpler radius-based or static vulnerability metrics. 

Finally, to address the third objective, the thesis discusses the conceptual and practical gaps to resolve in future research. Because the framework currently assumes a static social fabric, future extensions should capture how prolonged violence alters infrastructure, governance, and demographic distributions. Similarly, further work is required to calibrate model parameters—such as decay rates or thresholds for significant displacement—against field-based observations and humanitarian outcome data. These refinements require joined interdisciplinary efforts, drawing upon insights from physics, social science, and conflict analysis. By identifying such challenges in the discussion chapter, this thesis established a future outlook and possibilities for strengthening the model's empirical grounding, ethical safeguards, and, ultimately, its utility to practitioners. 

In sum, the research achieved its primary aim of showing how conflict events and local vulnerabilities can be analytically combined using a physics-based framework, highlighting how compound or protracted violence emerges from the interplay of event intensities and baseline fragilities. Through its proof-of-concept, it underscored that minor shocks in vulnerable areas could accumulate disproportionate harm, affirming the importance of integrating socio-economic data with conflict location and timing. While continuing validation and further model developments remain essential, this physics-based framework offers a novel lens for understanding and potentially mitigating the spatially varied and cumulative impacts of armed conflict.

%% file: content-folder/appendix.tex
\chapter{Social Indicator Mapping - Summary Table}\label{app:social-indicator-mapping-parameters}

\begin{table}[H]
    \centering
    \begin{tabular}{>{\raggedright\arraybackslash}p{75pt}|>{\raggedright\arraybackslash}p{75pt}|>{\raggedright\arraybackslash}p{75pt}|>{\raggedright\arraybackslash}p{75pt}|>{\raggedright\arraybackslash}p{75pt}|>{\raggedright\arraybackslash}p{50pt}} \hline
        \textbf{Indicator} & \textbf{Source} & \textbf{Parameter Type}& \textbf{Response Function}& \textbf{Function Parameter(s)}& \textbf{Weight} \\ \hline \hline
        \multicolumn{6}{l}{\textbf{Thickness $h$}} \\ \hline
        \ac{CISI}& \textcite{Nirandjan2022AInfrastructure} & Resilience & Power Law & $\gamma_r = 0.5$ & 0.15 \\ \hline
        \ac{SPI} Drought& \textcite{Funk2014AMonitoring} & Vulnerability & Power Law & $\gamma_v = 1.2$ & 0.20 \\ \hline
        \ac{SPI} Wetness& \textcite{Funk2014AMonitoring} & Vulnerability & Power Law & $\gamma_v = 1.2$ & 0.20 \\ \hline
        Health Infrastructure & \textcite{Nirandjan2022AInfrastructure} & Resilience & Logarithmic & $\alpha_r = 2.5$ & 0.30 \\ \hline
        Travel Time to Healthcare & \textcite{Weiss2020GlobalFacilities} & Prerequisite & Linear & $m_p = 1.0$& - \\ \hline
        Dependency Ratio & \textcite{WorldPop2016WorldPopAfrica} & Vulnerability & Exponential & $\beta_v = 0.8$ & 0.15 \\ \hline \hline
        \multicolumn{6}{l}{
\textbf{Young's Modulus $E$}} \\ \hline
        \ac{GDP} \ac{PPP}& \textcite{Kummu2018Gridded1990-2015} & Resilience & Linear & $m_r = 1.0$ & 0.60 \\ \hline
        Poverty Rate & \textcite{Tatem2013WorldPopPoverty} & Vulnerability & Power Law & $\gamma_v = 1.5$ & 0.20 \\ \hline
        Childhood Poverty & \textcite{Utazi2023ACountries} & Vulnerability & Power Law & $\gamma_v = 0.5$ & 0.20 \\ \hline \hline
        \multicolumn{6}{l}{\textbf{Poisson's Ratio $\nu$}} \\ \hline
        Population Density & \textcite{WorldPop2020WorldPopAdjusted} & Vulnerability & Linear & $m_v = 2.0$ & 0.60 \\ \hline
        Road Density & \textcite{Meijer2018GlobalInfrastructure} & Resilience & Logarithmic & $\alpha_r = 2.0$ & 0.40 \\ \hline
    \end{tabular}
    \caption[Summary of social indicator mapping parameters and their translation function]{Summary of social indicator mapping parameters and their translation functions. The table presents the complete set of indicators used to define the material properties in the proof-of-concept implementation. For each indicator, the source references, parameter type, selected response function, corresponding function parameters, and relative weights are specified. The parameter types determine whether the indicator contributes positively (resilience) or negatively (vulnerability) to the material properties.}
    \label{tab:social-indicator-mapping-parameters}
\end{table}

\chapter{ACLED Data Distribution Analysis}\label{app:acled-data-distribution-analysis}

\begin{table}[H]
    \centering
    \begin{tabular}{>{\raggedright\arraybackslash}p{150pt}|>{\raggedright\arraybackslash}p{60pt}|>{\raggedright\arraybackslash}p{80pt}} \hline
        \textbf{Event Type} & \textbf{Count} & \textbf{Percentage} \\ \hline \hline
 \textbf{Protests} & 404&32.22 \%\\ \hline
        \textbf{Violence against civilians} & 333& 26.56 \%\\ \hline
        \textbf{Battles} & 270& 21.53 \%\\ \hline
        \textbf{Riots} & 198& 15.79 \%\\ \hline
        \textbf{Explosions/Remote violence} & 49& 3.91 \%\\ \hline
    \end{tabular}
    \caption[Distribution of ACLED conflict events in Nigeria during 2018]{Distribution of \ac{ACLED} conflict events in Nigeria during 2018. The table presents the frequency of different event types, including only events with geo-precision code 1 and excluding strategic developments (1254 events total).}
    \label{tab:event-distribution}
\end{table}
\begin{table}[H]
    \centering
    \begin{tabular}{>{\raggedright\arraybackslash}p{140pt}|>{\raggedright\arraybackslash}p{140pt}|>{\raggedright\arraybackslash}p{80pt}|>{\raggedright\arraybackslash}p{90pt}} \hline
        \textbf{Event Type} & \textbf{Sub-event Type} & \textbf{Count} & \textbf{Fatalities} \\ \hline \hline
        \textbf{Violence against civilians}& Attack & 303& 1023\\ \cline{2-4}
        & Abduction/forced disappearance & 24& 0\\ \cline{2-4}
        & Sexual violence & 6& 4\\ \hline
        \textbf{Battles}& Armed clash & 242& 1049\\ \cline{2-4}
        & Government regains territory & 20& 135\\ \cline{2-4}
        & Non-state actor overtakes territory & 8& 95\\ \hline
        \textbf{Protests}& Peaceful protest & 375& 0\\ \cline{2-4}
        & Protest with intervention & 26& 0\\ \cline{2-4}
        & Excessive force against protesters & 3& 2\\ \hline
        \textbf{Explosions/Remote violence}& Remote explosive/landmine/IED & 7& 5\\ \cline{2-4}
        & Air/drone strike & 20& 150\\ \cline{2-4}
        & Suicide bomb & 20& 166\\ \cline{2-4}
        & Grenade & 1& 0\\ \cline{2-4}
        & Shelling/artillery/missile attack& 1& 10\\ \cline{2-4}
        & Chemical weapon & -& -\\ \hline
        \textbf{Riots}& Violent demonstration & 106& 71\\ \cline{2-4}
        & Mob violence & 92& 75\\ \hline
    \end{tabular}
    \caption[Detailed distribution of ACLED conflict events and associated fatalities in Nigeria during 2018.]{Detailed distribution of \ac{ACLED} conflict events and associated fatalities in Nigeria during 2018, broken down by event types and sub-event types (geo-precision code 1, without strategic developments (1254 events total)).}
    \label{tab:detailed-event-distribution}
\end{table}